# DUAL-FIT IMPERATIVE IN SECURITY LEADERSHIP: A GROUNDED THEORY INVESTIGATION OF CISO ROLE ENACTMENT IN MODERN ORGANISATIONS

Mazino Benson Onibere

*ORCID ID: 0000-0002-6197-0612*

submitted in total fulfillment for the

degree of Doctor of Philosophy

in the

School of Computing and Information Systems

Faculty of Engineering and Information Technology

THE UNIVERSITY OF MELBOURNE

March 2026

# ABSTRACT

The Chief Information Security Officer (CISO) has emerged as a strategically prominent executive role. The role confronts challenges that distinguish it fundamentally from other C-suite positions: an adversarial environment demanding continuous crisis readiness, irreconcilable accountability tensions between security and business enablement, and a prevention paradox in which the greatest successes remain invisible to those determining resource allocation. Despite these distinctive demands, systematic scholarly understanding of how the role operates across different organisational contexts has remained virtually absent, leaving consequential decisions about security leadership without empirical foundation.

This thesis addresses the research question: how is the CISO role enacted across different organisational contexts in modern organisations? The study employs constructivist grounded theory methodology with an abductive reasoning framework, adopting the Gioia method to move systematically from empirical observation to theoretical abstraction. Data were collected through twenty in-depth interviews with senior security executives across Australian organisations representing diverse industries and organisational sizes. Findings are grounded in the Australian context and offer theoretical propositions available for cross-cultural investigation.

The investigation reveals that the CISO role is not enacted as a stable configuration of responsibilities applied uniformly across settings. It is a continuously negotiated achievement, fundamentally constituted by the contexts within which it operates. Effectiveness emerges from dual-fit maintenance: the simultaneous management of alignment between the leader and the internal organisational context (CISO-Organisation Fit), and between the leader and the external environment (CISO-Environment Fit). What these alignment relationships require transforms systematically across three organisational maturity phases, Establishment, Maturation, and Strategic, with action-oriented, stewardship-oriented, and vision-oriented leaders generating effectiveness in each respective phase. Capabilities producing effectiveness in one phase may actively undermine it in another. Political capital operates as the central mechanism through which leaders navigate these shifting demands across phases of evolving complexity.

These patterns are synthesised in the Security Leadership Contingency Model (SLCM), the primary theoretical contribution of this thesis. Operational instruments developed from the SLCM translate these theoretical insights into immediately applicable guidance for leadership selection, development, and succession planning. The SLCM extends person-organisation fit theory, person-environment fit theory, and contingency theory to boundary-spanning security leadership, generating mid-range theory that advances understanding of executive effectiveness in roles characterised by irreconcilable accountability tensions, asymmetric visibility, and evolving contextual demands.

# DECLARATION

This is to certify that:

i. the thesis comprises only my original work towards the PhD,
ii. due acknowledgement has been made in the text to all other material used,
iii. the thesis is less than 100,000 words in length, exclusive of tables, maps, bibliographies, and appendices

..............................................................

Mazino Onibere

17 March 2026

# PREFACE

The following publications have been drawn from this thesis for the purpose of external validation. The thesis itself is not derived from these publications; rather, the publications were extracted from completed thesis work and submitted to peer-reviewed venues to subject the findings to independent scholarly scrutiny. In both publications, I was the primary author and was responsible for the research design, data analysis, and writing, with co-authors contributing in their supervisory capacity.

The initial findings, represented by the Gioia data structure developed in Chapter 5, were presented in the following publication:

- Onibere, M., Ahmad, A., & Maynard, S. B. (2025). The dual-fit imperative: A grounded theory of CISO effectiveness. *Proceedings of the Workshop on Information Security and Privacy (WISP 2025)*.

The Security Leadership Contingency Model developed in Chapter 6 was presented in the following publication:

- Onibere, M., Ahmad, A., & Maynard, S. B. (accepted). The Security Leadership Contingency Model: How Chief Information Security Officers adapt across organisational phases. *Proceedings of the European Conference on Information Systems (ECIS 2026)*.

**Other Publications Completed During Candidature**

The following publications were completed during the candidature but are not drawn from this thesis:

- Onibere, M., Ahmad, A., & Maynard, S. B. (2016). Information security strategy: Beyond protection, towards an organisational information security capability. Paper presented at the University of Melbourne Doctoral Colloquium, Melbourne, Australia.
- Onibere, M., Ahmad, A., & Maynard, S. B. (2017). Information security strategizing, capability, and organisational performance. Paper presented at the University of Melbourne Doctoral Colloquium, Melbourne, Australia.
- Onibere, M., Ahmad, A., & Maynard, S. B. (2017). The Chief Information Security Officer and the 5 dimensions of a strategist. *Proceedings of the Pacific Asia Conference on Information Systems (PACIS 2017)*.

- Maynard, S. B., Onibere, M., & Ahmad, A. (2018). Defining the strategic role of the Chief Information Security Officer. *Pacific Asia Journal of the Association for Information Systems, 10*(3), 61–85. https://doi.org/10.17705/1pais.10303
- Onibere, M., Ahmad, A., & Maynard, S. B. (2019). Dynamic information security management capability: Strategising for organisational performance. *Proceedings of the Australasian Conference on Information Systems (ACIS 2019)*.

This research was supported by the Commonwealth through an Australian Government Research Training Program Scholarship [DOI: doi.org/10.82133/C42F-K220].

# ACKNOWLEDGEMENT

I am grateful to all who made this possible. Every piece of scholarly work is, at its foundation, a collective achievement; this one is no different.

# TABLE OF CONTENTS

# LIST OF TABLES

# LIST OF FIGURES

# LIST OF ABBREVIATIONS

| Acronym | Full Term |
|---|---|
| AEMO | Australian Energy Market Operator |
| AI | Artificial Intelligence |
| APRA | Australian Prudential Regulation Authority |
| ASX | Australian Securities Exchange |
| ATT&CK | Adversarial Tactics, Techniques, and Common Knowledge |
| BAU | Business As Usual |
| CEO | Chief Executive Officer |
| CFO | Chief Financial Officer |
| CIO | Chief Information Officer |
| CISO | Chief Information Security Officer |
| CMMI | Capability Maturity Model Integration |
| CMO | Chief Marketing Officer |
| COBIT | Control Objectives for Information and Related Technologies |
| CPS | Cross-industry Prudential Standard |
| CRO | Chief Risk Officer |
| CTO | Chief Technology Officer |
| DDoS | Distributed Denial of Service |
| GDPR | General Data Protection Regulation |
| GRC | Governance, Risk, and Compliance |
| HIPAA | Health Insurance Portability and Accountability Act |
| ICS | Industrial Control Systems |
| ICT | Information and Communication Technology |
| IPS | Intrusion Prevention System |
| ISA | International Society of Automation |
| ISO/IEC | International Organization for Standardization and International Electrotechnical Commission |
| IT | Information Technology |
| KPI | Key Performance Indicator |
| NIST | National Institute of Standards and Technology |
| OT | Operational Technology |
| P&L | Profit and Loss |
| P-E | Person-Environment (fit) |
| P-O | Person-Organisation (fit) |
| P1 | Priority 1 |
| PCI | Payment Card Industry |

| Acronym | Full Term |
|---|---|
| ROI | Return on Investment |
| ROSI | Return on Security Investment |
| SLCM | Security Leadership Contingency Model |
| SMEs | Subject Matter Experts |
| SOC | Security Operations Centre |
| SOCI Act | Security of Critical Infrastructure Act 2018 |

# CHAPTER 1: INTRODUCTION

## 1.1 THE CONTEMPORARY SECURITY LEADERSHIP CHALLENGE

Organisations face a fundamental paradox in security leadership. Whilst cyber threats have evolved from technical nuisances to strategic risks commanding boardroom attention, the Chief Information Security Officer (CISO) role charged with managing these threats remains fundamentally undefined and poorly understood (Ramezan, 2025; Sahin & Vance, 2025). This disconnect between criticality and clarity creates profound organisational challenges that extend beyond technical security concerns to encompass strategic governance, executive dynamics, and business resilience.

Security leadership confronts distinctive challenges that set it apart from other executive domains and resist resolution through borrowed frameworks. First, security leaders operate in adversarial environments where intelligent opponents actively circumvent defensive measures, creating perpetual crisis conditions absent from competitive markets (Hannah et al., 2009). Traditional leadership frameworks treat crisis as episodic exception; security leaders face continuous crisis readiness as their operational baseline. Second, security leaders must simultaneously prevent losses and enable business innovation, objectives existing in fundamental tension. Unlike Chief Information Officers (CIOs) whose mandate centres on enabling business through technology, security leaders face irreconcilable dual accountability where improvement on one dimension necessarily degrades the other (Da Silva & Jensen, 2022). Third, security's primary value remains invisible. Security leaders' greatest successes, the breaches that never occurred, remain forever unobservable to stakeholders determining resource allocation. This inverts traditional executive visibility patterns where successful leaders demonstrate value through observable achievements. These three distinctions suggest that security leadership requires dedicated theoretical development; borrowed frameworks from adjacent domains prove insufficient for capturing its unique demands.

The CISO has emerged as arguably the most rapidly elevated yet least understood member of the contemporary C-suite. From obscurity in the early 2000s to boardroom prominence today, the role has ascended without the empirical foundation that underpins other executive positions (Da Silva & Jensen, 2022; Ferreira et al., 2024). Unlike Chief Financial Officers (CFOs) or Chief Marketing Officers (CMOs), whose effectiveness criteria have been systematically studied and validated, CISOs operate without established benchmarks for success, accepted competency frameworks, or even consensus on fundamental role responsibilities. This empirical vacuum persists despite organisations investing billions annually in security capabilities and recognising cyber risk as a major strategic financial threat (Abrardi et al., 2025; Ahmad et al., 2021; Franco et al., 2024).

The consequences of this empirical void manifest across multiple organisational dimensions. The profession experiences high turnover, with reported burnout rates and unclear career progression pathways contributing to succession challenges (da Veiga et al., 2020; Nepal et al., 2024; Pham et al., 2019). Organisations cannot optimally structure security functions when role requirements remain ambiguous, resulting in reporting relationships that span from Information Technology (IT) to risk to direct board accountability without empirical justification for any particular model. Security leaders themselves navigate career paths without validated development frameworks, often experiencing role transitions that prove unsuccessful despite technical excellence. Most critically, this leadership uncertainty compromises organisational security posture at a time when threat sophistication and regulatory requirements demand unprecedented coordination between technical controls and business strategy.

## 1.2 THE RESEARCH GAP

Despite extensive discourse about security leadership, systematic empirical investigation remains virtually absent from academic literature (Ramezan, 2025; Sahin & Vance, 2025). This gap persists across multiple knowledge sources, each failing to address the distinctive challenges that set security leadership apart from other executive domains.

Practitioner literature proliferates with prescriptive frameworks and maturity models, yet none rest on rigorous empirical foundations (Becker et al., 2009; Santos-Neto & Costa, 2019; Wendler, 2012). Consultancy reports offer confident recommendations about CISO capabilities, organisational structures, and success factors without systematic evidence supporting these claims. Industry surveys capture opinions and self-reported practices but lack the methodological rigour necessary for understanding causal mechanisms or contextual dependencies. Most critically, this practitioner discourse treats security leadership as analogous to other executive roles, applying generic leadership principles without accounting for the adversarial environment, irreconcilable accountability tensions, or invisible value paradox that fundamentally distinguish the CISO position. The absence of empirical grounding means that widely promoted best practices may actually diminish rather than enhance security leadership when applied without recognition of these unique contextual demands.

Academic security literature, whilst extensive in examining technical controls, governance processes, and incident response mechanisms, consistently overlooks the leadership dimension that orchestrates these elements (Ahmad et al., 2020; AlGhamdi et al., 2020; Naseer et al., 2021). This scholarship focuses on what security programmes should implement rather than how security leaders navigate the continuous crisis conditions created by intelligent adversaries. Research examines compliance frameworks without addressing how leaders manage the fundamental tension between preventing losses and enabling innovation. Studies measure security outcomes through breach rates and audit scores without confronting the attribution challenge posed by invisible successes. The systematic absence of leadership focus in security research leaves fundamental questions about how CISOs establish legitimacy, build influence, and demonstrate value entirely unaddressed.

Existing executive leadership literature, whilst rich in frameworks for understanding C-suite roles, proves inadequate for comprehending security leadership's distinctive challenges. Traditional theories of executive performance assume episodic crises that leaders manage before returning to steady-state operations; they cannot accommodate continuous crisis readiness as operational baseline (Hannah et al., 2009). Conventional frameworks presume executives optimise for unified organisational objectives; they lack mechanisms for navigating irreconcilable accountability tensions where improvement on one dimension necessarily degrades another (Da Silva & Jensen, 2022). Established models assume successful executives demonstrate value through observable achievements; they cannot explain effectiveness when greatest successes remain forever invisible to stakeholders determining resource allocation (Guhr et al., 2019; Schoemaker et al., 2018). The boundary-spanning nature of security leadership, requiring simultaneous credibility with technical teams and business executives whilst navigating regulatory requirements and threat landscapes, creates complexities that single-domain leadership theories fundamentally cannot capture.

This theoretical gap extends beyond academic curiosity to create profound practical organisational challenges. Without empirical understanding of how security leaders navigate adversarial environments, manage irreconcilable tensions, and demonstrate invisible value, organisations make critical decisions based on assumptions rather than evidence. Recruitment processes select CISOs using criteria borrowed from other executive roles without recognising that capabilities driving CFO or CIO success may prove irrelevant or counterproductive in security contexts. Development programmes emphasise technical certifications or generic leadership training without understanding which capabilities actually enable effectiveness when facing intelligent opponents, dual accountability, and value invisibility. Performance evaluation systems apply metrics designed for other roles, potentially rewarding behaviours that compromise security whilst penalising actions that genuinely protect organisations.

## 1.3 RESEARCH PROBLEM AND QUESTIONS

This thesis addresses the primary research question: How is the CISO role enacted across different organisational contexts in modern organisations? The question deliberately avoids presupposing specific role configurations or organisational outcomes, maintaining the breadth necessary for exploratory investigation whilst directing attention toward the contextual dynamics through which security leadership operates.

Three guiding themes emerged from the literature review and oriented the empirical inquiry. The first concerns the relationship between organisational context and security leadership, reflecting the literature's consistent acknowledgement that context matters without any systematic investigation of how or why. The second concerns how security leaders establish their standing and credibility within organisations where the role itself remains poorly understood and contested. The third concerns the longstanding debate about whether security leadership requires deep technical expertise or business acumen, a tension that pervades both practitioner discourse and academic literature without empirical resolution. These themes provided initial orientation for the investigation without constraining the discovery of unexpected patterns.

The research scope encompasses senior security executives in established organisations where digital dependencies create material security requirements. This boundary ensures investigation of mature security leadership rather than nascent functions, whilst maintaining sufficient diversity to identify patterns across contexts. The temporal scope captures contemporary security leadership during a period of unprecedented digital transformation and threat evolution, providing insights relevant to current organisational challenges whilst establishing foundations for longitudinal investigation.

## 1.4 THEORETICAL POSITIONING

This research adopted an exploratory stance that maintained theoretical openness whilst drawing on general awareness of executive leadership scholarship. The literature review established the boundaries of current knowledge across multiple domains: the contemporary cybersecurity context, the emergence of the CISO role in practitioner discourse, established executive leadership frameworks for understanding C-suite positions, security management and governance research, and organisational contextual factors shaping security requirements.

However, this review deliberately avoided committing to specific theoretical frameworks that might constrain empirical discovery. Existing executive leadership theories, whilst valuable for understanding traditional C-suite roles, appeared insufficient for explaining security leadership's unique challenges of adversarial environments, irreconcilable accountability tensions, and invisible value creation.

This theoretical positioning reflects grounded theory methodology's core principle: theories should emerge from systematic analysis of empirical data rather than being imposed a priori (Charmaz, 2014; Glaser & Strauss, 1967). The research began not with predetermined theoretical lenses but with open empirical investigation guided by the broad research question. The absence of established frameworks for security leadership, combined with fundamental differences from other executive roles, necessitated this abductive approach. Beginning with theoretical commitments risked forcing the phenomenon into ill-fitting conceptual moulds, potentially obscuring patterns that existing frameworks cannot accommodate.

The theoretical frameworks that ultimately inform the Security Leadership Contingency Model were not present at the outset of analysis. They were identified through targeted literature engagement at specific points when empirical patterns required explanation that the data alone could not provide. The methodological basis for this abductive process, and the specific analytical moments at which theoretical frameworks were recognised, are detailed in Chapter 4. The theoretical contributions that emerged from this process, and the manner in which multiple frameworks were integrated, are developed in Chapter 6, with broader implications explored in Chapter 7.

## 1.5 METHODOLOGICAL APPROACH

The exploratory nature of the research questions and the absence of established theoretical frameworks for security leadership necessitate a qualitative, abductive approach. This study employs grounded theory methodology, specifically adopting the Gioia method for its systematic approach to moving from empirical observation to theoretical abstraction (Gioia et al., 2013). This methodological choice reflects the need to understand security leadership from the perspective of those experiencing it rather than imposing predetermined frameworks that may not capture the phenomenon's complexity. The researcher's position as a practising CISO investigating peer security leaders introduces both methodological advantages and considerations that require explicit management; these are addressed through the reflexivity discussion in Section 4.6.5 and examined as a methodological contribution in Section 7.5.

The research design centres on in-depth, semi-structured interviews with 20 senior security executives in Australian organisations. This sample size, whilst modest by quantitative standards, proves appropriate for qualitative investigation seeking depth over breadth (Guest et al., 2006). Participants were purposefully selected to represent diversity across industries, organisation sizes, and security maturity levels whilst maintaining focus on senior executives with substantial leadership experience. The Australian context provides a bounded yet diverse research setting, with sophisticated regulatory requirements, mature security practices, and active threat landscapes that mirror global challenges whilst maintaining cultural coherence. The theoretical and methodological justification for this geographic focus is developed in Section 4.8 and Section 7.6 respectively.

Data collection proceeded iteratively, with early interviews informing subsequent questioning as patterns emerged. This approach enabled pursuit of emerging themes whilst maintaining consistency in core areas of investigation. Analysis followed the Gioia method's systematic progression from first-order concepts through second-order themes to aggregate dimensions, ultimately culminating in the Security Leadership Contingency Model. The analytical process involved constant comparison between data and emerging concepts, extensive memoing to capture analytical insights, and regular consultation with supervisory team members to challenge interpretations and ensure rigorous grounding in empirical evidence.

## 1.6 OVERVIEW OF CONTRIBUTIONS

The primary contribution of this research is the Security Leadership Contingency Model (SLCM), a comprehensive framework explaining how security leadership effectiveness emerges from dynamic alignment between leader capabilities, organisational contexts, and environmental demands. The SLCM challenges existing static conceptualisations of executive effectiveness by demonstrating that security leaders must maintain dual fits: simultaneously aligning with internal organisational requirements and external environmental pressures. The model reveals how these alignment requirements evolve through distinct organisational phases, each demanding different leadership configurations for optimal effectiveness.

Theoretical contributions extend established frameworks through mid-range theory development for security leadership contexts. The research extends person-organisation fit theory (P-O fit) to boundary-spanning security leadership by revealing that fit requirements transform systematically through organisational maturity phases, where capabilities creating fit in one phase may actively undermine fit in subsequent phases. This phase-based transformation creates more extreme fit transitions than traditional executive roles experience. Person-environment fit theory (P-E fit) is advanced by demonstrating how boundary-spanning security executives face unique environmental pressures that operate independently from organisational demands, requiring political capital for active fit maintenance over time. The integration of these perspectives with contingency theory and role theory generates new insights about how boundary-spanning security executives navigate complex, multi-stakeholder environments. Among these theoretical perspectives, contingency theory occupies a distinct position as the second core theoretical foundation. It is this principle, rather than fit theory alone, that most fundamentally distinguishes the SLCM and justifies its contingency designation; the full theoretical development of this choice is presented in Section 6.2. The concept of dual-fit maintenance emerges as a critical capability for security executives operating at organisational boundaries, providing theoretical language for phenomena previously recognised but not systematically conceptualised.

The research generates practical value through both conceptual and operational contributions. The conceptual framework identifies three distinct organisational phases, Establishment, Maturation, and Strategic, with associated leadership requirements, providing foundations for more sophisticated approaches to recruitment, development, and succession planning. The framework's emphasis on political capital as a critical enabler of security leadership effectiveness illuminates often-overlooked dimensions of executive success. To operationalise these insights, four practitioner-grade tools were developed: the Organisational Security Maturity Phase Diagnostic, the Leadership Orientation Self-Assessment, the Phase-Leadership Alignment Matrix, and the Dual-Fit Health Diagnostic. These operational tools translate theoretical insights into immediately usable instruments for leadership selection, development, and in-role effectiveness assessment. For the broader security profession, the research provides empirical foundations that can inform capability frameworks, career development pathways, and performance evaluation systems that reflect contextual realities rather than generic prescriptions.

## 1.7 THESIS STRUCTURE

This thesis unfolds through eight chapters that progressively build from problem identification through empirical investigation to theoretical contribution and practical application. Following this introduction, Chapter 2 presents a critical review of literature that establishes the boundaries of current knowledge whilst identifying specific gaps that justify exploratory investigation. The review examines the contemporary cybersecurity context, traces the emergence of the CISO role, evaluates existing executive leadership frameworks, and demonstrates why current theoretical approaches prove inadequate for understanding security leadership.

Chapter 3 synthesises the literature review findings, bridging from knowledge gaps to research design implications. This synthesis chapter clarifies how the patterns identified across multiple literatures inform the exploratory research approach and connects the empirical vacuum identified in existing scholarship to the methodological choices detailed subsequently.

Chapter 4 details the methodological approach, establishing philosophical foundations for the qualitative inquiry and justifying the selection of grounded theory methodology. The chapter provides transparency about research design decisions, sampling strategy, data collection procedures, and analytical processes, enabling readers to assess the rigour and trustworthiness of the investigation. Particular attention focuses on managing the researcher's dual identity as both academic investigator and industry practitioner, demonstrating how potential biases were acknowledged and mitigated through systematic procedures.

Chapter 5 presents the empirical findings emerging from systematic analysis of interview data. The chapter follows the Gioia methodology's progression from first-order concepts through second-order themes to aggregate dimensions, demonstrating the analytical journey from initial pattern recognition through theoretical abstraction. Rich participant quotations ground theoretical insights in lived experiences, whilst analytical narrative explains how patterns emerged and evolved through the investigation.

Chapter 6 develops the Security Leadership Contingency Model as the theoretical contribution emerging from empirical patterns. The chapter presents the model's dynamics, articulates four theoretical propositions, and addresses complexity through examination of negative cases and boundary conditions. This chapter maintains clear distinction between empirical observation and theoretical interpretation, establishing foundations for subsequent discussion.

Chapter 7 positions the empirical findings and theoretical model within broader scholarly conversations, demonstrating how the Security Leadership Contingency Model extends existing frameworks whilst addressing their limitations. The chapter explores practical implications for security leaders, organisations, and the profession, translating theoretical insights into actionable guidance through both conceptual frameworks and operational tools. Methodological contributions are examined, and limitations are acknowledged explicitly, with boundaries of applicability clearly defined whilst identifying opportunities for future investigation.

Chapter 8 concludes the thesis by synthesising the research journey and articulating its contributions to knowledge. The chapter provides comprehensive answers to the research questions, demonstrates how findings advance theoretical understanding, and offers specific recommendations for practice. The conclusion reflects on research quality whilst charting pathways for future investigation that build upon this foundational study.

For different audiences, the thesis offers multiple pathways through the material. Academic readers interested in theoretical contributions should focus on Chapters 2, 5, 6, and 7, where literature review, empirical findings, theoretical development, and theoretical positioning provide greatest depth. Practitioners seeking actionable insights might prioritise Chapters 1, 6, 7, and 8, which establish context, present the framework, explore practical implications, and provide recommendations. Methodology scholars will find Chapters 3 and 4's detailed exposition of grounded theory synthesis and qualitative research procedures of particular interest. Regardless of reading pathway, the thesis maintains coherent narrative that demonstrates how systematic empirical investigation can generate both theoretical advance and practical value in understanding complex organisational phenomena.

# CHAPTER 2: LITERATURE REVIEW

## 2.1 INTRODUCTION: THE LITERATURE REVIEW AS SENSITISING FRAMEWORK

This chapter explores existing knowledge across five domains, identifying the boundaries of current understanding and the empirical vacuum that justifies the present investigation.

### 2.1.1 Purpose and Philosophical Stance

This literature review serves a distinctive purpose within a grounded theory investigation: to sensitise rather than theorise, to illuminate domains of empirical uncertainty rather than impose predetermined conceptual structures (Birks et al., 2019; Cutcliffe, 2000; Turner & Astin, 2021). The review acknowledges that whilst the Chief Information Security Officer role has proliferated globally across organisations, systematic empirical investigation of this phenomenon remains virtually absent from academic discourse. This paradox, a critical organisational role emerging without scholarly foundation, necessitates an approach that maintains theoretical openness whilst identifying the boundaries of current knowledge (Ramezan, 2025; Sahin & Vance, 2025).

The philosophical stance adopted here recognises literature as a lens through which to identify what remains unknown rather than what is definitively established (Kraus et al., 2020; Linnenluecke et al., 2019; Paul & Criado, 2020). In grounded theory methodology, the literature review functions not to derive hypotheses for testing but to establish the legitimacy of exploratory inquiry (Charmaz & Thornberg, 2021; Noble & Mitchell, 2016; Walker & Myrick, 2006). This approach proves particularly essential when investigating phenomena where practitioner discourse has substantially outpaced academic investigation, creating a landscape dominated by prescription without description, advocacy without evidence. The review therefore systematically examines existing knowledge whilst consciously avoiding premature conceptual closure that might constrain the discovery of emergent patterns in subsequent empirical investigation (Paul et al., 2023; Rieger, 2019; Simsek et al., 2023).

The literature identification approach adopted here reflects grounded theory principles rather than systematic review protocols. Literature discovery proceeded iteratively and theoretically, driven by emerging conceptual needs rather than predetermined search protocols. Initial engagement with foundational security leadership sources led to forward and backward citation searching, following conceptual threads as they emerged. This iterative approach remained open to unexpected sources and perspectives that rigid systematic protocols might exclude, aligning with the exploratory nature of grounded theory investigation whilst maintaining analytical rigour in evaluating source quality and relevance.

### 2.1.2 Scope Definition

The literature reviewed spans publications from 2000 to 2025, reflecting the period of emerging scholarly and practitioner attention to security leadership roles. The disciplinary foundations draw from three primary domains: information systems, management studies, and security studies. Information systems literature provides the technological and organisational context within which security leadership operates (Haislip et al., 2021; Johnson & Goetz, 2007; Moon et al., 2018). Management studies, particularly executive leadership and organisational behaviour research, offers theoretical frameworks for understanding C-suite roles and their evolution (AlNuaimi et al., 2022; Guhr et al., 2019; Schoemaker et al., 2018). Security studies contributes perspectives on risk, governance, and the changing threat landscape that shapes security leadership requirements (Aldaajeh & Alrabaee, 2024; Garcia-Perez et al., 2023; Maynard et al., 2018).

The review draws from multiple source types, recognising that understanding of the CISO role requires engagement with diverse communities of practice. Academic literature provides theoretical frameworks and empirical studies, though as subsequent sections demonstrate, scholarly investigation of security leadership remains limited. Practitioner sources, including industry reports from major consultancies and professional body publications, offer insights into role perceptions and challenges, whilst requiring critical evaluation of methodological limitations and potential biases. Regulatory guidance and policy documents reveal how governmental and oversight bodies conceptualise security leadership requirements and responsibilities.

Geographic scope encompasses global perspectives with particular emphasis on developed economies where CISO roles first emerged and have achieved greatest maturity. However, this geographic concentration reflects limitations in available literature rather than deliberate exclusion. As Section 2.6.3 explores, the dominance of Anglo-American perspectives creates systematic blind spots in understanding how cultural and regional factors influence security leadership, a gap this review acknowledges critically whilst working within existing literature constraints.

### 2.1.3 Chapter Structure

Section 2.2 establishes the contemporary cybersecurity context, demonstrating how the evolution of threats has created unprecedented organisational challenges that traditional IT management approaches cannot address (Balzano & Marzi, 2025; Liu et al., 2022; Samtani et al., 2020). Section 2.3 examines the emergence and proliferation of the CISO role, critically analysing practitioner literature that attempts to define and describe this position. Section 2.4 turns to established executive leadership literature, exploring how existing C-suite research might inform understanding of security leadership whilst identifying why direct application of these frameworks may prove problematic.

Section 2.5 investigates security management and governance research, revealing a process-centric focus that systematically ignores leadership dimensions. Section 2.6 explores how organisational context shapes security requirements. Section 2.7 synthesises fundamental questions about the CISO role that emerge from literature analysis.

Throughout this structure, critical analysis takes precedence over description. The review does not merely catalogue existing knowledge but systematically evaluates its quality, identifies contradictions, and exposes assumptions masquerading as evidence. By maintaining this analytical perspective, the review establishes not what is known about security leadership but what remains to be discovered through systematic investigation.

## 2.2 THE CONTEMPORARY CYBERSECURITY CHALLENGE

The cybersecurity landscape has undergone fundamental transformation over the past two decades, evolving from a technical concern managed within information technology departments to a strategic risk demanding executive attention. This evolution reflects not merely quantitative changes in threat volume but qualitative shifts in attack sophistication, business impact, and organisational vulnerability. Understanding these changes provides essential context for examining how organisations structure their security leadership responses, particularly through the emergence of dedicated executive roles.

### 2.2.1 Evolution of the Threat Landscape

The contemporary threat environment differs markedly from the security challenges organisations faced at the turn of the millennium. Where early cyber threats primarily consisted of opportunistic attacks by individual actors seeking notoriety or causing disruption for its own sake (Li & Liu, 2021; Skopik et al., 2016), modern threats emanate from sophisticated criminal enterprises, nation-state actors, and well-resourced groups pursuing specific economic or strategic objectives (Council on Foreign Relations, 2024; Europol, 2023; Mandiant, 2024). This professionalisation of cybercrime has transformed both the nature and impact of security incidents.

Statistical evidence reveals the accelerating frequency of cyber attacks across all organisational sectors. Major breach notifications have increased exponentially, with reported incidents growing from hundreds annually in the early 2000s to tens of thousands per year by 2024 (Falowo et al., 2024; IBM Security, 2023; Verizon, 2024). However, these figures likely represent significant underreporting, as many organisations remain reluctant to disclose breaches unless legally compelled (Agrafiotis et al., 2018; Martin et al., 2017). The true scale of cyber attacks may exceed reported figures by orders of magnitude, particularly for incidents that do not trigger regulatory notification requirements.

Beyond frequency, the sophistication of modern attacks demonstrates qualitative advancement in threat actor capabilities. Advanced persistent threats employ multi-stage attack chains, leveraging zero-day vulnerabilities, sophisticated social engineering, and living-off-the-land techniques that evade traditional security controls (Alshamrani et al., 2019; Kazimierczak et al., 2024; Sharma et al., 2023). Supply chain compromises, exemplified by incidents affecting software providers and managed service providers, demonstrate threat actors' strategic thinking in targeting upstream dependencies to maximise impact across multiple organisations simultaneously (European Union Agency for Cybersecurity, 2021; Tan et al., 2025). The emergence of ransomware-as-a-service models has democratised access to sophisticated attack tools, enabling less technically proficient actors to launch devastating attacks (European Union Agency for Cybersecurity, 2023; Palo Alto Networks Unit 42, 2024).

Economic impacts from cyber incidents have reached levels that command board-level attention, as cyber risk is increasingly treated as a major strategic financial threat (Abrardi et al., 2025; Franco et al., 2024). Direct costs from major breaches now routinely exceed hundreds of millions of dollars, with extreme scenarios involving infrastructure or cloud outages resulting in aggregate losses in the billion-dollar range (Eling et al., 2023). These figures do not capture indirect costs such as intellectual property theft, competitive disadvantage from stolen trade secrets, or long-term reputational damage that persists years after an incident (Adekoya et al., 2025; Agrafiotis et al., 2018; Molok et al., 2010). For many organisations, a significant cyber incident now represents an existential threat capable of destroying shareholder value or causing permanent regulatory and financial exposure (Cheng et al., 2017).

The shift from nuisance to existential threat manifests clearly in critical infrastructure sectors where cyber attacks threaten not only economic losses but potential physical harm and societal disruption. Attacks on energy grids, water treatment facilities, healthcare systems, and transportation networks demonstrate threat actors' willingness and ability to target essential services (Humayun et al., 2020; Langner, 2011; McLaughlin et al., 2016; Riggs et al., 2023). These incidents elevate cybersecurity from a business continuity concern to a matter of national security and public safety, fundamentally altering the stakes for organisational security leaders.

### 2.2.2 Critical Analysis: Why These Changes Demand New Leadership

The evolution of the threat landscape creates leadership challenges that transcend traditional information technology management approaches. A critical pattern emerges across the literature: whilst technical solutions remain necessary, they prove insufficient for addressing threats that exploit human vulnerabilities and organisational processes (Ahmad et al., 2021; Parkin et al., 2023). Social engineering attacks bypass sophisticated technical controls by targeting human psychology, whilst business email compromise schemes exploit organisational procedures rather than technical vulnerabilities (Adekoya et al., 2025; Varga et al., 2021). This human-centric dimension of modern threats requires leadership capable of influencing organisational culture, behaviour, and risk awareness beyond implementing technical controls (Agrafiotis et al., 2018; Wasserman & Wasserman, 2022).

The literature reveals different perspectives regarding threat severity and appropriate responses, reflecting varying methodologies and stakeholder interests. Vendor-produced threat reports often emphasise emerging threats and worst-case scenarios, potentially influenced by commercial imperatives to demonstrate product value (CrowdStrike, 2024; Gartner, 2025). Conversely, academic researchers and independent analysts frequently present more measured evaluations, suggesting that whilst threats are serious, many organisations can achieve reasonable security through fundamental practices rather than cutting-edge technologies (Falowo et al., 2024; Ferdous et al., 2023). This divergence in risk assessment perspectives creates challenges for security leaders who must navigate varying claims whilst making evidence-based investment decisions.

The leadership implications of evolving threats extend beyond technical decision-making to encompass strategic risk management at the highest organisational levels. Security decisions increasingly involve trade-offs between business enablement and risk mitigation, requiring leaders who can articulate security concerns in business terms and participate effectively in executive strategy discussions (Kure et al., 2022; Luo, 2022; Saeed et al., 2023). The shift from technical management to strategic risk decisions demands security leaders with broader business acumen, communication skills, and political sophistication than traditional IT security managers typically possessed (Lehto & Limnéll, 2021; Loonam et al., 2022).

An unresolved tension permeates discussions of security's organisational role: whether security functions primarily as an enabler of business objectives or as a necessary constraint on business activities (Luo, 2022; Saeed et al., 2023). Practitioner-focused research often advocates for security as a business enabler, arguing that when built into strategic renewal (Warner & Wäger, 2019), it underwrites the trust required for digital innovation (Beck et al., 2017). However, operational reality often positions security as imposing constraints through heavy process demands and compliance requirements (Kamil et al., 2023; Wylde et al., 2022) that can slow business velocity (Perdana et al., 2022; Skafi et al., 2020). Security leaders must navigate this fundamental tension, balancing enablement aspirations with protection imperatives whilst maintaining credibility with both business and technical stakeholders (Ahmad et al., 2020; Soomro et al., 2016).

### 2.2.3 Organisational Vulnerability

Digital transformation initiatives, whilst delivering business benefits, simultaneously expand organisational attack surfaces in ways that many organisations fail to fully appreciate (Nadkarni & Prügl, 2020; Saeed et al., 2023). Cloud migration, whilst offering scalability and flexibility, introduces shared responsibility models that many organisations misunderstand, leading to security gaps (Hammi et al., 2023; Schmitt, 2023). Internet of Things deployments multiply connected endpoints, often with minimal security controls and limited visibility into device behaviour (Aidoo et al., 2022; Krishnan et al., 2022). Remote work arrangements, accelerated by global pandemic responses, extend corporate networks beyond traditional perimeters, challenging established security architectures (Fletcher & Griffiths, 2020; Qollakaj et al., 2025). Each digital transformation initiative potentially introduces new vulnerabilities whilst organisations struggle to maintain visibility and control across increasingly complex technology ecosystems (Reis & Melão, 2023).

Legacy system vulnerabilities compound modern security challenges, as organisations maintain ageing systems that cannot be easily patched or replaced. Critical business systems running on unsupported operating systems, applications with hard-coded credentials, and industrial control systems designed without security considerations create persistent vulnerabilities that threat actors actively exploit (Aloseel et al., 2021; Walker-Roberts et al., 2020). The technical debt accumulated through decades of technology deployments creates security risks that cannot be addressed through incremental improvements, requiring strategic decisions about system replacement, isolation, or risk acceptance (Axon et al., 2022; Ramač et al., 2022; Tom et al., 2013).

Supply chain and third-party risks have emerged as critical vulnerability vectors that traditional, internally focused security approaches inadequately address (Ghadge et al., 2020; Hammi et al., 2023). Modern organisations depend on complex webs of suppliers, partners, and service providers, each representing potential entry points where cyber risks can penetrate and then propagate across the entire network (Colicchia et al., 2018; Li & Xu, 2021). Traditional vendor risk assessment approaches, relying on questionnaires and annual reviews, often fail to capture the dynamic nature of these risks or the cascading effects of upstream compromises (Colicchia et al., 2018; Topping et al., 2021). Managing these extended enterprise risks requires security leaders to move toward network-centric governance, influencing stakeholders beyond organisational boundaries and negotiating security requirements with suppliers whilst maintaining vital business relationships (Creazza et al., 2021; Ghadge et al., 2020).

Human factors consistently emerge as the most significant vulnerability in organisational security, yet remain the most challenging to address through traditional security management approaches (Nifakos et al., 2021; Pollini et al., 2022). Despite extensive security awareness training investments, employees continue to fall victim to phishing attacks and circumvent security controls that are perceived as impediments to productivity (Jaeger & Eckhardt, 2021; Pollini et al., 2022). The literature reveals a persistent gap between security knowledge and security behaviour, suggesting that awareness alone does not translate to improved security outcomes (Khando et al., 2021; Prümmer et al., 2024). Insider threats, whether malicious or inadvertent, account for a significant proportion of security incidents, yet detection and prevention approaches often remain immature (Ghafir et al., 2018; Safa et al., 2019). Addressing human vulnerabilities requires security leaders capable of influencing organisational culture, designing usable security controls, and building security considerations into business processes rather than treating security as an add-on control (da Veiga et al., 2020; Xu et al., 2024).

### 2.2.4 Business Impact: Beyond Technical Metrics

Synthesis of impact studies reveals that organisations systematically underestimate the full business impact of security incidents, focusing on direct costs whilst overlooking longer-term consequences. Direct costs represent only the initial financial impact. Indirect costs often exceed direct costs by significant multiples, encompassing lost productivity, customer defection, and elevated cost of capital due to perceived risk (Perera et al., 2022; Varga et al., 2021). Reputational damage, whilst difficult to quantify, may persist for years following an incident, affecting customer trust, partner relationships, and talent acquisition (Perera et al., 2022).

Conflicting evidence emerges regarding optimal investment strategies for security, particularly the relative return on investment between prevention and response capabilities. Formal models of this trade-off find that optimal strategies generally require non-zero resources for both preventive and reactive measures, depending on attack probabilities and asset criticality (Fielder et al., 2016; Paarporn & Xu, 2024). Some research suggests that in interdependent systems, purely preventive strategies can be inefficient, as residual risk and spillovers require robust resilience and recovery capabilities (Abdallah et al., 2020; Baskerville et al., 2014; Fedele & Roner, 2022; Paul & Zhang, 2021). This prevention versus response debate reflects deeper uncertainties about security investment optimisation in environments where threat probabilities remain difficult to quantify accurately (Zhang & Malacaria, 2025).

A clear pattern emerges from longitudinal studies: board attention to cybersecurity correlates strongly with recent incidents rather than proactive risk assessment. Firm-level data show that higher breach costs and external discovery of incidents lead to subsequent increases in investment, indicating that salient, embarrassing events trigger budgeting and scrutiny (Fernandez De Arroyabe et al., 2023; Shaikh & Siponen, 2024). Organisations without recent incidents often maintain a minimal, compliance-oriented posture despite comparable risk exposures, effectively "chasing" incidents instead of investing to a higher target level in advance (Jalali & Kaiser, 2018; Patterson et al., 2023).

Critical observation reveals a significant gap in the empirical evidence base: whilst the assumption that security leadership directly impacts security outcomes appears intuitively logical, empirical evidence directly linking leadership characteristics to measurable security outcomes remains primarily correlational (Choi, 2016; Guhr et al., 2019; Moon et al., 2018). The literature frequently assumes that senior security executives improve breach prevention, yet rigorous causal studies demonstrating these relationships are remarkably sparse (Ahmad et al., 2020; Almuqrin, 2024; Loonam et al., 2022). This limited empirical validation represents a critical gap given the significant investments organisations make in these roles.

**Gap Identified:** Whilst the security challenge is well documented with technical and economic dimensions covered, limited empirical understanding exists of how these changes translate into leadership requirements. The literature documents what is happening but not how organisations should respond through leadership structures. Whilst frameworks for security leadership are beginning to emerge (Ganin et al., 2020; Goel et al., 2020; Sutton & Tompson, 2025), they remain largely theoretical with limited empirical validation in practice. The challenge has shifted to a lack of evidence demonstrating which approaches actually improve security outcomes and how to implement them effectively. The assumption that elevated security leadership represents an appropriate response to evolving threats lacks rigorous causal validation, leaving organisations to implement structures based on peer practices rather than empirically demonstrated effectiveness criteria.

## 2.3 THE EMERGENCE OF THE CISO ROLE

The Chief Information Security Officer role represents a relatively recent addition to the executive suite, emerging from the convergence of escalating cyber threats, regulatory pressures, and recognition that security requires dedicated leadership beyond traditional IT management. This section critically examines the proliferation of this role, the definitional challenges surrounding it, and the quality of existing literature attempting to describe and prescribe CISO practice.

### 2.3.1 Historical Development and Role Proliferation

The evolution of dedicated security leadership reflects broader transformations in how organisations conceptualise and manage information risk. Early security responsibilities typically resided within IT departments, where technical administrators managed access controls, implemented firewalls, and responded to viruses (Johnson & Goetz, 2007; Sahin & Vance, 2025). The first documented CISO appointments emerged in financial services during 1994-1995, with Steve Katz appointed at Citicorp (now Citigroup) following cyberattacks by Russian hackers, establishing the role that would proliferate across industries (Horkan & Baker, 2025). This initial adoption pattern, concentrated in highly regulated industries managing sensitive financial data, established precedents that would shape subsequent role development.

The trajectory from niche position to widespread adoption reveals dramatic growth patterns. From rare positions in 2000, CISO appointments experienced remarkable expansion across

Fortune 500 companies and beyond, though precise adoption rates remain undocumented in the academic literature (Ferreira et al., 2024; Karanja & Rosso, 2017). The proliferation extends beyond large enterprises, with mid-market organisations increasingly appointing security executives and even small businesses designating security leadership responsibilities (Karanja, 2017; Sahin & Vance, 2025). Geographic variations persist, with North American and European organisations leading adoption whilst Asia-Pacific markets demonstrate accelerating appointments (Maschmeyer et al., 2021; Ramezan, 2025). Industry analysis reveals differential adoption rates, with financial services achieving near-universal CISO presence whilst manufacturing and retail sectors demonstrate more variable patterns (Karanja & Rosso, 2017).

The drivers behind this proliferation merit critical examination. Practitioner literature consistently attributes role creation to breach events, regulatory requirements, and board pressure (Karanja & Rosso, 2017; Shahim, 2021). However, empirical validation of these claims remains absent. Surveys report that a majority of CISO appointments follow security incidents (Fernandez De Arroyabe et al., 2023; Shaikh & Siponen, 2024), yet without control groups or longitudinal analysis, causation cannot be established (Guhr et al., 2019; Loonam et al., 2022). Regulatory compliance emerges as another frequently cited driver, particularly following legislation such as General Data Protection Regulation (GDPR) and sector-specific requirements (Johnson & Goetz, 2007; Shahim, 2021). Nevertheless, the relationship between regulatory mandates and actual role creation lacks systematic investigation, with anecdotal evidence substituting for rigorous analysis.

### 2.3.2 The Definitional Challenge: What Is a CISO?

Fundamental definitional ambiguity plagues attempts to understand the CISO role, with varied conceptualisations across professional bodies, scholarly literature, and organisations themselves. Early academic characterisations often framed the CISO as a technical head of IT security operations, responsible for defining and delivering specific organisational security goals and running specialised security functions (Ashenden & Sasse, 2013; Whitten, 2008). This technology-centric framing contrasts with more recent business-oriented conceptualisations that position the CISO as a strategic leader responsible for translating cyber risk into business language and aligning security with organisational resilience (Hooper & McKissack, 2016; Luo, 2022). Such definitional variance extends beyond semantic differences to reflect fundamentally different conceptions of role purpose and scope.

Academic literature offers virtually no clarity, with scholarly definitions either absent entirely or borrowed uncritically from practitioner sources (Da Silva & Jensen, 2022; Sahin & Vance, 2025). The few academic papers addressing CISOs typically adopt operational definitions specific to their research context without engaging broader definitional debates (Ashenden & Sasse, 2013; Karanja, 2017). This theoretical vacuum leaves fundamental questions unaddressed: Is the CISO primarily a technical role elevated to executive status or a business role with security specialisation? Does the position require deep technical expertise or strategic business acumen? Recent evidence suggests that while technical depth remains desired, it is increasingly secondary to strategic and managerial focus (Ramezan, 2025; Whitten, 2008).

Title inflation further complicates definitional clarity. Organisations employ varied titles including Chief Security Officer, Vice President of Security, Director of Information Security, and Head of Cyber, often without clear differentiation in responsibilities or authority (Ramezan, 2025). Research reveals significant disconnects between titles and actual authority, with many "CISO" positions lacking budget authority, independent hiring decisions, or direct executive access (Da Silva & Jensen, 2022; Karanja, 2017). CISOs often grapple with a confused role identity, occupying an "expert system" position where they interpret a mystical domain for senior management; this is a precarious role akin to a "modern-day soothsayer" rather than a stable executive position (Da Silva & Jensen, 2022; Sahin & Vance, 2025).

The scope of responsibilities attributed to CISOs varies dramatically across sources. While practitioner literature routinely lists extensive responsibilities ranging from strategic planning to incident response, there is a lack of consensus on the core competencies required for the role (Ramezan, 2025). Scholarly efforts to categorise these duties often rely on theoretical frameworks or expert opinions rather than empirical data regarding which activities CISOs actually perform (Zwilling, 2022). Some empirical efforts to map security management activities do exist; early taxonomic work has demonstrated that such activities cluster into distinct practitioner domains (Alshaikh et al., 2014). These taxonomies offer partial empirical anchoring for understanding security practice but remain disconnected from executive role-level investigation, leaving the relationship between security management activity and CISO role enactment largely unaddressed. No existing study quantifies how CISOs allocate time across their portfolios or links discrete activities to objective security outcomes (Ramezan, 2025). This absence of empirical grounding reduces many role descriptions to aspirational wish lists rather than evidence-based professional requirements.

### 2.3.3 Critical Analysis of Practitioner Literature

The practitioner literature dominating CISO discourse exhibits systematic methodological weaknesses that undermine its credibility and utility. Industry reports from major consultancies typically rely on convenience samples with response rates between five and fifteen percent, raising fundamental questions about representativeness (Maschmeyer et al., 2021; Stavru, 2014). Gartner surveys, for instance, typically sample 100-500 participants from client organisations, introducing potential selection bias toward enterprises already investing in security guidance (Bethlehem, 2010; Chen et al., 2023). Self-selection bias pervades these studies, as CISOs facing challenges or seeking validation disproportionately participate whilst those managing effectively may remain uninvolved (Maschmeyer et al., 2021). Geographic and industry skews further limit generalisability, with North American financial services over-represented whilst emerging markets and diverse industries remain under-examined (Maschmeyer et al., 2021).

Vendor sponsorship introduces additional bias concerns. Research funded by security vendors consistently emphasises threats and advocates solutions aligned with sponsor offerings (Fabbri, 2020; Lundh et al., 2017, 2018). Studies sponsored by recruitment firms inflate salary expectations and emphasise talent shortages (Wang et al., 2018). Professional associations produce research supporting certification programmes and membership benefits (Dobson, 2003; Lexchin, 2012). Whilst not necessarily invalidating findings, these conflicts of interest demand critical evaluation rarely present in literature citations or practical applications.

The atheoretical nature of practitioner research represents perhaps its most fundamental limitation. Competency models lack grounding in leadership theory or empirical validation (Poeppelbuss et al., 2011; Wendler, 2012). Maturity frameworks borrow from other domains without adaptation or testing (Becker et al., 2009; Santos-Neto & Costa, 2019). Best practice recommendations derive from anecdote and opinion rather than systematic investigation (Buis et al., 2023; Kolukisa Tarhan et al., 2020). This theoretical vacuum prevents cumulative knowledge building, leaving each new report to reinvent concepts without advancing understanding.

Contradictions across sources further highlight quality concerns regarding the empirical grounding of security leadership guidance. Reporting structure recommendations vary widely, with different studies advocating Chief Executive Officer (CEO), CIO, Chief Risk Officer (CRO), or board reporting without robust empirical justification (Dupont, 2019; Magnusson et al., 2025; Szczepaniuk et al., 2020). Success metrics prove equally inconsistent, ranging from technical indicators such as mean time to detect through compliance scores to business alignment measures (Diesch et al., 2020; Liu et al., 2025; Philippou et al., 2020). Role priorities shift dramatically between studies, with operational, strategic, and transformation focuses alternating based on sample composition and organisational maturity contexts (Magnusson et al., 2025; Uchendu et al., 2021). These inconsistencies suggest that purported insights may reflect methodological artifacts rather than genuine patterns.

### 2.3.4 The Business-IT Divide: Unresolved Debates

The positioning of security leadership between technology and business domains represents a fundamental tension inadequately addressed in existing literature. Approximately forty-five percent of CISOs report through IT structures, typically to CIOs or Chief Technology Officers (CTOs) (Karanja, 2017; Peppard, 2010; Schobel & Denford, 2013). This arrangement raises questions about independence, particularly when security assessments must evaluate IT performance. Advocates for IT reporting cite technical synergies, shared infrastructure responsibilities, and operational efficiencies (Law & Ngai, 2007; Luftman & Brier, 1999). Critics argue that IT reporting subordinates security to technology priorities, limits strategic influence, and creates conflicts of interest (Patacsil & S. Tablatin, 2017; Peppard, 2007).

Alternative reporting structures to CEOs, boards, or risk committees reflect attempts to elevate security's organisational status (Karanja et al., 2021; Klaus et al., 2022). Direct CEO reporting theoretically provides strategic alignment, executive visibility, and organisational authority (Bajwa et al., 1998; Bansal & Agarwal, 2015). However, whether CEOs possess the expertise or bandwidth to effectively oversee security remains questionable. Board reporting suggests independence and enterprise-wide perspective, yet practical challenges include board meeting frequency, technical complexity, and spans of control (Barnes et al., 2021; Hekkala et al., 2022).

The skills debate further illustrates the business-IT divide. Technical advocates argue that deep security expertise remains essential for credibility and effective decision-making (Patacsil & S. Tablatin, 2017; Wang et al., 2018). Business proponents counter that strategic thinking, communication skills, and business acumen matter more than technical depth (Hu et al., 2007; Merhi & Ahluwalia, 2019). This debate reflects deeper questions about whether security leadership represents technical management elevated or business leadership specialised, a distinction with profound implications for recruitment, development, and evaluation.

Scholarly research reveals persistent challenges bridging business and technology domains. Communication gaps between security and business stakeholders appear in multiple academic reviews of security culture and governance (Soomro et al., 2016; Uchendu et al., 2021). CISOs often report struggling to translate technical risks into business impact, whilst executives express frustration with security's perceived negativity and complexity (Nifakos et al., 2021; Pollini et al., 2022). These communication challenges suggest fundamental misalignment in how different stakeholders conceptualise security's organisational role. Without empirical investigation of how successful security leaders navigate the business-IT divide, prescriptive guidance remains speculative (Ganin et al., 2020).

### 2.3.5 Challenges and Constraints: Consistent Patterns

Despite methodological limitations, certain challenges appear consistently across practitioner literature, suggesting genuine phenomena warranting investigation. Resource constraints dominate survey findings, with a high proportion of respondents reporting insufficient budgets and understaffing (Gao et al., 2023; Magnusson et al., 2025). The universality of resource complaints raises questions about whether these reflect objective shortfalls or endemic features of security management. Without baseline metrics or comparative analysis, the validity of resource inadequacy claims cannot be assessed, as current research lacks established objective adequacy thresholds (Żebrowski et al., 2022).

The influence without authority paradox emerges repeatedly, with CISOs expected to ensure enterprise security whilst lacking formal authority over business units, technology decisions, or resource allocation (Da Silva & Jensen, 2022). This structural contradiction creates fundamental role tensions that prescriptive frameworks fail to address. How security leaders navigate influence challenges, build coalitions, and achieve outcomes without positional power remains unexplored empirically.

Rapid technological change and evolving threat landscapes impose continuous adaptation requirements (Ashenden & Sasse, 2013; Balozian et al., 2023). CISOs report perpetual learning curves, with emerging technologies introducing new vulnerabilities faster than security capabilities develop. The cognitive and organisational demands of continuous adaptation receive minimal scholarly attention, leaving practitioners without evidence-based strategies for managing change velocity.

Stakeholder management complexity appears consistently, with CISOs balancing demands from boards, executives, regulators, auditors, technology teams, and business units (Jalali & Kaiser, 2018; Uchendu et al., 2021). Each stakeholder group brings different expectations, risk tolerances, and success metrics, creating competing pressures that complicate priority setting.

The absence of empirical research on stakeholder navigation strategies leaves security leaders reliant on intuition rather than evidence (Ganin et al., 2020).

Talent shortages represent another universal challenge, with organisations reporting difficulty recruiting and retaining qualified security professionals (Blažič, 2022; Furnell, 2021). Evidence suggests that skills mismatches and over-narrow role definitions contribute significantly to these gaps alongside genuine supply shortages (Graham & Lu, 2023; Polakova et al., 2023). The talent challenge extends beyond headcount to encompass a lack of professionals who combine technical depth with the social and methodological competencies needed to bridge business-technical divides (Bendler & Felderer, 2023).

Career progression uncertainties plague the profession, with unclear pathways to CISO positions and limited advancement opportunities beyond the role (da Veiga et al., 2020; Poon & Wagner, 2001; Sahin & Vance, 2025). This ambiguity complicates succession planning and professional development, potentially contributing to reported burnout rates (Nepal et al., 2024; Pham et al., 2019). This turnover pattern appears robust to sampling variations, though causation remains disputed. Resource constraints emerge universally, with studies assuming cybersecurity operates under tight budgets without providing cross-organisational benchmarks for adequacy (Magnusson et al., 2025).

The strategic aspiration versus operational reality tension surfaces repeatedly. While organisations formally define the CISO as a strategic, management-level role, perceptions of a heavy operational load persist (Hooper & McKissack, 2016; Ramezan, 2025). This pattern persists across industries and regions, yet without observational data or validated time-use studies, self-reported allocations may reflect perception more than reality. Industry reports suffer from multiple validity threats, including selection bias, vendor influence, and a lack of triangulation, which mirror broader concerns about reporting bias in cybersecurity secondary data (Ampatzoglou et al., 2019; Clougherty et al., 2016; Maschmeyer et al., 2021).

Critical evaluation reveals that industry reports provide valuable descriptive data whilst lacking analytical depth. Surveys document what CISOs say they do without observing actual behaviour. Reports identify challenges without exploring causation or solutions. Studies capture snapshots without longitudinal perspective. These limitations do not invalidate practitioner research but highlight the need for complementary academic investigation employing rigorous methods and theoretical frameworks. The contradiction between aspirational strategic positioning and operational reality, the communication gap between technical and business domains, and the persistent challenge of exercising influence without authority all point to fundamental role tensions requiring systematic empirical investigation.

**Gap Identified:** Despite the proliferation of practitioner literature attempting to define and describe the CISO role, no rigorous empirical research examines what CISOs actually do, how they contribute to organisational performance, or what makes them effective. The existing literature consists primarily of atheoretical surveys with methodological weaknesses, anecdotal accounts lacking systematic analysis, and prescriptive frameworks without empirical validation. This absence of scholarly investigation leaves fundamental questions about role purpose, activities, value creation, and effectiveness determinants completely unaddressed. The contrast between role prominence in practice and absence in academic research represents a critical knowledge gap requiring systematic empirical investigation.

## 2.4 EXECUTIVE LEADERSHIP LITERATURE

### 2.4.1 C-Suite Role Research: Patterns and Contrasts

The landscape of executive role research presents a stark contrast between deeply studied traditional positions and the theoretical vacuum surrounding security leadership. Established C-suite roles benefit from decades of scholarly investigation that has produced clear success metrics, well-understood career trajectories, and validated performance frameworks (Hambrick & Mason, 1984; Judge et al., 1995). The Chief Executive Officer role, for instance, has been subjected to more than fifty years of rigorous academic scrutiny, resulting in comprehensive understanding of strategic leadership requirements, decision-making processes, and performance indicators (Koch et al., 2015; Waldman et al., 2004). This extensive research foundation enables organisations to make informed decisions about CEO selection, development, and evaluation based on empirically validated criteria.

Similarly, the Chief Financial Officer role has evolved through systematic scholarly investigation that has documented its transformation from accounting oversight to strategic value creation (Hoitash et al., 2016; Zorn, 2004). Research has clearly established how CFOs contribute to organisational performance through capital allocation decisions, investor relations management, and strategic planning participation (Caglio et al., 2018; Liu et al., 2021). The clarity of financial metrics provides unambiguous measures of CFO effectiveness, enabling both scholars and practitioners to assess performance and identify success factors with considerable precision.

The Chief Information Officer role, whilst more recent than CEO or CFO positions, nonetheless benefits from four decades of sustained academic attention that has chronicled its evolution from technical management to business partnership (Chun & Mooney, 2009; Kratzer et al., 2023). Extensive research has examined how CIOs navigate the tension between operational excellence and strategic innovation, how they build credibility with non-technical executives, and how they demonstrate value through digital transformation initiatives (Chen et al., 2010; Peppard, 2010; Richardson et al., 2024). This body of work provides theoretical frameworks and empirical evidence that inform both CIO development and organisational structuring of the IT leadership function.

Emerging C-suite roles demonstrate predictable patterns in their organisational elevation and scholarly examination. The Chief Marketing Officer has transitioned from tactical campaign management to strategic growth leadership, with research documenting this evolution and its impact on firm performance, showing an average 15% performance boost when CMOs participate in strategic decision-making (Germann et al., 2015; Nath & Mahajan, 2008). The Chief Digital Officer represents organisations' response to digital transformation imperatives, with nascent but growing literature examining role requirements and success factors (Firk et al., 2021; Kunisch et al., 2022). The Chief Risk Officer emerged from regulatory requirements following financial crises, with research establishing connections between CRO presence and improved risk management outcomes (Bailey, 2019; Karanja, 2017).

Analysis across these diverse executive roles reveals consistent patterns in C-suite evolution. New executive positions typically emerge when functional domains experience crisis or transformation that elevates their strategic importance (Smith & Foti, 1998; van Kemenade, 2019). Initial role creation often responds to external pressures such as regulatory requirements, competitive dynamics, or stakeholder demands. Over time, roles mature from reactive compliance functions to proactive strategic positions, though this evolution requires both organisational learning and individual capability development. The trajectory from functional management to strategic leadership follows predictable stages, with research documenting common challenges and success factors at each phase.

However, the CISO role conspicuously lacks this theoretical foundation despite exhibiting many characteristics that should attract scholarly attention. While security breaches generate significant media attention, academic investigation remains limited and largely atheoretical (Karanja, 2017; Sahin & Vance, 2025). Regulatory requirements mandate security oversight similar to financial controls, yet stimulate minimal research into security executive effectiveness. Digital transformation creates security challenges paralleling IT complexity, yet triggers little examination of how security leaders should respond. This absence of scholarly attention creates a theoretical vacuum that leaves both practitioners and organisations without evidence-based guidance for one of their most critical executive positions.

### 2.4.2 Leadership Effectiveness: Lost in Translation

The challenge of measuring leadership effectiveness becomes particularly acute when examining boundary-spanning roles that operate across technical and business domains. Traditional executive effectiveness frameworks, developed primarily through studies of CEOs and other established C-suite positions, rely heavily on observable financial metrics and stakeholder satisfaction measures (Judge et al., 1995; Ng et al., 2005). These frameworks assume clear line-of-sight between executive actions and measurable outcomes, an assumption that proves problematic when applied to prevention-focused roles where success manifests as the absence of negative events rather than the presence of positive results.

Competency-based approaches to leadership effectiveness dominate practitioner literature, proposing comprehensive lists of skills, knowledge areas, and capabilities required for executive success (Shet et al., 2019). These frameworks typically enumerate technical proficiencies, business acumen requirements, and interpersonal skills deemed essential for role performance. However, empirical research consistently demonstrates limited predictive validity for competency-based models, with meta-analyses revealing weak correlations between assessed competencies and actual leadership outcomes (Bolden et al., 2006; Levenson et al., 2006; Stevens, 2013). The proliferation of competency frameworks without empirical validation creates an illusion of understanding whilst providing little practical guidance for leadership development or selection decisions.

Behavioural approaches attempt to identify specific actions and decision patterns that distinguish effective from ineffective leaders (Hooijberg, 1996; Yukl et al., 2019). This research stream examines how executives allocate time, interact with stakeholders, make decisions, and respond to challenges. Whilst behavioural research provides valuable insights into leadership practices, it systematically undervalues contextual factors that shape both the availability and effectiveness of different behavioural options (Russell, 2001; Willis et al., 2017). The assumption that effective behaviours transfer across contexts ignores fundamental differences in organisational cultures, industry dynamics, and stakeholder expectations that constrain or enable different leadership approaches.

Outcome-based effectiveness measures offer apparent objectivity by focusing on results rather than inputs or processes (Davies & Crombie, 1995; Heinrich, 2002). For traditional executive roles, outcomes such as profitability, growth, market share, or operational efficiency provide seemingly clear performance indicators. However, attribution problems plague outcome-based assessment, particularly in complex organisational environments where multiple factors influence results (Ling, 2012; Vaessen & Todd, 2008). The time lag between leadership decisions and observable outcomes further complicates assessment, especially for strategic initiatives with multi-year implementation horizons.

The unique characteristics of security leadership expose fundamental limitations in all three effectiveness frameworks. Competency models fail to address the paradox of technical expertise that must remain current whilst executives increasingly focus on strategic rather than technical activities (Bendler & Felderer, 2023; White, 2024). Behavioural frameworks cannot account for the adversarial context where security leaders face intelligent, adaptive opponents actively working to circumvent controls (Garnaev et al., 2016; Georgiadou et al., 2021). Outcome measures struggle with the counterfactual nature of security success, where the primary value proposition involves preventing events that might never have occurred regardless of security interventions (Lichtenthaler & Fischbach, 2018; Pearsall et al., 2023; Pfleeger & Cunningham, 2010).

The measurement challenge extends beyond individual frameworks to encompass the fundamental question of what constitutes security leadership success. Unlike revenue generation or cost reduction, security value creation resists simple quantification (Farrow & von Winterfeldt, 2020; Ryan & Ryan, 2006). Stakeholder satisfaction varies dramatically based on perspective, with technical teams valuing different outcomes than business executives or board members (Davis et al., 2009; van der Raadt et al., 2010). The temporal dimension adds further complexity, as security investments may show no return for years until preventing a single catastrophic incident, making traditional return on investment (ROI) calculations largely meaningless (Barik et al., 2023; Xu et al., 2017).

### 2.4.3 The CIO Analogue: Critical Lessons and Limitations

The Chief Information Officer role appears to offer the most relevant precedent for understanding CISO emergence and evolution, given apparent similarities in technical foundations, business alignment challenges, and credibility building requirements. Extensive CIO research documents the transition from technical management to business partnership, providing frameworks and insights that practitioners often attempt to apply to security leadership (Applegate & Elam, 1992;

Chun & Mooney, 2009). However, critical examination reveals both valuable lessons and fundamental limitations in using CIO research as a template for understanding CISO roles.

The parallels between CIO and CISO challenges appear compelling at first examination. Both roles emerged from technical functions that organisations initially viewed as operational necessities rather than strategic capabilities (Gerth & Peppard, 2016; Loonam et al., 2022). Both face the challenge of translating technical complexity into business language that non-technical executives can understand and act upon (Bushee et al., 2017; Volckmann, 2005). Both must balance operational excellence in running critical services with strategic innovation in applying technology to business challenges (Chawla et al., 2023; Chen et al., 2010). Both struggle to demonstrate value using traditional financial metrics whilst arguing for increased investment based on risk mitigation or opportunity enablement (Jones et al., 2020; Manfreda & Stemberger, 2019).

Particularly influential CIO research provides typologies and evolution models that practitioners frequently reference when discussing CISO development (Al-Taie et al., 2018; Barnes et al., 2021). The progression from utility IT director focused on operational efficiency, through evangelist promoting technology adoption, to innovator driving digital transformation, and ultimately facilitator enabling business-led technology initiatives offers an appealing framework for conceptualising security leadership evolution (Li & Tan, 2013; Peppard, 2010). Similarly, research examining CIO-CEO relationships, board presence, and organisational influence provides templates that security leaders attempt to replicate (Benlian & Haffke, 2016; Karahanna & Preston, 2013; Preston et al., 2008).

Yet critical distinctions undermine the transferability of CIO frameworks to security leadership contexts. The fundamental value propositions differ markedly: CIOs enable capabilities whilst CISOs prevent losses. This creates what has been termed the "prevention paradox," where success becomes invisible whilst failure generates intense scrutiny (Bendig et al., 2022; Smaltz et al., 2006). CIOs can demonstrate value through successful system implementations, efficiency improvements, or innovation initiatives that produce tangible business benefits. CISOs face the challenge of proving counterfactuals: attacks that did not succeed, breaches that did not occur, losses that were avoided.

The stakeholder dynamics and political contexts differ substantially between the roles. CIOs primarily manage relationships with business units seeking technology solutions, creating collaborative dynamics where shared success is possible (Davis et al., 2009; Manfreda & Stemberger, 2019). CISOs often find themselves in adversarial relationships with business units, imposing controls that slow processes, restrict capabilities, or increase costs in the name of risk reduction (Nath, 2017; Tassabehji et al., 2016). This fundamental tension between enabling business objectives and protecting against risks creates political challenges not captured in CIO research.

Security operates under asymmetric visibility where failures generate intense scrutiny whilst success remains largely invisible, with CISOs facing "alienation and scapegoating due to visibility only in failure" (Ashenden & Sasse, 2013; Da Silva & Jensen, 2022). This visibility pattern creates different political dynamics, stakeholder relationships, and communication requirements that CIO frameworks do not address.

The temporal dimension of role establishment creates another critical distinction. CIOs benefit from four decades of organisational experience and established legitimacy, with most organisations accepting technology leadership as essential for competitive success (Kratzer et al., 2023; Richardson et al., 2024). CISOs operate in a nascent space where role legitimacy remains contested, reporting relationships vary dramatically, and fundamental questions about strategic necessity persist (Ashenden & Sasse, 2013; Sahin & Vance, 2025). The assumption of organisational learning embedded in CIO evolution models may not apply to security leadership where external shocks rather than gradual maturation drive role development.

Perhaps most significantly, the contextual differences between enabling and protecting create distinct stakeholder dynamics that limit framework transferability. CIOs typically operate in collaborative contexts where business partners want technology to succeed because it enables their objectives (Chan, 2021; Wicks et al., 2015). CISOs frequently encounter resistance from business stakeholders who view security as impediments to agility, innovation, or customer experience (Cumps et al., 2006; Orts & Strudler, 2002). These different relational dynamics require distinct influencing strategies, political skills, and organisational navigation capabilities not captured in CIO research.

### 2.4.4 What General Leadership Theory Misses

The application of general leadership theory to security leadership contexts reveals significant blind spots that suggest the need for domain-specific theorising. Traditional leadership frameworks, developed primarily through studies of business leaders operating in competitive markets, embed assumptions about organisational contexts, success metrics, and stakeholder relationships that do not hold in security leadership situations (Oc, 2018; Willis et al., 2017).

Crisis leadership as routine rather than exceptional fundamentally challenges traditional leadership frameworks that treat crisis response as temporary departures from normal operations (Collins et al., 2023; Dirani et al., 2020). Security leaders operate in perpetual crisis readiness where incident response capabilities must remain constantly available, threat landscapes evolve continuously, and adversaries actively probe for weaknesses (Forster et al., 2022; Kersten, 2005). This continuous crisis orientation requires different cognitive capabilities, stress management approaches, and decision-making processes than traditional leadership theory addresses (Esmaeili et al., 2025; Okoli & Watt, 2018; Tabesh & Vera, 2020).

The adversarial context of security leadership introduces dynamics entirely absent from general leadership theory. Unlike competitive business contexts where rivals operate within legal and ethical boundaries, security leaders face malicious actors unconstrained by such limitations (Hannah et al., 2009; Korzhyk et al., 2011; Roponen et al., 2020). These adversaries employ game-theoretic strategies, actively attempting to deceive and exploit organisational vulnerabilities whilst adapting their tactics in response to defensive measures (Garnaev et al., 2016; Jakóbik, 2020; Pita et al., 2010). This adversarial dynamic requires what the literature describes as adaptive, game-theoretic approaches and continuous scenario planning (Parkin et al., 2023; Roponen et al., 2020), fundamentally different from the collaborative, trust-building emphasis of traditional leadership theory (Schinagl et al., 2022).

The measurement paradox of prevention-focused leadership challenges fundamental assumptions about performance assessment and value demonstration (Lichtenthaler & Fischbach, 2018; Pearsall et al., 2023). General leadership theory assumes observable connections between leadership actions and organisational outcomes, enabling performance evaluation and accountability (Davies & Crombie, 1995; Heinrich, 2002). Security leadership's value proposition centres on preventing negative events that may never manifest, creating counterfactual measurement challenges that traditional frameworks cannot address (Farrow & von Winterfeldt, 2020; Ryan & Ryan, 2006). The extreme difficulty of quantifying prevented incidents undermines conventional approaches to leadership effectiveness assessment and succession planning.

Technical expertise requirements in security leadership create tensions unaddressed by general leadership theory's assumption of functional knowledge sufficiency. Security leaders must maintain current technical understanding of rapidly evolving threat landscapes, attack techniques, and defensive technologies whilst simultaneously developing business acumen and strategic thinking capabilities (Bendler & Felderer, 2023; Dawson & Thomson, 2018). This dual expertise requirement differs qualitatively from other executive roles where functional knowledge becomes less critical as leaders advance to strategic positions. The technical complexity of security creates what some researchers term the "expert leadership paradox," where maintaining credibility with technical teams requires depth that strategic responsibilities preclude (Müller et al., 2024; Ramezan, 2025; Sriharan et al., 2024).

Research identifies the need for ambidextrous approaches that integrate both transitional, governance-focused leadership and transformational, culture-focused leadership, highlighting limitations of traditional frameworks that assume more linear leadership approaches (Loonam et al., 2022; Schinagl et al., 2022). The relational and social alignment dimensions critical for security effectiveness are often underemphasised in general leadership models, despite evidence that these factors significantly influence information security system outcomes (Guhr et al., 2019; Moon et al., 2018).

The combination of technical expertise, crisis management, and prevention focus creates leadership requirements not captured in general leadership literature. Attempts to apply existing frameworks without adaptation ignore fundamental differences in context, measurement, and stakeholder dynamics that require entirely new theoretical approaches (Stone & Jawahar, 2021; Uhl-Bien & Arena, 2018). While practitioner frameworks from consulting firms and professional bodies attempt to address these gaps through competency models and maturity frameworks, these often lack the theoretical grounding and empirical validation necessary for scholarly understanding of the role (Marican et al., 2023; Yeoh et al., 2023). The assumption that leadership principles transfer across domains regardless of context fails when confronted with security leadership's unique characteristics that challenge core theoretical assumptions about organisational life, success metrics, and leadership itself.

**Gap Identified**: Existing executive role research, developed for established positions with clear success metrics and well-understood career paths, fails to address the unique challenges of security leadership. The CISO role's combination of technical expertise, crisis management, adversarial context, and prevention focus creates requirements not captured in general leadership literature. Attempts to apply existing frameworks without adaptation ignore fundamental differences that necessitate distinct theoretical approaches to security leadership.

## 2.5 SECURITY MANAGEMENT AND GOVERNANCE RESEARCH

### 2.5.1 Security Governance: Process Focus, Leadership Absence

The security governance literature presents a paradox: whilst frameworks proliferate with increasing sophistication, the human dimension of who implements these frameworks and how remains systematically unexplored. Dominant frameworks have evolved to address technical and procedural requirements (De Cássia De Faria et al., 2021; Herath et al., 2020), yet fundamental questions about leadership and implementation persist.

The International Organization for Standardization and International Electrotechnical Commission (ISO/IEC) 27001, the most widely adopted information security management standard globally, exemplifies this process-centric orientation (Culot et al., 2021; De Haes et al., 2013). The ISO/IEC 27001:2013 version meticulously detailed 114 controls across 14 domains, specifying what organisations should implement (Diamantopoulou et al., 2020), subsequently consolidated to 93 controls in four themes in the 2022 revision (Badakhshan et al., 2022). However, the framework assumes rather than addresses the existence of capable leadership to drive implementation. The standard references "management" and "leadership" primarily as inputs to processes rather than as subjects requiring examination (Guhr et al., 2019; Kamil et al., 2023). This treatment reduces leadership to a checkbox requirement rather than recognising it as a complex organisational capability requiring development and support.

The National Institute of Standards and Technology (NIST) Cybersecurity Framework, whilst providing valuable technical and procedural guidance, similarly exhibited this leadership blind spot until the 2024 introduction of the Govern function, though implementation guidance remains limited (Dedeke, 2017; NIST, 2024). The framework's five core functions, Identify, Protect, Detect, Respond, and Recover, offer comprehensive coverage of security activities (Akter et al., 2025; Benz & Chatterjee, 2020). Yet even with the recent Govern addition, the framework provides limited guidance on the organisational capabilities, leadership structures, or human competencies required to execute these functions effectively. The implicit assumption appears to be that technical frameworks alone drive security outcomes, despite mounting evidence that human and organisational factors determine implementation success (Dalal et al., 2022; Pollini et al., 2022).

Control Objectives for Information and Related Technologies (COBIT)'s integration of IT governance with security components represents a more holistic approach, acknowledging the interconnection between technology management and security governance (De Haes et al., 2013; von Solms, 2005). Nevertheless, COBIT too focuses predominantly on processes, metrics, and maturity models rather than the leadership dynamics that enable or constrain governance effectiveness (McIntosh et al., 2024). The framework's emphasis on "what" and "how much" overshadows questions of "who" and "how" in terms of leadership execution.

A systematic analysis of these frameworks reveals a consistent pattern: governance literature treats leadership as an assumed input rather than a variable requiring investigation (Alhassan et al., 2016, 2018; Harvey et al., 2021; Horne et al., 2016; Kristensen & Andersen, 2023). This assumption becomes particularly problematic when considering the evidence from practitioner reports suggesting that governance framework adoption often fails due to leadership and organisational factors rather than technical deficiencies (Alshaikh et al., 2015; Jalali & Kaiser, 2018; Magnusson et al., 2025). The frameworks assume the existence of competent security leadership without addressing how such leadership develops, what competencies it requires, or how it navigates organisational complexities.

Furthermore, the proliferation of governance frameworks has created what some researchers term *framework fatigue* (Batyashe & Iyamu, 2017), where organisations struggle to select, adapt, and implement appropriate governance structures. This challenge inherently requires leadership judgement and organisational navigation skills, yet the frameworks themselves provide no guidance on these critical leadership dimensions. The result is a body of literature rich in process documentation but impoverished in leadership understanding.

### 2.5.2 The Structure Debate: Form Without Function

The security management literature engages in extensive debate about optimal organisational structures for security functions, yet this discourse occurs without empirical foundation linking structure to effectiveness. Various structural models have been proposed and advocated (AlGhamdi et al., 2020; Mattord et al., 2023), each with claimed advantages that remain largely unvalidated through systematic research.

Centralised security models dominate traditional thinking, with advocates arguing for economies of scale, standardisation benefits, and unified control (Liu et al., 2020; Zhao et al., 2019). Proponents suggest that centralisation enables consistent policy implementation, reduces duplication of effort, and provides clear accountability structures (Flowerday & Tuyikeze, 2016; Varadharajan et al., 2019). However, these arguments rely primarily on logical reasoning rather than empirical evidence. Critics of centralised models point to reduced business alignment, slower response times, and disconnection from operational realities (Alotaibi & Liu, 2017; Moon et al., 2018), though these criticisms similarly lack systematic empirical support.

Federated models have gained prominence as organisations seek to balance central coordination with business unit autonomy (Shen et al., 2022; Wen et al., 2023). The theoretical advantages include improved business alignment, faster local response capabilities, and better integration with business processes (KhoKhar et al., 2022; Qammar et al., 2023). Yet empirical studies examining whether federated structures actually deliver these benefits remain conspicuously absent from the literature. The advocacy for federated models appears driven more by dissatisfaction with centralised approaches than by demonstrated superiority of alternatives.

Hybrid structures, combining elements of centralised and federated models, represent attempts to capture the benefits of both approaches whilst mitigating their respective weaknesses (Javid et al., 2022; Venkatesan & Rahayu, 2024). These models typically involve central policy and governance functions with distributed implementation capabilities (Srivastava & Prakash, 2020). However, the complexity introduced by hybrid structures may create coordination challenges that offset theoretical benefits. Without empirical investigation, organisations adopting hybrid models operate on assumption rather than evidence.

The literature's preoccupation with structural forms obscures more fundamental questions about how leaders navigate different structures to achieve security outcomes. Regardless of formal structure, security leaders must influence without authority, coordinate across boundaries, and align diverse stakeholder interests (Guhr et al., 2019; Lehto & Limnéll, 2021; Moon et al., 2018). These leadership challenges persist across structural models, yet receive minimal attention in the structure-focused discourse.

Critical analysis reveals that structural recommendations often reflect consultant preferences or vendor architectures rather than empirical research (Humayun et al., 2020; Khan et al., 2021; Nguyen et al., 2015). The absence of longitudinal studies tracking structural changes and their impacts leaves organisations without evidence-based guidance for structural decisions. More problematically, the focus on structure as the primary variable ignores the possibility that leadership quality, organisational culture, or contextual factors may be more significant determinants of security effectiveness than structural form.

### 2.5.3 Security Value: The Measurement Paradox

The quest to demonstrate security value represents one of the field's most persistent challenges, generating extensive literature yet yielding limited practical solutions. The fundamental challenge of proving counterfactuals, demonstrating value through prevented incidents, creates a measurement paradox that confounds both practitioners and researchers (Cavusoglu et al., 2004; Ryan & Ryan, 2006; Schatz & Bashroush, 2017; Wang et al., 2008).

Traditional cost–benefit analyses struggle to capture security's value proposition. Security investments aim to prevent events that may never occur, making ROI calculations inherently speculative (Barik et al., 2023; Yaqoob et al., 2019). The literature documents numerous attempts to quantify security value through various frameworks and models, yet none have achieved widespread acceptance or demonstrated validity (Ramos et al., 2017; Schatz & Bashroush, 2017). This measurement challenge has profound implications for security leaders attempting to justify resources and demonstrate contribution.

The Gordon–Loeb model represents the most prominent theoretical attempt to optimise security investment (Gordon & Loeb, 2006; Hausken, 2006). The model suggests that organisations should not spend more than 37 percent of expected loss on security controls for any given asset (Gordon et al., 2020; Skeoch, 2022). However, empirical validation of the model remains elusive, with research noting a distinct lack of empirical tests on real firm data (Fedele & Roner, 2022; Haapamäki & Sihvonen, 2019). The model's elegant mathematics mask problematic assumptions about probability distributions, loss estimations, and control effectiveness that rarely hold in practice (Ebel & Mitra, 2024; Krutilla et al., 2021). Whilst offering valuable conceptual insights about diminishing returns, its direct application is hampered by the difficulty of accurately quantifying its core inputs.

Return on Security Investment (ROSI) calculations proliferate in the literature, offering formulas to quantify security value (Collier et al., 2023; Sonnenreich et al., 2006; Yaqoob et al., 2019). These calculations typically involve estimating annual loss expectancy, risk reduction percentages, and control costs to derive investment returns (Barik et al., 2023; Schatz & Bashroush, 2017). However, the assumptions required for these calculations often overwhelm any analytical value. Estimating breach probabilities, potential losses, and control effectiveness with sufficient accuracy to support investment decisions remains beyond current capabilities (Abrahamsen et al., 2021; Weishäupl et al., 2018). Scholarly assessments of economic valuation highlight that such models often face severe challenges in parameter estimation, creating an illusion of financial rigour not justified by the quality of the inputs (Schatz & Bashroush, 2017).

The precision implied by ROSI calculations masks fundamental uncertainties that can render results misleading for real investment decisions.

Maturity model approaches attempt to sidestep direct value measurement by correlating security maturity with organisational outcomes (Hasan et al., 2021; Park & Kim, 2014). These models suggest that higher maturity levels correlate with better security outcomes and, by extension, business value (Saeed et al., 2023). However, research indicates that maturity scores primarily reflect the implementation level of processes and controls useful for demonstrating compliance, rather than providing a direct measure of realised risk reduction or business value (Koolen et al., 2024; Marican et al., 2023). Correlation studies suffer from numerous methodological limitations, including selection bias, confounding variables, and the absence of clear causal mechanisms (Diesch et al., 2020; Schmitz et al., 2021; Uchendu et al., 2021). Organisations with higher security maturity may simply be better managed overall, making security maturity a proxy for general management capability rather than a driver of specific security value (Diesch et al., 2020; Schmitz et al., 2021).

The value demonstration challenge extends beyond measurement technicalities to fundamental questions about security's organisational role. The traditional framing of security as a cost centre perpetuates zero-sum thinking about security investment (Cavusoglu et al., 2015; Dutta & McCrohan, 2002). Alternative framings positioning security as a business enabler offer more positive narratives but often lack concrete substantiation (Dhillon & Torkzadeh, 2006; Gordon et al., 2020). Claims that security enables digital transformation, supports innovation, or provides competitive advantage remain largely rhetorical rather than empirically demonstrated (AlGhamdi et al., 2020; Fielder et al., 2016).

Risk reduction represents the most commonly cited security value proposition, yet quantifying risk reduction proves equally challenging (Kraude et al., 2022; Krutilla et al., 2021). The dynamic nature of threats, the complexity of modern IT environments, and the human factors in security incidents make risk quantification highly uncertain (Bentley et al., 2020; Qazi et al., 2018). As systematic reviews of human factors demonstrate, the human element introduces a level of unpredictability that undermines many quantitative models (Nifakos et al., 2021; Pollini et al., 2022). Security leaders must navigate this measurement impossibility whilst maintaining organisational support and resources, a leadership challenge the technical literature largely ignores (Collier et al., 2023; Papathanasiou & Adey, 2020; Rios et al., 2020).

### 2.5.4 Strategic Versus Operational: False Dichotomy?

The security management literature consistently distinguishes between strategic and operational security activities, yet this separation may represent an artificial construct that obscures rather than illuminates security leadership realities. The literature's treatment of strategy and operations as distinct domains (Anderson & Choobineh, 2008; Purser, 2004; von Solms, 2005; White, 2009; White, 2024) fails to capture the dynamic interplay between these dimensions in security practice.

Strategic security activities, as characterised in the literature, encompass policy development, risk assessment, governance design, and executive engagement (Da Silva & Jensen, 2022; Horne et al., 2017; Loonam et al., 2022; Ramezan, 2025; Sahin & Vance, 2025). These activities supposedly require different skills, operate on longer time horizons, and engage senior stakeholders (AlGhamdi et al., 2020; Donalds & Barclay, 2022; Ganin et al., 2020). The literature positions strategic activities as the proper domain of senior security leadership, particularly the CISO role (Da Silva & Jensen, 2022; Ramezan, 2025; Steinbart et al., 2018). This characterisation implies that strategic work represents higher-value contribution and justifies executive positioning.

Operational security activities, by contrast, include incident response, vulnerability management, security monitoring, and system patching (Bhatt et al., 2014; Demertzis et al., 2019; Vielberth et al., 2020). The literature often characterises these activities as tactical and technical, though the characterisation as 'routine' is increasingly contested (Chowdhury & Gkioulos, 2021b; Haqaf & Koyuncu, 2018). Operational work supposedly prioritises technical skills over leadership capabilities and focuses on immediate rather than long-term concerns. This framing positions operational activities as necessary but insufficient for security leadership, work that is often delegated to enable leaders to balance strategic and operational responsibilities.

However, synthesis of empirical research reveals a different reality where strategic and operational dimensions interweave inextricably (Ahmad et al., 2012; Ahmad et al., 2019; Baskerville et al., 2014; Efthymiopoulos, 2019; Vielberth et al., 2020). Incident response, whilst operationally focused, provides critical intelligence that informs strategic planning (Ahmad et al., 2012; Ahmad et al., 2021; Ahmad et al., 2015; Ahmad et al., 2019; Naseer et al., 2021). Major incidents often trigger strategic reviews, policy changes, and investment decisions that shape long-term security posture (Ahmad et al., 2020; Baskerville et al., 2014). Conversely, strategic decisions manifest through operational implementation, with strategy quality revealed through operational effectiveness (Friesl et al., 2021; Tawse & Tabesh, 2021; Tu et al., 2018; Verhagen et al., 2023).

The artificial separation between strategy and operations creates problematic implications for security leadership development and evaluation. Leaders who lack operational credibility struggle to gain and maintain technical team respect and make contextually informed strategic decisions (Ashenden & Sasse, 2013; Choi, 2016; Dhillon & Torkzadeh, 2006; Ramezan, 2025). Conversely, leaders consumed by operational demands risk becoming reactive, missing opportunities for organisational learning and failing to align security initiatives with business objectives (Ahmad et al., 2020; Loonam et al., 2022; Naseer et al., 2024; Soomro et al., 2016). Whilst the literature's dichotomous treatment provides conceptual clarity, it offers limited practical guidance for navigating this tension, with empirical accounts consistently describing the need for 'bilingual' leaders who integrate both domains (Loonam et al., 2022; Ramezan, 2025).

Furthermore, the pace of change in cybersecurity challenges traditional strategic planning horizons. Threat landscapes evolve rapidly, new vulnerabilities emerge continuously, and technology shifts alter risk profiles fundamentally (Aftabi et al., 2025; AlGhamdi et al., 2020; Soomro et al., 2016). In this context, the distinction between strategic planning and operational adaptation becomes increasingly meaningless. Security leaders must simultaneously maintain long-term vision whilst remaining responsive to immediate threats, a capability the literature's strategic-operational dichotomy fails to address.

Critical analysis suggests that the strategic-operational separation may serve rhetorical rather than analytical purposes, supporting arguments for CISO elevation without addressing practical leadership challenges (Ahmad et al., 2019; Baskerville et al., 2014; Soomro et al., 2016). The dichotomy allows advocates to position CISOs as strategic executives whilst acknowledging operational realities that consume most security leaders' time (Sallos et al., 2019). However, this rhetorical device provides limited value for understanding or developing security leadership capabilities. Whilst Solms (von Solms) argues for maintaining some separation for governance and compliance purposes, the weight of evidence suggests that effective security leadership requires integration rather than separation of these domains.

**Gap Identified**: Security management literature provides extensive guidance on processes, technologies, and governance structures but systematically ignores the leadership dimension. The literature describes what should be done but not who does it or how. This process-centric focus leaves fundamental questions about security leadership roles, their strategic positioning, and effectiveness completely unaddressed. The absence of empirical research examining how security leaders navigate governance frameworks, organisational structures, value demonstration challenges, and strategic-operational tensions represents a critical knowledge gap requiring investigation.

## 2.6 ORGANISATIONAL CONTEXT AND SECURITY

The examination of security leadership cannot be separated from the organisational contexts within which it operates. Whilst the cybersecurity literature universally acknowledges the importance of context (Soomro et al., 2016; Uchendu et al., 2021), systematic investigation of how organisational characteristics shape security leadership requirements remains remarkably absent.

### 2.6.1 Industry Patterns: Same Threats, Different Responses

Cross-industry synthesis reveals striking variations in security approach despite facing fundamentally similar threat landscapes (Mishra et al., 2022; Riggs et al., 2023). Whilst sophisticated threat actors demonstrate clear industry preferences aligned with their motivations (financial, strategic, or ideological), organisations across sectors face similar technical vulnerabilities that can be exploited opportunistically (Ahmad et al., 2019; Holt et al., 2020; Smaga, 2025; Walker-Roberts et al., 2020). Technical vulnerabilities and broad threat types, including ransomware, Distributed Denial of Service (DDoS), and social engineering, remain remarkably consistent across critical sectors, often exploiting weak authentication and poor patching protocols (Riggs et al., 2023; Xenofontos et al., 2022). This targeted nature of sophisticated attacks, combined with the ubiquity of opportunistic threats, creates a complex threat landscape that industries navigate through markedly different approaches.

Financial services organisations demonstrate regulation-driven maturity, with security functions often representing the most resource-rich departments within the enterprise (Ducas & Wilner, 2017; Yang & Li, 2018). These organisations typically maintain dedicated security operations centres, threat intelligence capabilities, and specialised incident response teams (Ahmad et al., 2021; Khayat et al., 2025; Vielberth et al., 2020). Regulatory baselines, such as the Australian Prudential Regulation Authority (APRA)'s Cross-industry Prudential Standard (CPS) 234, serve to harmonise expectations and trigger standardised minimum investment levels in governance and technical controls (McIntosh et al., 2024; Mishra et al., 2022). Financial institutions are repeatedly highlighted as early adopters of cyber-resilience concepts, pushed by supervisors to move beyond simple prevention toward withstanding and rapidly recovering from attacks (Darem et al., 2023; Dupont, 2019). This reflects the sector's fundamental dependence on trust and the maturity of its risk management frameworks, with external requirements often acting as the primary motivator for formalising security spending (Culot et al., 2021; Lee, 2021; Wang et al., 2024).

Healthcare organisations present a contrasting profile, characterised by privacy-focused security programmes that often struggle with technological lag (Paul et al., 2023; Sahi et al., 2018; Tazi et al., 2024). Healthcare security functions operate with demonstrably fewer resources and historically lax cybersecurity compared with the financial sector, despite handling comparably sensitive data (Almalawi et al., 2023; Coventry & Branley, 2018; Kruse et al., 2017; Wasserman & Wasserman, 2022). The emphasis on patient privacy under frameworks such as Health Insurance Portability and Accountability Act (HIPAA) has created security programmes oriented toward compliance documentation and confidentiality rather than robust operational security capabilities (Abouelmehdi et al., 2018; Choi et al., 2006; Kwon & Johnson, 2013; Mbonihankuye et al., 2019; Paul et al., 2023). This privacy-centric approach has created documented operational vulnerabilities, particularly regarding legacy IT and connected medical devices which often run outdated software and lack strong authentication (Newaz et al., 2021; Wasserman & Wasserman, 2022). The consequences are severe; ransomware attacks in healthcare have direct operational and clinical consequences, including service disruptions and risks to patient outcomes (Argaw et al., 2020; He et al., 2021).

Manufacturing organisations face unique challenges arising from operational technology (OT) and information technology (IT) convergence (Felser et al., 2019; Kok et al., 2024; Mantravadi et al., 2020). The integration of industrial control systems (ICS) with corporate networks has created attack surfaces that traditional IT security professionals struggle to address, as ICS were originally designed for availability rather than security (Bhamare et al., 2020; Dhirani et al., 2021). Security leadership in manufacturing must navigate not only data protection requirements but also safety implications of cyber attacks on physical processes, where compromise can cause physical damage or environmental hazards (Pop et al., 2021; Pospisil et al., 2021; Riggs et al., 2023). Manufacturing organisations increasingly recognise the inadequacy of pure IT frameworks, adopting hybrid approaches that combine IT governance with OT-specific standards like International Society of Automation (ISA)/IEC 62443 to capture real-time constraints and legacy controller requirements (Mantravadi et al., 2020; Mullet et al., 2021; Toussaint et al., 2024).

Technology companies confront complexity through product security requirements that extend beyond organisational boundaries (Aidoo et al., 2022; Iqbal et al., 2020). These organisations must consider not only their own security posture but also the security implications of their products in customer environments, where perceived security significantly shapes purchasing decisions and user trust (Handoyo, 2024; Jafri et al., 2024; Valdez-Juárez et al., 2021). The dual mandate of securing internal operations whilst ensuring product security creates role complexity not addressed in traditional security leadership frameworks (Akbar et al., 2022; Spiekermann et al., 2019). Technology CISOs operate as supply-chain security orchestrators, expected to embed secure-by-design practices and manage risks that can propagate across partner networks (Hammi et al., 2023; Loonam et al., 2022).

Critical analysis reveals a fundamental paradox: whilst threat actors demonstrate industry-specific targeting strategies (Kaloudi & Li, 2020; Sailio et al., 2020), organisational responses vary dramatically by sector. This variation appears driven more by regulatory requirements and historical precedent than by empirical assessment of threat landscapes or security effectiveness (Bouraffa & Hui, 2025; Mishra et al., 2022). The literature provides extensive description of these industry differences but fails to examine whether sector-specific approaches actually improve security outcomes (Ali Milaat & Lubell, 2024; Burke et al., 2024; Chowdhury & Gkioulos, 2021a). Moreover, the assumption that industry context determines optimal security leadership approaches remains untested. The literature describes surface-level differences without investigating underlying capability requirements or examining the transferability of security leadership competencies across industries (Hooper & McKissack, 2016; Zwilling, 2022).

### 2.6.2 Size and Scale: The Security Paradox

Synthesis of size-related findings reveals a fundamental security paradox that challenges conventional wisdom about organisational scale advantages (Heidt et al., 2019; Mijnhardt et al., 2016). Large organisations possess greater resources for security investment, typically maintaining substantial security teams and larger absolute budgets (Fedele & Roner, 2022; Shaikh & Siponen, 2024). These resources enable sophisticated security operations centres, dedicated threat intelligence capabilities, and specialised security functions (Schlette et al., 2021; Vielberth et al., 2020). However, the complexity inherent in large organisations often overwhelms these resource advantages. Endpoint and organisational complexity are identified as dominant drivers of cyber-attack risk, frequently outweighing raw resource availability (Jalali & Kaiser, 2018). This “complexity tax” demonstrates that substantial portions of security budgets are consumed simply managing the friction of scale, internal politics, and integration challenges rather than advancing defensive posture (Corradi et al., 2022; Offner et al., 2020; Patera et al., 2021).

Medium-sized organisations navigate distinctive challenges that defy simple categorisation. These organisations face enterprise-grade threats without enterprise-scale resources, creating a documented "security gap" where budgets are perpetually constrained relative to threat exposure (Gozman & Currie, 2014; Mansfield-Devine, 2017). However, reduced complexity and shorter decision-making chains can enable more agile security responses; such organisations are often able to adapt quickly through iterative processes and tailored security practices rather than rigid, monolithic frameworks (Chan et al., 2018; Mihelic et al., 2024; Valdés-Rodríguez et al., 2024).

Small organisations face the starkest challenges, with security often representing an additional responsibility for IT generalists rather than a dedicated function (Chidukwani et al., 2022; Tam et al., 2021). Resource constraints force these organisations to rely heavily on outsourced security services and cloud-based security solutions to access built-in protections they cannot afford in-house (Compastié et al., 2023; Saeed et al., 2023). Whilst agility and simplified infrastructure provide theoretical advantages, the lack of dedicated expertise and structured controls leaves small organisations highly vulnerable to basic attacks (Armenia et al., 2021; Chidukwani et al., 2024). Small businesses experience significant gaps in security awareness and social engineering defences, with ransomware attacks causing disproportionate disruption and cash-flow stress compared to larger counterparts (Chaudhary et al., 2023; Fernandez De Arroyabe & Fernandez de Arroyabe, 2023; Romanosky, 2016).

Critical examination reveals that the relationship between organisational size and security leadership effectiveness remains poorly understood. The literature assumes that resource availability translates to security capability, yet evidence suggests that architectural complexity may negate resource advantages by increasing the attack surface faster than controls can mature (Jalali & Kaiser, 2018; Saeed et al., 2023). Large organisations experience breaches despite exponentially greater security investments; however, the typical incident cost is often modest relative to total revenue, which can inadvertently encourage "good enough" security rather than optimised risk-based investment (Fedele & Roner, 2022; Romanosky, 2016). This paradox suggests that scale effects in security differ fundamentally from other organisational functions.

The impact of organisational size on security leadership requirements receives even less attention. Whilst some literature suggests that security leaders in smaller organisations require broader, more hybrid skill sets, there is almost no direct empirical evidence on how leadership approaches vary with size (Guhr et al., 2019; Müller et al., 2024). Qualitative research highlights that CISOs in large organisations struggle primarily with role ambiguity, limited power, and organisational blockages (Ashenden & Sasse, 2013). The assumption that security leadership requirements scale linearly with organisational size remains untested and likely oversimplified.

### 2.6.3 Cultural and Geographic Blind Spots

Critical analysis of the literature reveals profound cultural and geographic biases that limit understanding of security leadership in global contexts. The vast majority of published security leadership studies originate from Western countries, particularly the United States, United Kingdom, and Australia (Maschmeyer et al., 2021; Sutton & Tompson, 2025; Uchendu et al., 2021), with minimal representation from Asia, Africa, Latin America, or even continental Europe. This geographic concentration creates systematic blind spots in understanding how cultural factors influence security leadership effectiveness.

The dominance of Anglo-American perspectives imposes Western governance assumptions on security leadership discussions (Zhang et al., 2023). Concepts such as independent oversight, transparent reporting, and challenge-oriented governance reflect cultural values that may not translate to high power distance cultures (Gupta et al., 2022; Liu et al., 2016). Security frameworks developed in low power distance environments assume that subordinates will challenge authority and report problems upward, yet these assumptions fail in hierarchical cultures where deference to authority prevails (Moon et al., 2018).

Language barriers further compound these limitations, with the vast majority of security leadership research published exclusively in English (Brummer et al., 2020). Non-English literature that might provide alternative perspectives on security leadership remains largely inaccessible to English-speaking researchers and practitioners (Pellegrini et al., 2020). This linguistic limitation not only restricts access but also creates translation challenges where core security and governance concepts lose critical cultural context when crossing linguistic boundaries. The dominance of Anglo-American perspectives is thus perpetuated, preventing cross-pollination of ideas from different cultural contexts.

Risk tolerance variations across national cultures receive minimal attention despite their obvious relevance to security leadership (Ameen et al., 2020; Rieger et al., 2015). Cultures with high uncertainty avoidance may demand different security leadership approaches than those comfortable with ambiguity (Hassandoust & Johnston, 2023). The balance between security controls and business enablement likely varies with cultural risk appetites, yet no studies examine these relationships empirically (Bansal & Axelton, 2025).

Regulatory environments, whilst acknowledged as important, are treated as technical requirements rather than cultural artefacts (Benamati et al., 2021; Meso et al., 2021). The emphasis on prescriptive compliance in some jurisdictions versus principle-based regulation in others reflects deeper cultural attitudes toward rule-following and professional judgement (Zhang et al., 2018). These regulatory philosophies fundamentally shape security leadership requirements in ways that remain unexplored, with evidence suggesting that cultural values have a stronger effect on regulatory attitudes than the regulatory structures themselves (Meso et al., 2021).

The implications of cultural factors for security leadership effectiveness remain almost entirely unexamined. Questions about whether Western-trained CISOs can effectively lead security functions in different cultural contexts, how security leaders navigate cultural differences in global organisations, and whether security leadership competencies are culturally universal or specific receive no empirical attention (Da Silva & Jensen, 2022; Ramezan, 2025). The literature acknowledges cultural diversity exists but treats it as demographic noise rather than a variable requiring systematic investigation.

### 2.6.4 Context Matters, But How?

Synthesis attempts across the contextual literature reveal a troubling pattern of superficiality that undermines claims about context's importance. Every study examined acknowledges that organisational context influences security requirements and challenges (Diesch et al., 2020; Heidt et al., 2019; Loonam et al., 2022), yet none systematically investigate how contextual factors actually shape security leadership. This universal acknowledgement, coupled with the absence of systematic investigation, represents one of the most significant gaps in security leadership literature. Context is routinely invoked as important but is rarely modelled or tested as an explanatory mechanism (AlGhamdi et al., 2020; Khando et al., 2021; Uchendu et al., 2021).

Industry context typically appears as a demographic variable in survey research; it is noted as a descriptor but not analysed as a predictor or moderator (da Veiga et al., 2020; Guhr et al., 2019; Onumo et al., 2021). Studies often report the percentage of respondents from different industries without examining whether industry membership predicts different security leadership requirements, challenges, or success factors (Saeed et al., 2023; Wang et al., 2019). The implicit assumption that industry matters remains largely speculative, with almost no robust comparative tests of sector-specific security leadership approaches (Awan et al., 2022; Wang et al., 2019).

Organisational size receives similar treatment, recorded as employee count or revenue but rarely theorised as a variable affecting security leadership (Heidt et al., 2019; Imran et al., 2021). Studies acknowledge size differences without exploring how scale fundamentally alters security leadership dynamics. The complexity paradox identified earlier suggests non-linear relationships between size and security leadership effectiveness, yet current research focuses on what security leaders do rather than how organisational size changes what they must be (Jalali & Kaiser, 2018; Müller et al., 2024). The contingency work that does exist focuses on transformation type or strategic positioning rather than size as an explanatory variable (Müller et al., 2024; Volberda et al., 2021).

Cultural factors suffer the most profound neglect. Whilst globalisation creates multi-national security challenges, the literature remains overwhelmingly Western-centric (Pellegrini et al., 2020). The absence of cross-cultural security leadership research creates fundamental gaps in understanding how security leaders navigate diverse cultural contexts within global organisations. Systematic reviews explicitly call for more attention to contextual variation, such as national culture and external factors, acknowledging that existing work treats context as a broad, decontextualised construct (AlGhamdi et al., 2020; Uchendu et al., 2021).

This pattern of superficial acknowledgement without systematic investigation suggests that contextual influences on security leadership remain one of the field's most significant unexplored territories. The universal recognition that "context matters", coupled with the complete absence of empirical investigation into how it matters, represents a critical knowledge gap. Without understanding how industry, size, culture, and other contextual factors shape security leadership requirements and effectiveness, prescriptive frameworks remain untethered from organisational realities. Security leadership requirements are context-contingent, yet the literature offers almost no direct evidence on how they vary across these dimensions (Imran et al., 2021; Müller et al., 2024).

## 2.7 THE CISO ROLE: FUNDAMENTAL QUESTIONS FROM LITERATURE

The extensive practitioner discourse surrounding the Chief Information Security Officer role stands in stark contrast to the empirical evidence supporting its fundamental claims. This section synthesises the literature's assertions, contradictions, and gaps to identify the critical questions that remain unanswered about this increasingly prominent executive position. Through thematic analysis of existing sources, we expose the chasm between what is claimed and what is known, revealing a role characterised more by assumption than evidence.

### 2.7.1 Thematic Analysis: What Literature Claims but Cannot Prove

#### *2.7.1.1 Theme 1: Role Creation and Requirements*

The literature presents confident assertions about why organisations establish CISO positions, yet these claims rest on anecdotal foundations rather than systematic investigation. Practitioners consistently claim that breach-driven appointments dominate role creation (Karanja, 2017; Karanja & Rosso, 2017), suggesting that security incidents catalyse executive-level security leadership establishment. However, no empirical studies trace actual appointment patterns or validate this reactive narrative. Similarly, the assertion that compliance requirements drive role creation appears throughout academic discussions (da Veiga et al., 2020; Johnson & Goetz, 2007; Shahim, 2021; Uchendu et al., 2021), yet lacks substantiation through organisational data or longitudinal analysis.

The contradictions across sources further complicate understanding. Whilst some advocate for proactive role creation as strategic vision (Ferreira et al., 2024; Karanja, 2017), others frame the position as a compliance checkbox (Sahin & Vance, 2025). These competing narratives reveal fundamental uncertainty about organisational motivations. The critical question remains unanswered: why do organisations really create CISO roles? Without empirical investigation of actual appointment decisions, role creation drivers remain speculative rather than understood.

#### *2.7.1.2 Theme 2: Value and Performance*

Perhaps no aspect of the CISO role generates more discussion yet less evidence than questions of value and performance measurement. The literature proposes numerous measurement approaches, each failing to capture the essence of security leadership effectiveness. Incident metrics cannot account for prevented breaches (Ahmad et al., 2021; Ahmad et al., 2015), compliance scores measure activity rather than outcomes (AlGhamdi et al., 2020; Herath et al., 2023), and maturity assessments document processes without demonstrating performance impact (Becker et al., 2009; Santos-Neto & Costa, 2019). These measurement limitations create a fundamental attribution problem for security leadership.

Value claims proliferate without validation throughout the literature. Assertions that CISOs reduce breach probability (Haislip et al., 2021; Smith et al., 2021) lack counterfactual analysis, whilst statements about security leadership improving resilience (Ahmad et al., 2020; Naseer et al., 2021) assume causation from correlation. The absence of control groups or comparative studies means these claims remain untested hypotheses rather than demonstrated relationships. Academic studies attempting to quantify security returns (Gordon & Loeb, 2006;

Hausken, 2006) cannot isolate leadership contribution from technological or process improvements. The critical question of how CISO value can be demonstrated remains not merely unanswered but potentially unanswerable given current methodological approaches.

#### *2.7.1.3 Theme 3: The Work Itself*

Despite extensive discussion of CISO responsibilities, the literature provides remarkably little empirical evidence about actual role activities. Time allocation debates illustrate this gap clearly. Strategic aspirations suggest CISOs should spend 20% to 40% of time on strategic activities (Hooper & McKissack, 2016), whilst self-reported data suggests significant operational involvement, though such reports often fail to align with actual logged behaviour (Parry et al., 2021). However, no observational studies validate these self-reports or examine how time allocation varies across contexts. The disconnect between aspiration and reported reality suggests either widespread role failure or fundamental misunderstanding of role requirements.

Activity lists proliferate across the literature, with academic reviews commonly enumerating fifty or more CISO responsibilities (Ramezan, 2025; Zwilling, 2022). Yet these lists lack empirical ranking or weighting, presenting all activities as equally important without evidence of actual priority or time investment. Academic frameworks attempt to categorise responsibilities (Karanja, 2017; Zwilling, 2022), but do so prescriptively rather than descriptively. The fundamental question of what CISOs actually do all day remains surprisingly opaque, with the literature substituting prescription for description.

### 2.7.2 Constraints and Politics: The Unnamed Challenges

Resource constraints represent the most universally claimed yet least systematically studied aspect of the CISO role. Every industry survey reports budget limitations as a primary challenge (Jalali & Kaiser, 2018; Shaikh & Siponen, 2024), whilst talent shortage narratives dominate professional discourse (Heidt et al., 2019; Mijnhardt et al., 2016). However, these constraints remain quantitatively undefined and qualitatively unexplored. No research examines how resource limitations shape role priorities, force trade-offs, or influence effectiveness. The assumption that more resources equal better security lacks empirical validation, whilst the strategies successful CISOs employ to overcome constraints remain undocumented.

Political dimensions of the CISO role receive acknowledgement without investigation. The literature references organisational politics, stakeholder management, and influence dynamics (Ashenden & Sasse, 2013; Da Silva & Jensen, 2022), yet provides no systematic analysis of these factors. How CISOs build coalitions, navigate resistance, or translate technical risks into business language remains unexplored. The critical skill of managing up appears frequently in practitioner advice; however, there is no empirical examination of what this actually entails or how it affects outcomes (Da Silva & Jensen, 2022; Sahin & Vance, 2025). The fundamental political nature of security leadership, operating at the intersection of technical expertise and organisational power, remains theoretically undeveloped.

### 2.7.3 Structure and Reporting: Debates Without Data

Reporting relationship debates dominate structural discussions of the CISO role, yet these arguments proceed without empirical foundation. Advocates for CEO reporting emphasise strategic elevation and enterprise-wide perspective (Peppard, 2010; Schobel & Denford, 2013), whilst those favouring CIO reporting cite operational efficiency and technical alignment (Karanja, 2017; Peppard, 2007). Emerging arguments for Chief Risk Officer reporting stress risk management integration (Kristensen & Andersen, 2023), whilst aspirational proposals for direct board reporting emphasise independence (De Haes et al., 2013). These positions reflect ideological preferences rather than evidence-based conclusions.

The evidence base for structural recommendations consists entirely of opinion and advocacy. No studies systematically compare effectiveness across different reporting structures, examine how structure affects role execution, or validate claimed benefits of particular arrangements. Empirical research documenting reporting relationships describe current state without evaluating effectiveness (Karanja, 2017; Uchendu et al., 2021). The fundamental question of whether structure actually affects security leadership effectiveness remains completely unexamined.

Beyond reporting relationships, broader structural questions receive even less attention. Matrix organisations, dotted-line relationships, and federal models appear in practice but not in research (Mattord et al., 2023; von Solms, 2005). The literature assumes traditional hierarchical structures whilst practitioners navigate increasingly complex organisational forms. How CISOs operate within different structural contexts or adapt their approach to organisational design remains unexplored territory.

### 2.7.4 Evolution Claims: Change Without Documentation

The transformation narrative pervades CISO literature, consistently claiming fundamental role evolution without providing supporting evidence. The progression from technical to strategic appears throughout academic and practitioner sources (Hooper & McKissack, 2016; Ramezan, 2025; Zwilling, 2022), suggesting a maturation from operational focus to business alignment. Similarly, evolution from reactive to proactive (Ahmad et al., 2020; Naseer et al., 2021) implies a shift from incident response to risk prevention. The transformation from cost centre to business enabler (Badakhshan et al., 2022; Kraus et al., 2022; von Solms, 2005) advocates for value creation rather than loss prevention. These narratives present evolution as both inevitable and beneficial.

However, no longitudinal evidence supports these transformation claims. The absence of studies tracking role evolution over time means these narratives remain unvalidated assertions (Ferreira et al., 2024; Sahin & Vance, 2025). No documentation exists of actual transition patterns, whether at individual or organisational levels. The literature cannot demonstrate whether roles actually evolve, what triggers transformation, or whether evolution improves effectiveness. Without baseline measurements or longitudinal data, claims of role transformation remain rhetorical rather than empirical.

The evolution narrative also assumes unidirectional progress, ignoring potential regression or cyclical patterns. Crisis events might force strategic CISOs back to operational focus, whilst organisational changes could reverse achieved progress (Kristensen & Andersen, 2023). The literature's teleological assumption of inevitable advancement prevents examination of complex, non-linear development patterns. The critical question remains: has the CISO role actually changed, or do we simply talk about it differently?

### 2.7.5 Questions That Demand Answers: Implications for Critical Inquiry

The synthesis of thematic patterns reveals fundamental questions about the CISO role that remain empirically unanswered despite extensive practitioner discussion:

**Role Purpose and Creation**: Why do organisations create CISO positions? What triggers appointment decisions? How do organisations determine need for security leadership? What differentiates organisations with CISOs from those without?

**Value and Effectiveness**: How can CISO value be demonstrated? What constitutes effectiveness in security leadership? How do prevention and enablement create measurable value? What performance indicators actually matter for security leadership effectiveness?

**Activities and Time Allocation**: What do CISOs actually do? How do they allocate time across competing demands? Which activities contribute most to effectiveness? How do successful CISOs prioritise responsibilities?

**Political and Organisational Dynamics**: How do CISOs navigate organisational politics? What influence mechanisms prove effective? How do they build coalitions and overcome resistance? What role does political capital play in security leadership?

**Structure and Context**: Does reporting structure affect effectiveness? How do different organisational contexts shape role requirements? What structural arrangements enable success? How do matrix and federal structures impact security leadership?

**Evolution and Development**: Do CISO roles actually evolve? What triggers role transformation? How do individuals and organisations navigate transitions? What development pathways lead to effectiveness?

These questions expose the empirical vacuum at the heart of security leadership discourse. The absence of systematic investigation means that organisations create, structure, and evaluate these critical roles based on assumption rather than evidence. This gap between practitioner discussion and empirical validation demands exploratory research capable of generating foundational understanding.

**Critical Gap**: The literature surrounding the CISO role exhibits a profound disconnect between assertion and evidence. Claims proliferate about role creation, value, activities, challenges, structure, and evolution, yet systematic empirical investigation remains virtually absent. This is not a gap in coverage but in evidence: much is said, little is known. The fundamental questions about this critical organisational role remain unanswered, creating an empirical vacuum that demands exploratory investigation. The absence of evidence-based understanding impedes both theoretical development and practical guidance, leaving organisations and practitioners to navigate complex security leadership challenges without empirical foundation.

Chapter 3 synthesises these patterns to establish the research requirements and question guiding the empirical investigation that follows.

# CHAPTER 3: SYNTHESIS AND RESEARCH DESIGN IMPLICATIONS

## 3.1 INTRODUCTION

The preceding literature review has exposed a paradoxical landscape: the Chief Information Security Officer role has achieved global prominence whilst remaining fundamentally under-researched. Practitioner discourse dominates with assertion replacing empirical validation, prescription substituting for description, and advocacy masquerading as evidence. This chapter synthesises the critical patterns emerging from this examination to establish the foundation for empirical investigation.

The synthesis reveals not merely isolated knowledge gaps but rather a systematic absence of empirical grounding for an increasingly critical organisational role. What we know about the CISO role derives primarily from methodologically limited surveys, anecdotal practitioner accounts, and advocacy-driven frameworks rather than rigorous scholarly investigation. What we assume far exceeds what we have demonstrated through systematic research. This disconnect between the role's organisational prominence and its empirical foundations demands explicit acknowledgement and methodological response.

The analysis proceeds from assessment of the current state of knowledge through mapping of literature patterns to research requirements, culminating in articulation of the research question that emerges from this synthesis. By explicitly connecting literature patterns to research design requirements, the chapter justifies the exploratory approach detailed in Chapter 4 whilst maintaining the theoretical openness essential for grounded theory investigation.

## 3.2 THE STATE OF KNOWLEDGE: MUCH ASSUMED, LITTLE KNOWN

The literature establishes several uncontested observations about the contemporary security leadership landscape. CISO roles are proliferating globally, with appointment rates accelerating across industries and geographies (Hooper & McKissack, 2016; Karanja, 2017; Maynard et al., 2018). Security challenges are intensifying in both frequency and sophistication, creating unprecedented organisational vulnerabilities (Dritsas & Trigka, 2025; Falowo et al., 2024; Ferdous et al., 2023; Kazimierczak et al., 2024). Practitioners consistently perceive the role as important, evidenced through survey responses, professional body advocacy, and compensation trends (Choi, 2016; Guhr et al., 2019; Loonam et al., 2022). Challenges and constraints facing security leaders are widely reported, though predominantly through anecdotal accounts rather than systematic investigation (Ashenden & Sasse, 2013; Balozian et al., 2023; Da Silva & Jensen, 2022).

These surface-level observations, however, mask profound uncertainties about fundamental aspects of the role. The proliferation of CISO appointments proceeds without empirical understanding of what drives organisational decisions to create these positions. The perceived importance lacks validation through measurable outcomes or demonstrated value creation. The reported challenges remain unexamined through systematic research that could identify patterns, causes, or mitigation strategies.

### 3.2.1 What We Assume but Do Not Know

The literature reveals extensive assumptions operating without empirical validation. How the CISO role creates value remains unexplored beyond assertion and advocacy (Fielder et al., 2016; Gordon & Loeb, 2006; Gordon et al., 2020; Hausken, 2006). The mechanisms through which security leaders influence organisational outcomes lack theoretical grounding or empirical investigation. What makes CISOs effective constitutes perhaps the most critical unanswered question, with no validated competency models, success factors, or performance criteria established through research (Guhr et al., 2019; Haqaf & Koyuncu, 2018; Karanja, 2017; White, 2024).

Contextual influences receive universal acknowledgement but no systematic examination. The few academic studies that mention CISOs typically treat the role as context rather than phenomenon, assumption rather than variable. Information systems research has extensively examined other C-suite roles, particularly the Chief Information Officer, yet has largely ignored the security leadership dimension (Gerth & Peppard, 2016; Jones et al., 2020; Karahanna & Preston, 2013; Peppard, 2010; Peppard et al., 2011). Management research on executive roles and upper echelons theory provides relevant frameworks but has not been applied to security leadership contexts.

This combination results in prescription without description, advocacy without evidence, and assumption without validation. Fundamental questions about role creation, activities, value, challenges, and evolution remain not merely unanswered but largely unasked through systematic inquiry.

## 3.3 MAPPING LITERATURE PATTERNS TO RESEARCH NEEDS

Analysis of the literature reveals distinct patterns that indicate specific research requirements. These patterns do not represent simple gaps but rather systematic biases and blind spots that shape current understanding whilst limiting advancement of knowledge.

### 3.3.1 Pattern 1: Description Before Prescription Needed

The literature exhibits extensive prescription of best practices without empirical description of current practices (Culot et al., 2021; De Haes et al., 2013; Diamantopoulou et al., 2020; Herath et al., 2023). Frameworks proliferate for what CISOs should do without investigation of what they actually do. Maturity models define aspirational states without documenting current states. Competency frameworks specify required capabilities without validating which capabilities actually influence outcomes.

This pattern indicates a fundamental research need for empirical investigation of actual CISO activities, time allocation, decision-making processes, and daily work realities. Understanding what security leaders do precedes meaningful guidance about what they should do. Description of current practice provides the foundation for prescription of improved practice.

### 3.3.2 Pattern 2: Context Acknowledged but Not Examined

Every study acknowledges that organisational context influences security requirements and leadership approaches, yet none systematically examine how contextual factors operate (Mijnhardt et al., 2016; Mohammed et al., 2025; Uchendu et al., 2021). Industry differences receive mention without analysis. Organisational size appears as a demographic variable rather than a theoretical construct. Cultural factors gain acknowledgement before being ignored in analysis.

This pattern reveals the need for systematic exploration of contextual influences on security leadership. Research must move beyond noting context to examining how specific contextual factors shape role requirements, constrain leadership options, and influence effectiveness criteria. Understanding contextual variation enables development of contingency approaches rather than universal prescriptions.

### 3.3.3 Pattern 3: Evolution Claimed but Not Documented

Transformation narratives dominate practitioner discourse about the CISO role evolving from technical to strategic, reactive to proactive, operational to executive (Hooper & McKissack, 2016; Karanja, 2017; Maynard et al., 2018). These evolution claims lack longitudinal evidence documenting actual role transitions. No studies track how individual roles change over time, how organisational expectations shift, or how role holders adapt to changing requirements.

This pattern indicates the need for understanding how roles actually develop rather than how advocates claim they should develop. Longitudinal or retrospective research could examine role evolution patterns, identify transition triggers, and understand adaptation mechanisms. Documentation of actual evolution provides grounding for transformation guidance.

### 3.3.4 Pattern 4: Politics Mentioned but Not Studied

Organisational dynamics and political factors appear consistently in practitioner accounts but receive no systematic investigation (Ashenden & Sasse, 2013; Balozian et al., 2023; Da Silva & Jensen, 2022). The need to influence without formal authority becomes a recurring theme without examination of influence mechanisms. Executive buy-in gains universal endorsement without understanding of how such support develops. Organisational resistance receives frequent mention without analysis of resistance sources or mitigation approaches.

This pattern reveals the need for exploration of political and organisational factors shaping security leadership effectiveness. Research must examine how CISOs navigate organisational politics, build coalitions, manage stakeholder relationships, and overcome resistance. Understanding political dynamics moves beyond technical competence to organisational effectiveness.

### 3.3.5 Pattern 5: Value Assumed but Not Demonstrated

Return on investment discussions and value creation claims proceed without empirical foundation (Gordon & Loeb, 2006; Gordon et al., 2020; Hausken, 2006; Xu et al., 2017). The assumption that CISOs create value lacks validation through measurable outcomes or demonstrated impact. Prevention value remains particularly problematic, requiring proof of counterfactuals. Business enablement claims lack connection to business outcomes.

This pattern indicates the need for understanding how CISOs conceptualise and demonstrate value rather than assuming value creation. Research must explore value definition, measurement challenges, and communication strategies. Understanding value demonstration enables more effective positioning and resource justification.

Figure 3.1 maps each of the five literature patterns to its corresponding research requirement, consolidating the systematic relationship between what the literature omits and what empirical investigation must address.

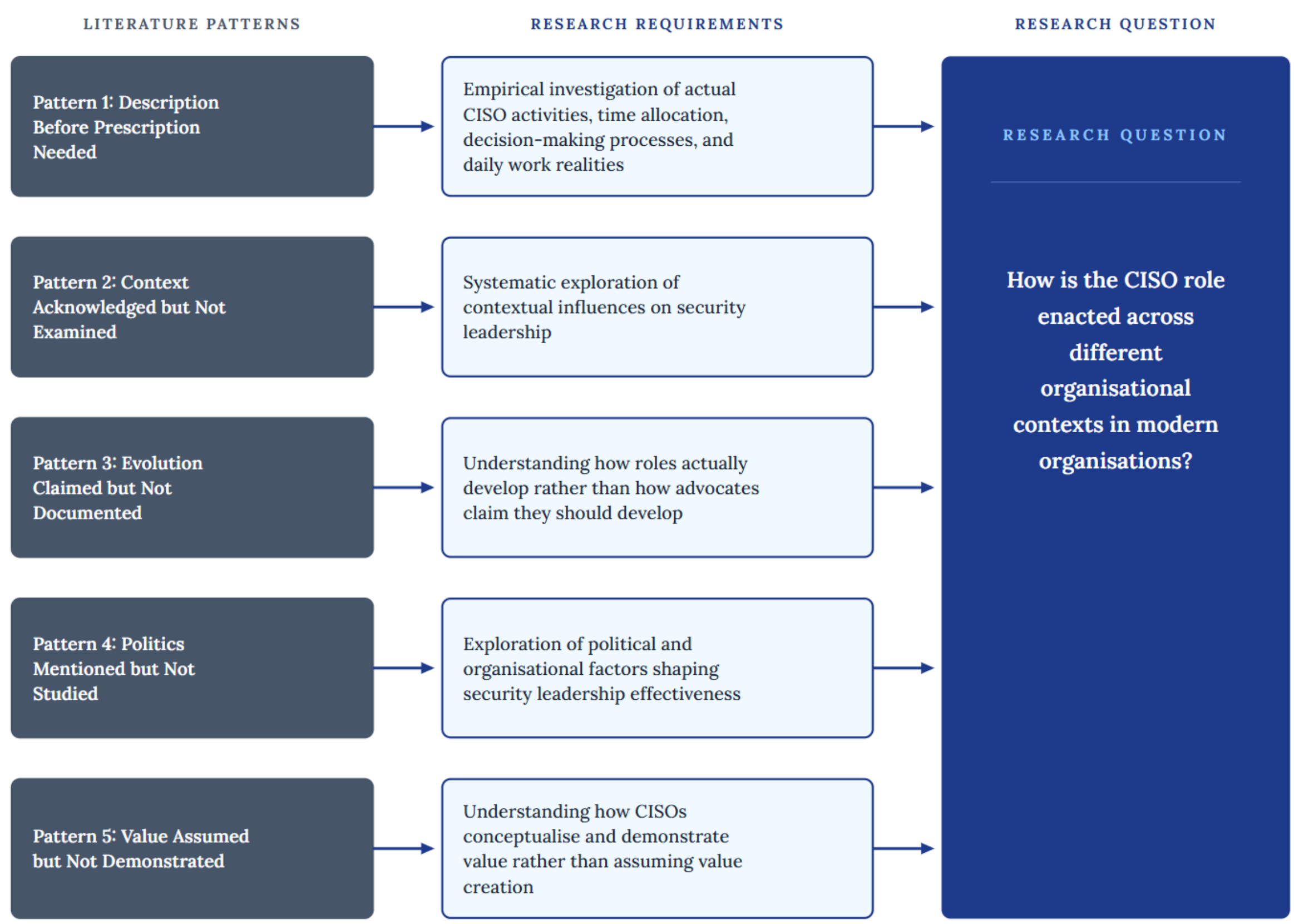

*Figure 3.1. Mapping of literature patterns to research requirements.*

## 3.4 RESEARCH QUESTION: EMBRACING THE UNKNOWN

The synthesis of literature patterns and knowledge gaps leads to a research question that embraces rather than constrains the empirical uncertainty surrounding the CISO role:

**How is the CISO role enacted across different organisational contexts in modern organisations?**

Three guiding themes emerged from the literature synthesis and oriented the empirical inquiry. The first asks how organisational context shapes the demands placed on security leadership, responding to the literature's consistent acknowledgement that context matters without systematic investigation of how or why. The second asks how security leaders establish their standing and credibility within organisations where the role remains poorly understood and contested, responding to the persistent appearance of political and stakeholder dynamics in practitioner accounts without empirical examination. The third asks whether security leadership requires deep technical expertise or business acumen, responding to a tension that pervades both practitioner discourse and academic literature without empirical resolution. These themes provided initial orientation for the investigation without constraining the discovery of unexpected patterns. Figure 3.2 presents the research question alongside its three guiding themes.

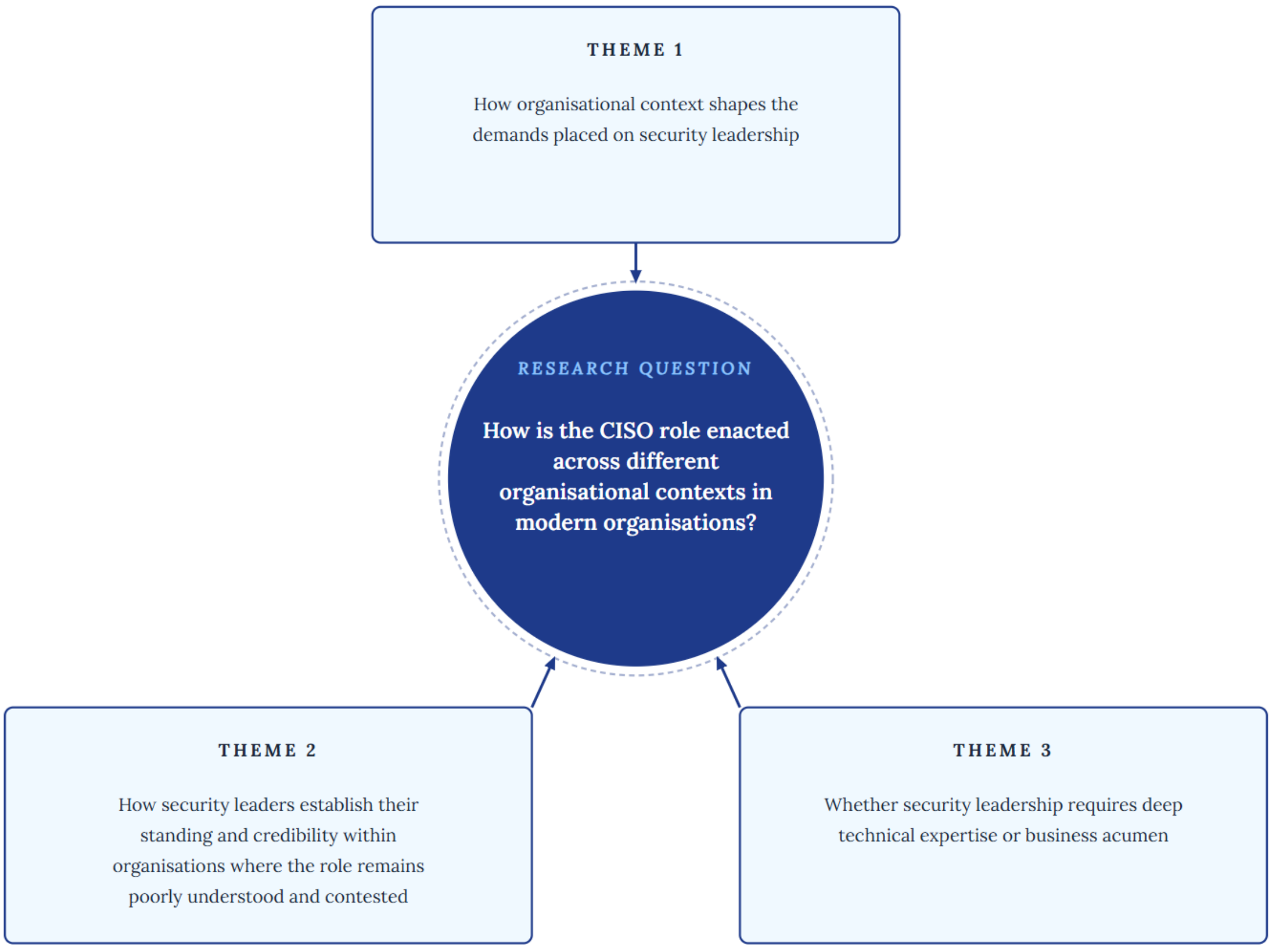

*Figure 3.2. Research question and guiding themes.*

### 3.4.1 What This Question Enables

The research question enables discovery of unexpected patterns that more focused investigation might overlook (Turner & Astin, 2021; Walker & Myrick, 2006). Security leadership may involve dimensions not anticipated by researchers unfamiliar with the role's daily realities. Relationships between factors may operate differently than in other executive contexts. New theoretical constructs may be necessary to explain observed phenomena.

Contextual variation can be captured rather than controlled, allowing understanding of how different organisational settings shape role requirements and effectiveness criteria (Fiedler, 1964). The question permits examination of multiple role dimensions simultaneously, recognising that isolation of variables may be neither possible nor desirable in complex organisational phenomena.

Practitioner perspectives can be understood authentically without forcing responses into predetermined categories. The emic perspective becomes paramount, allowing those who inhabit the role to define its nature, challenges, and requirements. This insider knowledge provides the foundation for theoretical development grounded in empirical reality rather than researcher speculation.

Theory development from an empirical foundation becomes possible through this open approach (Charmaz & Thornberg, 2021; Cutcliffe, 2000; Noble & Mitchell, 2016). Rather than testing existing theories that may not apply to this unique context, new theoretical frameworks can emerge from systematic analysis of empirical data. This grounded approach ensures that resulting theory reflects the phenomenon rather than forcing the phenomenon into existing theoretical moulds.

The research question thus positions the investigation to address the fundamental gap identified throughout the literature review: the absence of systematic empirical research on a critical and emerging executive role. By embracing the unknown rather than assuming knowledge, the research can provide the empirical foundation necessary for both theoretical advancement and practical guidance.

## 3.5 CRITICAL ANALYSIS OF LITERATURE QUALITY

The quality assessment of reviewed literature reveals systematic limitations that further justify the exploratory approach. Academic studies mentioning CISOs typically exhibit methodological weaknesses including small sample sizes, limited geographic scope, and superficial treatment of the security leadership dimension. Practitioner literature, whilst offering rich descriptive accounts, lacks the systematic rigour necessary for theory development. The absence of longitudinal studies prevents understanding of role evolution, whilst the dominance of survey methods limits depth of insight into complex organisational dynamics.

These quality limitations do not diminish the value of existing literature but rather highlight the nascent state of security leadership research. The literature serves its purpose in establishing the importance of the phenomenon and identifying areas requiring investigation. However, the systematic absence of rigorous empirical research on fundamental aspects of the CISO role necessitates grounded theory approaches that can build knowledge from empirical foundations rather than testing premature hypotheses.

## 3.6 CONCLUSION: TOWARDS EMPIRICAL INVESTIGATION

The patterns identified through this synthesis directly inform the methodological choices detailed in Chapter 4. The absence of empirical description necessitates qualitative approaches capable of capturing rich, contextual data about actual security leadership practice. The lack of theoretical frameworks requires inductive methodologies that can generate theory from empirical observation rather than testing existing models. The diversity of contexts and implementations demands research designs that embrace rather than control variation.

Grounded theory methodology emerges as the appropriate response to these requirements, offering systematic procedures for developing theoretical understanding from empirical data. The Gioia methodology specifically provides the rigour necessary for doctoral research whilst maintaining the flexibility required for exploratory investigation. This methodological approach, detailed in Chapter 4, enables the research to address the fundamental gaps identified through the preceding literature review.

By mapping specific knowledge gaps to research design requirements, this synthesis creates the bridge from literature review to empirical investigation. The research question, grounded in systematic analysis of literature patterns, provides the foundation for exploratory investigation whilst maintaining the theoretical openness essential for discovery of unexpected patterns and development of grounded theory. Security leadership research requires not merely incremental addition to existing frameworks but fundamental empirical investigation of a role that has proliferated in practice whilst remaining largely unexplored in scholarship.

# CHAPTER 4: METHODOLOGY

## 4.1 INTRODUCTION: PHILOSOPHICAL FOUNDATIONS AND METHODOLOGICAL CHOICES

### 4.1.1 Chapter Purpose and Structure

This chapter presents the methodological approach undertaken to investigate the role of the Chief Information Security Officer in contemporary organisations. The research design employs a rigorous qualitative methodology grounded in systematic empirical investigation, selected to address the exploratory nature of the research questions and the nascent state of scholarly understanding regarding security leadership (Flemming & Noyes, 2021; Gioia et al., 2013; Hennink & Kaiser, 2022). The chapter articulates the philosophical foundations underpinning the research, justifies the selection of grounded theory methodology, details the data collection and analytical procedures, and demonstrates the measures taken to ensure research quality and trustworthiness.

The qualitative approach adopted for this investigation reflects the complex, socially embedded nature of executive security leadership (Cassell et al., 2020; Da Silva & Jensen, 2022). Given the absence of established theoretical frameworks for understanding the CISO role, as identified in Chapters 2 and 3, an inductive methodology proves essential for theory development rather than hypothesis testing (Charmaz, 2014; Gioia et al., 2013; Locke, 2007). The research design facilitates deep exploration of participant experiences, enabling the emergence of patterns and relationships not readily apparent through deductive approaches. This methodological choice aligns with established practices in management research when investigating emerging organisational phenomena (Einola & Alvesson, 2021; Fisher & To, 2012; Pass, 2020).

The structure of this chapter progresses from philosophical foundations through to practical implementation. Following this introduction, Section 4.2 examines the research design, focusing on the grounded theory methodology and specific adoption of the Gioia method. Section 4.3 establishes the research context and sampling strategy, situating the study within the Australian organisational landscape. Section 4.4 details the data collection procedures, while Section 4.5 presents the systematic analytical approach employed. Section 4.6 addresses research quality and rigour, incorporating trustworthiness criteria and reflexivity considerations. Sections 4.7 and 4.8 examine ethical considerations and research limitations respectively, before Section 4.9 provides a summary that bridges to the empirical findings presented in Chapter 5 and the theoretical model developed in Chapter 6.

### 4.1.2 Research Questions Revisited

The primary research question guiding this investigation asks: "How is the CISO role enacted across different organisational contexts in modern organisations?" Three guiding themes emerged from the literature review and oriented the empirical inquiry. The first asks how organisational context shapes the demands placed on security leadership. The second asks how security leaders establish their standing and credibility within organisations where the role remains poorly understood and contested. The third asks whether security leadership requires deep technical expertise or business acumen, a tension that pervades both practitioner discourse and academic literature without empirical resolution. These themes provided initial focus for the investigation without presupposing particular relationships or outcomes. The primary question and guiding themes are addressed through the empirical findings presented in Chapter 5, culminating in the development of the Security Leadership Contingency Model (SLCM) detailed in Chapter 6.

The exploratory nature of this investigation reflects the current state of knowledge regarding security leadership, characterised by limited empirical research and an absence of validated theoretical frameworks (Corbin & Strauss, 2015; Glaser & Strauss, 1967). Whilst practitioner literature offers numerous prescriptive models for CISO success, these lack the theoretical grounding and empirical validation required for scholarly advancement (Gambrill, 2011; Mura & Wijesinghe, 2023; Shaw, 2005). The research question therefore demands a methodology that can bridge this theory-practice divide, generating insights that are both theoretically robust and practically relevant. This dual imperative shapes the methodological choices presented throughout this chapter.

The formulation of the research question acknowledges that security leadership operates within complex organisational systems, influenced by technological change, regulatory requirements, threat landscapes, and organisational dynamics (Georgiadou et al., 2021; Hannousse & Yahiouche, 2021). Understanding the CISO role therefore requires methodological approaches that can accommodate this complexity whilst maintaining analytical rigour without presuming particular effectiveness criteria. The question avoids presupposing particular relationships or outcomes, instead maintaining the openness necessary for genuine discovery. This stance aligns with the principles of qualitative inquiry, particularly grounded theory approaches that prioritise empirical observation over theoretical preconceptions (Chapman et al., 2015; Charmaz, 2014).

### 4.1.3 Philosophical Positioning

The philosophical foundations of this research rest upon a critical realist ontology, acknowledging both the objective reality of security threats and organisational structures, and the subjective interpretations through which these realities are understood and enacted (Brock, 2025; Fleetwood, 2005; Yucel, 2018). This position recognises that while cyber threats exist independently of human perception, the organisational responses to these threats, including the construction of the CISO role, emerge through social processes of interpretation and negotiation. Critical realism provides an appropriate foundation for investigating security leadership as it accommodates both the material aspects of technology and threats, and the social dimensions of organisational life (Crabtree, 2025; Sætra, 2023).

Epistemologically, the research adopts a moderate constructionist stance, recognising that knowledge about security leadership emerges through the interaction between researcher and participants (Denzin & Lincoln, 1994; Mir & Watson, 2000). This approach acknowledges that while objective organisational structures and security challenges exist, understanding of these phenomena is mediated through individual experiences and interpretations. The epistemological position influences both data collection and analysis, emphasising the importance of capturing multiple perspectives and remaining sensitive to contextual variations in how security leadership is enacted and understood (O'Neil & Koekemoer, 2016; Timonen et al., 2018).

The adoption of an abductive reasoning approach enables movement between empirical observation and theoretical abstraction, facilitating theory development grounded in empirical data while remaining open to theoretical insights (Folger & Stein, 2017; Timmermans & Tavory, 2012). Abduction proves particularly valuable when investigating emerging phenomena where existing theory provides insufficient explanation. This reasoning approach aligns with the grounded theory methodology employed, supporting the iterative movement between data collection, analysis, and theoretical development that characterises rigorous qualitative research (Kovács & Spens, 2005; Mirza et al., 2014).

## 4.2 RESEARCH DESIGN: GROUNDED THEORY METHODOLOGY

### 4.2.1 Justification for Grounded Theory Approach

The selection of grounded theory methodology represents a deliberate response to the empirical vacuum identified in Chapters 2 and 3, where extensive practitioner discourse about security leadership exists without corresponding scholarly investigation. This methodological choice aligns directly with the research objectives of theory development rather than theory testing, acknowledging that the absence of established theoretical frameworks for understanding the CISO role necessitates an approach that allows patterns to emerge from empirical data (Charmaz & Thornberg, 2021; Noble & Mitchell, 2016). The exploratory nature of the research question, "How is the CISO role enacted across different organisational contexts in modern organisations?", demands a methodology capable of capturing the complexity and nuance of an emerging executive position without imposing predetermined conceptual structures that might obscure unexpected findings (Paul et al., 2023; Rieger, 2019).

The alignment between grounded theory and the phenomenon under investigation proves particularly compelling given the need to capture practitioner perspectives authentically. Security leaders operate within complex organisational environments where technical expertise intersects with business strategy, regulatory compliance meets risk management, and operational necessity confronts resource constraints. These multifaceted dynamics require a methodological approach that can accommodate complexity whilst maintaining analytical rigour (Turner & Astin, 2021; Walker & Myrick, 2006). Grounded theory provides the systematic procedures necessary for moving from rich descriptive data to abstract theoretical understanding, enabling the development of substantive theory that remains grounded in the lived experiences of security leaders whilst offering explanatory power beyond individual cases (Birks et al., 2019; Cutcliffe, 2000).

The methodological fit between grounded theory and the security leadership phenomenon extends beyond mere alignment with research objectives. The emerging nature of the CISO role, lacking established constructs or validated frameworks, creates ideal conditions for grounded theory investigation. Unlike phenomena with extensive theoretical foundations where hypothesis testing might prove appropriate, security leadership represents what Sahin and Vance (2025) describe as a substantive area requiring fresh theoretical development. The complex, context-dependent processes through which security leaders navigate organisational dynamics, manage stakeholder relationships, and balance competing priorities demand processual understanding that grounded theory excels at capturing. This methodology enables the researcher to trace how security leaders make sense of their evolving responsibilities, adapt to organisational contexts, and develop strategies for their organisational roles.

The specific selection of constructivist grounded theory (Charmaz, 2014) reflects recognition of the researcher's interpretive role in knowledge construction. Unlike classical grounded theory's claims to objectivity, constructivist grounded theory acknowledges that understanding emerges through the interaction between researcher and participants, with meaning co-constructed rather than discovered. This philosophical stance proves particularly appropriate given the researcher's position as a practising CISO investigating the experiences of peer security leaders. The constructivist approach acknowledges that the researcher's professional experience constitutes a form of theoretical sensitivity, whilst systematic procedures ensure that this sensitivity sharpens analytical perception rather than predetermining theoretical conclusions (Paul & Criado, 2020; Simsek et al., 2023).

Two distinct forms of prior knowledge operated within this investigation, and their respective roles require clarification. The researcher's professional experience as a practising CISO constituted what Charmaz (2014) terms a sensitising resource: heuristic knowledge that sharpened pattern recognition, enabled sophisticated questioning, and facilitated participant candour through shared professional identity. This form of prior knowledge orients perception of a phenomenon without predetermining its theoretical explanation. It is categorically different from prior familiarity with the formal theoretical frameworks that ultimately inform the Security Leadership Contingency Model. No prior acquaintance with person-organisation fit theory, person-environment fit theory, contingency theory, or political capital scholarship existed at the outset of data collection. These frameworks entered the analysis because empirical patterns demanded theoretical explanation, not because they had been anticipated in advance.

This study operated within the abductive tradition of constructivist grounded theory (Gioia et al., 2013; Kelle, 2007; Timmermans & Tavory, 2012). In this tradition, iterative movement between empirical observation and scholarly literature constitutes rigorous theory building, not methodological compromise. Practitioner experience oriented the inquiry and sharpened analytical perception. The theoretical frameworks that ultimately inform the SLCM earned their explanatory role by demonstrating resonance with patterns the data revealed, consistent with Thornberg (2012) principle that prior knowledge sensitises rather than predetermines.

### 4.2.2 The Gioia Methodology

The adoption of the Gioia methodology (Gioia et al., 2013) within the broader grounded theory framework represents a strategic choice to enhance methodological rigour and transparency. Developed specifically to address criticisms of qualitative research in management studies, the Gioia methodology provides systematic procedures for demonstrating the progression from raw data to theoretical contributions. This approach has established credibility within top-tier management journals, offering a proven framework for qualitative rigour that satisfies the expectations of scholarly audiences whilst maintaining the flexibility necessary for exploratory investigation (Gehman et al., 2018; Magnani & Gioia, 2023).

The selection rationale for the Gioia methodology extends beyond mere credibility to encompass its specific affordances for investigating executive phenomena. The methodology's emphasis on preserving participant voice through first-order concepts ensures that the unique perspectives and language of security leaders remain visible throughout the analytical process. This preservation proves essential when investigating a role where practitioner understanding may diverge significantly from academic conceptualisation (da Veiga et al., 2020; Dhillon et al., 2021). The systematic progression from first-order concepts to second-order themes enables researcher interpretation whilst maintaining clear connections to empirical data, creating what Langlois et al. (2020) describe as a *semi-grounded* approach that balances emergence with structure.

The balance between emergence and structure inherent in the Gioia methodology addresses a fundamental tension in grounded theory research: the need for openness to unexpected patterns whilst maintaining systematic analytical procedures. The methodology's structured approach to data analysis, progressing through clearly defined stages from initial coding to theoretical abstraction, provides the analytical scaffolding necessary for managing complex qualitative data whilst allowing space for creative theoretical development (Gehman et al., 2018; Murphy et al., 2017). This structure proves particularly valuable when investigating multifaceted phenomena like security leadership, where multiple theoretical perspectives might illuminate different aspects of the empirical patterns. However, scholars have noted that this structured approach risks becoming overly template-driven, potentially constraining interpretive depth (Mees-Buss et al., 2020).

The key principles adopted from the Gioia methodology shape both analytical procedures and presentation of findings. First-order concepts preserve the authentic voice of participants, capturing how security leaders describe their experiences, challenges, and strategies in their own terms. These concepts emerge directly from interview data through systematic coding that maintains fidelity to participant language (Braun & Clarke, 2019). Second-order themes represent researcher interpretation, where patterns across first-order concepts are identified and abstracted to more theoretical language. This interpretive move requires the researcher to step back from descriptive detail to identify underlying patterns and relationships (Bouncken et al., 2021; Gehman et al., 2018). Aggregate dimensions enable theoretical abstraction, grouping related second-order themes into broader theoretical categories that form the building blocks of the emergent theoretical framework.

The data structure visualisation central to the Gioia methodology provides transparency in analytical progression, allowing readers to trace the path from empirical observation to theoretical contribution. This visual representation demonstrates the systematic nature of the analysis, showing how multiple first-order concepts support each second-order theme and how themes combine to form aggregate dimensions (Magnani & Gioia, 2023; Sankaran et al., 2024). Such transparency proves essential for establishing trustworthiness in qualitative research, particularly when investigating phenomena where scepticism about qualitative methods might exist. However, this visualisation approach has been critiqued for potentially becoming formulaic, with some scholars arguing it may privilege structural clarity over interpretive nuance (Mees-Buss et al., 2020).

Adaptations of the Gioia methodology for the security leadership context reflect the unique characteristics of the phenomenon under investigation. Extended attention to technical-business translation acknowledges the distinctive challenge facing security leaders who must communicate complex technical risks in business language, translating between disparate professional domains (da Veiga et al., 2020; Dhillon et al., 2021). This adaptation ensures the methodology captures not merely what security leaders do but how they bridge the persistent divide between technical and business perspectives that characterises contemporary organisations.

The incorporation of regulatory and threat dynamics into the analytical framework recognises that security leadership operates within a distinctive environmental context characterised by evolving threats and expanding regulatory requirements. Unlike other executive roles that might face relatively stable external environments, security leaders must continuously adapt to emerging threats, new attack vectors, and changing regulatory landscapes (Brass & Sowell, 2021; Radanliev, 2024). The methodology must therefore accommodate this dynamism, capturing how security leaders make sense of and respond to environmental turbulence.

Sensitivity to political and power dimensions reflects recognition that security leadership involves navigating complex organisational politics where competing priorities, resource constraints, and stakeholder interests create challenging dynamics. Research demonstrates that security leaders frequently rely on political acumen and influence rather than formal authority, with expertise, relational leadership, and soft power often proving effective in achieving security objectives (Dang-Pham et al., 2022; Kammersgaard, 2021; Moon et al., 2018). However, the application of these influence-based strategies is context-dependent, varying with organisational culture, hierarchical structures, and crisis situations (Garretsen et al., 2022). The methodology must therefore attend to these political dimensions, examining how security leaders develop and deploy political capabilities to achieve organisational security outcomes.

### 4.2.3 Ensuring Methodological Rigour

The establishment of methodological rigour requires explicit attention to quality criteria appropriate for qualitative inquiry. The trustworthiness criteria established by Lincoln and Guba (1985) provide a comprehensive framework for evaluating qualitative research quality, offering alternatives to positivist notions of validity and reliability. Credibility, achieved through prolonged engagement with participants and triangulation across data sources, ensures that findings accurately represent the phenomenon under investigation. The sixty-minute interview duration enabled movement beyond surface descriptions to explore the complexities and contradictions inherent in security leadership experiences. Transferability, facilitated through thick description of contexts and cases, enables readers to assess the applicability of findings to their own settings.

Dependability, maintained through careful documentation of research procedures and decisions, creates an audit trail that allows evaluation of the research process. The use of NVivo software for initial coding provided systematic organisation of analytical procedures, whilst supervisory oversight ensured that analytical decisions remained grounded in data rather than researcher assumptions (Castleberry, 2014; Dalkin et al., 2021). Confirmability, established through reflexive practices that acknowledge and manage researcher influence, proves particularly critical given the researcher’s dual position as practitioner and investigator. The systematic employment of supervisory challenge as the primary reflexivity mechanism provided external scrutiny that maintained analytical distance whilst leveraging insider knowledge appropriately.

Beyond general trustworthiness criteria, specific quality criteria for grounded theory research demand attention (Charmaz, 2014). Fit between the emergent theory and the data ensures that theoretical categories accurately represent participant experiences rather than researcher impositions. The systematic coding procedures and constant comparative analysis employed throughout this research maintained this fit, with supervisory validation providing additional assurance. Workability of the emergent theory, its capacity to explain and predict patterns within the phenomenon, emerged through careful attention to relationships between categories rather than mere description of isolated concepts. Relevance to both academic and practitioner communities guided theoretical development, ensuring that the Security Leadership Contingency Model addresses genuine concerns whilst contributing to scholarly understanding.

The iterative nature of grounded theory analysis creates particular quality challenges that require explicit management (Foley et al., 2021). The movement between data collection and analysis, whilst enabling theoretical sampling and refinement, risks premature theoretical closure or endless iteration without resolution. This research addressed these risks through clear saturation criteria, supervisory oversight, and systematic documentation of analytical decisions. The temporal boundaries of doctoral research provided pragmatic constraints that prevented endless iteration whilst ensuring sufficient analytical depth.

Specific procedural safeguards enhanced methodological rigour throughout the research process. The interview protocol, whilst maintaining flexibility for emergent themes, ensured consistency in core topics across participants. The recording and professional transcription of interviews created accurate data records for analysis, eliminating reliance on researcher memory or notes. The use of participant validation, sharing preliminary findings with interested participants, provided member checking opportunities that enhanced credibility. However, this validation process recognised that participants might not recognise their individual experiences within abstracted theoretical categories, a common challenge in grounded theory research (Urquhart et al., 2010).

The application of grounded theory methodology, operationalised through the Gioia method, provided the systematic framework necessary for investigating the complex phenomenon of security leadership. The combination of methodological structure and interpretive flexibility enabled the development of theoretical insights that remain grounded in empirical data whilst achieving the abstraction necessary for scholarly contribution. The Security Leadership Contingency Model presented in Chapter 6 represents the culmination of this methodological approach, demonstrating how systematic qualitative research can generate novel theoretical understanding of emerging organisational phenomena.

## 4.3 RESEARCH CONTEXT AND SAMPLING STRATEGY

### 4.3.1 Australian Organisational Context

The selection of Australia as the research context for examining CISO leadership represents a deliberate methodological choice grounded in both theoretical and practical considerations. The Australian organisational landscape provides a critical case for security leadership research, characterised by mature regulatory frameworks, advanced digital infrastructure, and sophisticated threat environments that mirror global security challenges whilst maintaining distinct contextual features (Ahmad et al., 2021; Ahmad et al., 2019).

#### *4.3.1.1 Contextual Boundaries and Justification*

The research focused specifically on Australian Securities Exchange (ASX)-listed and large private organisations operating within the post-2020 timeframe, capturing the period of pandemic-accelerated digitalisation that fundamentally transformed security leadership requirements. This temporal boundary proved particularly significant as organisations rapidly shifted to distributed work models, cloud-based infrastructures, and digital service delivery, creating unprecedented security leadership challenges (Garcia-Perez et al., 2023). The Australian regulatory environment, encompassing the Privacy Act 1988 (with subsequent amendments), the Security of Critical Infrastructure Act 2018 (SOCI Act), and the Notifiable Data Breaches scheme, establishes a sophisticated compliance landscape that demands strategic security leadership rather than merely technical management (Alazab et al., 2021; Ali et al., 2021; Kure et al., 2022).

The decision to concentrate on established organisations rather than start-ups reflects the research focus on mature security leadership roles where the CISO position has evolved beyond initial establishment phases. These organisations typically demonstrate formalised governance structures, established reporting relationships, and defined security functions, enabling examination of security leadership within structured organisational contexts (Poeppelbuss et al., 2011; Santos-Neto & Costa, 2019). The exclusion of vendor organisations ensures that participating CISOs bear direct responsibility for internal security outcomes rather than customer-facing consultancy roles, maintaining focus on organisational security leadership rather than commercial security services.

#### *4.3.1.2 Industry Diversity and Cross-Sector Patterns*

The deliberate pursuit of industry diversity within the sample responds to the need to distinguish between sector-specific phenomena and universal security leadership patterns. The Australian economy's sectoral composition, spanning financial services, healthcare, manufacturing, technology, retail, and aviation, provides sufficient variation in regulatory requirements, threat profiles, and digital maturity to enable robust cross-sector analysis. This diversity proved essential for the development of the Security Leadership Contingency Model (SLCM), which transcends industry boundaries whilst acknowledging contextual variations.

Furthermore, Australia's governance structures, based on the Westminster system, share fundamental characteristics with other Commonwealth nations including the United Kingdom, Canada, and New Zealand. This similarity in corporate governance models, regulatory approaches, and business cultures enhances the potential transferability of findings to comparable contexts. The Australian context thus serves as a representative case for security leadership in mature digital economies whilst maintaining sufficient specificity for deep contextual understanding.

### 4.3.2 Sampling Strategy

The sampling strategy employed purposive sampling techniques aligned with the exploratory nature of the research and the requirements of the methodology detailed in Section 4.2. This approach enabled deliberate selection of information-rich cases that could illuminate the phenomenon of security leadership whilst maintaining sufficient diversity for theoretical development.

#### *4.3.2.1 Purposive Sampling Criteria*

The establishment of clear sampling criteria ensured consistency in participant selection whilst allowing for emergent theoretical sampling as patterns developed. The primary criterion required participants to occupy the top-ranking security position within their organisation, whether titled CISO, Chief Security Officer, Head of Security, or equivalent designation. This focus on the most senior security decision-maker ensured access to strategic perspectives on security leadership rather than operational or technical viewpoints (Campbell et al., 2020; Palinkas et al., 2015).

The requirement for established organisations rather than start-ups reflects the research focus on mature security functions where leadership dynamics have stabilised beyond initial formation stages. Start-up environments, whilst presenting interesting security challenges, typically lack the organisational complexity, formalised governance structures, and established stakeholder relationships necessary for examining security leadership in its full complexity. The exclusion of vendor organisations eliminated potential confusion between internal security leadership responsibilities and customer-facing security consultancy roles, maintaining clear focus on organisational security accountability.

Participants needed to hold direct accountability for organisational security outcomes, distinguishing true security leaders from advisory or consultancy positions. This criterion ensured that participants experienced the full weight of security leadership responsibilities, including budget management, team leadership, stakeholder engagement, and accountability for security incidents (Tongco, 2007).

#### *4.3.2.2 Sampling Rationale and Theoretical Considerations*

The purposive sampling approach aligns with established qualitative research practices for exploring complex organisational phenomena where random sampling would fail to capture necessary variation and depth (Campbell et al., 2020; Palinkas et al., 2015; Tongco, 2007). By focusing on the most senior security decision-makers, the research captures strategic rather than operational perspectives, accessing insights into executive dynamics, board interactions, and organisational politics that shape security leadership.

The emphasis on cross-industry comparison reflects theoretical interest in identifying patterns that transcend sectoral boundaries whilst recognising context-specific variations. This approach enables development of a robust theoretical framework applicable across diverse organisational contexts rather than industry-specific models with limited transferability. The sampling strategy thus balances the need for depth within cases against breadth across contexts, supporting both detailed understanding and theoretical abstraction.

In practice, the relationship-based recruitment approach and the elite nature of the target population meant that participant selection remained primarily purposive throughout the study, guided by the predetermined criteria outlined above. The theoretical aspect manifested more in the iterative refinement of interview protocol emphasis and analytical focus rather than in strategically selecting participants to test emergent concepts. This represents a pragmatic adaptation of grounded theory appropriate for elite interviewing contexts where access constraints limit strict theoretical sampling (Charmaz, 2014). Nonetheless, the constant comparative analysis conducted between interviews ensured that emerging patterns informed subsequent data collection, with later interviews exploring themes and relationships identified in earlier conversations whilst maintaining the overarching purposive sampling framework.

### *4.3.2.3 Sample Composition Achieved*

The final sample comprised twenty senior security executives representing diverse industries and organisational contexts. The distribution across sectors included financial services (6 participants), manufacturing (4), professional services (2), higher education (2), food services (1), government (1), aviation (1), health and aged care (1), retail (1), and information technology services (1), providing sufficient variation for cross-industry analysis whilst maintaining adequate representation within key sectors. Organisational sizes ranged from about 500 to over 120,000 employees, capturing security leadership across different scales of operation and complexity.

Table 4.1 presents the composition of the final sample, listing participants in interview order and summarising each participant's industry sector, organisational size, and year of entry into the security profession.

*Table 4.1. Participant characteristics.*

| Interview Order | Participant ID | Industry Sector | Organisation Size | Year Started in Security |
|---|---|---|---|---|
| 1 | AgedCare01 | Health & Aged Care Services | Large (5000+) | 2015 |
| 2 | Insurance01 | Financial Services - Health Insurance | Mid-size (3000+) | 2001 |
| 3 | Beverage01 | Food & Beverage Manufacturing | Mid-size (4000+) | 2016 |
| 4 | MgtConsult01 | Management Consulting | Mid-size (700+) | 2012 |
| 5 | Bank01 | Financial Services - Digital Banking | Small (<500) | 2004 |
| 6 | Super01 | Financial Services - Superannuation | Large (7000+) | 2003 |
| 7 | Food01 | Food Service - Fast Food / Restaurants | Large (22000+) | 2012 |
| 8 | Aviation01 | Aviation - Airport Operations | Mid-size (600+) | 2004 |
| 9 | Beverage02 | Manufacturing | Large (7000+) | 2000 |
| 10 | BuildingMat01 | Construction Materials - Building | Large (7000+) | 2003 |
| 11 | Super02 | Financial Services - Superannuation | Small (<500) | 2004 |
| 12 | GovAgency01 | Government | Mid-size (600+) | 2000 |
| 13 | HigherEd01 | Higer Education | Mid-size (3000+) | 2003 |
| 14 | Food02 | Food Manufacturing | Large (38000+) | 2003 |
| 15 | MgtConsult02 | Accounting, Audit & Advisory | Mid-size (1800+) | 2008 |
| 16 | FinServ01 | Financial Services - Mortgage Broking | Mid-size (2400+) | 2006 |
| 17 | Tech01 | Information Technology Services | Small (<500) | 2004 |

| 18 | Retail01 | Retail Supermarket | Large (120000+) | 2000 |
|---|---|---|---|---|
| 19 | HigherEd02 | Higer Education | Mid-size (2000+) | 2003 |
| 20 | Auto01 | Financial Services - Automotive | Small (<500) | 2002 |

This composition enabled examination of security leadership across varying regulatory intensities, from highly regulated financial services to less prescribed manufacturing contexts. The range of organisational sizes captured different resource constraints, team structures, and governance complexities that influence security leadership. The diversity achieved supports theoretical development beyond single-industry or single-size limitations whilst maintaining sufficient commonality for meaningful comparison.

#### *4.3.2.4 Theoretical Saturation Assessment*

Data collection continued until theoretical saturation was achieved, defined as the point where additional interviews ceased to generate new theoretical insights relevant to the emerging framework (Guest et al., 2006; Guest et al., 2020; Nelson, 2017). The saturation assessment involved continuous monitoring of concept emergence throughout data collection, with particular attention to the stabilisation of core categories and relationships.

Analysis revealed that core concepts related to leadership-organisation fit stabilised after approximately fifteen interviews, with no fundamentally new patterns emerging in this domain. The theoretical categories of organisational maturity phases and corresponding leadership types had crystallised by interview seventeen, with subsequent interviews confirming rather than extending these patterns. The final interviews (eighteen through twenty) provided valuable confirmation of the emerging theoretical framework whilst enriching understanding through negative cases and boundary conditions rather than introducing entirely new dimensions.

This saturation pattern aligns with established guidance for grounded theory studies, where fifteen to twenty interviews typically suffice for achieving theoretical completeness in focused investigations (Guest et al., 2006; Nelson, 2017). The decision to continue to twenty interviews, despite apparent saturation around interview fifteen, reflects commitment to theoretical rigour and ensures confidence in the stability of emerged patterns. The additional interviews proved valuable in confirming theoretical relationships and identifying boundary conditions, strengthening the final theoretical framework.

### 4.3.3 Access and Recruitment

The recruitment process evolved significantly throughout the research, shifting from initial cold outreach strategies to relationship-based approaches that proved more effective for accessing senior security executives. This evolution reflects both the practical challenges of elite interviewing and the importance of trust in security research contexts (Guest et al., 2020; Marsh et al., 2020).

#### *4.3.3.1 Recruitment Strategy Evolution*

Initial recruitment attempts employed LinkedIn outreach to identify and contact CISOs meeting the sampling criteria. This approach, whilst providing access to a broad population of potential participants, yielded limited success with low response rates and minimal conversion to actual interviews. The impersonal nature of cold outreach proved particularly ineffective for accessing senior executives who receive numerous unsolicited requests and maintain careful boundaries around their time and information sharing.

Recognition of these limitations prompted a strategic shift toward relationship-based recruitment leveraging existing professional networks and conference interactions. The researcher's position within the CISO community, established through years of professional practice, provided crucial access to potential participants. This insider status facilitated trust-building essential for candid discussions about sensitive topics including security failures, political challenges, and career vulnerabilities (Choi et al., 2018; Posey et al., 2014).

Face-to-face recruitment at professional events proved particularly effective, especially when the researcher presented at conferences, establishing credibility and visibility within the target population. These interactions enabled informal relationship building that preceded formal recruitment, with many participants expressing interest in the research before being formally approached. Conference networking provided opportunities for explaining the research purpose, addressing concerns about confidentiality, and establishing the academic rigour that distinguished this research from commercial consultancy studies.

#### *4.3.3.2 Leveraging Professional Networks*

The researcher's existing position within the CISO community provided significant advantages for recruitment whilst requiring careful management of dual practitioner-researcher roles. Most participants knew the researcher from professional contexts, whether through direct collaboration, conference interactions, or community engagement. This familiarity facilitated access to senior executives who might otherwise decline participation, whilst establishing baseline trust that enabled frank discussions about sensitive leadership challenges (Choi et al., 2018; Denner et al., 2019). However, this insider status also necessitated reflexive awareness of potential biases and the implementation of strategies to maintain analytical distance, as discussed in Section 4.6.

Referrals from interviewed participants expanded the sample beyond immediate professional networks, with several participants recommending colleagues who met sampling criteria and might provide valuable perspectives. These warm introductions proved significantly more effective than cold outreach, with referred participants demonstrating higher engagement levels and greater willingness to share detailed experiences. The referral process also enabled access to CISOs outside the researcher's immediate network, enhancing sample diversity whilst maintaining the trust foundation essential for quality data collection.

The relationship-based recruitment approach aligned with established practices for elite interviewing, where personal connections and professional credibility prove essential for accessing senior executives (Guest et al., 2020; Marsh et al., 2020). The approach also reflects the close-knit nature of the Australian CISO community, where professional relationships and peer recommendations carry significant weight in decision-making about research participation.

#### *4.3.3.3 Participation Criteria and Engagement*

Beyond meeting the formal sampling criteria, successful recruitment required addressing practical and ethical considerations that influenced participation decisions. All participants provided voluntary informed consent, understanding their right to withdraw at any stage without consequence. The agreement to sixty-minute interviews acknowledged executives' time constraints whilst providing sufficient duration for in-depth exploration of complex leadership dynamics.

Permission for recording and transcription proved uncontentious, with all participants agreeing to audio recording for accuracy whilst receiving assurances about confidentiality and data security. Notably, whilst transcription review was not offered given practical constraints, participants expressed primary interest in the final research outcomes rather than reviewing individual transcripts. This orientation toward research findings rather than transcript accuracy reflects participants' strategic focus and interest in comparative insights rather than individual narrative control.

Participants' expressed interest in receiving final research outcomes highlights the perceived value of the research to practitioner communities. Many participants explicitly stated their motivation as contributing to professionalisation of the CISO role and advancing understanding of security leadership. This alignment between research objectives and participant interests facilitated recruitment whilst establishing foundations for potential research impact within practitioner communities.

## 4.4 DATA COLLECTION PROCEDURES

The data collection procedures for this study were designed to capture the rich, contextualised experiences of senior security leaders whilst maintaining the systematic rigour demanded by grounded theory methodology. This section details the development of the interview protocol, the management of the interview process, and the assessment of data saturation, demonstrating how these procedures enabled the emergence of theoretical insights whilst ensuring methodological transparency.

### 4.4.1 Semi-Structured Interview Protocol

The selection of semi-structured interviews as the primary data collection method aligned with the exploratory nature of this research and the constructivist grounded theory approach adopted for the study, following the methodology detailed in Section 4.2. This method provided the flexibility to pursue emergent themes whilst maintaining sufficient structure to ensure comprehensive coverage of the research domain. The balance between consistency and adaptability proved particularly valuable given the diverse organisational contexts and career trajectories of the participating security leaders.

#### *4.4.1.1 Protocol Development*

The interview protocol was developed through careful collaboration with supervisors, drawing directly from the fundamental gaps identified in the literature review regarding role creation, value demonstration, core activities, structural positioning, and role evolution. This systematic development process ensured that questions would address the central research concerns whilst allowing participants sufficient flexibility to articulate their experiences in their own terms (Dahlin, 2021). The protocol structure deliberately commenced with broad, open questions about the role and its organisational context, allowing participants to frame their experiences before progressing to more specific domains (Wood & Ford, 1993). This approach, consistent with recommendations for elite interviewing, recognised that senior executives possess sophisticated understanding of their domains and benefit from the opportunity to articulate their perspectives without premature constraint (Wood & Ford, 1993).

The protocol remained consistent throughout all 20 interviews to ensure comparability across participants and maintain methodological rigour. This consistency proved essential for identifying patterns across diverse organisational contexts and industry sectors. While the questions themselves did not change, the emphasis placed on different domains naturally evolved as certain themes emerged as particularly significant. Political dynamics and stakeholder relationships, for instance, received increased attention in later interviews as their importance to security leadership became increasingly apparent through early data collection.

The protocol organised questions into six core domains, each designed to elicit different aspects of the security leadership experience. These domains comprised: (1) role requirements and organisational drivers, exploring why the CISO position was established and what organisations expected from it; (2) organisational value and contribution, examining how CISOs perceived and demonstrated their impact on organisational performance; (3) operational constraints and challenges, investigating the obstacles participants faced in fulfilling their responsibilities; (4) business-IT perceptions, probing how security was positioned relative to technical and business domains; (5) organisational structure and political dynamics, examining reporting relationships, positional power, and political navigation; and (6) strategic versus operational orientation, understanding how participants allocated their attention across different types of activities. Additional questions explored role evolution in response to changing threat landscapes and invited participants to reflect on their multifaceted responsibilities using a conceptual framework. This approach consistently generated rich narratives about role scope, reporting relationships, and organisational expectations whilst revealing important contextual information about how the position had been established and evolved. The average interview duration of 60 minutes provided sufficient time to explore topics in depth whilst respecting executive time constraints. Several participants, engaged by the discussion, extended slightly beyond the scheduled time.

The semi-structured format proved particularly valuable in managing the diverse expertise and experiences within the sample. Whilst core questions remained consistent, the flexibility to pursue emergent themes enabled exploration of industry-specific challenges in financial services, unique regulatory pressures in healthcare, and distinctive operational requirements in aviation. This adaptability, fundamental to grounded theory methodology, allowed theoretical categories to emerge from the data rather than being imposed through rigid questioning (Foley et al., 2021).

### 4.4.2 Interview Process Management

The management of the interview process required careful attention to both practical logistics and methodological considerations, particularly given the seniority of participants and the sensitive nature of security leadership discussions.

#### *4.4.2.1 Pre-Interview Preparation*

Prior to each interview, basic organisational background research was conducted to understand the industry context and publicly available information about the organisation's security posture. This preparation enabled more sophisticated questioning and demonstrated respect for participants' time by avoiding requests for basic factual information readily available elsewhere. However, this research was deliberately limited to public sources to avoid biasing the interview process with assumptions about organisational security challenges or leadership requirements.

The plain language statement, distributed at least 48 hours before each interview, outlined the research objectives, interview process, and data handling procedures. This document emphasised the academic nature of the research and the steps taken to ensure confidentiality, addressing concerns particularly acute for security leaders discussing organisational vulnerabilities and leadership challenges.

#### *4.4.2.2 During Interview Management*

The interview process itself required careful management of multiple competing demands: maintaining conversational flow whilst ensuring comprehensive coverage, pursuing emergent themes whilst respecting time constraints, and building rapport whilst maintaining analytical distance. The practitioner-researcher position created both advantages and challenges in this regard. Shared professional background facilitated rapid establishment of credibility and enabled sophisticated discussion of technical and strategic issues. However, this same familiarity required conscious effort to avoid assumptions and to probe for explicit articulation of concepts that might otherwise remain implicit between security professionals.

Active listening techniques proved essential, with minimal interruption allowing participants to develop complex narratives about their leadership experiences (Lavee & Itzchakov, 2023). Note-taking during interviews served multiple purposes: capturing non-verbal observations not evident in audio recordings, identifying themes for follow-up questions, and maintaining engagement during lengthy explanations. These notes, whilst not subjected to formal analysis, provided valuable context during the transcription review and initial coding phases.

The progression from broad to specific questioning followed a natural conversational arc whilst ensuring coverage of key domains. Initial questions about the current role and its organisational positioning consistently generated narratives that touched upon multiple theoretical domains, providing organic transitions to more focused exploration. This approach aligned with recommendations for executive interviewing, where participants expect sophisticated engagement with complex organisational issues (Wood & Ford, 1993).

#### *4.4.2.3 Post-Interview Procedures*

Professional transcription services were engaged to ensure accurate and timely conversion of audio recordings to text. The transcription service, selected for their experience with business and technical content and their privacy protocols, ensured appropriate handling of the sensitive security discussions. Transcripts were reviewed for accuracy, with particular attention to technical terminology and organisational references that automated transcription might misinterpret. This review process, whilst time-consuming, ensured data quality and provided an initial immersion in the interview content.

Communication with participants following interviews maintained engagement and demonstrated appreciation for their contribution. Thank you emails, sent within 48 hours, reiterated the commitment to share research findings upon completion. This commitment reflected both ethical research practice and pragmatic recognition that participants, as senior security leaders, represented valuable stakeholders in the research outcomes. Several participants expressed particular interest in the comparative insights across industries and organisations, viewing the research as an opportunity to understand broader patterns in security leadership evolution.

### 4.4.3 Data Saturation Assessment

The assessment of data saturation represented a critical methodological decision point, requiring systematic evaluation of whether additional data collection would yield new theoretical insights or merely confirm existing patterns (Nelson, 2017). This assessment occurred iteratively throughout the data collection process rather than as a single determination point.

#### *4.4.3.1 Theoretical Saturation Indicators*

The primary indication of approaching saturation emerged around interview 15, when discussions of leadership-organisation fit began yielding consistent patterns rather than new conceptual categories. The core theoretical categories of organisational security phases and corresponding leadership types had stabilised, with subsequent interviews providing elaboration and confirmation rather than fundamental reconceptualisation. This stabilisation was particularly evident in the recurring descriptions of transition challenges when security leaders moved between organisations at different maturity levels, suggesting robust theoretical categories rather than idiosyncratic experiences.

By interview 17, the relationship between environmental factors and leadership adaptation had crystallised into consistent patterns. Participants described similar pressures from regulatory change, threat evolution, and technological transformation, with variations in specific examples but consistency in underlying dynamics. The final interviews (18-20) served primarily confirmatory purposes, validating the theoretical framework without adding new dimensions. These interviews did, however, provide valuable negative cases that enriched understanding of boundary conditions and exceptional circumstances where typical patterns might not apply.

The assessment of saturation extended beyond simple category repetition to consider the depth and dimensionality of theoretical development (Rowlands et al., 2015). Whilst new anecdotes and examples continued to emerge, these illustrated rather than extended existing theoretical categories.

#### *4.4.3.2 Saturation Documentation*

The documentation of saturation followed a systematic approach, with theoretical memos maintained throughout the data collection process. These memos tracked the emergence and stabilisation of key categories, noting when new interviews confirmed rather than challenged existing patterns. The supervisory review process provided external validation of saturation assessment, with independent evaluation confirming that additional data collection would likely yield diminishing theoretical returns.

The iterative nature of data collection and analysis, following the methodology detailed in Section 4.2, enabled real-time saturation assessment. Preliminary coding of each interview before conducting the next allowed for evaluation of whether new theoretical insights were emerging. This approach revealed that whilst surface-level variety continued (different industries, specific incidents, individual career paths), the underlying theoretical patterns had stabilised by interview 20.

#### *4.4.3.3 Implications for Theoretical Development*

The achievement of theoretical saturation enabled confident progression to focused theoretical development, knowing that the empirical foundation was comprehensive and robust. The patterns identified across the 20 interviews provided sufficient variation to understand both typical manifestations and boundary conditions of security leadership. The diversity of industries, organisational sizes, and leadership backgrounds within the sample strengthened confidence that saturation represented genuine theoretical completeness rather than sample homogeneity.

The saturation assessment also revealed the importance of theoretical sensitivity in recognising emergent patterns. Early interviews that initially seemed divergent later revealed themselves as variations within consistent theoretical categories. This recognition required iterative engagement with the data, moving between individual cases and cross-case patterns to identify underlying commonalities. The supervisory challenge process proved particularly valuable in testing whether apparent saturation reflected genuine theoretical completeness or analytical fatigue.

The documentation and assessment of saturation ultimately provided the empirical foundation for the Security Leadership Contingency Model presented in Chapter 6. The confidence that data collection had captured the full range of relevant phenomena enabled theoretical development that extended beyond description to explanation, addressing not only what patterns exist but why they emerge and how they evolve. This theoretical depth, grounded in comprehensive data collection, represents a key contribution of this research to understanding security leadership in contemporary organisations.

## 4.5 DATA ANALYSIS PROCEDURES

The analysis of interview data followed the Gioia methodology (Section 4.2.2) to transform raw interview data into theoretical understanding, with empirical findings presented in Chapter 5 and the resulting Security Leadership Contingency Model detailed in Chapter 6. This section details the analytical procedures employed to systematically abstract participant experiences into theoretical constructs whilst maintaining clear linkages to empirical foundations.

### 4.5.1 Gioia Method Implementation

The implementation of the Gioia methodology (Section 4.2.2) proceeded through three distinct analytical phases, each building upon the previous to create progressively abstract theoretical understanding.

#### *4.5.1.1 First-Order Analysis*

The initial analytical phase focused on preserving participant voice through detailed coding of interview transcripts. Each transcript underwent line-by-line analysis to identify concepts directly expressed by participants. This granular approach ensured comprehensive capture of participant experiences before any theoretical abstraction occurred.

The coding process generated over sixty initial first-order concepts, reflecting the richness and complexity of CISO experiences across diverse organisational contexts. These codes retained participants' language wherever possible, creating informant-centric concepts that preserved the authentic voice of security leaders.

NVivo software facilitated the management of this extensive code set, enabling systematic organisation whilst maintaining clear audit trails from codes to source data. The software's capabilities proved particularly valuable for managing the volume of codes generated from twenty hour-long interviews, allowing for efficient retrieval and comparison of coded segments across participants.

#### *4.5.1.2 Second-Order Analysis*

The transition from first-order to second-order analysis involved identifying patterns and relationships amongst the initial codes. This phase required shifting from participant language to researcher interpretation, seeking theoretical patterns that could explain the phenomena described by participants.

Through constant comparison across cases, the initial first-order concepts were progressively grouped into sixteen second-order themes. These themes emerged through identifying conceptual similarities whilst respecting meaningful differences. The themes naturally clustered into three contextual groupings: Organisational Context (encompassing Security Needs, Culture, Structure, and Maturity), CISO Characteristics (including three leadership types alongside Technical Skills, Experience, and Personality), and Environmental Context (capturing external pressures and constraints).

The development of second-order themes required theoretical sensitivity to both emergent patterns and existing literature. For example, various first-order codes describing different approaches to stakeholder engagement were grouped under broader themes related to leadership orientations. The three distinct leadership types that emerged, Action-Oriented, Stewardship, and Vision-Oriented, represented different approaches CISOs adopted in response to their organisational contexts.

#### *4.5.1.3 Aggregate Dimension Development*

The final analytical phase involved abstracting second-order themes into aggregate dimensions representing the highest level of theoretical abstraction. Through iterative analysis and supervisory discussion, two aggregate dimensions emerged that captured the essential components of the Security Leadership Contingency Model: CISO-Organisation Fit and CISO-Environment Fit.

The progression from sixteen second-order themes to two aggregate dimensions required careful theoretical consideration. The recognition that both CISO-Organisation fit and CISO-Environment fit must be maintained simultaneously emerged as a critical insight, representing a significant departure from traditional fit theories that examine these dimensions in isolation. This dual-fit dynamic, emerging from the data analysis presented in Chapter 5, became the theoretical cornerstone of the Security Leadership Contingency Model developed in Chapter 6.

### 4.5.2 Analytical Procedures

Beyond the Gioia method's structural framework, several complementary analytical procedures enhanced the rigour and depth of analysis in this study.

#### *4.5.2.1 Constant Comparative Method*

The constant comparative method operated throughout the analysis at multiple levels. Within-case analysis examined internal coherence of individual participant narratives, ensuring that emerging themes accurately reflected each CISO's complete experience rather than selective excerpts. This involved reviewing each transcript multiple times, checking that coded segments aligned with the participant's overall narrative.

Cross-case comparison identified patterns transcending individual experiences. By systematically comparing responses across participants, the analysis distinguished between idiosyncratic experiences and recurring phenomena. Industry comparisons revealed how sectoral contexts influenced CISO experiences, whilst temporal comparisons examined how role requirements evolved across different career stages.

The constant comparative approach proved particularly valuable for identifying boundary conditions of emerging concepts. For instance, whilst most participants described increasing business engagement as essential, comparison across industries revealed variations in how this manifested. Financial services CISOs emphasised regulatory compliance discussions, whilst technology sector CISOs focused more on product security integration. These comparative insights enriched theoretical development by specifying contextual conditions influencing concept expression.

#### *4.5.2.2 Theoretical Coding*

Theoretical coding focused on identifying relationships between categories rather than simply grouping similar concepts. This involved mapping connections between second-order themes and exploring how different elements of the emerging model interacted. The analysis examined causal relationships, temporal sequences, and conditional factors influencing the CISO role.

Process mapping revealed temporal dynamics in leadership evolution, demonstrating how role requirements changed as organisations matured through different phases. The identification of phase-based patterns in organisational needs emerged as a pivotal analytical moment. Rather than treating organisational contexts as static, the data revealed dynamic progression through identifiable phases, each creating distinct leadership requirements.

Conditional matrix development identified factors moderating relationships between leadership characteristics and outcomes. For example, the analysis revealed that technical expertise's importance varied depending on organisational maturity phase and industry context. The theoretical coding process benefited significantly from supervisory discussions, which challenged initial interpretations and pushed for deeper theoretical engagement.

#### *4.5.2.3 Negative Case Analysis*

Deliberate attention to contradictory evidence and outlier cases enhanced theoretical refinement. Rather than dismissing cases that did not fit emerging patterns, the analysis examined these instances to understand boundary conditions and refine theoretical claims. This approach proved essential for developing nuanced theory capable of explaining variation rather than oversimplifying complex phenomena. The significance of negative case analysis for validating and refining the Security Leadership Contingency Model is detailed further in Chapter 6, where these exceptional cases illuminate the moderating role of political capital and formal authority.

Several participants succeeded despite apparent misalignment between their leadership orientation and organisational phase. These negative cases revealed the moderating influence of political capital and formal authority. Rather than undermining the emerging pattern, these exceptions illuminated boundary conditions and contingency factors that enrich theoretical understanding. The analysis revealed that political capital could temporarily buffer misalignment, providing time for adaptation, whilst formal authority structures could compensate for certain capability gaps.

### 4.5.3 Theoretical Development Process

The progression from empirical patterns to theoretical concepts involved multiple iterations of analysis, reflection, and refinement. This section details how raw data transformed into the Security Leadership Contingency Model through systematic analytical procedures.

#### *4.5.3.1 Analytical Evolution*

The analytical journey began with immersion in participant narratives, seeking to understand CISO experiences on their own terms before imposing theoretical structure. Initial coding sessions focused purely on capturing participant meanings without theoretical interpretation. As patterns emerged through constant comparison, the analysis progressively moved toward greater abstraction whilst maintaining clear links to empirical foundations. The software facilitated initial coding and systematic organisation of interview data, with documentation evolving organically as patterns emerged from the data.

Regular supervisory meetings provided crucial external perspective throughout this evolution. Supervisors challenged interpretations, suggested alternative explanations, and ensured that theoretical development remained grounded in participant data rather than researcher preconceptions. These discussions often identified instances where personal experience as a practising CISO might influence interpretation, prompting renewed examination of the data to ensure participant voices remained paramount.

The entry of theoretical frameworks into the analysis followed the abductive logic of inquiry described in Section 4.2.1. Theoretical frameworks were absent at the outset; they were identified through targeted literature engagement at specific points when empirical patterns required explanation that the data alone could not provide.

The first such moment arose from persistent alignment and misalignment patterns across participant accounts. Participants consistently described their effectiveness as contingent on a form of correspondence between their own capabilities and orientations and the characteristics of their organisations. The pattern was clear and recurring. A targeted search of the executive leadership literature, prompted by this observation, identified person-organisation fit theory as offering explanatory purchase. The framework was not imposed on the data; the data created the analytical need that led to its discovery.

A second analytical moment emerged as the data revealed that organisational phase systematically altered what effective security leadership required. Participants who described strong alignment with their organisations during early security programme development often reported changes in their effectiveness as the organisation matured, despite unchanged personal capabilities. This pattern prompted engagement with contingency theory, which provided a framework for understanding how contextual factors shape leadership requirements. The theoretical framework was identified because the empirical pattern demanded explanation. This sequence, from observed pattern to analytical puzzle to targeted literature engagement, exemplifies the abductive reasoning characteristic of constructivist grounded theory (Gioia et al., 2013; Thornberg, 2012; Timmermans & Tavory, 2012).

Analytical memos captured evolving interpretations, theoretical insights, and decision rationales, creating a comprehensive record of the analytical journey. These memos documented critical analytical moments, such as the recognition of phase-based patterns in organisational needs and the emergence of political capital as a moderating factor. This documentation proved invaluable during later stages of analysis when reviewing earlier coding decisions or tracing the evolution of particular concepts.

#### *4.5.3.2 Theoretical Saturation*

Evidence of theoretical saturation manifested through diminishing conceptual emergence as the analysis progressed (as detailed in Section 4.4.3). By interview seventeen, no new concepts emerged related to the core theoretical framework, and the relationships between concepts had stabilised. The final three interviews served as validation exercises, confirming saturation across all categories and strengthening confidence in the emergent structure.

The saturation assessment focused particularly on the core theoretical categories of organisational phases and leadership types. Whilst individual variations continued to emerge, these represented instantiations of established patterns rather than fundamentally new theoretical insights. Supervisory validation of saturation claims ensured that the decision to conclude data collection at twenty interviews was theoretically justified rather than pragmatically driven.

The analytical procedures detailed in this section provided the systematic foundation for transforming rich interview data into theoretical understanding. Through careful implementation of the Gioia methodology (Section 4.2.2), complemented by established grounded theory techniques, the analysis maintained rigorous standards whilst remaining open to emergent insights. The resulting empirical findings, presented in Chapter 5, enabled development of the Security Leadership Contingency Model detailed in Chapter 6, representing the culmination of this systematic analytical journey from empirical observation to theoretical contribution.

## 4.6 ENSURING RESEARCH QUALITY AND RIGOUR

The quality and rigour of qualitative research demand systematic attention to trustworthiness criteria that establish the credibility, dependability, confirmability, and transferability of findings (Lincoln & Guba, 1985). This section explicates the strategies employed to ensure methodological rigour throughout the research process, addressing both the opportunities and challenges inherent in practitioner-researcher positioning whilst demonstrating adherence to established quality criteria for grounded theory research.

### 4.6.1 Credibility Strategies

Credibility in qualitative research corresponds to internal validity in quantitative studies, concerning the truth value and accuracy of findings (Cassell et al., 2020; Denzin & Lincoln, 1994). Multiple strategies were implemented to enhance credibility, each contributing to confidence in the authenticity and accuracy of the emergent theoretical model.

**Prolonged engagement** characterised the data collection process, with sixty-minute in-depth interviews providing sufficient time to move beyond surface descriptions to deeper exploration of participants' experiences. The extended duration enabled participants to reflect thoughtfully on their roles, with all participants expressing genuine interest in the outcomes of this research. This depth of engagement suggested that the interviews facilitated meaningful reflection beyond routine professional discourse.

**Supervisory involvement** provided crucial external perspective and methodological guidance throughout the research process. One supervisor participated directly in several early interviews, offering real-time coaching on interview technique and ensuring consistency in approach. This involvement served multiple purposes: it provided quality assurance for interview conduct, enabled immediate feedback on emerging themes, and established a shared understanding of the data that enriched subsequent analytical discussions. The supervisor's participation proved particularly valuable in refining the balance between allowing participants to direct conversation flow whilst ensuring coverage of key theoretical domains.

**Triangulation approaches** strengthened credibility through multiple forms of comparison and cross-validation (Jack & Raturi, 2006; Thurmond, 2001). Data source triangulation occurred through sampling participants across diverse industries, enabling identification of patterns that transcended sector-specific contexts whilst also revealing industry-contingent variations. Temporal triangulation emerged through participants' reflections on both current roles and previous positions, providing longitudinal perspective on leadership evolution. Perspective triangulation involved exploring phenomena from multiple angles within each interview, examining the same issues through technical, business, and political lenses to develop comprehensive understanding.

**Peer debriefing** through regular supervisory meetings provided systematic external challenge to emerging interpretations. Six-monthly presentations to the full supervisory panel created formal opportunities for scrutiny of analytical decisions and theoretical development. These sessions proved invaluable for identifying potential biases, challenging preliminary interpretations, and ensuring that theoretical abstractions remained grounded in empirical data. The supervisory panel's diverse expertise, spanning information systems, organisational behaviour, and security management, enriched the analytical process through multidisciplinary perspectives.

### 4.6.2 Analytical Documentation and Transparency

Systematic documentation of analytical decisions enhanced confirmability by creating an audit trail from raw data through to theoretical conclusions (Dalkin et al., 2021). This documentation evolved organically as patterns emerged, reflecting the iterative nature of grounded theory analysis whilst maintaining sufficient structure to demonstrate rigour.

**NVivo utilisation** provided systematic support for initial coding and data organisation. The software facilitated management of the extensive initial coding framework, enabling efficient comparison across interviews and identification of emerging patterns. However, software use remained instrumental rather than deterministic; NVivo organised data but did not drive analytical decisions. As analysis progressed toward theoretical abstraction, physical whiteboards and visual mapping supplemented digital tools, reflecting the non-linear nature of theory development.

**Coding consistency strategies** developed progressively as understanding deepened. Initial coding remained deliberately open, using participants' own language wherever possible to preserve authentic meaning. As patterns emerged, code definitions were refined through supervisory discussion, creating increasingly precise conceptual boundaries. Regular supervisory review of coding decisions provided external validation whilst maintaining flexibility for emergent insights. This balance between consistency and emergence proved crucial for developing robust yet grounded theoretical categories.

The evolution from first-order concepts through second-order themes to aggregate dimensions was meticulously documented, though this documentation process itself evolved as understanding deepened. Early attempts at rigid documentation protocols proved constraining; the final approach balanced systematic recording with flexibility for creative theoretical development. Each analytical decision point was captured through memos that recorded not just what decisions were made but why, preserving reasoning for later scrutiny.

**Analytical transparency** extended beyond documentation to include clear representation of the abstraction process in research outputs. The data structure diagram presented in Chapter 5 explicitly shows the progression from raw data through theoretical constructs, enabling readers to trace conceptual development. Representative quotes for each first-order concept demonstrate empirical grounding, whilst the articulation of abstraction logic from first-order to second-order themes reveals analytical reasoning. This transparency allows readers to assess the credibility of theoretical claims independently.

### 4.6.3 Dependability Through Systematic Procedures

Dependability, paralleling reliability in quantitative research, concerns the consistency and stability of findings over time and across researchers (Lincoln & Guba, 1985). Whilst acknowledging that qualitative research cannot achieve mechanical replicability, several strategies enhanced dependability through systematic and traceable procedures.

**Consistent interview protocols** provided structure whilst maintaining flexibility for emergence. The semi-structured interview guide ensured coverage of core domains across all participants whilst allowing conversation to follow unexpected directions when theoretically interesting themes emerged. This balance proved crucial; overly rigid protocols would have prevented discovery of unanticipated patterns, whilst completely unstructured interviews would have compromised comparability.

**Analytical procedure standardisation** occurred through development of clear processes for moving from raw data to theoretical abstraction. The Gioia methodology (detailed in Section 4.2.2) provided systematic guidelines that were consistently applied across all interviews. This standardisation did not impose mechanical rigidity but rather established clear analytical stages that could be documented and reviewed. Regular supervisory meetings ensured consistent application of analytical procedures whilst maintaining sensitivity to emergent patterns.

**Temporal stability** was addressed through the theoretical saturation process. The stabilisation of core categories by interview fifteen, with confirmation through interviews seventeen to twenty, demonstrated that findings reflected robust patterns rather than idiosyncratic observations. The consistency of themes across diverse organisational contexts further enhanced confidence in the stability of findings.

### 4.6.4 Confirmability Through Audit Trails

Confirmability addresses the degree to which findings reflect participants' experiences rather than researcher preferences or biases (Lincoln & Guba, 1985; Tufford & Newman, 2012). The maintenance of comprehensive audit trails enabled external review of analytical decisions and theoretical development.

**Decision documentation** captured key analytical choice points throughout the research process. Supervisory meeting notes recorded discussions about coding decisions, theoretical development, and potential bias identification. These records demonstrate how interpretations evolved through systematic analysis rather than predetermined expectations. When analytical directions shifted, such as the recognition of political dynamics as central rather than peripheral, documentation captured both the evidence prompting reconsideration and the reasoning supporting new interpretations.

**Chain of evidence** from raw data to theoretical propositions remained traceable throughout analysis. Interview transcripts were preserved in original form, with NVivo coding maintaining links between theoretical concepts and supporting data. This chain of evidence enables independent assessment of whether theoretical claims are warranted by empirical foundations. The ability to trace any theoretical proposition back to multiple participant quotes demonstrates that findings emerged from data rather than being imposed upon it.

### 4.6.5 Reflexivity and Practitioner-Researcher Positioning

The dual identity as both practising CISO and researcher investigating CISOs created unique methodological considerations requiring careful reflexive management (Berger, 2015; Dwyer & Buckle, 2009). This positioning offered advantages through insider knowledge and access whilst simultaneously introducing potential for practitioner bias that could compromise analytical rigour.

**Researcher positionality** as an active member of the professional community under study provided several advantages. Insider status facilitated access to senior executives who might otherwise decline participation, with several participants explicitly noting they agreed to interviews because of shared professional background. Technical fluency enabled rapid comprehension of complex security scenarios without requiring extensive explanation, preserving interview time for exploring leadership rather than technical dimensions. Perhaps most significantly, shared professional experience appeared to enhance participant candour, with many expressing views about organisational politics and leadership challenges they indicated they would not share with non-practitioner researchers.

However, this insider position also introduced risks of assumed understanding and practitioner bias. The potential for interpreting data through personal experience rather than participant meaning required constant vigilance and systematic strategies to maintain analytical distance throughout the research process.

**Reflexivity management** centred primarily on supervisory challenge as the key mechanism for identifying and addressing practitioner bias. Regular supervisory sessions specifically examined potential influences of researcher experience on data interpretation. Supervisors consistently challenged assumptions that appeared to derive from personal experience rather than participant data, demanding explicit grounding of all interpretations in interview transcripts. When analytical memos referenced concepts familiar from practice, supervisors required demonstration that these concepts emerged from participants rather than being imposed by researcher preconceptions.

This supervisory challenge process proved particularly valuable during theoretical development phases. Initial tendencies to draw parallels with CIO literature, reflecting researcher familiarity with that domain, were identified and corrected through supervisory intervention. This external scrutiny ensured that CISO-specific patterns could emerge independently rather than being forced into existing theoretical frameworks from adjacent domains. The process of defending interpretations to supervisors unfamiliar with security practice demanded clear articulation of reasoning, enhancing analytical transparency.

**Conscious bracketing** of preliminary concepts represented another reflexivity strategy, though its implementation evolved throughout the research process. Rather than attempting complete suspension of prior knowledge, neither feasible nor desirable given the technical nature of many discussions, bracketing focused on maintaining openness to unexpected patterns. This involved deliberately seeking disconfirming evidence and remaining alert to findings that contradicted professional assumptions.

The practitioner-researcher position, whilst requiring careful reflexive management, ultimately enhanced rather than compromised research quality. Insider knowledge enabled recognition of subtle patterns that might escape external researchers, whilst systematic reflexivity strategies prevented this knowledge from overwhelming participant voices. The combination of insider sensitivity and analytical distance achieved through supervisory challenge produced findings that are both theoretically robust and practically relevant.

### 4.6.6 Transferability Through Thick Description

Transferability in qualitative research does not claim statistical generalisation but rather provides sufficient detail for readers to assess applicability to other contexts (Lincoln & Guba, 1985). Rich contextual description enables informed judgements about whether findings might apply to different settings.

**Contextual detail** throughout data presentation preserves the situated nature of participants' experiences. Industry contexts, organisational characteristics, and regulatory environments are described sufficiently to enable readers to assess similarity to their own contexts. This thick description extends beyond surface features to include cultural dynamics, political contexts, and historical trajectories that shaped security leadership in studied organisations.

**Boundary specification** explicitly delineates the scope conditions for theoretical claims. The focus on Australian organisations within mature regulatory environments provides clear context for findings. The emphasis on established organisations rather than start-ups, and on senior security leaders rather than technical managers, creates defined boundaries for theoretical application. These boundaries do not limit relevance but rather enable appropriate translation to different contexts.

The Australian focus, whilst creating geographic boundaries, offers particular value for understanding security leadership in comparable contexts. The Westminster system of governance, similar regulatory maturity to other developed nations, and mix of local and multinational organisations create conditions likely replicated in markets such as the United Kingdom, Canada, and New Zealand. Rather than claiming universal applicability, the research provides sufficient detail for readers to assess transferability to their specific contexts.

### 4.6.7 Quality Criteria for Grounded Theory

Beyond general qualitative quality criteria, this research addressed specific quality standards for grounded theory studies (Charmaz, 2014; Corbin & Strauss, 2015). These criteria ensure that the research achieves the theory-building objectives of grounded methodology whilst maintaining empirical grounding.

**Theoretical saturation** was systematically assessed throughout data collection. The emergence of P-O fit concepts around interview eight, followed by P-E fit themes in interview ten, demonstrated progressive theoretical development. The integration of fit theories in interviews thirteen and fifteen marked theoretical crystallisation. Interviews seventeen through twenty confirmed saturation by reinforcing existing patterns without introducing new theoretical dimensions. This saturation assessment focused not merely on repetition of surface themes but on theoretical completeness, whether additional data would enhance theoretical understanding.

**Conceptual depth** achieved through the analytical process moved beyond descriptive categorisation to develop abstract theoretical relationships. The progression from over sixty first-order concepts through sixteen second-order themes to two aggregate dimensions demonstrates systematic abstraction whilst maintaining empirical grounding. Each level of abstraction added theoretical value: initial codes preserved participant meanings, second-order themes identified patterns, and aggregate dimensions enabled theoretical explanation.

**Theoretical resonance** with both participants and the broader practitioner community validates the practical relevance of findings (Charmaz, 2014; Dodgson, 2019). Informal discussions with security leaders beyond the participant sample confirmed that the theoretical model resonated with their experiences. The identification of political dynamics as central to security leadership, whilst potentially controversial in technical communities, was consistently validated by practitioners who recognised this reality in their own experience.

### 4.6.8 Integration of Quality Strategies

The multiple quality strategies employed throughout this research operated synergistically rather than independently. Supervisory challenge enhanced reflexivity whilst also contributing to credibility through external scrutiny. Documentation for confirmability simultaneously supported dependability through systematic procedures. This integration of quality strategies created multiple overlapping safeguards against threats to rigour.

The evolution of quality strategies throughout the research process reflected the emergent nature of grounded theory methodology. Rather than implementing rigid quality protocols that might constrain discovery, strategies developed responsively as analytical needs became apparent. This adaptive approach to quality assurance maintained rigour whilst preserving the flexibility essential for theoretical discovery.

The Security Leadership Contingency Model (SLCM) emerged through these rigorous quality strategies, ensuring that theoretical contributions rest on solid empirical foundations whilst maintaining relevance to practitioner experience. The systematic attention to trustworthiness criteria throughout the research process provides confidence that findings represent authentic patterns in security leadership rather than artefacts of methodological limitations or researcher bias.

## 4.7 ETHICAL CONSIDERATIONS

The ethical dimensions of this research extended beyond procedural compliance to encompass the substantive responsibilities inherent in studying elite security leaders navigating sensitive organisational contexts. This section details the ethical framework governing the research, protective measures implemented for participants and organisations, and strategies employed to manage the unique sensitivities of security leadership research.

### 4.7.1 Institutional Ethics Approval

This research operated under formal ethics approval from the University Human Research Ethics Committee (approval reference numbers: 2021-11429-13523-3 and 2023-26041-42767-3), ensuring alignment with the National Statement on Ethical Conduct in Human Research (NHMRC, 2018 edition). The formal approval letters are provided in Appendix A. The approval process recognised the research as minimal risk, acknowledging that whilst participants occupied senior positions with sophisticated understanding of research implications, the sensitive nature of security leadership discussions warranted careful ethical consideration.

The ethics review process examined several dimensions specific to elite security research. First, the power dynamics between researcher and participants required consideration, particularly given the researcher’s practitioner status within the CISO community. Second, the potential for discussing organisational vulnerabilities or security incidents necessitated clear boundaries around information disclosure. Third, the interconnected nature of the Australian CISO community raised considerations around confidentiality protection across a relatively small professional network. The committee's approval validated the research design's capacity to address these ethical complexities whilst maintaining methodological rigour.

Annual progress reporting to the ethics committee maintained ongoing oversight throughout the data collection period. These reports documented participant recruitment progress, confirmed adherence to approved protocols, and reported any emerging ethical considerations. No adverse events or protocol deviations occurred during the research period, with all data collection proceeding according to approved procedures.

### 4.7.2 Participant Protection

Comprehensive participant protection mechanisms operated throughout the research process, recognising the vulnerability that even senior executives face when discussing leadership challenges, organisational politics, and career experiences. These protections balanced the need for rich, authentic data with participants' legitimate concerns about professional exposure.

#### *4.7.2.1 Informed Consent Processes*

The informed consent process extended beyond procedural compliance to ensure genuine understanding and voluntary participation (Christians, 2018). Each participant received a detailed Plain Language Statement and Consent Form via email alongside their Zoom interview invitation, outlining the research purpose, data collection procedures, confidentiality measures, and their rights as participants. This documentation explicitly addressed concerns specific to security leadership research, including how sensitive security information would be handled and the boundaries around discussing specific incidents or vulnerabilities.

Acceptance of the calendar invitation constituted initial consent to participate, with participants having reviewed the provided documentation at their convenience before the scheduled interview. At the commencement of each session, prior to activating the recording, participants were asked to confirm their consent to proceed with the recorded interview. This two-stage approach accommodated the practical constraints of conducting research with senior executives via video conferencing whilst ensuring fully informed participation. The process proved particularly valuable given the relationship-based recruitment approach, as it clearly delineated the shift from professional interaction to research participation. Participants explicitly confirmed their understanding that whilst the researcher was a fellow CISO, the interview constituted formal research requiring candid responses rather than professional courtesy.

The voluntary nature of participation was emphasised throughout, with participants reminded of their right to decline specific questions or redirect discussions away from sensitive areas. Withdrawal rights were clearly explained, though no participant chose to withdraw following their interview.

#### *4.7.2.2 Confidentiality Safeguards*

Protecting participant and organisational identities required sophisticated approaches given the Australian CISO community's interconnected nature. Pseudonym assignment followed a systematic protocol ensuring no inadvertent connections to actual identities through naming patterns. Industry codes (such as Financial01, Manufacturing02) replaced organisation names whilst preserving analytical capacity to examine sector-specific patterns.

Organisational de-identification extended beyond simple name replacement to include removal of identifying characteristics such as specific employee numbers, unique security incidents, or distinctive organisational features that might enable identification. This process required careful balance between preserving analytical richness and ensuring confidentiality. For instance, whilst maintaining that a participant worked in financial services provided important context, specific details about their institution's size, location, or recent publicised incidents were obscured or generalised.

Industry-level reporting further protected individual identities by aggregating insights across multiple participants within sectors. This approach enabled discussion of sector-specific patterns whilst preventing association of specific comments with individual organisations. The presentation of findings employs representative quotes that capture essential insights whilst removing identifying speech patterns or distinctive terminology that might reveal participants' identities.

Beyond identity protection, the research maintained clear boundaries regarding sensitive content. Security vulnerabilities and specific technical weaknesses were explicitly excluded from discussion scope, with participants guided toward leadership and organisational dynamics rather than technical details when such topics arose. This approach maintained research focus whilst avoiding creation of documents potentially useful to threat actors. Throughout analysis and writing, each quote selection and example required careful evaluation to balance analytical richness with protective responsibility, ensuring the research's contribution to knowledge did not compromise participants who had shared sensitive professional experiences.

#### *4.7.2.3 Data Security Measures*

Rigorous data security protocols protected participant information throughout the research process, recognising that security leaders would reasonably expect exemplary data protection practices (Dalkin et al., 2021). All interviews were conducted and recorded using the University's institutional Zoom platform with password-protected recordings. Recordings were immediately transferred to secure University OneDrive storage with enterprise-level encryption and appropriate access controls aligned with University data governance policies.

The de-identification process occurred immediately upon transcription completion, with original recordings retained in secure storage for verification purposes whilst de-identified transcripts were created for analysis. Only de-identified transcripts were shared with the supervisory team through secure University channels, ensuring that supervisors could not inadvertently re-identify participants without deliberate cross-referencing of secured documents. The professional transcription service was selected based on demonstrated compliance with Australian privacy laws and robust data security measures, operating under strict commercial confidentiality agreements.

Research team access followed the principle of least privilege, with only the primary researcher having access to identified recordings and the complete de-identification matrix. Supervisors received only de-identified transcripts necessary for methodological guidance and analytical oversight. This tiered access model protected participant confidentiality whilst enabling appropriate supervisory support throughout the analytical process.

## 4.8 LIMITATIONS AND DELIMITATIONS

The transparency required of rigorous qualitative research demands explicit acknowledgement of both methodological limitations and deliberate boundary decisions that shaped this investigation. This section delineates the constraints inherent in the research design, the conscious delimitations established to maintain focus, and the mitigation strategies employed to enhance the study's credibility and transferability.

### 4.8.1 Methodological Limitations

Several limitations emerge from the qualitative, interpretive approach adopted for this investigation, each requiring careful consideration regarding its impact on the findings and their applicability.

#### *4.8.1.1 Single-Country Context*

The decision to focus exclusively on Australian organisations introduces geographic constraints to the generalisability of findings. Australia's regulatory environment, characterised by the Privacy Act amendments and the SOCI Act, creates a specific context that may not translate directly to other jurisdictions. The Westminster governance system prevalent in Australian corporations shapes board-executive relationships in ways that differ from alternative governance models. Furthermore, Australia's relatively mature cybersecurity market and established CISO community may not reflect the developmental stage of security leadership in emerging economies or regions with different technological adoption patterns (Campbell et al., 2020; Palinkas et al., 2015).

The Australian focus also means that cultural factors specific to Australian business practices influence the findings. The relatively egalitarian corporate culture, emphasis on regulatory compliance, and particular risk appetite characteristics of Australian organisations colour the experiences and perspectives captured in the data. These contextual factors inevitably shape how security leadership manifests and evolves within the studied organisations.

#### *4.8.1.2 Cross-Sectional Design*

Whilst the research captures evolution and change through participants' retrospective accounts, the cross-sectional nature of data collection presents temporal limitations. Career trajectories and organisational transformations were understood through participant recollection rather than longitudinal observation, introducing potential recall bias and retrospective sense-making that may smooth over contradictions or uncertainties experienced in real-time (Charmaz, 2014; Wolgemuth et al., 2015).

The snapshot nature of interviews, conducted between 2021 and 2024, captures a specific moment in the evolution of security leadership. This period, marked by pandemic-driven digital acceleration and heightened cyber threats, represents a particular inflection point that may amplify certain aspects of the CISO role whilst diminishing others. This temporal specificity necessarily influences the theoretical model developed, potentially emphasising crisis-responsive capabilities over steady-state leadership requirements.

#### *4.8.1.3 Sample Composition and Model Generalisability*

The composition of the sample introduces particular considerations for the generalisability of the Security Leadership Contingency Model across different regulatory and organisational contexts. The distribution of participants across industries reveals a notable, though not dominant, representation from heavily regulated sectors.

Specifically, nine of twenty participants (45%) came from industries characterised by cybersecurity-specific regulatory frameworks beyond the Privacy Act 1988 that applies universally across Australian organisations. Table 4.2 summarises the regulatory frameworks applicable to this group. These industries share characteristics that distinguish them from less regulated sectors: compliance requirements drive systematic security investment, regulatory frameworks establish minimum security capabilities, and external mandates elevate security to strategic importance.

*Table 4.2. Participants Subject to Cybersecurity-Specific Regulatory Frameworks.*

| Industry | Participants (n) | Applicable Regulatory Framework(s) |
|---|---|---|
| Financial services | 6 | APRA CPS 234 (Information Security) and sector-specific security requirements |
| Health and aged care | 1 | SOCI Act (designated critical infrastructure sector) |
| Government | 1 | Protective Security Policy Framework; SOCI Act |
| Aviation | 1 | SOCI Act |

The remaining eleven participants (55%) operated in sectors without cybersecurity-specific regulatory requirements beyond privacy legislation. This majority included manufacturing (4 participants), professional services (2), higher education (2), food services (1), retail (1), and information technology services (1). In these contexts, security investment decisions derive primarily from business risk assessment, competitive positioning, and internal governance rather than regulatory mandate. This diversity proved methodologically valuable, enabling examination of security leadership across varying drivers for the CISO role.

The regulatory distribution within the sample creates both potential limitations and analytical strengths. The presence of participants from heavily regulated industries may influence certain aspects of the findings. The stewardship leadership orientation, emphasising compliance and risk management, may resonate particularly strongly in regulated contexts where these capabilities directly address external requirements. Similarly, the organisational maturity progression identified in the model (establishment, maturation, strategic integration) may follow more predictable trajectories in regulated industries where compliance deadlines and regulatory expectations create structured development paths.

However, the fact that the majority of participants operated outside heavily regulated environments strengthens rather than undermines confidence in the model's core propositions. The dual-fit framework (CISO-Organisation fit and CISO-Environment fit) emerged consistently across both regulatory contexts, with participants from manufacturing, consulting, education, and technology sectors describing fundamentally similar dynamics of alignment between leadership approach and organisational needs. The necessity of maintaining simultaneous person-organisation and person-environment fit transcended regulatory intensity, appearing equally relevant whether security investment stemmed from compliance requirements or business imperatives.

Furthermore, several participants had worked across multiple industries during their careers, including transitions between heavily regulated and minimally regulated sectors. These participants explicitly reflected on how their leadership approaches needed adaptation when moving between regulatory contexts, whilst the fundamental importance of dual-fit alignment remained constant. This within-participant variation provides additional confidence that regulatory context moderates specific manifestations of security leadership rather than determining whether the dual-fit dynamic operates.

The sample composition does suggest boundary conditions requiring acknowledgement. The Security Leadership Contingency Model, whilst grounded in diverse organisational contexts, may not fully capture dynamics in environments with characteristics underrepresented in the sample. Industries facing rapid technological disruption without corresponding regulatory drivers, organisations in jurisdictions with minimal security compliance requirements, or sectors where security remains purely technical rather than strategic may exhibit patterns requiring model extension. Future research examining security leadership in minimally regulated industries across different jurisdictions, technology start-ups prioritising innovation over compliance, or emerging markets with less developed security governance could test whether the dual-fit framework holds whilst potentially revealing additional contingencies.

#### *4.8.1.4 Sample Characteristics*

The purposive sampling strategy, whilst appropriate for accessing elite perspectives, creates inherent limitations in representativeness. The focus on senior security executives in established organisations excludes early-career security leaders, those in smaller enterprises, and professionals who may have left security leadership roles due to various challenges. This survivor bias potentially skews findings toward successful navigation strategies whilst obscuring the full spectrum of security leadership experiences, including failures and departures (Palinkas et al., 2015; Tongco, 2007).

The requirement for current or recent (post-2020) CISO experience further constrains the sample. Security leaders who transitioned to other executive roles, moved to consulting, or left the profession entirely remain unrepresented. Their perspectives might reveal different dynamics, particularly regarding the sustainability of security leadership careers and the factors contributing to role exit.

##### *4.8.1.5 Power Dynamics and Disclosure*

The elite nature of participants introduces considerations of power and disclosure that may influence data richness (Cassell et al., 2020; Einola & Alvesson, 2021). Senior executives, acutely aware of professional reputation and organisational sensitivities, may self-censor when discussing failures, conflicts, or politically sensitive navigation strategies. Despite assurances of confidentiality, the relatively small Australian CISO community heightens concerns about identification, potentially limiting candour about controversial or challenging experiences.

The researcher's position as a practising CISO, whilst facilitating access and rapport, may paradoxically constrain disclosure. Participants might assume shared understanding, leaving important assumptions unexamined, or avoid admitting uncertainties that could affect professional standing within the peer community. This peer dynamic differs from researcher-participant relationships where professional distance might encourage fuller disclosure.

##### *4.8.1.6 Self-Reported Data*

The reliance on self-reported experiences and perceptions, without triangulation through organisational documents or stakeholder perspectives, presents another limitation. Participants' accounts of their influence and organisational impact remain unverified against objective measures or alternative viewpoints. The absence of board members', executives', or team members' perspectives means the construction of security leadership relies solely on CISOs' own narratives and interpretations.

This self-reporting limitation proves particularly relevant when participants discuss successful influence attempts or strategic victories. Social desirability bias may lead to overemphasis on successes whilst minimising failures or ongoing challenges. The confidential nature of many security incidents further complicates verification, as participants cannot provide documentary evidence without breaching organisational confidentiality. This necessary discretion potentially limits the depth of insight into the more challenging aspects of security leadership.

#### *4.8.1.7 Researcher Interpretation*

The interpretive nature of qualitative analysis introduces subjectivity that, whilst managed through systematic procedures, cannot be entirely eliminated. As a practising CISO conducting research on fellow security leaders, the researcher brings both valuable insider knowledge and potential interpretive biases. The analytical process, whilst following established procedures and benefiting from supervisory challenge, ultimately involves researcher judgement in identifying patterns, developing themes, and constructing theoretical relationships (Charmaz, 2014; Gioia et al., 2013; Lincoln & Guba, 1985).

The translation from first-order concepts to second-order themes and ultimately to aggregate dimensions involves progressive abstraction that moves away from participants' direct language. Each level of abstraction introduces interpretive decisions that, whilst grounded in data and subjected to scrutiny, reflect the researcher's theoretical sensitivity and analytical choices.

### 4.8.2 Delimitations and Boundaries

Beyond inherent limitations, several deliberate boundaries were established to maintain research focus and enable depth of investigation within manageable scope.

#### *4.8.2.1 Scope Boundaries*

The exclusive focus on executive-level security leaders represents a conscious delimitation. Middle management security professionals, security architects, and operational security staff were excluded despite their important roles in security programme delivery. This boundary ensures that the investigation captures strategic leadership perspectives rather than operational management challenges, though it necessarily excludes insights about how security leadership translates through organisational layers.

The restriction to established organisations, explicitly excluding start-ups and small enterprises, creates another important boundary. Start-up environments, with their distinct risk appetites, resource constraints, and organisational fluidity, likely produce different security leadership dynamics. Similarly, the exclusion of vendor organisations removes the commercial dimension of security leadership, where customer-facing responsibilities and product development considerations alter the fundamental nature of the role.

The post-2020 timeframe for participant experiences ensures currency but limits historical perspective. Whilst participants discussed career histories extending earlier, the focus on current and recent roles means the findings primarily reflect contemporary security leadership rather than its historical evolution. This temporal boundary proves particularly relevant given the rapid evolution of cyber threats and regulatory requirements.

The English-language interview requirement, whilst practical for transcription and analysis, potentially excludes perspectives from non-English-speaking security leaders who might bring different cultural or operational insights. This linguistic boundary intersects with the geographic limitation to create a relatively homogeneous cultural sample.

#### *4.8.2.2 Theoretical Boundaries*

The investigation maintains focus on leadership at individual and organisational levels, deliberately excluding team dynamics and departmental structures. Whilst security teams undoubtedly influence leadership outcomes, the complexity of team-level analysis would have diluted the focus on executive leadership. This boundary means the findings do not address how CISOs build, manage, and leverage their teams for organisational impact.

The emphasis on leadership and influence processes deliberately minimises attention to technical competencies. Whilst technical knowledge emerged as relevant for credibility, the research does not systematically investigate the relationship between technical expertise and leadership. This delimitation reflects the study's positioning within leadership rather than information security literature.

The exclusion of specific security frameworks, maturity models, and technical architectures represents another theoretical boundary. Whilst participants referenced various frameworks, the research does not evaluate their relative merits or implementation approaches. This boundary maintains focus on leadership processes rather than security programme content.

### 4.8.3 Mitigation Strategies

Acknowledging these limitations and delimitations, several strategies were employed to enhance the credibility and potential transferability of findings.

#### *4.8.3.1 Enhancing Transferability*

To address geographic limitations, thick description provides sufficient detail for readers to assess transferability to their contexts (Gioia et al., 2013; Lincoln & Guba, 1985). The comprehensive presentation of participant perspectives, organisational contexts, and environmental conditions enables informed judgement about applicability beyond Australian settings. The Australian context, whilst specific, shares important characteristics with other Commonwealth nations, particularly the United Kingdom, Canada, and New Zealand, where similar governance structures and regulatory approaches prevail.

The detailed specification of boundary conditions throughout the analysis enables appropriate theoretical claims. Rather than asserting universal applicability, the findings explicitly acknowledge their grounding in particular organisational and cultural contexts, inviting future research to test and extend the theoretical framework in different settings.

#### *4.8.3.2 Cross-Industry Comparison as Mitigation*

The deliberate pursuit of industry diversity within the sample, whilst revealing the regulatory concentration discussed above, also served as a mitigation strategy for identifying universal versus context-specific patterns. The cross-case analytical approach systematically compared experiences across regulatory environments, explicitly examining whether patterns transcended or were contingent upon regulatory intensity.

This comparative analysis revealed that the core dual-fit framework (CISO-Organisation fit and CISO-Environment fit) operated consistently across all industries represented, including less regulated sectors. Participants from technology and manufacturing contexts, despite operating under lighter regulatory requirements, still described the fundamental importance of alignment between their leadership approach and organisational needs. The consistency of this pattern across diverse regulatory contexts strengthens confidence in the model's core theoretical propositions whilst acknowledging that specific manifestations may vary.

Furthermore, the inclusion of participants who had worked across multiple industries during their careers provided within-participant comparison of regulatory contexts. These participants explicitly reflected on how their leadership approaches needed to adapt when moving between heavily regulated and minimally regulated environments, offering insights into regulatory context as a moderating variable rather than a determinant of the core dual-fit dynamic.

#### *4.8.3.3 Ensuring Analytical Rigour*

Multiple strategies addressed the limitations of researcher interpretation. The systematic application of the Gioia methodology provided structured analytical procedures that enhance transparency and replicability. The maintenance of a clear audit trail, from raw data through analytical decisions to theoretical claims, enables external scrutiny of the interpretive process.

Regular supervisory meetings throughout analysis provided critical challenge to emerging interpretations. Supervisors, less embedded in the phenomenon than the researcher, questioned assumptions, challenged interpretations, and demanded evidence for theoretical claims. This supervisory involvement proved particularly valuable in maintaining analytical distance from phenomena familiar to the researcher through professional practice.

The participation of one supervisor in early interviews provided additional quality assurance, ensuring that questioning techniques avoided leading participants toward predetermined concepts. This direct supervisory involvement in data collection helped establish rigorous interviewing practices that maintained openness to emergent insights.

#### *4.8.3.4 Managing Scope*

The clear specification of boundaries enables appropriate theoretical claims whilst acknowledging what remains unexplored. By explicitly stating what the research does and does not address, the findings avoid overreach whilst identifying areas requiring future investigation. This boundary clarity proves essential for positioning the contributions within existing literature whilst highlighting opportunities for extension.

The focus on senior executive perspectives, whilst limiting breadth, enables depth of insight into strategic leadership challenges. This depth reveals nuanced dynamics that broader sampling might have obscured, justifying the elite focus whilst acknowledging its constraints.

### 4.8.4 Implications for Interpretation

These limitations and delimitations necessarily influence how the findings should be interpreted and applied. The Security Leadership Contingency Model (SLCM) emerging from this research represents a theoretical framework grounded in the experiences of successful senior security executives in established Australian organisations during a specific historical period. Its applicability beyond these boundaries requires empirical testing and potential adaptation. Additional limitations relating to the theoretical model are discussed in Chapter 7.

The model's emphasis on dual-fit maintenance (CISO-Organisation fit and CISO-Environment fit) and political navigation reflects the realities of executive leadership in mature organisational contexts. Start-ups, rapidly growing companies, or organisations in crisis may experience different dynamics requiring modified theoretical frameworks. Similarly, security leaders at different career stages or organisational levels may face distinct challenges not captured in this executive-focused investigation.

The temporal specificity of the research, conducted during a period of heightened cyber awareness and accelerated digital transformation, may amplify certain aspects of security leadership whilst understating others. As the security landscape continues evolving, the relative importance of different model components may shift, requiring ongoing theoretical refinement.

Despite these boundaries, the systematic methodology, rigorous analysis, and theoretical grounding provide confidence in the findings within their specified scope. The limitations do not invalidate the contributions but rather define their appropriate application domain. By acknowledging what this research can and cannot claim, the study maintains the intellectual honesty required of doctoral scholarship whilst offering valuable insights into an under-researched phenomenon of increasing organisational importance.

## 4.9 CHAPTER SUMMARY

This chapter has established the methodological foundation for investigating the role of the Chief Information Security Officer in contemporary organisations. The research design responds to the empirical vacuum identified in Chapters 2 and 3, employing constructivist grounded theory operationalised through the Gioia methodology to enable systematic theory development from practitioner experiences. This methodological choice proves particularly appropriate for exploring an emerging executive position where established theoretical frameworks remain absent.

The combination of purposive sampling, semi-structured interviews, and systematic analytical procedures transformed twenty interviews with senior security leaders into robust theoretical understanding. Research quality strategies including supervisory challenge, triangulation, and reflexive management of the practitioner-researcher position ensure that emergent theory remains grounded in empirical data whilst achieving necessary analytical abstraction. The transparent acknowledgement of methodological limitations and deliberate boundaries establishes appropriate scope for theoretical claims.

The methodological rigour detailed in this chapter provides the foundation for the empirical findings presented in Chapter 5 and the Security Leadership Contingency Model developed in Chapter 6, demonstrating how systematic qualitative research can generate novel theoretical insights into complex organisational phenomena.

# CHAPTER 5: EMPIRICAL FINDINGS

## 5.1 INTRODUCTION: POSITIONING THE EMPIRICAL JOURNEY

This chapter represents the empirical heart of our investigation into security leadership effectiveness. Through systematic analysis of twenty in-depth interviews with Chief Information Security Officers across diverse Australian organisations, we uncover the complex, contingent nature of the CISO role. Our analysis reveals how security leadership effectiveness emerges not from universal characteristics but from dynamic alignment between leaders and their contexts. This introductory section establishes the foundation for understanding how we developed the Security Leadership Contingency Model through rigorous grounded theory analysis, providing a clear pathway from empirical observation to theoretical contribution.

### 5.1.1 Chapter Purpose and Scope

The research question guiding this investigation, “How is the CISO role enacted across different organisational contexts in modern organisations?”, yields an answer of considerable theoretical depth. Our empirical journey reveals that the CISO role cannot be understood through static descriptions or universal prescriptions. Instead, effectiveness emerges from maintaining dual dynamic fits: between CISOs and their organisational contexts, and between CISOs and their environmental demands. This chapter presents the systematic uncovering of these relationships through grounded theory analysis, culminating in the Security Leadership Contingency Model that reconceptualises security leadership as requiring continuous dual-fit maintenance.

Our interpretivist stance recognises that understanding the CISO role requires deep engagement with practitioners’ lived experiences. We prioritise participants’ own interpretations and meaning-making processes, allowing theoretical insights to emerge from their narratives rather than imposing predetermined frameworks. This approach proves essential for capturing the dynamic, contextual nature of security leadership effectiveness. The transparency of our analytical process enables readers to evaluate the empirical grounding of each theoretical claim, ensuring the integrity of our theory-building endeavour.

The chapter follows a carefully structured progression. We begin with first-order concepts drawn directly from participant accounts, advance through second-order themes that reveal cross-case patterns, and ultimately arrive at two aggregate dimensions: CISO-Organisation Fit and CISO-Environment Fit, that form our theoretical framework. This systematic presentation demonstrates how abstract theoretical constructs remain firmly grounded in empirical reality.

### 5.1.2 Contextualising the Inquiry

Our investigation unfolds within the Australian organisational landscape during an era of rapid digital transformation. The study encompasses twenty CISOs and heads of cybersecurity from ASX-listed companies and comparable organisations, representing financial services, manufacturing, healthcare, aviation, retail, and technology sectors. This sectoral diversity proves crucial; our findings reveal how industry context profoundly shapes security leadership requirements and effectiveness patterns.

The temporal context, the post-COVID digital acceleration era, captures security leadership at a pivotal moment. Participants navigate rapidly evolving threat landscapes, accelerated cloud adoption, distributed workforce challenges, and shifting regulatory expectations. These environmental pressures create natural variation in leadership contexts, enabling rich insights into how security leaders adapt to dynamic conditions. As one participant noted, "The pandemic changed everything; suddenly we were defending a perimeter that no longer existed."

The participant profile, presented in Table 4.1 (Section 4.3.2), reflects the breadth of perspectives captured. Participants range from newly appointed leaders transforming nascent security functions to seasoned executives driving strategic integration in mature organisations. This variation in experience levels, organisational contexts, and industry settings enables comprehensive exploration of the CISO role across its full spectrum.

The Australian context provides a bounded yet diverse setting for theory development. Whilst this geographic focus represents a limitation, it enables deep understanding of how security leadership operates within a coherent regulatory framework and business culture. The theoretical propositions we develop from this context offer foundations for comparative studies in other national settings, contributing to cumulative knowledge development.

### 5.1.3 Bridging from Methodology

The application of the Gioia methodology (detailed in Section 4.2.2) proves particularly suited to exploring the CISO role's emergent nature. The systematic yet flexible approach honours participants' voices whilst developing theoretically rigorous abstractions. Through iterative cycles of data collection, coding, and theoretical reflection, initial interviews generate provisional concepts that inform subsequent data collection, whilst constant comparison across cases reveals patterns and contradictions demanding theoretical explanation.

Through this process, we made a pivotal discovery: effectiveness depends not on leadership characteristics alone but on maintaining dual fits with both organisational and environmental contexts. Participant narratives consistently emphasise alignment and misalignment experiences. Phrases such as “I’m not really a steady state one” and “the organisation evolved beyond my sweet spot” signal the inadequacy of static role conceptualisations. Through systematic analysis, we develop a dynamic model capturing how security leaders must continuously maintain both fit dimensions.

The progression from empirical observation to theoretical abstraction involves multiple analytical pivots. Initial coding focuses on activities and responsibilities, yet participants redirect attention to questions of timing, evolution, and fit. This empirical redirection proves theoretically generative, leading to insights that predetermined frameworks would have obscured. The transparency of this analytical journey ensures our theoretical contributions bridge the oft-lamented divide between academic theory and practitioner experience.

## 5.2 DEMONSTRATING ANALYTICAL RIGOUR

We developed the Security Leadership Contingency Model through systematic analytical processes grounded in established qualitative research principles. This section provides transparency into the analytical journey, demonstrating how we systematically derived theoretical insights from empirical data. We maintained the interpretive integrity essential to understanding CISO leadership effectiveness throughout this process.

### 5.2.1 The Iterative Analytical Process

#### *5.2.1.1 Timeline and Progression*

The analytical journey spanned the duration of this part-time doctoral study, with data collection occurring between 2021 and 2024. This extended timeframe presented both challenges and opportunities. The challenges included maintaining analytical consistency across years; however, the opportunities for deep reflection between interviews enriched the theoretical development. The longitudinal nature of data collection coincided with significant cybersecurity landscape evolution, including emerging threat patterns and regulatory developments, thus providing rich contextual understanding of CISO role transformation.

#### *5.2.1.2 Critical Analytical Turning Points*

The analysis underwent several transformative moments that fundamentally shaped theoretical outcomes:

**After Interviews 4-6: Initial Pattern Recognition.** Early analysis catalogued role descriptions and responsibilities, generating numerous first-order concepts. A critical shift emerged when participants consistently employed alignment language. Phrases such as *growing beyond the role*, *not the right fit anymore*, and *evolving with the organisation* appeared across transcripts. These patterns signalled deeper dynamics requiring exploration beyond surface-level role descriptions.

**After Interview 8: The 'Fit' Revelation.** Interview 8 provided a transformative articulation when the participant stated: "I'm more of an agent of change… I'm someone you bring in to uplift or change directions. I'm not really a steady state one" (Aviation01). This statement crystallised the matching pattern between leadership orientations and organisational contexts. The recognition triggered fundamental analytical reorientation from descriptive categorisation toward dynamic relationship exploration.

**After Interviews 10-12: Temporal Dynamics Discovery.** Our analysis revealed that alignment evolved temporally rather than remaining static. Participants described effectiveness changes as organisations matured, despite consistent leadership approaches. One participant reflected: "So my role probably changed a lot in the 18 months that I've been in my position" (BuildingMat01). This insight led us to incorporate temporal dimensions into the emerging framework, moving beyond static fit conceptualisations.

**After Interview 15: Environmental Complexity.** Through negative case analysis, we identified environmental factors as critical dimensions. Participants in highly regulated industries described external pressures overriding natural phase-leadership alignment. These cases revealed that regulatory requirements could force particular approaches regardless of organisational maturity, adding essential complexity to the model.

#### *5.2.1.3 Constant Comparison and Theoretical Sampling*

The analytical process employed constant comparison across multiple levels. Within-case analysis examined each interview for internal consistency and contradictions. Through cross-case comparison, we identified patterns across participants from similar industries and organisational phases. Cross-industry validation tested emerging patterns against diverse sectoral contexts. Through negative case analysis, we deliberately sought disconfirming evidence to refine theoretical boundaries.

Theoretical sampling guided later interviews, with participants selected to explore emerging dimensions. Following our identification of three leadership orientations, we specifically recruited participants representing potential hybrid approaches. This purposive sampling challenged discrete categorisations and enriched theoretical understanding.

### 5.2.2 Critical Incidents in Theoretical Development

Several critical incidents fundamentally shaped theoretical sensitivity:

**The “Agent of Change” Moment (After Interview 8).** When a participant articulated *I’m more of an agent of change*, this triggered our recognition of leadership typologies beyond technical-strategic dichotomies. This moment illustrates how participant language can crystallise emerging patterns.

**The Misalignment Revelation (After Interview 10).** Analysis of participants’ descriptions of role evolution highlighted the temporal dynamics of fit. Several participants described how their effectiveness changed as organisations matured, or how they preferred transformation over steady-state operations. This pattern of evolving alignment requirements led us to move beyond static fit concepts.

**The Political Capital Insight (After Interview 13).** Our analysis of a negative case, where a visionary CISO succeeded in an establishment phase, revealed political capital as a critical moderating factor, adding complexity to the emerging model.

### 5.2.3 Theoretical Frameworks: Recognition and Integration

#### *5.2.3.1 Emergence of Theoretical Resonance*

We identified relevant theoretical frameworks organically through analytical progression rather than predetermined selection. This emergence process exemplified grounded theory principles.

Our recognition of patterns triggered theoretical connections at specific analytical junctures. After interview 8, persistent *matching* and *misalignment* patterns evoked P-O fit theory relevance; however, traditional static conceptualisations proved insufficient for observed dynamics. After interview 10, participant descriptions of context-dependent leadership effectiveness suggested contingency theory’s explanatory power. After interview 13, capability development narratives pointed toward dynamic capabilities frameworks. After interview 15, multiple role transition references evoked role theory, particularly regarding ambiguity during organisational transitions.

#### *5.2.3.2 Integration Process*

Integrating multiple theoretical lenses required careful analytical discipline. We ensured frameworks were recognised for explanatory power rather than forced onto data. Iterative literature consultation occurred as patterns emerged, with supervisory discussions challenging theoretical abstractions. The decision to integrate multiple theories reflected the phenomenon's complexity, as single frameworks proved insufficient for explaining observed dual-fit dynamics.

This analytical journey exemplifies systematic rigour in qualitative research. The transparency provided enables assessment of finding credibility and potential transferability whilst acknowledging applicable boundaries. The resulting Security Leadership Contingency Model represents theoretical innovation grounded in empirical observation, demonstrating how rigorous analysis generates novel contributions whilst maintaining strong empirical foundations.

This analytical journey, from initial pattern recognition through theoretical integration, exemplifies the systematic rigour required for trustworthy qualitative research. The transparency provided here enables readers to assess both the credibility of findings and their potential transferability to other contexts, whilst acknowledging the boundaries within which these insights apply.

## 5.3 THE ARCHITECTURE OF FINDINGS: DATA STRUCTURE PRESENTATION

### 5.3.1 Introducing the Emergent Structure

Through constant comparative analysis, we organised first-order concepts into sixteen second-order themes, then organised these themes into three primary contextual groupings: Organisational Context, CISO Characteristics, and Environmental Context. At the highest level of abstraction, we constructed two aggregate dimensions through systematic analysis of relational patterns across these groupings: CISO-Organisation Fit and CISO-Environment Fit.

We constructed the aggregate dimensions not as hierarchical containers for the contextual groupings but rather as theoretical constructs that explain relational dynamics we observed across them. Whilst the three contextual groupings (Organisational Context, CISO Characteristics, Environmental Context) describe what elements exist and how they cluster, we developed the two aggregate dimensions to capture dynamic relationships between these elements. The groupings provide the substantive content, the constituent factors that matter for security leadership. We constructed the aggregate dimensions to conceptualise how these factors interact to produce effectiveness outcomes through dual-fit dynamics. This distinction between descriptive taxonomy (groupings) and relational dynamics (dimensions) proved essential for theoretical development, enabling us to move from cataloguing what matters to understanding how effectiveness emerges.

We developed this structure through iterative engagement with the data rather than from predetermined theoretical categories. Multiple rounds of coding, comparison, and refinement ensured that the emergent framework remained grounded in participant experiences whilst achieving theoretical abstraction.

Theoretical saturation manifested through diminishing conceptual emergence as the analysis progressed. Early interviews yielded numerous new concepts and relationships, whilst later interviews primarily confirmed and refined existing categories. By interview seventeen, no new concepts emerged, and the relationships between concepts had stabilised. The final three interviews served as validation exercises, confirming saturation across all categories and strengthening confidence in the emergent structure.

### 5.3.2 Visual Presentation: The Complete Data Structure

The complete data structure encompasses over sixty first-order concepts systematically organised into sixteen second-order themes, culminating in two aggregate dimensions that form the foundation of the Security Leadership Contingency Model. This visual representation, shown in figure 5.1, demonstrates the analytical progression from empirical observation to theoretical abstraction whilst maintaining clear connections to participant voice.

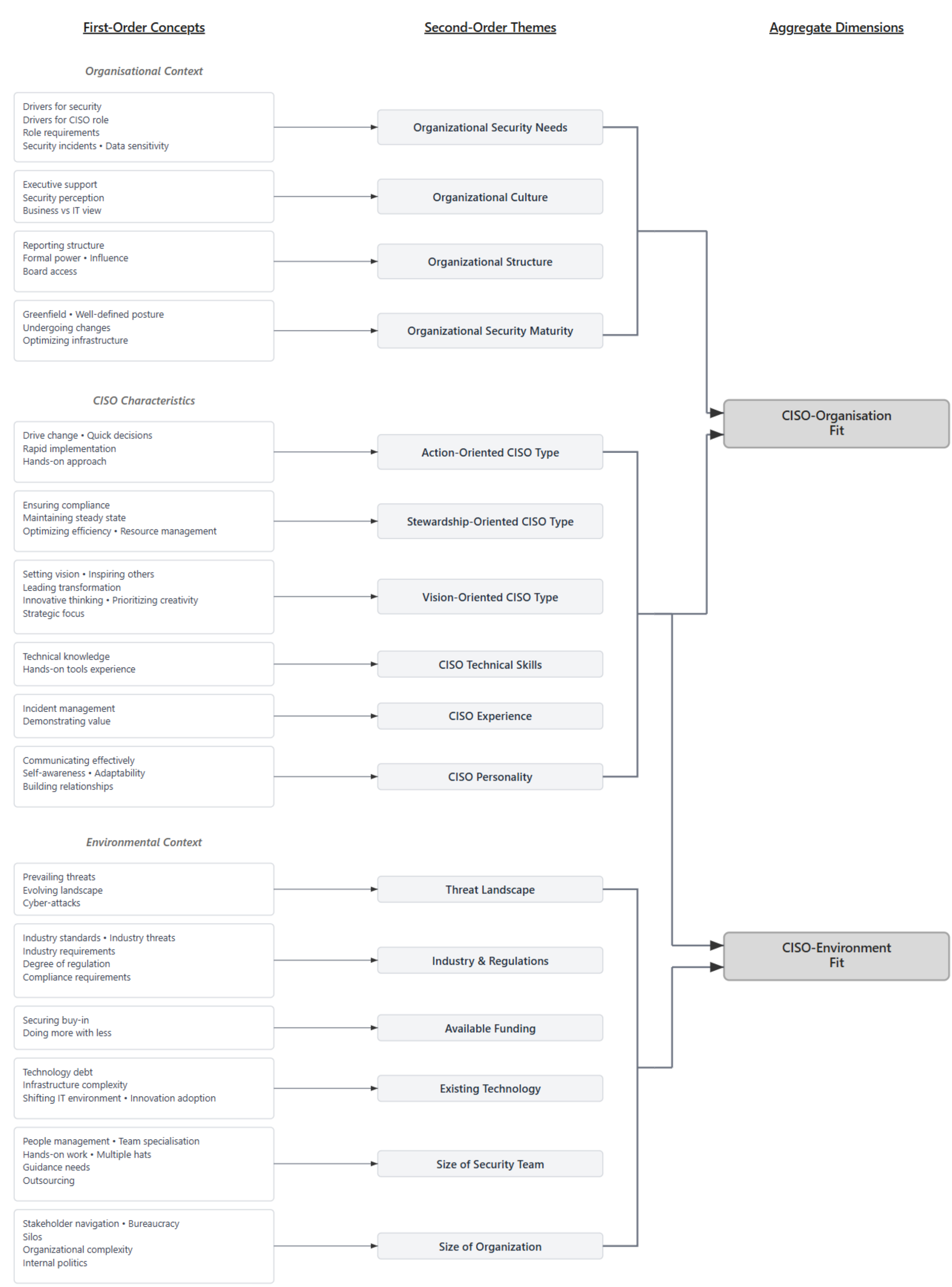

*Figure 5.1. Complete Gioia Data Structure - CISO Leadership Effectiveness.*

### *5.3.2.1 Organisational Context Grouping*

The organisational context grouping comprises four second-order themes capturing institutional factors that shape security leadership requirements.

**Organisational Security Needs.** We developed this theme from first-order concepts including drivers for security, drivers for CISO role, role requirements, security incidents, and data sensitivity. Participants consistently described how fundamental organisational imperatives created specific demands for security leadership. As one CISO noted: "The evolving threat landscape and the risk that cyber posed the organisation" (BuildingMat01) drove the establishment of their role.

**Organisational Culture.** We constructed this theme from concepts addressing executive support patterns, security perception, and the critical distinction between viewing security as a business versus IT concern. This theme captures the shared beliefs and values shaping how organisations understand and value security leadership. The contrast emerged clearly when participants described their experiences: "I've been quite fortunate in that I got senior leadership buy in from day one" (Aviation01) versus "I report into a CIO and whilst a CISO reports into a CIO, it's always going to be seen as technology problem" (BuildingMat01).

**Organisational Structure.** We developed this theme from concepts examining reporting structures, formal power arrangements, influence mechanisms, and board access. These structural elements define the formal positioning and authority granted to security leadership, directly affecting their capacity to enact change.

**Organisational Security Maturity.** We constructed this theme from participant descriptions ranging from greenfield implementations to well-defined programmes. Organisations were characterised as establishing basic capabilities, undergoing transformation, or optimising existing infrastructure, with each state creating distinct leadership requirements.

### *5.3.2.2 CISO Characteristics Grouping*

The CISO characteristics grouping encompasses six second-order themes capturing individual leadership attributes and orientations.

**Leadership Orientation.** We identified three distinct leadership types inductively from the data:

*Action-Oriented CISO Type* characterised by transformation focus, hands-on implementation, and rapid change orientation. These leaders described themselves in active terms: "I'm more of an agent of change, head of cyber. I'm someone you bring in to uplift or change directions" (Aviation01).

*Stewardship-Oriented CISO Type* distinguished by compliance focus, operational excellence, process optimisation, and measurement orientation. Their descriptions emphasised systematic approaches: "Identifying what our current level of capability maturity is helps us to identify what our gaps are" (AgedCare1).

*Vision-Oriented CISO Type* defined by strategic focus, business alignment, innovation enablement, and future-state orientation. These leaders articulated broader organisational impact: "I think that's one of the things as security leaders that we should be able to articulate. The 'so what' factor" (MgtConsult01).

**CISO Technical Skills.** We consolidated concepts addressing technical knowledge depth and hands-on tools experience.

**CISO Experience.** We constructed this theme from practical wisdom accumulated through incident management and value demonstration.

**CISO Personality.** We developed this theme from concepts addressing communication effectiveness and self-awareness of personality type's impact on leadership approach.

#### *5.3.2.3 Environmental Context Grouping*

The environmental context grouping comprises six second-order themes capturing external forces shaping security leadership contexts.

**External Threat Landscape.** We developed this theme from participant descriptions of evolving threats and expanding attack surfaces.

**Industry & Regulations.** We constructed this theme to capture compliance requirements and sector-specific pressures.

**Available Funding**, **Existing Technology**, **Size of Security Team**, and **Size of Organisation.** Each represented contextual factors that participants identified as significantly influencing their leadership approaches and effectiveness.

#### *5.3.2.4 Aggregate Dimensions*

The three contextual groupings converge to form two aggregate dimensions capturing the dynamic nature of security leadership effectiveness:

**Aggregate Dimension 1: CISO-Organisation Fit.** We constructed this dimension to represent alignment between CISO characteristics and organisational context. This dimension emerged from observed patterns of “matching” and “misalignment” between leader capabilities and organisational requirements. We extended person-organisation fit theory by incorporating temporal dynamics, as both leaders and organisations evolve.

**Aggregate Dimension 2: CISO-Environment Fit.** We constructed this dimension to represent alignment between CISO capabilities and environmental demands. This dimension captures how leaders adapt to external pressures including threats, regulations, and resource constraints. We extended person-environment fit theory to the security leadership context, recognising that effectiveness requires continuous environmental adaptation.

### 5.3.3 Structural Genesis and Refinement

The journey from initial coding to final data structure involved multiple iterations of refinement and abstraction. Initial coding generated extensive concepts through line-by-line analysis of interview transcripts. This expansive conceptual landscape underwent systematic reduction through constant comparison, identifying patterns of similarity and difference across participant accounts.

A pivotal analytical moment occurred when phase-based patterns in organisational needs became apparent. Rather than treating organisational contexts as static, the data revealed dynamic progression through identifiable phases, each creating distinct leadership requirements. This insight fundamentally shaped the emerging theoretical framework, suggesting that leadership effectiveness depends on dynamic alignment as organisations evolve.

Negative cases proved particularly valuable in refining the structure. Several participants succeeded despite apparent misalignment between their leadership orientation and organisational phase. These cases revealed the moderating influence of political capital and formal authority. Rather than undermining the emerging pattern, these exceptions illuminated boundary conditions and contingency factors that enrich theoretical understanding.

The validation phase with final participants served dual purposes: confirming theoretical saturation and validating the interpretive framework. These participants, selected for diverse organisational contexts and leadership experiences, reviewed and confirmed the resonance of the emergent structure with their lived experiences.

### 5.3.4 Methodological Rigour in Structure Development

Development of this data structure exemplified methodological rigour through systematic application of Gioia principles. First-order concepts maintained fidelity to participant voice through extensive use of in vivo codes. The transition to second-order themes required theoretical reasoning whilst remaining empirically grounded. We constructed aggregate dimensions through highest-level abstraction, connecting empirical patterns to theoretical constructs without abandoning the phenomenological foundation.

Supervisory oversight and peer debriefing provided critical validation throughout the analytical journey. Regular analytical memos documented decision-making processes, creating an audit trail enhancing trustworthiness. NVivo software facilitated systematic coding procedures whilst enabling complex queries to identify patterns and test emerging relationships.

The recognition that both CISO-Organisation fit and CISO-Environment fit must be maintained simultaneously emerged as a critical insight. This dual-fit dynamic represents a significant departure from traditional fit theories that examine these dimensions in isolation. Furthermore, the pervasive role of power and politics emerged throughout all relationships, suggesting that organisational politics constitute the medium through which alignments are negotiated and maintained.

### 5.3.5 From Structure to Process: Implications for Theory Building

The data structure provides a comprehensive taxonomy of relevant concepts, yet analysis revealed that static categorisation cannot fully capture security leadership effectiveness dynamics. The structure establishes what elements matter, but relationships between elements, particularly temporal and contextual dependencies, necessitate a process model.

Our construction of two aggregate dimensions rather than multiple discrete categories reflects sophisticated understanding of how security leadership operates. The dual-fit framework suggests leaders must navigate multiple alignment challenges simultaneously. Our recognition that both fits require continuous maintenance rather than one-time achievement represents fundamental reconceptualisation of fit theory in security leadership contexts.

This structural presentation serves as both culmination and transition. It culminates the categorisation phase whilst transitioning toward dynamic theoretical modelling. The progression from structure to process, from taxonomy to theory, demonstrates the Gioia method's capacity for building grounded theory that honours empirical complexity whilst achieving theoretical elegance. The two aggregate dimensions provide the conceptual foundation for the Security Leadership Contingency Model, demonstrating how these fits operate dynamically across evolving organisational and environmental contexts.

## 5.4 UNPACKING THE EMPIRICAL FOUNDATION: FIRST-ORDER DISCOVERIES

### 5.4.1 Navigating the First-Order Landscape

The empirical journey begins with the authentic voices of twenty Australian CISOs, each contributing distinct perspectives on their roles, challenges, and organisational contexts. This section presents the foundational layer of the Gioia data structure: first-order concepts emerging directly from participants' words and experiences. Rather than imposing theoretical frameworks prematurely, these concepts preserve the language and meaning-making of security leaders navigating complex organisational environments.

The presentation follows a deliberate progression from organisational context through CISO characteristics to environmental factors. This structure emerged naturally from the data, reflecting participants' conceptualisation of effectiveness as contingent upon the interplay between organisational setting, personal leadership approaches, and broader environmental pressures. Direct quotations maintain the Australian vernacular that characterised many interviews, preserving both content and cultural context. Representative quotes supporting each first-order concept are compiled in Appendix C: Representative Quotes

The twenty security leaders represented in this analysis operated across markedly different organisational and environmental contexts. Organisational scale ranged from about 500 employees to more than 120,000, creating vastly different stakeholder landscapes, political dynamics, and resource constraints. Security team sizes varied from solo practitioners building initial capability to established functions exceeding 85 professionals. Seven participants joined greenfield environments requiring foundational security programmes; others inherited mature capabilities demanding optimisation and business integration. Industry contexts spanned from minimally regulated manufacturing and hospitality to heavily scrutinised financial services where regulatory compliance dominated priorities. Reporting structures placed some security leaders three or four organisational levels removed from chief executives, whilst others maintained direct board access and executive committee membership. Role tenure ranged from nine months to more than six years, with several participants at inflection points: newly appointed leaders establishing credibility, mid-tenure practitioners navigating organisational change, and departing leaders reflecting on lessons learned. Geographic scope varied from single-site operations to international responsibilities spanning 140 countries across multiple business units. This contextual diversity, whilst creating analytical complexity, proved essential for theoretical development: patterns emerging across such varied circumstances demonstrated robustness whilst contextual variations refined boundary conditions and contingent relationships. The first-order concepts presented below preserve participants' language and meaning-making across this diverse landscape.

## 5.4.2 Organisational Context Concepts

The organisational context emerged as a multifaceted construct, with participants describing various aspects that shaped their role expectations and effectiveness boundaries. From these participant accounts, we developed four second-order themes: organisational security needs, organisational culture, organisational structure, and organisational security maturity.

### *5.4.2.1 Organisational Security Needs - First-Order Concepts*

**Drivers for security**. Participants articulated diverse motivations underlying their organisations' security investments. One CISO who joined as the first security executive to build capability from a three-person junior team noted the threat landscape emerged as a primary driver:

*"... it was all purely based around, I guess, the evolving threat landscape and the risk that cyber posed the organization. There was very little focus there or drive from a compliance perspective"* (BuildingMat01).

Yet customer demands often proved equally influential. A CISO operating as a solo function through facilitation and project management explained:

*"One of the primary reasons is that, when customers ask and if a regulator asks, they can be completely transparent and proud as an organization of how they're going about protecting the information"* (MgtConsult01).

This diversity revealed fundamental differences in how organisations conceptualised security value.

**Drivers for CISO role**. The impetus for establishing CISO positions varied considerably. Risk elevation drove many appointments:

*"I think the role is ultimately responsible and in some ways accountable for managing cyber risk. I think cyber risk in our organization sits on the enterprise risk register. And it's within our top ten risks"* (BuildingMat01).

Organisational growth created different pressures:

*"But as the business is growing and... we have businesses across 140 countries... that is a driver for an organization to think about a CISO function..."* (Food02).

**Role requirements**. Expectations for CISO roles demonstrated remarkable variation. A security leader who had built capability over two years in a government agency with new law enforcement powers emphasised that technical leadership remained fundamental:

*"I am... critical about... people who race to that level without having a really good technical understanding of security threats and vulnerabilities who are just pure managers. I don't think that is a good trend for the industry."* (GovAgency01).

However, business partnership increasingly dominated:

*"I guess, like anything else, it's probably more on the lines of ensuring that they have a trusted partner in security who can advise and consult with them and work with them in their business and be at the table..."* (AgedCare1).

**Security incidents**. Prior security events profoundly shaped organisational approaches. Post-incident organisations demonstrated heightened awareness:

*"I know coming into [an organization] that was hit with ransomware... they said, 'we thought we had the risk managed...' but when everything everywhere all at once stops working, that's when you realize just how significant cybersecurity threats are."* (Beverage02).

Conversely, incident-free organisations often struggled for attention:

*"But the board, they want to hear more about cyber, but they don't always understand the risks, or they just, you know it's not a priority. You know It's about increasing profitability and fattening up the bottom line and all those sorts of things, market share."* (Super02).

**Data sensitivity**. The nature of organisational data fundamentally influenced security approaches. One healthcare insurer particularly emphasised this aspect:

*"Well, I think the heart of everything in Insurance01 is focused on customers and customer information"* (Insurance01).

A CISO whose role had evolved from officer to head to CISO as their organisation grew through major acquisitions noted healthcare data created particular imperatives:

*"We've got a lot of personal information. In fact, we've got 100-point data. So we've got medical records, IDs, bank accounts. We've got everything."* (Super01).

#### *5.4.2.2 Organisational Culture - First-Order Concepts*

**Executive support**. We identified leadership championship as perhaps the most critical cultural factor. A security leader in a lean aviation organisation focused on availability and disruption noted that strong support transformed possibilities:

*"I've been quite fortunate in that I got senior leadership buy in from day one"* (Aviation01).

Yet structural impediments often undermined individual support:

*"...whilst a CISO reports into a CIO, it's always going to be seen as technology problem or issue or a function of technology only until you bring that role out into wherever it might sit in the organization..."* (BuildingMat01).

**Security perception**. Organisational conceptualisation of cybersecurity created fundamental boundaries. Mature perspectives enabled strategic engagement:

*"Not at all. I've worked in a lot of places, and I think Insurance01 is probably one of the most mature in terms of understanding it as an operational risk, and it's not necessarily linked to anything specific to IT"* (Insurance01).

IT-centric views limited influence:

*"I think they think of it as a business issue. They treat it as an IT issue... But they ended up putting the cybersecurity function within an IT department... organization thinks it's a business issue. They treat it like a technical issue."* (HigherEd02).

**Business vs IT view**. The evolution from technical to business perspectives marked organisational maturity:

*"If you asked me this question probably 12 months ago, I would have a different response to you. So, in a nutshell, I would say it's evolving and it's evolving more positively"* (AgedCare1).

This transition required sustained effort:

*"I would say 18 months ago, yes, definitely. I would say 18 months on, i.e., now, to a certain extent, yes, but not as much. But there's still a long way to go"* (BuildingMat01).

**Risk culture**. Organisations demonstrated varying risk management sophistication:

*"As a business, our risk appetite is low for cybersecurity, but we treat all risks down to a medium level. So if you've got a critical or high inherent risk, you apply controls, which will reduce the residual risk down to a medium."* (Super01).

Less mature risk cultures struggled with security integration:

*"...when they are working towards that goal, then I also try to chip in. I try to understand where we have a need for IT security control implementation or where we have risks related to ID security that is not kind of captured or not kind of discussed with senior leadership."* (Food02).

#### *5.4.2.3 Organisational Structure - First-Order Concepts*

**Reporting structure**. CISO positioning varied dramatically with profound implications:

*"The structure of where the CISO role fits within Insurance01 is basically there's a board, then underneath the board, there are basically functional groups. And one of those functional groups is led by a senior executive who's part of the executive leadership group"* (Insurance01).

Alternative structures created different dynamics:

*"So my role probably changed a lot in the 18 months that I've been in my position... My role reports into the CIO, which forms part of the finance function. So reporting line is CEO, CFO, CIO, CISO"* (BuildingMat01).

**Formal power/influence**. Authority levels ranged from advisory to decisive:

*"Absolutely. I think you need a level of decision-making authority, right? So whether that's through setting a policy, having that endorsed, and then being able to enforce that policy without having to have everything approved every single time... then yes, right?"* (MgtConsult02).

Limited authority constrained effectiveness. The first security hire at a global food and hospitality company focused on closing gaps and Payment Card Industry (PCI) compliance noted reporting challenges:

*"Particularly that I report to the national IT operations manager, who reports to the CIO that said, it would be much more beneficial and, I mean, easier to deal directly with the CIO"* (Food01).

**Board access**. Direct board engagement transformed CISO influence:

*"I regularly meet with two board members every month outside of cycle. And it just makes my key messaging really clear and crystal"* (AgedCare1).

Mediated access limited strategic impact:

*"I've had situations where, in some instances, the CIO has been really happy for me to present to Audit risk, for example. And then in other situations where they're not, they want to do it themselves."* (HigherEd01).

### *5.4.2.4 Organisational Security Maturity - First-Order Concepts*

**Greenfield/Well-defined posture**. Starting points varied dramatically:

*"I'm probably in a quite unique position as a CISO because I came into Super01 just up to nine years ago. And it was no secret, but there was no security...I had a green field there."* (Super01)

Mature organisations presented different challenges:

*"For entities or organizations that are not regulated... their cybersecurity strategy is... very based on operational requirements. Things are broken. Things are fixed... whereas financial institutions, A, they already had a head start... For them it's been a case of, okay... we got some of these gaps that we need to fix..."* (Beverage02).

**Undergoing changes**. Transitional states created unique challenges:

*"We acquire businesses. Each two to three years, we acquire some large businesses... And so that's how we kind of morphed, that the business grew, the industry grew, and the landscape changed drastically to create this role really."* (Super01).

Rapid evolution demanded adaptability:

*"So my role probably changed a lot in the 18 months that I've been in my position... But we have since divested all of our US and Asian asset businesses and a number of local businesses, so it's now very much an Australian based role."* (BuildingMat01).

**Optimising infrastructure**. Advanced organisations focused on refinement. A CISO leading a brand new cybersecurity function less than one year old articulated success simply:

*"The bottom line is that if we have avoided a major cyber incident, then we've been successful. If we haven't, then we've been unsuccessful. That's the bottom line."* (Beverage01).

Optimisation required different leadership:

*"I think a good strategy is really around clear articulation of how we're going to manage and control risk going forward as it relates to where the business is going"* (Beverage02).

#### *5.4.2.5 Cross-Case Patterns in Organisational Context*

Cross-case analysis of organisational context revealed systematic variations in how organisations conceptualised and supported security leadership. We observed strong clustering patterns: organisations with high regulatory requirements consistently articulated security needs in compliance terms, whilst minimally regulated organisations emphasised threat response and business enablement. Cultural maturity varied dramatically, with some CISOs describing security as deeply integrated into business thinking whilst others struggled against persistent IT-centric perceptions. These cultural differences profoundly shaped what security leaders could accomplish, regardless of individual capability.

Our analysis revealed structural positioning as particularly consequential. CISOs reporting directly to chief executives or with regular board access described fundamentally different operational realities from those buried within IT hierarchies. The former group articulated security in business risk language and influenced enterprise-wide decisions, whilst the latter fought for visibility and struggled to elevate security beyond technical concerns. Maturity phases created distinct leadership demands: establishment-phase organisations needed capability building, maturation-phase organisations required systematisation, whilst strategic-phase organisations sought business integration. These organisational context dimensions operated interdependently rather than independently, creating complex alignment challenges for security leaders.

We identified temporal instability in organisational contexts as particularly significant. Several participants described how their organisations evolved rapidly through mergers, divestitures, regulatory changes, or security incidents. These transitions fundamentally altered organisational security needs, cultural expectations, and structural positioning, sometimes creating acute misalignment between previously well-fitted leaders and their evolving contexts. This dynamism suggested that organisational context cannot be treated as static background but rather as continuously evolving conditions requiring ongoing leadership adaptation.

### 5.4.3 CISO Characteristic Concepts

Participants' descriptions revealed distinct patterns in leadership approaches, technical capabilities, experience profiles, and personality traits. These characteristics interacted with organisational contexts to produce varying effectiveness outcomes.

##### *5.4.3.1 Action-Oriented CISO Type - First-Order Concepts*

**Drive change**. Action-oriented CISOs explicitly identified as transformation catalysts:

*"I'm more of an agent of change, head of cyber. I'm someone you bring in to uplift or change directions"* (Aviation01).

This identity shaped strategic approaches:

*"I usually go to organizations when they have little maturity in terms of information security and I look to turn that around, and turn that around all within using a risk-based approach"* (AgedCare1).

**Quick decisions**. Decisive action characterised this leadership style:

*"On Wednesday, I meet vendors. So Wednesday is my vendor day. I just listen to everyone. A lot of CISO sound too busy, 'Go away,' but actually, you learn from the vendors"* (Super01).

Crisis situations demanded rapid response:

*"...one area where it's critical that they have formal power is that explicit agreement to do whatever needs to be done at speed during a major incident."* (Beverage01).

**Rapid implementation**. Speed trumped perfection for action-oriented leaders.

*"...there was a bunch of certain technical technology and technical controls that we should be implementing that I don't need a strategy to tell me we should be doing...we went really hard in terms of implementing that program."* (AgedCare1).

**Hands-on approach**. Despite senior positions, these leaders maintained operational involvement:

*"I came into the role about 18 months ago... the team was three and three quite junior resources without any real structure... So one of the first things I did initially was build that operating model"* (BuildingMat01).

Direct engagement accelerated progress:

*"When I joined, there were still quite a few vacant roles... But I also restructured the team when I started because I felt like it wasn't really focused on secure by design"* (Retail01).

##### *5.4.3.2 Stewardship-Oriented CISO Type - First-Order Concepts*

**Ensuring compliance**. Regulatory adherence provided organising principles:

*"PCI compliance was a big one. Our businesses revolves around taking payment. One of the compliance items was security, definitely"* (Food01).

Even in less regulated environments, compliance frameworks guided programmes:

*"So working in the industry that we work in… there are very little regulatory compliance requirements for cybersecurity outside of ASX obligations"* (BuildingMat01).

**Maintaining steady state**. Operational stability motivated stewardship leaders:

*"The business wants to do lots of things. We need sufficient prioritization of the particular supply chain engineers and also the IT infrastructure staff to improve our security posture"* (Beverage01).

Consistency mattered more than transformation:

*"I'm not really a steady state one, but I mean, some of it's simplistic, don't get breached, mitigate any impact of any attack"* (Aviation01).

**Optimising efficiency**. Systematic improvement characterised this approach:

*"I worked very much on it using a maturity model… it's easier for the executive and the board to understand-- and palatable, of talking risk, but also talking about maturity and what maturity levels that you want to go, and between what domains."* (AgedCare1).

Measurement enabled optimisation:

*"And that's where I've actually gone and put OKRs in for myself, so objectives and key results… And then I use CMMI to measure about 35 metrics in terms of security… I'll be able to show where the organisation was and then where the organisation is now, and then show a tangible uplift."* (Super02).

**Resource management**. Stewardship leaders excelled at maximising limited resources:

*"…people would look at the security team and say, 'Look, just protect us. Just do whatever you can, but live within your means.' You've only got a percentage of budgets."* (Super01).

Strategic prioritisation guided investments:

*"Funding and resources is always a problem. Look but I wouldn't say that it is preventing me from delivering what I want to deliver…Could I have had more money and do more things? Yeah."* (GovAgency01).

#### *5.4.3.3 Vision-Oriented CISO Type - First-Order Concepts*

**Setting vision**. Strategic articulation distinguished vision-oriented leaders:

*"A successful security leader articulates a clear strategy... you've got to have a strategy and a vision and you've got to be able to articulate that vision... be able to clearly articulate that strategy piece to management, to the board, to peers, etc."* (Beverage02).

Business value communication proved essential:

*"My role as CISO really is to break it all down in layman's terms for the executive team because information security is an operational risk. They don't really know what a firewall is, IPS. They don't understand all of the technical jargon. They simply need to know where are the concerns..."* (Super01).

**Inspiring others**. Vision-oriented CISOs recognised influence beyond authority:

*"My role in that from my previous experience was to really sell it to the executive board... So I really am responsible for the entire information security strategy and roadmap, as we call it."* (MgtConsult02).

Coalition building enabled success:

*"I'd fail without [other executives' support]. There's no point in my role without it. I strongly believe that"* (AgedCare1).

**Leading transformation**. These leaders repositioned security strategically:

*"...you needed some helicopter view, someone to give that strategic advice... the business grew, the industry grew, and the landscape changed drastically to create this role really."* (Super01)

Innovation drove approaches. A CISO whose role was created primarily to satisfy customer security questionnaires emphasised the need for automation:

*"...just like the bad guys are really good at innovating and automating, we can only survive if we go through a similar process ourselves, right? So anywhere we can look to not rely on people to do repetitive tasks, and they can be automated at scale..."* (Tech01).

**Innovative thinking**. Creative approaches distinguished vision-oriented leaders:

*"On Wednesdays, I meet vendors... I just listen to everyone. A lot of CISO sound too busy, 'Go away,' but actually, you learn from the vendors. You learn, 'Well, why are you actually working on that product?'"* (Super01).

Technology adoption enabled business:

*"I think it's about enabling the business. Enabling the business to do what they need to do, but doing it securely."* (HigherEd01).

**Prioritising creativity**. Vision-oriented CISOs encouraged experimentation. A security leader recently elevated from manager to head level who operated with high autonomy in a fast-paced organisation described innovative approaches:

*"We just built a little side project... to give exemptions to certain security controls... We built a little chatbot into Slack... Whereas... most organisations... want a Word document where the Word document is emailed to the manager."* (FinServ01).

Cultural transformation required persistence:

*"Because of how shiny cyber is at the moment, we've actively gone out and done, 'We're here to help. We've got some threat warning. How can we help?'"* (Aviation01).

**Strategic focus**. Business alignment dominated thinking:

*"I keep close contacts with elements of board. I regularly meet with two board members every month outside of cycle"* (AgedCare1).

Long-term perspective guided decisions:

*"I think we're coming into end of financial year... So right now, I'd say I'm probably 80% strategic, 20% operational"* (BuildingMat01).

#### *5.4.3.4 CISO Technical Skills - First-Order Concepts*

**Technical knowledge**. Deep expertise remained essential despite leadership focus:

*"...security people who are not really strong in technology...don't last very long... if you don't understand technology...you'll struggle."* (FinServ01)

Technical credibility enabled influence:

*"As a security person, you need to be fairly well abreast in all other aspects of IT to really be able to have enough insight into how to secure things."* (Tech01).

**Hands-on tools experience**. Operational proficiency varied by context, with participants describing direct engagement with specific security technologies suited to their environment:

*"I've got products like Trend Micro that I actually have to touch on, visit, see their loading, what's happening, etc. And also dealing with ad hoc requests that come through, whether it be someone wants to get out of geo-blocking because they're going overseas."* (Food01)

*"The tool that we are using to enable real-time analysis in our cybersecurity incident response process is SPLUNK... search and monitoring of cybersecurity events."* (MgtConsult01)

Tool expertise enabled rapid response:

*"And then what we do in general is we do benchmark our security capabilities... And then we benchmark against some of the best of industry frameworks."* (Food02).

#### 5.4.3.5 CISO Experience - First-Order Concepts

**Incident management**. Crisis experience proved invaluable:

*"My role in responding to a major cyber incident would be the Incident Manager or the Incident Commander, using the example of a severity 1, P1, critical ransomware event"* (BuildingMat01).

Prior incidents shaped approaches:

*"I've sat across my career... across about 150 incidents, right? So I'm, I'm used to dealing with a big incident, dealing with incidents at scale, dealing with incidents that go on for weeks, sometimes months. And so I would be actively in that."* (Super02).

**Demonstrating value**. Experienced CISOs understood value articulation:

*"If a CISO is not reporting into the into the C suite or to the executive without having that seat at the board. I don't think you will be, you won't be truly. You can't be effective"* (BuildingMat01).

Metrics mattered strategically:

*"The bottom line is that if we have avoided a major cyber incident, then we've been successful."* (Beverage01).

#### 5.4.3.6 CISO Personality - First-Order Concepts

**Communicating effectively**. Translation skills proved fundamental:

*"Fundamentally, the business requires someone who can translate the threats that are obviously active in the environment, the capabilities of the organization at an operational level and align them based on risk to the organization's objectives. And usually to be a good CISO, in my opinion, means that you need to be able to translate between those two layers."* (Tech01).

Stakeholder communication required finesse:

*"Because the CISO in the C Suite might have the board appointed responsibility, unless they've actually got everyone else in the C Suite on board, they can't actually implement what they're trying to achieve"* (MgtConsult01).

**Self-awareness**. Successful CISOs understood their impact:

*"So I do little things, I wear a tie, as you can see, I'm one of the few people who does wear a tie. I'm quite vocal, I'm quite obvious and present because, to be in it– not so much an attention seeker, but a nexus point around that cultural change"* (Aviation01).

Personal style influenced effectiveness:

*"I'm a massive believer of– if we're making a big decision or making a change in some space that's important to the business, then the executive leadership group needs to understand it"* (Insurance01).

**Adaptability**. Context sensitivity distinguished effective leaders:

*"CISOs need to be humble to understand that everyone's got their own pressures, everyone's got their own KPIs, and their own challenges. And the CISOs can help those executives with their challenges"* (Super01).

Flexibility enabled navigation:

*"...being able to have, I guess, a successful role, it's all about managing expectations of a very difficult audience. And it's across to your entire business."* (BuildingMat01).

**Building relationships**. Collaboration underpinned success:

*"It is very much a strategic and a role which requires building partnerships with, I guess, key stakeholders across the business, from the corporate side to the operational side of the business, to the executive to the board"* (BuildingMat01).

Trust creation proved essential:

*"I met with all of them. Each of the offices have a managing partner as well. So I met with all the office managing partners so that they would understand that cybersecurity, me, I'm just a person as well"* (MgtConsult02).

#### *5.4.3.7 Cross-Case Patterns in CISO Characteristics*

Our analysis of CISO characteristics revealed three distinct leadership orientations that extended beyond simple technical versus strategic dichotomies. Action-oriented CISOs explicitly identified as change agents, describing themselves in transformation terms and preferring rapid implementation over extended planning. Stewardship-oriented CISOs emphasised compliance, process optimisation, and operational stability, demonstrating systematic thinking and measurement focus. Vision-oriented CISOs prioritised strategic integration, innovation enablement, and future-state transformation, articulating security in business value terms. These orientations appeared relatively stable across contexts, with several participants explicitly acknowledging their preferred operating mode and recognising mismatches with organisational needs.

We observed that leadership orientation interacted dynamically with technical skills, experience, and personality. Highly technical action-oriented leaders operated differently from less technical ones, with the former maintaining hands-on involvement whilst the latter focused on team building. Experience provided crucial credibility, particularly incident management experience that enabled confident crisis response. Personality traits, particularly communication effectiveness and self-awareness, moderated how leadership orientations manifested in practice. Adaptable personalities partially compensated for orientation mismatches, whilst rigid personalities struggled even with good alignment.

Cross-case comparison revealed that no single CISO characteristic profile guaranteed effectiveness across all contexts. Instead, we identified configuration patterns where certain characteristic combinations proved more effective in specific organisational contexts. Action-oriented leaders with strong technical skills succeeded in establishment-phase organisations, stewardship-oriented leaders with compliance expertise thrived in highly regulated mature organisations, whilst vision-oriented leaders with strong stakeholder management skills excelled in strategic-phase contexts seeking business integration. This configurational finding suggested that effectiveness emerged from alignment patterns rather than universal leader attributes.

### 5.4.4 Environmental Factor Concepts

External environmental factors created the broader context within which CISOs operated, shaping both organisational priorities and individual effectiveness. These factors operated at multiple levels, from global threat evolution to local resource constraints.

#### *5.4.4.1 Threat Landscape - First-Order Concepts*

**Prevailing threats**. Current threat profiles varied by industry and exposure. A CISO building an entirely digital startup fintech whilst balancing security controls with usability observed the evolution:

*"The threat has completely changed to a point now that we are against things like nation-states if you're in an important sort of organization, but that hacker name has changed, right, because now it's actually not driven by one person... We're talking about organized crime..."* (Bank01).

Threat sophistication demanded continuous adaptation:

*"...threat landscape has definitely played a part. We know that's always ever evolving and whatnot, and with that, the CISO has to pivot and sort of move with that, as does an organization with that ever-changing threat landscape."* (Beverage01).

**Evolving landscape**. Technology adoption expanded attack surfaces:

*"CISO are having to evolve because of the changing threat landscape around, be it automation, be it advanced persistent threat, be it vulnerabilities because we've got people leaping to the cloud or other digital transformation means as well."* (BuildingMat01).

Attack complexity increased continuously:

*"...you're not just dealing with one person. You're actually dealing with an entity that's actually putting money in to do what they need to do to extract information from you or make it into a business on their end..."* (Bank01).

**Cyber attacks**. Direct attack experience transformed organisational approaches:

*"...organizations that have been through the war room... have gone through the battle scars in terms of the lessons learned... those organizations are more susceptible... to ensure that all the continuity plans are well understood... it usually takes an event like this to wake people up."* (AgedCare1).

Near-misses created learning opportunities:

*"We've done tabletop exercises in the past as well. And certainly, that has revealed some gaps in our incident response."* (Food01).

#### *5.4.4.2 Industry & Regulations - First-Order Concepts*

**Industry standards**. Sector-specific requirements shaped programmes:

*"So as a service provider, we have a lot of clients. Sometimes what they'll do is get me to go out in my CISO capacity to speak with those clients. Sometimes it's to help provide assurance to those clients that we are secure."* (Tech01).

Industry norms created baselines:

*"Industries that have high levels of regulation and oversight, such as the banking sector is a classic, right? ...what that has driven over the last decade has been an information security function that is very focused and orientated towards satisfying the regulator's requirements for cybersecurity."* (Beverage02).

**Industry threats**. Sector exposure created different risk profiles:

*"So disruption is our key cyber risk."* (Aviation01).

Industry targeting influenced priorities:

*"The environment is very volatile today. We have a lot of political. We have a lot of other motives for the people to use the non-physical parts I mean, to launch attacks"* (Food02).

**Industry requirements**. Sector expectations varied dramatically:

*"...there are very little regulatory compliance requirements for cybersecurity outside of ASX obligations, the Australian privacy act, PCI, DSS, where there isn't a regulator like APRA in finance or AEMO in energy."* (BuildingMat01).

Customer demands substituted for regulation:

*"One of the primary reasons is that, when customers ask and if a regulator asks, they can be completely transparent"* (MgtConsult01).

**Degree of regulation**. Regulatory intensity shaped entire programmes:

*"Industries that have high levels of regulation... banking sector is a classic"* (Beverage02).

Compliance burden varied significantly:

*"Being an organization where you have data protection legislations across Europe, across Asia, like China, and in APAC, and also in Latin America and the US"* (Food02).

**Compliance requirements**. Specific frameworks dominated attention:

*"Being in the financial services industry, that's a requirement."* (Super01).

Multiple requirements created complexity:

*"In the bank, it was regulated and it's been heavily regulated for a long time... people understand if you don't properly meet the regulations and obligations and you don't protect, then you will lose your banking license. There's a very clear consequence."* (Retail01).

#### *5.4.4.3 Available Funding - First-Order Concepts*

**Securing buy-in**. Resource acquisition required sophisticated approaches:

*"We are not profit-oriented. We are a public sector. You know, we don't have deep pockets, but we do have enough to make sure that we are able to run our regulatory business..."* (GovAgency01).

Business case development proved critical:

*"I've been quite fortunate in that I got senior leadership buy in from day one. So the traditional constraints around budget and time from the top and cut through weren't there." (Aviation01)*.

**Doing more with less**. Resource constraints forced creativity:

*"I want to make sure that any decision around buying things or getting more headcount is very, very calculated...we've maximized, we've got these five tools...We don't need to go and spend a million dollars on another one..."* (Insurance01)

Efficiency became paramount:

*"We are definitely entering a cycle of tightening our belts and you know low margins and things... we've got to make sure that every dollar we spend equates to security improvement, risk reduction." (Retail01)*

#### *5.4.4.4 Existing Technology - First-Order Concepts*

**Technology debt**. Legacy systems created unique challenges:

*"You've got legacy. You've got the business demands. You've got the collaboration tools that you need to make sure that staff are using and everything works"* (Super01).

Historical decisions constrained options:

*"When I picked up this role, for example, we had different ways of managing, for example, endpoints on our laptops, different types of agents"* (FinServ01).

**Infrastructure complexity**. Modern environments challenged traditional approaches:

*"...we have a very different sort of landscape and threats and risks that we're exposed to."* (Bank01).

Complexity demanded new capabilities:

*"There's never a simplistic utopian sort of thing. And certainly across organisation and organisation, there's no uniformity... what you're faced with as a CISO is lots of complexity..."* (Tech01).

**Shifting IT environment**. Continuous change required adaptation:

*"You should always be looking both forward and in your rearview mirror. If you're not, you're just going to get caught. Because it's constantly changing."* (Super01)

Technology velocity challenged security:

*"Just when you think that maybe you've gotten a handle on it and you got under control, suddenly something new comes along like AI"* (Tech01).

**Innovation adoption**. Technology advancement created opportunities and risks:

*"...we can take processes or even security capabilities that we had previously, which were quite manual and cumbersome, and turn them into automated, cost-effective, and far more effective approaches for managing security."* (Tech01).

Balance proved challenging:

*"...if you're a CISO and you've secured the business, but the business is now slower, less agile...you've effectively hindered the organisation."* (Tech01).

#### *5.4.4.5 Size of Security Team - First-Order Concepts*

**People management**. Team size influenced role focus:

*"Being in large organizations, for instance, at the airport, I had about 160 people in my team. And my role, I would say 80% of it was strategic"* (AgedCare1).

Small teams demanded different approaches:

*"When I started, there was nobody, including IT security. Now, I think we're about 12 people, and with plans to grow next year"* (HigherEd01).

**Team specialisation**. Scale enabled depth:

*"I do have an outsourced SOC as well, which we're just onboarding at the moment."* (Bank01).

Limited teams required breadth:

*“And now we're leading other projects...which may not be a traditional technology-led...but because of the risk that we manage and our approach, we've been asked to do more or to lead more of these...”* (Aviation01).

**Hands-on work**. Small teams necessitated operational involvement:

*“I think the smaller the company, probably the more technical the CISO needs to be.”* (Super02).

Scale allowed strategic focus:

*“I could not tell you what was happening on the ground. I had 200 security people under me. I didn’t know any of their names. You know I knew the next layer below me and that’s it.”* (Super02).

**Multiple hats**. Resource limitations forced role multiplication:

*“So my role two hats I actually have in there in an incident. I'm in the IT process side of things, providing recommendations, but I'm in the conductor in the actual incident response.”* (AgedCare1).

Wearing multiple hats challenged focus:

*“That's been a real struggle, to be honest with you, especially when we've had some of that attrition and had some gaps in quite a small team. I feel like I've been dragged into sort of more tactical tasks more often.”* (HigherEd01).

**Guidance needs**. Team maturity influenced leadership requirements:

*“But not in healthcare and certainly Aged care, the biggest challenge for me is people. Not the technology, not the process, not the government, it’s not the board, not the executive, it’s the people.”* (AgedCare1).

Junior teams required mentoring:

*“...three quite junior resources without any real structure...one of the first things I did initially was build...” (BuildingMat01)*.

**Outsourcing**. External support supplemented internal capabilities:

*“I do have an outsourced SOC as well, which we're just onboarding at the moment”* (Bank01).

Vendor management became critical:

*“I also have a bunch of vendors who provide lots of services.”* (Beverage01).

##### *5.4.4.6 Size of Organisation - First-Order Concepts*

**Stakeholder navigation**. Organisational scale multiplied stakeholder complexity:

*"I regularly meet with two board members every month outside of cycle. And it just makes my key messaging really clear and crystal"* (AgedCare1).

Large organisations demanded sophisticated engagement:

*"Given the spread of people on Aviation01 at any day could be up to 40,000 people working there"* (Aviation01).

**Bureaucracy**. Size correlated with procedural complexity:

*"We're only less than 600 people. Given the spread of people on Aviation01 at any day could be up to 40,000 people working there. It's a one to many type set up"* (Aviation01).

Process navigation consumed energy:

*"We're not really trying to address the inherent risk of cybersecurity and information security, which would be the vast volume of attacks"* (Super01).

**Silos**. Functional separation challenged collaboration:

*"Instead of influencing and working with your other technology managers as a peer, right, and within the– effectively the technology org unit, as opposed to being an outside party coming in"* (Beverage02).

Breaking silos required persistence:

*"Well, actually, that's been one of our success stories because we've broken down the silos. Because of how shiny cyber is at the moment"* (Aviation01).

**Organisational complexity**. Multiple dimensions created challenges:

*"I've worked in two organizations which are large family-owned businesses. And in large family-owned businesses, politics– I'm going to be very careful what I say here. The political agenda is very strong"* (BuildingMat01).

Complexity demanded navigation skills:

*"If we ever get breached, we're really prepared... at least having people aware of what to do if something happens."* (MgtConsult01).

**Internal politics**. Political dynamics pervaded large organisations:

*"There's politics everywhere, right? That's corporate world. To be honest, the corporate world is pretty brutal, especially if you're an ASX listed company"* (Super01).

Power structures shaped security programmes:

*"I've worked in two organizations which are large family-owned businesses. And in large family-owned businesses, politics– I'm going to be very careful what I say here. The political agenda is very strong"* (BuildingMat01).

#### *5.4.4.7 Cross-Case Patterns in Environmental Factors*

Environmental factors operated as external forces creating adaptive pressures on both organisations and security leaders. Cross-case analysis revealed how threat landscape evolution forced continuous capability development regardless of organisational preferences or leader orientations. Participants consistently described accelerating threat sophistication, from individual hackers to organised crime and nation-state actors, creating universal pressure for enhanced security postures. However, we observed that industry and regulatory contexts mediated how organisations responded to these threats, with heavily regulated industries demonstrating more systematic and resourced approaches than minimally regulated sectors.

We identified resource availability as a critical moderating factor that shaped what security leaders could achieve. We identified a complex dynamic where funding followed incidents or regulatory pressure rather than proactive risk assessment. This reactive pattern created ongoing challenges for security leaders attempting to build capabilities before crises occurred. Existing technology infrastructure and organisational scale further constrained or enabled security approaches, with legacy environments and large organisational complexity multiplying leadership challenges. The size of security teams particularly influenced whether CISOs could focus strategically or remained operationally involved, creating fundamentally different role experiences.

Political dynamics pervaded all environmental factors, operating as the medium through which external pressures translated into organisational action. We observed that political skill moderated environmental factor impacts: politically adept CISOs leveraged threat intelligence and regulatory requirements to secure resources and influence, whilst those lacking political acumen struggled despite similar environmental pressures. The interaction between organisational size and political complexity proved particularly significant, with larger organisations requiring sophisticated stakeholder navigation that consumed substantial leadership energy. These environmental factors did not operate independently but interacted dynamically, creating complex adaptive challenges requiring continuous leader attention and organisational responsiveness.

### 5.4.5 First-Order Synthesis

The first-order concepts reveal security leadership effectiveness as emerging from dynamic interactions between organisational context, CISO characteristics, and environmental factors. This empirical foundation challenges simplistic views of CISO success as dependent on individual competence alone.

**Concept emergence**. We identified numerous distinct first-order concepts before achieving theoretical saturation. These concepts captured the rich complexity of CISO experiences across diverse organisational settings, with saturation occurring after seventeen interviews when no new concepts emerged.

**Natural clustering**. The concepts grouped naturally into three major areas without forcing: organisational descriptors capturing what organisations need and provide; CISO descriptors revealing what leaders bring and do; and environmental descriptors highlighting external pressures and constraints. This tripartite structure emerged organically from the data rather than being imposed.

**Frequency patterns**. Analysis revealed varying concept frequency across participants. High-frequency concepts like executive support and compliance requirements were mentioned by most participants, suggesting their universal importance. Medium-frequency concepts such as board access and team size constraints appeared in many interviews. Low-frequency concepts including innovation adoption and greenfield security emerged in specific contexts, indicating situational relevance.

**Cross-industry variation**. The same concepts manifested differently across sectors, revealing contextual nuance. Financial services participants emphasised that "regulatory compliance is everything," whilst manufacturing counterparts noted "compliance is minimal, it's about operational risk." This variation highlighted how industry context shapes concept expression whilst core meanings remain consistent.

**Participant language**. The mix of technical and business terminology in participant responses reflected the dual nature of the CISO role. This linguistic diversity, from "threat vectors" to "business enablement", captured the boundary-spanning requirements of modern security leadership.

This empirical foundation, grounded in participants' authentic experiences, provides the basis for theoretical abstraction in subsequent sections. By preserving practitioner voices whilst identifying patterns, the analysis maintains both rigour and relevance: essential for developing theory that bridges academic and practitioner domains.

## 5.5 THEORETICAL ELEVATION: SECOND-ORDER THEMES

### 5.5.1 The Abstraction Process

Through systematic analysis of first-order concepts, we developed sixteen distinct second-order themes, revealing complex structures within the security leadership phenomenon.

For instance, descriptions of *driving change*, *rapid implementation*, and *hands-on problem-solving* coalesced into broader patterns of action-oriented leadership. This process required constant movement between the empirical data and emerging theoretical insights, maintaining the delicate balance between staying grounded in participants' experiences whilst elevating to theoretical significance.

The temporal dimension pervaded participants' narratives. CISOs consistently described dynamic processes of adaptation, evolution, and changing requirements rather than static roles or fixed organisational states. This discovery shaped our theoretical elevation, distinguishing our findings from traditional static conceptualisations of leadership roles and pointing toward a more dynamic understanding of security leadership effectiveness.

### 5.5.2 Theme: Organisational Security Needs

**Definition and Theoretical Boundaries.** We developed Organisational Security Needs as a foundational second-order theme encompassing the fundamental drivers that create demand for particular forms of security leadership. This theme transcends simple functional requirements to capture the complex interplay between business context, risk profile, and security maturity that shapes what organisations need from their security leaders. The theme represents the "demand side" of the leadership equation, establishing what organisations seek in their security leadership.

**Constituent First-Order Concepts.** We developed this theme from the convergence of five interrelated first-order concepts: security drivers, CISO role drivers, role requirements, security incidents, and data sensitivity. These concepts, whilst distinct in their expression, collectively revealed how organisations conceptualise and articulate their security leadership needs.

**Pattern Recognition and Evidence.** Analysis revealed that organisational security needs vary systematically with maturity levels. In establishment-phase organisations, participants consistently described fundamental capability needs. As one CISO explained their organisation's drivers:

*"The evolving threat landscape and the risk that cyber posed the organisation"* (BuildingMat01).

This focus on basic threat response contrasted sharply with more mature organisations where needs centred on optimisation and strategic alignment.

The variation in compliance drivers proved particularly revealing. Some organisations operated with:

*"Very little focus there or drive from a compliance perspective"* (BuildingMat01),

whilst others were driven by external requirements where:

*"Customers ask and if a regulator asks"* (MgtConsult01).

This variation suggested that security needs are shaped not just by internal factors but by the broader ecosystem in which organisations operate.

Our analysis identified data sensitivity as another critical driver, with participants in data-intensive industries emphasising this aspect:

*"...the heart of everything in Insurance01 is focused on customers and customer information"* (Insurance01).

This highlights how industry context and business model fundamentally shape security leadership requirements.

**Key Insight.** Security needs evolve predictably through organisational phases unless disrupted by external events such as security incidents or regulatory changes. This predictability suggests potential for anticipatory leadership development and succession planning aligned with organisational evolution.

### 5.5.3 Theme: Organisational Culture

**Conceptual Definition and Scope.** Organisational Culture as a second-order theme captures the shared beliefs, values, and assumptions about security's role and importance within the organisation. This extends beyond formal policies or stated positions to encompass the lived reality of how security is perceived, valued, and integrated into organisational decision-making processes. Culture shapes not just what security leaders can do, but what they are expected to do.

**Constituent Elements and Their Interplay.** We developed this theme from first-order concepts addressing executive support patterns, security perception, and the business versus IT view of security. This theme captures the shared beliefs and values shaping how organisations understand and value security leadership. The contrast emerged clearly when participants described their experiences:

*"I've been quite fortunate in that I got senior leadership buy in from day one"* (Aviation01)

versus:

*"I report into a CIO and whilst a CISO reports into a CIO, it's always going to be seen as technology problem"* (BuildingMat01).

Executive support patterns varied dramatically across participants. Some enjoyed strong backing:

*"I've been quite fortunate in that I got senior leadership buy in from day one"* (Aviation01).

Others faced structural challenges that reflected deeper cultural issues:

*"I report into a CIO and whilst a CISO reports into a CIO, it's always going to be seen as technology problem"* (BuildingMat01).

This variation in support directly influenced what security leaders could achieve.

**Cultural Archetypes and Evolution.** Our cross-case analysis revealed three distinct cultural archetypes regarding security. In organisations viewing security as a technical function, the perception limited security's strategic influence. As one CISO noted about their organisation's perception:

*"I think it is perceived as an IT problem"* (Beverage01).

Conversely, mature organisations demonstrated integrated security cultures:

*"Not at all. I've worked in a lot of places, and I think Insurance01 is probably one of the most mature"* (Insurance01).

This maturity manifested in security being viewed as a business enabler rather than a technical constraint.

**Boundary Conditions and Key Insight.** Strong compliance cultures in heavily regulated industries could override natural maturity considerations, forcing rapid cultural evolution. This suggests that whilst culture typically evolves slowly, external pressures can accelerate cultural transformation. The key insight is that culture shapes what is possible regardless of CISO capabilities; even highly skilled leaders struggle in cultures that view security as purely technical.

### 5.5.4 Theme: Organisational Structure

**Definition and Significance.** Organisational Structure encompasses the formal positioning and authority arrangements of the security function within the broader organisational hierarchy. This theme captures how reporting relationships, formal power arrangements, and board access collectively shape the CISO's ability to influence organisational security outcomes.

**Structural Variations and Their Implications.** Participants described diverse structural arrangements that profoundly influenced their effectiveness. One CISO outlined a complex structure:

*"The structure of where the CISO role fits within Insurance01 is basically there's a board, then underneath the board, there are basically functional groups"* (Insurance01).

Another described significant structural evolution:

*"So my role probably changed a lot in the 18 months that I've been in my position… My role reports into the CIO, which forms part of the finance function"* (BuildingMat01).

Our analysis revealed the reporting relationship as particularly significant. Some CISOs enjoyed direct executive access whilst others navigated multiple layers:

*"… Organizationally, I sit under the chief technology officer"* (Beverage02).

These structural differences created vastly different operating contexts for security leadership.

**Pattern Recognition and Key Insight.** Analysis revealed that structure reflects organisational views about security's importance. CISOs with board access and executive reporting relationships generally demonstrated greater strategic influence than those buried within IT hierarchies. However, structure alone did not determine effectiveness; political skills could compensate for suboptimal positioning. The key insight is that whilst structure enables or constrains CISO effectiveness, skilled leaders can work within structural limitations through political navigation and relationship building.

### 5.5.5 Theme: Organisational Security Maturity

**Theoretical Definition.** Organisational Security Maturity represents the evolution of security capability development and integration within the organisation. This theme captures not just the technical maturity of security controls but the organisational sophistication in understanding, resourcing, and leveraging security as a business capability.

**Maturity Phases and Characteristics.** We identified three distinct maturity phases from the data, each creating specific leadership requirements. Establishment-phase organisations demonstrated greenfield or underdeveloped security postures requiring foundational capability building. Maturation-phase organisations were undergoing significant changes to formalise and systematise security. Strategic-phase organisations focused on optimising existing infrastructure and integrating security with business strategy.

The importance of maturity alignment was consistently emphasised by participants, with phrases like "where the organisation is at" recurring throughout interviews. This suggests that CISOs intuitively understand the importance of phase-appropriate leadership approaches.

**Key Pattern and Insight.** Organisational maturity influences CISO role expectations more than any other single factor. The maturity phase creates specific leadership requirements that, when matched with appropriate leadership styles, enhance effectiveness. Critically, maturity represents a dynamic rather than static state; organisations continuously evolve, requiring adaptive leadership approaches.

### 5.5.6 Theme: Action-Oriented CISO Type

**Theoretical Definition and Characteristics.** Action-Oriented CISOs represent leaders who prioritise rapid implementation, transformational change, and hands-on problem-solving. This leadership style is characterised by a bias toward action over extended analysis, comfort with ambiguity, and the ability to build capabilities from minimal foundations. These leaders thrive in uncertainty and excel at creating momentum in stalled security programmes.

**Behavioural Manifestations.** A substantial portion of participants strongly identified with action-oriented characteristics. One captured the essence of this orientation:

*"I'm more of an agent of change, head of cyber. I'm someone you bring in to uplift or change directions"* (Aviation01).

This self-identification as a change agent permeated action-oriented leaders' narratives.

Another participant described their typical engagement pattern:

*"I usually go to organizations when they have little maturity in terms of information security and I look to turn that around, and turn that around all within using a risk-based approach when it comes to security"* (AgedCare1).

We observed this preference for transformation challenges over steady-state operations as a defining characteristic.

The hands-on nature of action-oriented leaders manifested in their operational approach. One CISO described their vendor engagement strategy:

*"On Wednesday, I meet vendors. So Wednesday is my vendor day. I just listen to everyone. A lot of CISO sound too busy, 'Go away,' but actually, you learn from the vendors"* (Super01).

This direct engagement with the ecosystem exemplifies the action-oriented preference for learning through doing.

**Effectiveness Patterns and Context.** Action-oriented CISOs demonstrated highest effectiveness in establishment-phase contexts where rapid capability building was essential. One participant detailed their transformation impact:

*"I came into the role about 18 months ago… the team was three and three quite junior resources without any real structure, function in absence of a defined operating model. So one of the first things I did initially was build that operating model, deploy that operating model, and obtain approved funding"* (BuildingMat01).

The ability to make quick decisions in ambiguous situations proved critical. As one CISO noted:

*"...one area where it's critical that they have formal power is that explicit agreement to do whatever needs to be done at speed during a major incident"* (Beverage01).

This mandate for rapid action aligns with the action-oriented leader's natural tendencies.

However, these leaders often struggled in mature environments requiring careful optimisation rather than radical change. One action-oriented CISO acknowledged this limitation:

*"I'm not really a steady state one"* (Aviation01),

recognising the mismatch between their change-driving nature and maintenance-focused requirements.

### 5.5.7 Theme: Stewardship-Oriented CISO Type

**Conceptual Framework and Definition.** Stewardship-Oriented CISOs focus on process optimisation, compliance achievement, and operational excellence within established frameworks. This leadership style emphasises systematic thinking, measurement rigour, and risk mitigation through proven methodologies. Stewardship leaders excel at bringing order to chaos and ensuring sustainable security operations.

**Core Characteristics and Evidence.** The majority of participants demonstrated some stewardship orientation, with many emphasising compliance and process maturity. We observed a strong focus on regulatory requirements:

*"Working in the industry that we work in… there are very little regulatory compliance requirements"* (BuildingMat01),

with this CISO noting the contrast with other industries. Another emphasised the centrality of compliance:

*"PCI compliance was a big one. Our business revolves around taking payment"* (Food01).

The systematic approach to security management characterised stewardship thinking. One CISO described their measurement focus:

*"Identifying what our current level of capability maturity is helps us to identify what our gaps are"* (AgedCare1).

This emphasis on assessment, gap analysis, and systematic improvement typified the stewardship approach.

**Contextual Effectiveness.** Stewardship-oriented leaders excelled in mature organisational contexts with established security programmes requiring optimisation. Their systematic approach proved particularly valuable in regulated industries where demonstrable compliance was paramount. The focus on process and measurement aligned with organisational needs for predictability and control in mature security environments.

### 5.5.8 Theme: Vision-Oriented CISO Type

**Definition and Strategic Focus.** Vision-Oriented CISOs represent security leaders who focus on strategic business integration, innovation enablement, and future-state transformation. This leadership style transcends traditional security boundaries to position security as a business differentiator and strategic capability. These leaders think beyond protection to enablement.

**Manifestations and Evidence.** Several participants in mature organisations exemplified vision-oriented leadership. One articulated the importance of formal strategic planning:

*"So a successful security leader articulates a clear strategy...you've got to have a strategy and a vision and you've got to be able to articulate that vision"* (Beverage02).

This emphasis on vision-setting distinguished these leaders from their more operationally focused counterparts.

We observed the ability to communicate strategic value as critical. Another participant emphasised:

*"You need now to be able to articulate what it means to the business"* (Insurance01).

This focus on articulating business relevance rather than technical details characterised vision-oriented communication.

**Strategic Integration Pattern.** We identified vision-oriented leaders exclusively in organisationally mature contexts where basic security capabilities were already established. Their effectiveness derived from ability to position security as business enabler rather than cost centre, creating competitive advantage through security excellence.

### 5.5.9 Theme: CISO Technical Skills

**Definition and Scope.** CISO Technical Skills encompasses the depth of technical knowledge and hands-on experience with security tools and technologies. This theme captures the ongoing debate about technical depth requirements for security leadership effectiveness.

**Technical Depth Variations.** Participants demonstrated wide variation in technical emphasis. Some maintained deep technical engagement:

*"Actually I lead up and head up all the security matters, both information security or the GRC elements, security ops, and all things information and cybersecurity-related"* (AgedCare1).

Others had evolved beyond technical roles whilst maintaining sufficient understanding for credibility.

We observed a clear relationship between organisational size and technical requirements. As one participant observed:

*"I think the technologist is probably more important in smaller to medium-sized companies. I think the smaller the company, probably the more technical the CISO needs to be"* (Super02).

This size-skill relationship reflected resource constraints and role breadth in smaller organisations.

**Key Pattern and Insight.** Technical skills proved necessary but insufficient for effectiveness across all contexts. The required technical depth varied by organisational phase and team size, with smaller teams and earlier phases demanding greater hands-on technical involvement from CISOs.

### 5.5.10 Theme: CISO Experience

**Conceptual Boundaries.** CISO Experience captures prior experience managing security incidents and demonstrating security value to organisations. This theme encompasses both the breadth and depth of security leadership experience, recognising that different experiences prove valuable in different contexts.

**Experience Types and Their Value.** Our analysis revealed incident management experience as particularly valuable, providing credibility and practical knowledge. The ability to demonstrate value, developed through experience across various organisational contexts, proved equally important. Participants with diverse industry experience brought valuable perspectives on security challenges and solutions.

**Pattern Recognition.** Experience type mattered more than duration. Alignment between prior experience and current organisational needs enhanced effectiveness more than years of experience alone. This suggests that organisations should prioritise relevant experience over tenure when selecting security leaders.

### 5.5.11 Theme: CISO Personality

**Definition and Components.** CISO Personality encompasses personal attributes that enable effective security leadership beyond technical or strategic capabilities. This theme captures how individual characteristics moderate leadership effectiveness across different contexts.

**Key Personality Dimensions.** We identified communication effectiveness as critical across all contexts. Self-awareness manifested in deliberate behavioural choices, as one CISO illustrated:

*"So I do little things, I wear a tie, as you can see, I'm one of the few people who does wear a tie"* (Aviation01).

This conscious attention to professional presentation reflected broader self-awareness about organisational fit.

Adaptability proved essential given the dynamic nature of security challenges and organisational evolution. The ability to build relationships across diverse stakeholder groups, from technical teams to board members, proved to be a universal success factor we observed.

**Personality as Moderator.** Personality traits moderated the effectiveness of different leadership styles. Highly adaptable personalities could partially compensate for style-context misalignment, whilst rigid personalities struggled even with good alignment. This suggests that personality flexibility enhances leadership effectiveness across varying contexts.

### 5.5.12 Theme: Threat Landscape

**Environmental Context Definition.** The Threat Landscape theme captures the external threat environment creating pressure on organisations and shaping security leadership requirements. This environmental factor operates independently of organisational control, forcing adaptive responses.

**Evolution and Sophistication.** Participants consistently emphasised the rapidly evolving threat environment. One noted:

*“The threat has completely changed to a point now that we are against things like nation-states... we’re talking about organized crime...”* (Bank01).

This constant evolution created pressure for continuous adaptation in security approaches and leadership.

We identified the expansion of attack surfaces through digital transformation as a critical concern:

*“...as we start at how the trend has evolved where everything is connected, cyber risk becomes very relevant.”* (Auto01).

This technological expansion multiplied the complexity of security leadership, requiring broader perspectives and capabilities.

**Acceleration Effect.** Threat sophistication drives security maturity acceleration, often forcing organisations to evolve faster than naturally inclined. Major incidents create “burning platforms” for change, enabling security leaders to drive transformations that might otherwise face resistance.

### 5.5.13 Theme: Industry & Regulations

**Sector-Specific Context.** Industry & Regulations encompasses sector-specific requirements and compliance demands that shape CISO roles. This theme recognises that industry context creates distinct security leadership requirements beyond general organisational factors.

**Regulatory Intensity Variations.** The degree of regulatory pressure varied dramatically across industries. Some operated with minimal requirements:

*"Very little regulatory compliance requirements... outside of ASX obligations"* (BuildingMat01).

Others faced intense regulatory scrutiny:

*"Industries that have high levels of regulation... banking sector is a classic"* (Beverage02).

This variation in regulatory intensity fundamentally shaped CISO priorities and approaches. In highly regulated industries, compliance considerations dominated security strategies, whilst less regulated sectors allowed greater flexibility in security approaches.

**Industry Override Pattern.** Industry context could override organisational preferences for security approaches. Even organisations preferring innovation-focused security might require compliance-oriented leadership in heavily regulated sectors. This suggests that industry context represents a powerful constraint on security leadership options.

### 5.5.14 Theme: Available Funding

**Resource Context Definition.** Available Funding captures the financial resources available for security initiatives and their impact on leadership approaches. This theme recognises that resource constraints fundamentally shape what security leaders can achieve.

**Funding Patterns and Triggers.** Funding availability often followed predictable patterns, with increases typically triggered by incidents or regulatory pressure. We observed the challenge of securing resources in the absence of visible threats as a common frustration. Security leaders had to excel at building business cases and demonstrating value to secure necessary resources.

The need to *do more with less* forced creativity and prioritisation. Resource constraints shaped leadership approaches, with limited funding requiring more strategic, risk-based approaches rather than comprehensive security programmes.

**Resource-Leadership Alignment.** Available funding needed to align with leadership style and organisational phase. Action-oriented leaders in establishment phases required sufficient resources for transformation, whilst stewardship-oriented leaders could optimise within constraints. Misalignment between resources and requirements created significant leadership challenges.

### 5.5.15 Theme: Existing Technology

**Technical Environment Context.** Existing Technology encompasses the current technology environment, including legacy systems and technical debt, that shapes security leadership challenges and opportunities. This theme recognises that inherited technology environments constrain security options.

**Legacy Challenges and Modernisation Opportunities.** We identified technical debt as a significant constraint on security effectiveness. Legacy environments required different leadership approaches than modern, cloud-native architectures. The pace of technology change, particularly cloud transformation, continuously altered security requirements and leadership demands.

Some organisations embraced innovation whilst others struggled with decades-old systems. This variation in technology environments created different security challenges requiring adapted leadership approaches.

**Technology as Enabler and Constraint.** Modern technology environments enabled advanced security approaches whilst legacy environments constrained options. Security leaders needed to balance ideal security architectures with technological realities, requiring pragmatic approaches to security implementation.

### 5.5.16 Theme: Size of Security Team

**Team Scale Implications.** Size of Security Team captures how team scale and structure shape CISO role requirements. This theme recognises that team size fundamentally determines whether CISOs can focus strategically or must remain operationally involved.

**Small Team Dynamics.** In small teams, CISOs necessarily maintained hands-on involvement. The need to *wear multiple hats* and provide direct technical guidance characterised small-team leadership. These constraints forced efficiency and prioritisation but limited strategic focus.

**Large Team Orchestration.** Larger teams enabled greater specialisation and strategic focus but required sophisticated people management skills. The shift from doing to directing represented a significant transition that not all CISOs navigated successfully.

**Size-Style Interaction.** Team size mediated the effectiveness of different leadership styles. Action-oriented leaders thrived in small teams where hands-on involvement was necessary, whilst vision-oriented leaders required larger teams to execute strategic initiatives. This interaction suggests that team size should align with leadership style for optimal effectiveness.

### 5.5.17 Theme: Size of Organisation

**Organisational Scale Effects.** Size of Organisation encompasses how organisational scale affects security leadership requirements through increased complexity, stakeholder management demands, and political dynamics. This theme recognises that scale fundamentally alters leadership challenges.

**Complexity and Navigation.** Larger organisations required sophisticated stakeholder navigation. As one CISO noted:

*"I regularly meet with two board members every month outside of cycle"* (AgedCare1),

illustrating the intensive relationship management required in large organisations.

The contrast between organisations of different scales was stark:

*"We're only less than 600 people. Given the spread of people on Aviation01 at any day could be up to 40,000 people working there"* (Aviation01).

This scale difference created vastly different operating contexts.

**Political Amplification.** Organisational size amplified political requirements. Larger organisations featured more complex stakeholder landscapes, deeper silos, and more intense resource competition. Security leaders in large organisations spent considerably more time on political navigation than their small-organisation counterparts.

### 5.5.18 Thematic Interrelationships

**Organisational Context Clustering.** The organisational themes, Security Needs, Culture, Structure, and Maturity, demonstrated strong interrelationships, collectively shaping the organisational context for security leadership. These themes operated as a system rather than independent factors, with changes in one affecting others.

**Leadership Type-Context Matching.** We identified clear patterns between leadership types and organisational phases. Action-oriented leaders consistently thrived in establishment phases where transformation was needed. Stewardship-oriented leaders excelled during maturation phases requiring systematisation. Vision-oriented leaders proved most effective in strategic phases demanding business integration.

**Environmental Catalysis.** Environmental pressures, particularly threats and regulations, consistently accelerated maturity progression. These external forces could override natural organisational evolution, forcing rapid adaptation. Security leaders who recognised and leveraged these environmental pressures achieved greater transformation success.

**Size as Universal Moderator.** Both organisational and team size moderated all other relationships. Larger scales required more sophisticated political navigation and stakeholder management, whilst smaller scales demanded greater technical involvement and operational flexibility. Size considerations permeated every aspect of security leadership effectiveness.

**Political Capital as Meta-Capability.** We identified political skills as a meta-capability enabling navigation across all themes. Leaders with strong political capital could succeed despite style-context misalignment, whilst those lacking political skills struggled even with perfect alignment. This finding suggests that political acumen may be the most critical security leadership capability.

**Pattern Synthesis and Theoretical Foundation.** Our cross-case analysis revealed consistent patterns. The majority of cases demonstrated phase-aligned leadership success. Several cases revealed how strong political capital could compensate for style-phase misalignment. A notable minority exhibited hybrid leadership styles, combining elements of multiple orientations to address complex or transitioning contexts.

Environmental factors occasionally overrode maturity-based expectations, forcing stewardship approaches in establishment-phase organisations facing regulatory pressure. We observed the reciprocal relationship between organisational culture and leadership effectiveness consistently, with cultural alignment amplifying leadership impact.

These sixteen themes form the foundation for understanding security leadership as a dynamic phenomenon requiring continuous navigation of organisational, environmental, and political landscapes. The three-way interaction between Organisation-Leader-Environment becomes clear, with fit emerging from alignment patterns whilst remaining inherently dynamic due to continuous contextual evolution.

## 5.6 THEORETICAL INTEGRATION: AGGREGATE DIMENSIONS

### 5.6.1 Ascending to Aggregate Dimensions

We constructed two aggregate dimensions through analysis: CISO-Organisation Fit and CISO-Environment Fit, representing the highest level of theoretical abstraction in this study.

The construction of these aggregate dimensions marked a critical juncture in our analysis. After identifying over 60 first-order concepts and distilling them into 16 second-order themes, we recognised a higher-order pattern. The data revealed not merely discrete characteristics or contextual factors, but a fundamental dynamic of fit between CISOs and their contexts.

The core insight crystallising through this process was that security leadership effectiveness emerges from the dynamic interplay between what CISOs bring to their roles and what their contexts demand of them. This dual-fit dynamic became the theoretical cornerstone of the Security Leadership Contingency Model.

### 5.6.2 Contextual Groupings: The Foundation for Fit

The journey to aggregate dimensions began with recognising that the second-order themes naturally clustered into three contextual groupings, each representing a distinct domain of influence on security leadership effectiveness.

#### *5.6.2.1 Organisational Context Grouping*

The first grouping encompassed four second-order themes that collectively define the organisational context within which CISOs operate: Security Needs, Culture, Structure, and Maturity. This grouping represents the organisational "demand side" of the fit equation, defining what the organisation needs from security leadership at any given point in time.

The multidimensional nature of organisational requirements emerged clearly from the data. As one CISO explained:

*"So my role probably changed a lot in the 18 months that I've been in my position"* (BuildingMat01).

This evolution of role requirements within the same organisation illustrates how organisational contexts shift over time, demanding different capabilities from security leadership.

#### *5.6.2.2 CISO Characteristics Grouping*

The second grouping consolidated six second-order themes that capture the individual attributes CISOs bring to their roles: three leadership types (Action-Oriented, Stewardship-Oriented, Vision-Oriented), Technical Skills, Experience, and Personality. This grouping represents the individual "supply side" of the fit equation.

The inclusion of three distinct leadership orientations alongside traditional attributes reflects a nuanced understanding emerging from the data. As one action-oriented CISO articulated:

*"I'm more of an agent of change, head of cyber. I'm someone you bring in to uplift or change directions"* (Aviation01).

In contrast, a more stewardship-oriented leader focused on optimisation:

*"I worked very much on it using a maturity model... it's easier for the executive and the board to understand–and palatable, of talking risk, but also talking about maturity and what maturity levels that you want to go, and between what domains."* (AgedCare1).

These distinct approaches fundamentally shape effectiveness in different contexts.

#### *5.6.2.3 Environmental Context Grouping*

The third grouping brought together six second-order themes capturing external forces: Threat Landscape, Industry & Regulations, Available Funding, Existing Technology, Size of Security Team, and Size of Organisation. This grouping represents factors creating adaptation pressures on the fit relationship.

The pervasive nature of environmental pressures was captured by participants. One noted the evolving threat environment:

*"The threat has completely changed to a point now that we are against things like nation-states if you're in an important sort of organization, but that hacker name has changed, right, because now it's actually not driven by one person or two people or five people. We're talking about organized crime..."* (Bank01).

Another highlighted the impact of technology shifts:

*"I would probably say technology hygiene. So technology hygiene, and that probably includes legacy technology across infrastructure, networks, applications, and even going into more detail on that, probably lifecycle management, hardware, software, application, lifecycle management..."* (BuildingMat01).

### 5.6.3 Aggregate Dimension 1: CISO-Organisation Fit

We observed that effectiveness patterns consistently arose from alignment between CISO characteristics and organisational contexts, leading us to conceptualise the first aggregate dimension: CISO-Organisation Fit.

#### *5.6.3.1 Definition and Emergence*

CISO-Organisation Fit represents the multidimensional alignment between what CISOs bring to their roles (leadership orientation, skills, experience, personality) and what their organisations require (based on security needs, culture, structure, maturity phase). This conceptualisation emerged from consistent patterns of "matching" between CISO characteristics and organisational needs across the interview data.

The importance of this alignment was articulated by participants. One CISO reflected on the challenge of misalignment:

*"I report into a CIO and whilst a CISO reports into a CIO, it's always going to be seen as technology problem"* (BuildingMat01).

This structural misalignment created ongoing friction between the CISO's strategic aspirations and organisational perceptions.

#### *5.6.3.2 Properties of CISO-Organisation Fit*

Our analysis revealed several key properties of this aggregate dimension:

**Multi-dimensional Nature.** Fit operates across multiple intersections simultaneously. One participant illustrated the complexity of achieving alignment across dimensions:

*"I've seen places that have every piece of technology that you can imagine. Or they've got ISO certification... But really, it doesn't really mean anything if the culture and people that sit around it and the culture of the business don't understand what security looks like"* (Insurance01).

This highlights how technical alignment without cultural fit proves insufficient.

**Dynamic Evolution.** The fit relationship changes as organisations evolve. Several participants described how their roles transformed over time within the same organisation. One noted:

*"So my role probably changed a lot in the 18 months that I've been in my position."* (BuildingMat01)

Another explained the shift in focus over time:

*"Like I said, because of the maturity level of the organization at the time of me joining, I found myself dedicating possibly 70 to 80 percent of my time in strategic undertakings. Whereas now, it's slowly starting to even out. As time goes on and we're becoming more mature as an organization, I see that percentage shifting a bit."* (Food01)

This preference for transformation over maintenance illustrates how fit requirements shift as organisations mature.

**Reciprocal Influence.** The data revealed bidirectional relationships where CISOs shape their organisations while being shaped by them. One leader explained the delicate balance:

*"Instead of influencing and working with your other technology managers as a peer... as opposed to being an outside party coming in and telling you what to do"* (Beverage02).

This highlights the importance of working within organisational dynamics rather than imposing external standards.

#### *5.6.3.3 Evidence of Effectiveness Differences*

We observed the impact of CISO-Organisation fit on effectiveness through contrasting experiences. Aligned leaders reported greater influence and success. One described the benefits of strong executive alignment:

*"I've been quite fortunate in that I got senior leadership buy in from day one"* (Aviation01).

This alignment enabled rapid progress on security initiatives.

Conversely, misalignment created significant challenges. As one CISO noted about organisational politics:

*"There's politics everywhere, right? That's corporate world. To be honest, the corporate world is pretty brutal, especially if you're an ASX listed company"* (Super01).

Navigating these political dynamics while maintaining security effectiveness required careful balance.

### 5.6.4 Aggregate Dimension 2: CISO-Environment Fit

The second aggregate dimension emerged from patterns showing how CISO effectiveness was shaped by alignment with environmental demands beyond organisational boundaries.

#### 5.6.4.1 Definition and Distinction

CISO-Environment Fit represents the alignment between CISO capabilities and demands imposed by the external environment, including threat landscapes, regulatory requirements, resource constraints, and technological contexts. While CISO-Organisation fit captured internal alignment, the data revealed that effectiveness also required alignment with external forces.

The distinction between organisational and environmental fit became clear through participant experiences. One CISO described managing regulatory complexity:

*"Obviously, as a health insurer, there's requirements around PCI compliance because credit cards are taken. So there's some obligations out that we have to be really mindful of our own storage and use and management of credit card information."* (Insurance01).

This external requirement operated independently of internal organisational preferences.

#### 5.6.4.2 Properties of CISO-Environment Fit

Our analysis identified several distinctive properties:

**Responsive Adaptation.** Environmental fit requires continuous responsiveness to external changes. One participant explained the challenge of keeping pace:

*"...you're not just dealing with one person. You're actually dealing with an entity that's actually putting money in to do what they need to do to extract information from you or make it into a business on their end, so."* (Bank01).

This dynamic threat environment demanded constant adaptation.

**Boundary-Spanning Nature.** CISOs serve as critical interfaces between organisations and external security environments. One leader described this translation role:

*"My role as CISO really is to break it all down in layman's terms for the executive team because information security is an operational risk. They don't really know what a firewall is, IPS. They don't understand all of the technical jargon. They simply need to know where are the concerns, where are our weaknesses..."* (Super01).

This involved making external threats meaningful to internal stakeholders.

**Resource Dependency**. Environmental fit heavily depends on resource availability. The contrast between sectors was stark. As one CISO observed about industry differences:

*"Industries that have high levels of regulation... banking sector is a classic"* (Beverage02).

These regulatory requirements created different resource demands across industries, affecting CISOs' ability to maintain environmental fit.

#### *5.6.4.3 Evidence of Environmental Fit Impact*

The data provided rich evidence of environmental fit's impact on effectiveness. CISOs who successfully navigated environmental challenges reported enhanced credibility and influence. One participant linked certification to organisational confidence:

*"And certainly, having the 27001 certification as well has helped with that"* (MgtConsult02).

Those struggling with environmental demands faced significant challenges. One CISO reflected on the evolution of threats:

*"When I started out in security like 20 years ago, we were worried about script kiddies, right? And you weren't worried about your files being encrypted or wiped. You were worried about your website being defaced"* (Super02).

The dramatic escalation in threat sophistication required continuous capability development to maintain environmental fit.

### 5.6.5 The Dual-Fit Dynamic

The most significant theoretical insight emerged from recognising that both types of fit must be maintained simultaneously for sustained effectiveness. This dual-fit dynamic represents a fundamental reconceptualisation of how security leadership effectiveness emerges.

#### *5.6.5.1 Core Finding: Simultaneous Fit Requirements*

Our analysis revealed that neither CISO-Organisation fit nor CISO-Environment fit alone suffices for effectiveness. CISOs must maintain both fits simultaneously, creating a complex balancing act. One participant captured this duality:

*"If a CISO is not reporting into the into the C suite or to the executive without having that seat at the board. I don't think you will be, you won't be truly. You can't be effective"* (BuildingMat01).

This statement reflects both organisational positioning (internal fit) and the need for strategic influence to address external challenges (environmental fit).

The simultaneous maintenance of both fits emerged as fundamentally different from achieving either fit independently. Participants' experiences revealed layered complexity: maintaining

CISO-Organisation fit enabled them to navigate internal dynamics and secure resources, while CISO-Environment fit ensured they could respond effectively to external pressures. However, the data revealed a crucial temporal dimension; misalignment in either dimension created cascading effects over time.

The wearing down process participants described was not merely personal dissatisfaction but manifested as decreased effectiveness over time. CISOs reported a progression from initial enthusiasm through increasing strain to eventual departure, even when organisational metrics showed continued security improvements. As participants repeatedly emphasised:

*"There really is a different CISO for the different time of an organisation… it's good for organisations… that there's quite a lot of clarity in terms of what they actually need at that particular time."* (Tech01)

*"If you're in the early days of building up a security function, you need a support of the CIO. If you are in the later days in the mature function, you need a support directly of a CEO… So it depends on where you are on your journey, how mature you are…"* (HigherEd02)

alignment is not a fixed state but a dynamic relationship that must evolve as contexts change. This pattern indicates that dual-fit maintenance is necessary not only for immediate organisational security outcomes but also for the CISO's continued capacity to deliver those outcomes. The sustainability of security leadership, therefore, emerges as inseparable from the sustainability of security performance.

#### *5.6.5.2 Interaction Effects*

We observed that the relationship between fit dimensions proved interactive rather than additive:

**Compensatory Effects**. Strong fit in one dimension could partially compensate for weaker fit in the other. One technically proficient CISO explained:

*"Because I am technical, I would be diving into technical solutions. So, you know, I have in incidents taken over from the incident handler because they were struggling"* (Super02).

This deep technical expertise helped overcome some organisational challenges.

**Amplification Effects.** When both fits aligned strongly, effectiveness amplified. One leader described achieving this synergy:

*“Well, actually, that’s been one of our success stories because we’ve broken down the silos. Because of how shiny cyber is at the moment, we’ve actively gone out and done, ‘We’re here to help. We’ve got some threat warning. How can we help?’”* (Aviation01).

This demonstrates successful navigation of both internal silos and external threats.

**Deterioration Effects**. Weakness in both dimensions created downward spirals. One CISO described the challenge of limited resources impacting both fits:

*“Working in the industry that we work in… there are very little regulatory compliance requirements”* (BuildingMat01),

yet still facing sophisticated threats with constrained budgets.

#### *5.6.5.3 Temporal Dynamics of Dual-Fit*

Both fits require continuous negotiation rather than one-time achievement. The temporal challenges were captured by one participant reflecting on role evolution:

*“The initial leadership team that took me in, they’re no longer here. The CTO is different. The executive director is different. CEO is different. Everybody is different… Now all your contacts and interpersonal relationships and influence, everything is reset. You have to start from the scratch…”* (GovAgency01).

This organisational churn required constant recalibration of internal fit while maintaining vigilance against evolving external threats.

#### *5.6.5.4 Political Capital as Moderator*

Political capital emerged as a key moderator of dual-fit maintenance ability. One experienced CISO explained the protective effect of accumulated credibility:

*“We’ve probably got one of the most mature control frameworks in the country, and that’s simply because of our client base. Our client base is the top end of town”* (Super01).

This established reputation provided buffer during periods of misalignment.

The importance of political navigation was emphasised repeatedly. As one participant noted:

*“I’ve worked in two organizations which are large family-owned businesses. And in large family-owned businesses, politics– I’m going to be very careful what I say here. The political agenda is very strong”* (BuildingMat01).

Successfully maintaining dual-fit required careful political management alongside technical competence.

#### *5.6.5.5 The Pervasive Role of Power and Politics*

Perhaps the most profound insight was recognising that power dynamics and organisational politics permeate all fit relationships rather than existing as separate variables.

Every successful alignment story included political navigation. One CISO articulated the reality:

*"There's politics everywhere, right? That's corporate world"* (Super01).

This was not an occasional challenge but a constant consideration in maintaining both organisational and environmental fit.

The critical role of political acuity was further emphasised:

*"And while the CISO in the C Suite might have the board appointed responsibility, unless they've actually got everyone else in the C Suite on board, they can't actually implement what they're trying to achieve or they implement something that's more on a silo that is not as integrated with the rest of the organization."* (MgtConsult01).

This highlights how formal authority without political alignment proves insufficient for effectiveness.

### 5.6.6 Theoretical Integration: From Empirical Patterns to Theoretical Framework

The journey from empirical patterns to theoretical framework required recognising how established theories could explain observed phenomena while acknowledging where new theoretical development was needed.

#### *5.6.6.1 Recognition of Theoretical Resonances*

As patterns solidified through analysis, connections to established theories became apparent. Person-Organisation fit theory emerged from consistent patterns of alignment and misalignment between CISOs and their organisations. One participant's reflection captured this dynamic:

*"...we are going about doing the right thing based on best practices to continuously improve our cybersecurity controls in the best interests of our clients."* (MgtConsult01),

highlighting the need for mutual adaptation between CISO and organisation.

Person-Environment fit theory arose from patterns of CISO adaptation to external pressures. The challenge of environmental alignment was captured by one participant:

*"...more contemporary sort of architectures that we have, the cloud and everything else. And that's a big one... The reality is not. It's actually a bit more complicated than that. You actually need to start thinking about controls in a completely different way."* (Bank01).

This technological evolution required continuous adaptation to maintain environmental fit.

Contingency theory became relevant through phase-dependent effectiveness patterns. Different organisational phases required different approaches, as one CISO noted:

*"Once you have a level of maturity of the organization to a certain level from cybersecurity and the organizational maturity of a different level, then you can start seeing how the CISO is going to add value to the performance"* (HigherEd02).

Dynamic capabilities theory resonated with the need for continuous evolution. One participant captured this imperative:

*"The framework we use for preventive is the MITRE ATT&CK framework, whereby we understand what techniques are the ones that the threat actors that we're most concerned about are using... And we test ourselves against each of those techniques to see whether we can both prevent them and or detect them."* (Beverage01).

This systematic capability development reflected dynamic adaptation to evolving threats.

Role theory emerged through discussions of changing expectations. The tension between role stability and evolution was captured by one CISO:

*"My role probably changed a lot in the 18 months that I've been in my position"* (BuildingMat01),

illustrating how role expectations shift even within stable organisational structures.

#### *5.6.6.2 Theoretical Synthesis Decision*

Our analysis revealed that no single theory adequately explained the dual-fit phenomenon. Each illuminated aspects of the findings but none captured full complexity. This theoretical insufficiency led us to synthesise multiple theories into an integrated framework.

The pervasive nature of these dual-fit dynamics across our cases suggested that neither P-O fit nor P-E fit theory alone could explain the observed patterns. Participants who achieved strong organisational fit but struggled with environmental demands described feeling protected internally but overwhelmed by external pressures. As one CISO noted about the evolving threat landscape:

*"The threat has completely changed to a point now that we are against things like nation-states... We're talking about organized crime..."* (Bank01).

Conversely, those adept at managing environmental challenges but misaligned organisationally described being constrained by internal limitations. The political realities of organisational life permeated these dynamics, with one participant simply noting,

*"There's politics everywhere, right? That's corporate world"* (Super01),

acknowledging the unavoidable influence of organisational politics on their effectiveness. Only when both fits were maintained did participants describe sustainable effectiveness: the ability to deliver security value while maintaining personal engagement and professional growth.

The Security Leadership Contingency Model emerged as a meta-framework showing how dual fits evolve through interaction of multiple theoretical mechanisms. This synthesis recognised that the phenomenon's complexity required multiple lenses for complete understanding.

#### *5.6.6.3 Theoretical Innovation: Dynamic Dual-Fit*

The key theoretical innovation involved reconceptualising fit as inherently dynamic and necessarily dual. Traditional fit theories often treat fit as an outcome to achieve rather than a process to maintain. The dynamic nature was captured by one participant using a powerful metaphor:

*"...when an effective leader has successfully done their job, people believed it happened naturally."* (Tech01).

This suggests that effective fit appears effortless but requires continuous, often invisible, work to maintain.

This reconceptualisation provides more nuanced understanding of how person-context alignment operates in dynamic, complex organisational environments. The dual-fit imperative means CISOs must simultaneously navigate internal organisational dynamics while responding to external environmental pressures, with success requiring continuous adaptation in both dimensions.

#### *5.6.6.4 Building Toward the Security Leadership Contingency Model*

The theoretical integration accomplished in this section established the conceptual foundation for the Security Leadership Contingency Model. We identified three contextual groupings that shape security leadership: organisational factors, individual CISO characteristics, and environmental pressures. We constructed two aggregate dimensions, CISO-Organisation Fit and CISO-Environment Fit, that together define effectiveness through their dynamic interaction. We recognised that political navigation permeates these fit relationships, operating as the medium through which alignment is negotiated and sustained. Finally, we synthesised multiple theoretical perspectives to explain these complex dynamics.

These empirical and theoretical foundations set the stage for Chapter 6, where we present the Security Leadership Contingency Model as a comprehensive process framework. The model integrates these elements to explain how security leadership effectiveness emerges from continuous dual-fit maintenance across evolving organisational and environmental contexts.

## 5.7 CHAPTER SYNTHESIS: FROM EMPIRICAL PATTERNS TO THEORETICAL FRAMEWORK

This chapter presented the empirical heart of the investigation into security leadership effectiveness. Through systematic application of the Gioia methodology to twenty in-depth interviews with Australian CISOs, our analysis progressed from over sixty first-order concepts preserving participant voice, through sixteen second-order themes capturing cross-case patterns, to two aggregate dimensions representing the highest level of theoretical abstraction. The analytical journey revealed that security leadership effectiveness emerges not from universal characteristics or best practices, but from dynamic alignment between CISOs and their contexts. This synthesis reflects on the empirical foundations established and bridges to the theoretical development presented in Chapter 6.

### 5.7.1 The Analytical Progression

We began with open coding that generated first-order concepts drawn directly from participant language. Through constant comparative analysis, we abstracted these into second-order themes capturing patterns across diverse organisational settings. Three leadership orientations emerged inductively: action-oriented leaders focused on building, stewardship-oriented leaders emphasising optimisation, and vision-oriented leaders driving strategic integration. Organisational contexts revealed systematic variation across security needs, culture, structure, and maturity phases. Environmental factors created adaptation pressures that participants navigated with varying success.

As we identified first-order concepts, we organised them into three contextual groupings based on what aspect of security leadership they addressed: Organisational Context, CISO Characteristics, and Environmental Context. Within each grouping, we then abstracted similar first-order concepts into second-order themes, ultimately developing sixteen themes that maintained the three-grouping structure. However, static categorisation could not adequately explain the effectiveness patterns we observed. The critical insight emerged when we recognised that effectiveness arose not from characteristics within any single grouping, but from dynamic alignment relationships across groupings. This led us to construct two aggregate dimensions: CISO-Organisation Fit and CISO-Environment Fit. We developed these not as hierarchical containers but as relational constructs capturing how alignment between CISOs and their contexts produces effectiveness outcomes.

Throughout this progression, we maintained grounded theory principles. Theoretical saturation occurred at interview seventeen. Multiple analytical turning points shaped theoretical development: the fit revelation at interview eight, the temporal dynamics discovery after interviews ten through twelve, the political capital insight at interview thirteen, and the environmental complexity recognition at interview fifteen.

### 5.7.2 What the Empirical Analysis Revealed

Our analysis revealed patterns that challenge conventional assumptions about security leadership. Participants who succeeded in one context struggled in others despite unchanged capabilities. Effectiveness declined when organisations evolved beyond phases matching a CISO's natural orientation, even when technical competence remained constant. Both organisational contexts and environmental demands evolved continuously, requiring ongoing fit recalibration. Political navigation pervaded all successful alignment stories. Most significantly, we discovered that both fits must be maintained simultaneously: strong organisational fit without environmental fit produced internally focused leadership, whilst strong environmental fit without organisational fit created politically ineffective leadership.

These patterns point toward a fundamental reconceptualisation. Effectiveness cannot be understood through static role descriptions, universal competency models, or acontextual best practices. Rather, our empirical findings indicate that effectiveness emerges as a dynamic phenomenon requiring continuous adaptation across multiple dimensions. The data structures presented in this chapter provide essential taxonomy, yet the relationships between elements demand theoretical development beyond description.

### 5.7.3 From Empirical Patterns to Theoretical Need

The empirical patterns we identified reveal theoretical requirements that existing frameworks cannot adequately address. Static categorisation, whilst useful for organising findings, proves insufficient for explaining dynamic effectiveness relationships. Participants consistently described evolution, adaptation, and transition. Their experiences revealed temporal dependencies and contextual contingencies that demand process-oriented theorising.

No single established theory adequately explained the dual-fit phenomenon we discovered. Person-Organisation fit theory illuminated internal alignment but neglected environmental pressures. Person-Environment fit theory captured external adaptation but underspecified organisational dynamics. Contingency theory explained phase-dependent patterns but lacked mechanisms for continuous evolution. Each theoretical perspective illuminated aspects of our findings whilst leaving others unexplained.

This theoretical insufficiency necessitates integration and innovation. The empirical patterns demand frameworks capable of explaining how CISOs and contexts co-evolve, how fits deteriorate and regenerate, how political capital moderates alignment relationships, and how environmental disruptions cascade through both fit dimensions. Chapter 6 addresses these theoretical requirements through development of the Security Leadership Contingency Model, synthesising multiple theoretical perspectives whilst introducing novel constructs to explain the complex, dynamic nature of security leadership effectiveness revealed through our empirical analysis.

# CHAPTER 6: THEORETICAL DEVELOPMENT - THE SECURITY LEADERSHIP CONTINGENCY MODEL

## 6.1 INTRODUCTION: FROM PATTERNS TO THEORY

Chapter 5 presented the empirical foundation revealing how security leadership effectiveness emerges from maintaining dual dynamic fits: alignment between CISOs and their organisational contexts, and alignment between CISOs and their environmental demands. This chapter builds on that empirical foundation to present the Security Leadership Contingency Model (SLCM), a theoretical framework synthesising multiple theoretical lenses to explain how security leadership effectiveness is achieved and maintained over time. The model represents this study's primary theoretical contribution, offering both scholarly advancement in understanding organisational security leadership and practical guidance for security leader selection, development, and transition planning.

### 6.1.1 Chapter Purpose and Scope

This chapter transforms empirical patterns into theoretical contribution. Whilst Chapter 5 revealed what matters for security leadership effectiveness through systematic grounded theory analysis, this chapter explains why these patterns occur, how they interact dynamically, and what implications they hold for theory and practice. We present the SLCM as a process model explaining the complex, contingent, and political nature of security leadership effectiveness.

The chapter progresses through four key stages. First, we present the model's conceptual architecture, showing how three contextual groupings generate two aggregate dimensions that define effectiveness through their dynamic interaction. Second, we articulate the model's theoretical logic, demonstrating how different leadership orientations align with specific organisational phases and environmental conditions. Third, we formalise four theoretical propositions capturing core relationships discovered through empirical analysis. Fourth, we engage with complexity by examining negative cases, alternative explanations, and contextual variations that refine the model's boundary conditions.

This theoretical development maintains strong empirical grounding whilst achieving conceptual abstraction necessary for theoretical contribution. The model emerged from participant experiences rather than predetermined frameworks, ensuring relevance to security leadership practice whilst advancing scholarly understanding of how contextual alignment produces effectiveness in dynamic organisational environments.

### 6.1.2 Why Theoretical Development Was Necessary

The empirical patterns revealed through grounded theory analysis pointed beyond descriptive categorisation toward theoretical explanation. Three inadequacies in existing frameworks necessitated new theoretical development.

First, static categorisation proved insufficient for explaining dynamic effectiveness relationships. Our data structure successfully organised over sixty first-order concepts into sixteen second-order themes grouped into three contextual domains. However, participants consistently described evolution, adaptation, and transition. Their experiences revealed temporal dependencies and contextual contingencies that taxonomies cannot capture. Effectiveness emerged not as a fixed state but as continuous accomplishment requiring ongoing adaptation. This dynamic nature demanded process-oriented theorising rather than structural categorisation.

Second, single theoretical perspectives could not adequately explain the dual-fit phenomenon. Person-Organisation fit theory illuminated internal alignment but neglected environmental pressures. Person-Environment fit theory captured external adaptation but underspecified organisational dynamics. Contingency theory explained phase-dependent patterns but lacked mechanisms for continuous evolution. Dynamic capabilities perspective addressed adaptation imperatives but insufficiently theorised political dimensions. Role theory clarified transition challenges but required integration with fit concepts. Each perspective illuminated aspects of our findings whilst leaving others unexplained, necessitating theoretical synthesis.

Third, the pervasive role of organisational politics demanded explicit theorisation. Every successful alignment story included sophisticated political navigation. CISOs accumulated and deployed political capital to maintain fits during transitions, build coalitions for security initiatives, and manage competing stakeholder interests. Political acumen emerged not as peripheral skill but as meta-capability enabling all other aspects of security leadership. Political capital refers to the accumulated influence a CISO derives from relationships, reputation, and track record, treated in its moderating function rather than as an independently theorised construct. Existing fit theories acknowledge contextual factors but inadequately theorise the political medium through which alignment is negotiated and sustained.

These theoretical inadequacies required developing a process model capable of explaining how CISOs and contexts co-evolve, how fits deteriorate and regenerate, how political capital moderates alignment relationships, and how environmental disruptions cascade through both fit dimensions.

### 6.1.3 The Security Leadership Contingency Model Preview

The Security Leadership Contingency Model reconceptualises effectiveness as the continuous maintenance of dual dynamic fits between security leaders and their contexts. The model integrates three contextual groupings, Organisational Context, CISO Characteristics, and Environmental Context, into two aggregate dimensions: CISO-Organisation Fit and CISO-Environment Fit. Effectiveness emerges when both fits are maintained simultaneously through sophisticated political navigation across evolving organisational and environmental landscapes.

The model's theoretical innovation lies in reconceptualising fit as inherently dynamic and necessarily dual. Traditional fit theories treat alignment as an outcome to achieve. Our empirical analysis revealed fit as continuous process requiring active maintenance. This reconceptualisation transforms theoretical understanding: effectiveness becomes not a state of being but ongoing accomplishment. The dual-fit imperative compounds this complexity, as CISOs must simultaneously maintain two distinct alignment relationships, each with its own dynamics and demands.

The model operates through four primary mechanisms. First, contingent alignment patterns match leadership orientations with organisational phases: action-oriented leaders excel during establishment, stewardship-oriented leaders during maturation, and vision-oriented leaders during strategic integration. Second, environmental factors moderate both fit dimensions, creating adaptation pressures that leaders must navigate whilst maintaining organisational alignment. Third, political capital buffers temporary misalignment during transitions, enabling adaptive leaders to recalibrate fits as contexts evolve. Fourth, when dual-fit maintenance becomes untenable, strategic succession preserves organisational effectiveness through leadership transitions.

The following sections present the model in detail, demonstrating how these mechanisms operate, formalising theoretical propositions that capture core relationships, and engaging with complexity through examination of negative cases and boundary conditions. The model synthesises multiple theoretical perspectives whilst remaining firmly grounded in the empirical patterns revealed through grounded theory analysis.

## 6.2 THEORETICAL CRYSTALLISATION: THE SECURITY LEADERSHIP CONTINGENCY MODEL

### 6.2.1 From Structure to Process: Theoretical Framework Development

Our empirical analysis revealed patterns that static categorisation could not adequately explain. Whilst the data structure identified discrete organisational phases, leadership types, and environmental factors, participants' lived experiences demonstrated continuous evolution and adaptation. This recognition necessitated theoretical development beyond descriptive typologies toward a dynamic process model.

#### *6.2.1.1 Building the Theoretical Framework*

We developed the Security Leadership Contingency Model through systematic integration of multiple theoretical perspectives. Each theoretical lens addressed distinct empirical patterns whilst contributing to a coherent theoretical whole.

**Dual-Fit Foundation.** The model's foundation comprises two aggregate dimensions: CISO-Organisation Fit and CISO-Environment Fit. Data analysis consistently demonstrated both fits as necessary for effectiveness. However, traditional fit theories conceptualise alignment as a static achievement (Kristof-Brown et al., 2005), a perspective insufficient for explaining participants' dynamic experiences. One CISO articulated this temporal dimension:

*"My role probably changed a lot in the 18 months that I've been in my position... it's now very much an Australian based role"* (BuildingMat01).

This dynamism demanded theoretical innovation beyond established fit frameworks.

**Contingency Theory Layer.** Contingency theory (Fiedler, 1967; Lawrence & Lorsch, 1967) accounts for the situational effectiveness patterns observed across interviews. Different organisational phases require fundamentally different leadership approaches; environmental conditions create varying leadership demands. The data rejected notions of universal "best" CISO characteristics, instead revealing effectiveness as inherently context-dependent. This layer explains why action-oriented leaders excel during establishment phases whilst struggling in strategic contexts.

**Dynamic Capabilities Integration.** The dynamic capabilities perspective (Eisenhardt & Martin, 2000; Teece et al., 1997) illuminates how CISOs and organisations co-evolve. Leaders must develop new capabilities as contexts shift, technical CISOs learning business strategy, operational leaders developing transformation skills. Simultaneously, organisations must evolve security capabilities from basic controls to integrated risk management. The environment continuously presents novel challenges requiring capability development. This theoretical lens explains the “growing beyond the role” phenomenon participants repeatedly described.

**Role Theory Lens.** Role theory (Biddle, 1986; Katz & Kahn, 1978) illuminates critical transition challenges CISOs face. Role expectations shift dramatically with organisational maturity, from hands-on implementer to strategic advisor. Environmental changes redefine role requirements: new regulations demanding compliance expertise, emerging threats requiring technical depth. These role transitions represent critical junctures for maintaining dual fits, explaining why many participants described leaving roles when organisations evolved beyond their preferred operating mode.

These four theoretical perspectives contribute differently to the model's architecture. Person-organisation fit theory and person-environment fit theory provide the foundational alignment requirements; dynamic capabilities theory and role theory illuminate specific mechanisms of adaptation and transition. Contingency theory occupies a distinct position as the second core theoretical foundation. It transforms the model beyond static fit conceptualisations by establishing that alignment requirements are not universal but contingent upon organisational phase and environmental conditions. The model is therefore designated a contingency model rather than a fit model; the name reflects what is theoretically distinctive rather than what is foundational.

Figure 6.1 illustrates this theoretical framework, showing how the four theoretical perspectives combine across the three tiers described above to inform the SLCM.

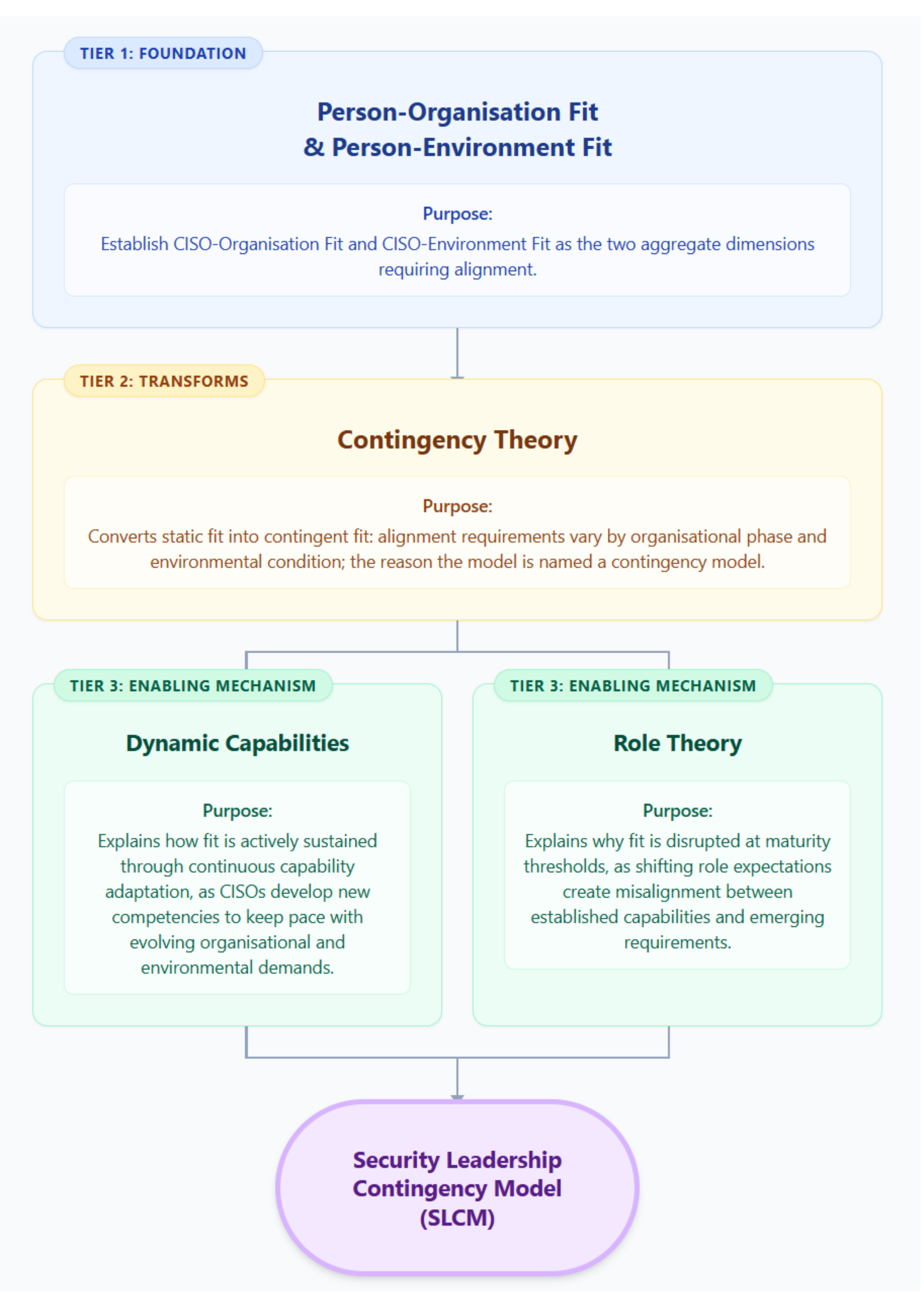

*Figure 6.1. Theoretical Framework Underpinning the SLCM*

#### *6.2.1.2 Theoretical Innovation: Dynamic Dual-Fit*

Our theoretical contribution synthesises these perspectives into the concept of "Dynamic Dual-Fit." This synthesis represents more than theoretical eclecticism; it constitutes a novel reconceptualisation of fit as inherently processual. The integration operates through multiple mechanisms:

- **Dual Fits** (CISO-Organisation + CISO-Environment) establish fundamental alignment requirements
- **Contingency** logic explains how these fits vary by organisational and environmental context
- **Dynamic Capabilities** account for continuous evolution required to maintain fits
- **Role Theory** illuminates transition challenges that threaten fit maintenance

The key insight emerging from this synthesis: both fits are not achieved but continuously maintained through ongoing adaptation. This reconceptualises dual fits as dynamic properties requiring active management across multiple dimensions simultaneously.

### 6.2.2 Presenting the Conceptual Model

Figure 6.2 presents the Security Leadership Contingency Model, visualising dynamic relationships between contextual factors, leadership characteristics, and effectiveness outcomes.

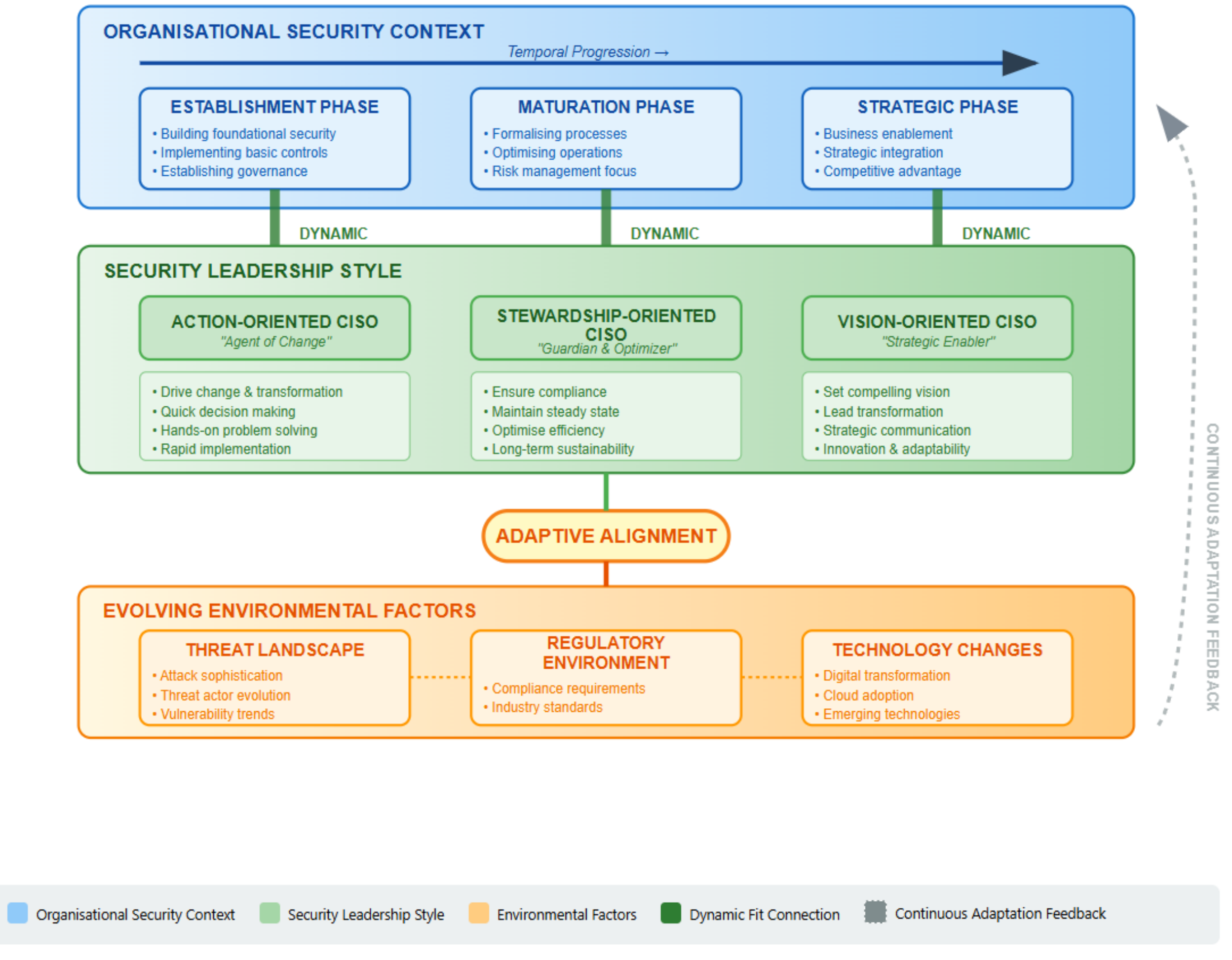

*Figure 6.2. The Security Leadership Contingency Model.*

### 6.2.2.1 Model Components

**Three Contextual Groupings.** First, **Organisational Context** encompasses internal factors creating leadership demands. This grouping progresses through three phases: Establishment → Maturation → Strategic. Each phase generates distinct leadership requirements, with transitions between phases creating adaptation imperatives.

Second, **CISO Characteristics** represents individual leader attributes and orientations. We identified three leadership types: Action → Stewardship → Vision. Additionally, technical skills, experience, and personality traits contribute to overall leadership capability. These characteristics evolve through deliberate development and experiential learning.

Third, **Environmental Context** captures external factors creating adaptation pressures. These include threat landscape evolution, regulatory requirements, technological change, and resource constraints. Unlike organisational phases, environmental factors operate continuously rather than sequentially.

**Two Aggregate Dimensions.** The model's core comprises two fit dimensions. **CISO-Organisation Fit** represents alignment between leader characteristics and organisational needs. This multi-dimensional construct encompasses surface-level skill matching and deep-level value alignment. The fit exhibits three key properties: it operates across multiple theme intersections, changes as organisations evolve, and involves reciprocal influence between leader and organisation.

**CISO-Environment Fit** captures alignment between leader capabilities and environmental demands. This construct acknowledges the boundary-spanning nature of security leadership, requiring external legitimacy alongside internal effectiveness. Environmental fit demands continuous adaptation to external changes whilst managing resource constraints.

**Dynamic Dual-Fit Process.** The model's innovation lies in conceptualising both fits as continuously negotiated rather than achieved. Environmental disruptions affect both fit dimensions simultaneously, whilst organisational evolution alters parameters of effective alignment. Political capital and formal authority moderate the ability to maintain fits during turbulent periods.

**Adaptive Alignment Mechanisms.** Four mechanisms enable ongoing fit maintenance. Individual adaptation occurs through capability development. Organisational adaptation involves structural evolution. Environmental navigation requires proactive positioning. Strategic succession preserves effectiveness when dual-fit becomes untenable.

### 6.2.3 Model Dynamics and Theoretical Logic

#### *6.2.3.1 Dual-Fit Alignment Patterns*

**CISO-Organisation Fit Patterns.** Through systematic cross-case analysis, we identified specific alignment patterns between leadership types and contextual phases.

*Establishment Phase and Action-Oriented Leadership.* Action-oriented CISOs excel when organisations require rapid security capability building. Their bias toward implementation over analysis matches organisational needs for tangible progress. One participant exemplified this alignment:

*"I came into the role about 18 months ago… the team was three"* (BuildingMat01).

These leaders tolerate ambiguity whilst driving foundational changes.

*Maturation Phase and Stewardship-Oriented Leadership.* Stewardship-oriented leaders align with organisations seeking process refinement and compliance. Their systematic thinking and measurement focus match organisational needs for standardisation. A CISO in this phase described:

*"We use the NIST framework and we are assessed independently by a third party every 12 months"* (Retail01).

*Strategic Phase and Vision-Oriented Leadership.* Vision-oriented CISOs succeed when organisations view security as strategic enabler. Their stakeholder influence skills and business acumen match needs for enterprise-wide integration. One participant captured this alignment:

*"We've been seen as an enabler and as a business enabler rather than just a technology enabler"* (Aviation01).

**CISO-Environment Fit Patterns.** Environmental demands create distinct capability requirements:

- High threat environments demand technical depth for effective defence
- Regulatory environments require compliance expertise for governance
- Innovation environments need strategic thinking for enablement

These patterns interact with organisational fit, creating complex alignment challenges. A technically skilled CISO might achieve strong environment fit in high-threat contexts whilst experiencing organisational misalignment if the company prioritises compliance over defence.

#### *6.2.3.2 Environmental Disruption Effects*

Environmental factors moderate both fit dimensions simultaneously, creating cascading adaptation requirements. We identified three primary disruption patterns.

**Threat-Driven Acceleration.** Major security incidents accelerate organisational progression through maturity phases. One participant described this compression:

*"So in the event of a ransomware threat, the loss of critical ICT services and the impact that that would have on our wider supply chains. So the ability to be able to order ticket all the way through to dispatch the business ultimately comes to a screaming halt"* (BuildingMat01).

Such events compress typical evolution timelines, potentially creating acute misalignment as organisations suddenly require different leadership capabilities.

**Regulatory-Forced Maturation.** New compliance requirements force standardisation regardless of organisational readiness. A participant noted:

*"PCI compliance was a big one. Our business revolves around taking payment"* (Food01).

This external pressure creates tension when action-oriented leaders must suddenly deliver process documentation and audit evidence.

**Technology-Enabled Leapfrogging.** Technological shifts enable organisations to bypass traditional maturity progression. One participant observed:

*"Because we are a startup and fintech"* (Bank01).

This requires leaders capable of operating across multiple maturity levels simultaneously.

#### *6.2.3.3 Adaptive Mechanisms for Maintaining Dual Fits*

Four interconnected mechanisms enable ongoing fit maintenance:

**Individual Adaptation.** Leaders actively develop capabilities to maintain alignment as contexts evolve. This includes formal education, mentoring relationships, and deliberate skill building. Successful CISOs described continuous learning as essential:

*"So when I joined, I guess I was– this was back in September. So I was still figuring out how the organization worked... I had to figure out how to get the board engaged, how to come up with a strategy very quickly"* (MgtConsult01).

Individual adaptation encompasses both technical capability development and softer skills like political navigation.

**Organisational Adaptation.** Organisations evolve structures and processes to support leader effectiveness. This includes adjusting reporting relationships, modifying role scope, and building complementary teams. Several participants described how their organisations created new governance structures to maintain alignment as complexity increased.

**Environmental Navigation.** Proactive environmental scanning and positioning help maintain external fit. Leaders build regulatory relationships, participate in threat intelligence sharing, and influence technology decisions. This mechanism involves shaping the environment rather than merely reacting to changes.

**Strategic Succession.** When dual-fit maintenance becomes untenable, strategic leadership transitions preserve organisational effectiveness. Several participants recognised when they had "grown beyond the role" and facilitated smooth succession to leaders better suited to evolved contexts. This mechanism acknowledges that not all misalignments can be resolved through adaptation.

#### *6.2.3.4 Temporal Complexity of Dual-Fit Maintenance*

A critical pattern emerging from our data reveals the temporal complexity of dual-fit maintenance. Participants consistently described scenarios where short-term organisational security objectives could be achieved despite misalignment, but at significant personal cost. One CISO who left after a successful transformation reflected:

*"As long as you can keep your, your mental health and your well-being in a good state, then for me, the job is really satisfying and it's really rewarding. But if that starts to get out of kilter, I think the answer's gotta be no… I would say probably in the last four or five months, that balance is just maybe that's why I decided well, it is a big part why I decided to take another opportunity"* (HigherEd01).

This pattern of *burnout despite delivering organisational value* illuminates why both fits must be maintained for sustainable effectiveness.

The data suggests that organisations often mistake short-term security achievements for effective security leadership. They fail to recognise that without proper dual-fit alignment, such achievements come at the cost of leadership sustainability and ultimately organisational security continuity. This temporal dimension distinguishes between episodic security success and sustainable security leadership effectiveness. Several participants described delivering transformational outcomes whilst experiencing progressive misalignment, ultimately leading to their departure despite organisational success metrics. This pattern underscores the model's central premise: dual-fit maintenance is not merely optimal but essential for sustainable security leadership effectiveness.

### 6.2.4 Theoretical Propositions

The Security Leadership Contingency Model generates four theoretical propositions, each empirically grounded and theoretically significant.

#### *6.2.4.1 Proposition 1: Dual-Fit Contingency*

**Security leadership effectiveness is contingent upon maintaining dual fits, both CISO-Organisation fit and CISO-Environment fit simultaneously**

This proposition challenges single-fit perspectives by demonstrating interdependence. Strong organisational alignment without environmental fit produces internally focused leadership that misses external threats. Strong environmental alignment without organisational fit creates technically sound but politically ineffective leadership. Only simultaneous achievement of both fits enables sustained effectiveness.

#### *6.2.4.2 Proposition 2: Environmental Moderation*

**Environmental factors moderate both types of fit, creating dynamic adaptation requirements**

Environmental changes do not merely affect external fit; they cascade through organisational contexts, altering internal fit parameters. A major breach shifts organisational priorities, potentially misaligning previously well-fitted leaders. Regulatory changes force organisational evolution, requiring leaders to adapt both externally and internally.

#### *6.2.4.3 Proposition 3: Continuous Recalibration*

**Sustained effectiveness requires continuous recalibration of both fits as contexts evolve**

Fit represents not an achievement but an ongoing process of alignment maintenance. This distinguishes our model from static fit conceptualisations. Leaders must simultaneously track organisational evolution, environmental shifts, and personal capability development, continuously recalibrating their approach to maintain dual alignment.

#### *6.2.4.4 Proposition 4: Political Capital Moderation*

**Political capital and formal authority moderate the ability to maintain dual fits over time**

Whilst dual-fit alignment creates effectiveness conditions, political dynamics determine sustainability. Leaders with strong political capital can buffer temporary misalignment during transitions. Formal authority provides structural power to influence fit parameters rather than merely adapting to them. This proposition acknowledges the pervasive role of organisational politics in enacting fit relationships.

#### *6.2.4.5 Construct Definitions for Clarity*

Theoretical precision requires explicit construct definitions:

**CISO-Organisation Fit**: The alignment between CISO characteristics (leadership style, skills, experience, personality) and organisational context (security needs, culture, structure, maturity phase). This multi-dimensional construct encompasses both surface-level skill matching and deep-level value alignment.

**CISO-Environment Fit**: The alignment between CISO capabilities and environmental demands (threat landscape, regulatory requirements, technological change, resource constraints). This construct captures the boundary-spanning nature of security leadership, requiring external legitimacy alongside internal effectiveness.

**Action-Oriented CISO**: Leaders who prioritise rapid implementation, change management, and hands-on problem-solving. Distinguished by bias toward action over analysis, comfort with ambiguity, and focus on building foundational capabilities. These leaders excel at establishing security programmes but may struggle with standardisation requirements.

**Stewardship-Oriented CISO**: Leaders who emphasise process optimisation, compliance, and operational stability. Characterised by systematic thinking, measurement focus, and risk mitigation orientation. These leaders excel at refinement but may struggle with transformational mandates.

**Vision-Oriented CISO**: Leaders who focus on strategic integration, innovation, and business enablement. Identified by strategic thinking, stakeholder influence, and future-state orientation. These leaders excel at business alignment but may struggle with operational demands.

Whilst ideal types are presented for analytical clarity, several participants demonstrated hybrid orientations, typically combining primary and secondary styles. This suggests the typology represents orientations along continua rather than discrete categories, with leaders potentially developing repertoires spanning multiple types.

Having presented the Security Leadership Contingency Model, its theoretical foundations, dynamics, and core propositions, Section 6.3 examines how the model addresses complexity through negative cases, alternative explanations, and contextual variations.

## 6.3 ADDRESSING COMPLEXITY: NUANCES AND CONTRADICTIONS

The pursuit of theoretical rigour demands explicit engagement with complexity. This section systematically examines negative cases, alternative explanations, and contextual variations that emerged during analysis. Rather than weakening the Security Leadership Contingency Model, these nuances strengthen its explanatory power by establishing clear boundary conditions and acknowledging the multifaceted nature of security leadership effectiveness.

### 6.3.1 Negative Cases and Boundary Conditions

#### *6.3.1.1 Systematic Analysis of Negative Cases*

We identified a small but significant number of participants representing clear negative cases where CISOs achieved effectiveness despite apparent misalignment with the model's core propositions. These anomalous cases proved instrumental in refining theoretical boundaries and identifying compensatory mechanisms.

**Case 1: Action-Oriented Leadership in Mature Phase.** Aviation01 exemplified successful misalignment, operating as an action-oriented leader within a mature organisation. Traditional model predictions would suggest friction between change-driven approaches and steady-state requirements. However, this CISO created continuous transformation initiatives within the stable environment:

*"I'm not a business as usual CISO, like to get the organization to a position where things are getting to a steady state, but not interested in BAU, running the same thing over and over."* (Aviation01)

Additionally, they leveraged exceptional political capital from board relationships:

*"I've been quite fortunate in that I got senior leadership buy in from day one... I knew the chairman personally from [previous role], which meant I could bypass the usual prove-your-worth phase."* (Aviation01)

This case revealed how action-oriented leaders can maintain effectiveness in mature contexts by identifying transformation opportunities within stable operations, using political capital to create space for change initiatives that would typically be resisted in steady-state environments.

**Case 2: Vision-Oriented Leadership Signaling Phase Transition.** Super01 presented what initially appeared as a negative case but actually illustrated phase transition dynamics. Operating within a highly mature financial services environment with established frameworks, this vision-oriented CISO's success signalled organisational evolution toward strategic phase:

*"We've probably got one of the most mature control frameworks in the country, and that's simply because of our client base. Our client base is the top end of town."* (Super01)

Yet rather than maintaining steady state, they drove strategic innovation:

*"We've kind of got into quite a unique situation that security is truly an enabler. We run our own P&L for financial crimes, so we've got our own revenue, profits… We've saved over 70 million in superannuation fraud in two years through that product."* (Super01)

This case reveals how vision-oriented leadership success in mature contexts may indicate organisational readiness for strategic phase transition. When mature organisations begin viewing security as a revenue generator and competitive differentiator rather than just a compliance requirement, they require vision-oriented leadership to navigate this evolution. The case demonstrates the model's dynamic nature: phases are not static states but evolving processes.

**Case 3: Stewardship Leadership Driving Transformation.** One participant challenged assumptions about change leadership, successfully driving digital transformation through stewardship-oriented approaches. This CISO assembled coalitions of action-oriented deputies whilst maintaining governance stability. As Bank01 explained:

*"To be honest, right, there is no single blueprint for a security team. The security team needs to be built based on the experience of the CISO. So if the CISO is more GRC focused, then you get more technical people. If the CISO is more technical focus, you get more sort of GRC compliance type of people, and then there's a mix in between."* (Bank01)

This approach allowed the CISO to compensate for their own stewardship orientation by building a team with complementary capabilities. Another participant echoed this strategy:

*"I'll delegate based on the actual– whatever the technology may be that is affected, I'll delegate that to the SMEs."* (Food01)

This distributed leadership model enabled transformation without requiring personal style adaptation, highlighting the importance of complementary team composition.

#### *6.3.1.2 Theoretical Refinement Through Negative Case Analysis*

Initial model conceptualisation assumed direct correspondence between leadership style and organisational phase. Our negative case analysis revealed greater complexity, identifying three compensatory mechanisms:

**Political Capital as Buffer.** Pre-existing relationships and reputation create temporary flexibility in style-phase alignment. However, longitudinal pressure for alignment intensifies over time. Multiple participants described the exhausting nature of maintaining misalignment, with several noting that sustained misfit eventually forces either adaptation or departure from the role.

**Structural Workarounds.** Misaligned leaders create organisational structures that complement their limitations. These mechanisms include deputisation strategies, committee structures, and selective domain focus that enable effectiveness despite personal misalignment. As GovAgency01 described their team structure:

*"The way how I structure my team is one responsible for GRC kind of function, policy frameworks, risk management, gap assessment, and audits... And there is another role about security architect role, which actually interacts with the business projects or even IT projects."* (GovAgency01)

This structural approach allows leaders to maintain effectiveness by distributing responsibilities according to team members' strengths.

**Temporal Limitations.** Compensatory mechanisms prove unsustainable long-term. Multiple participants described "exhaustion" from maintaining misalignment, suggesting biological and psychological limits to adaptation. As HigherEd01 reflected on finding the right fit:

*"Get the right fit for what you need which really does require, you know, whoever's hiring a CISO to think about, 'Well, what do I need at this point in time?' Because what they need now might not be what they need in a couple of years' time."* (HigherEd01)

#### *6.3.1.3 Specified Boundary Conditions*

The Security Leadership Contingency Model operates within specific boundary conditions that emerged through theoretical saturation and negative case analysis:

**Organisational Type.** The model applies to established organisations with existing security functions. Pure greenfield contexts exhibit different dynamics where founder effects and rapid growth override phase-based requirements. Start-up environments may require fundamentally different leadership approaches not captured by the model.

**Geographic Context.** Developed economies with mature governance expectations provide the institutional backdrop. The model emerged from Australian organisational contexts characterised by strong regulatory frameworks and governance traditions. Emerging markets may exhibit compressed or alternative phase progressions.

**Temporal Context.** The model describes evolutionary periods where normal organisational dynamics operate. However, the adversarial nature of cybersecurity creates frequent disruptions to these normal dynamics. Crisis situations, whether from major incidents, sophisticated attacks, or threat actor campaigns, fundamentally alter leadership requirements. As participants noted:

*"The second one is really around OT and ICS and the potential for disruption to the manufacturing process."* (BuildingMat01)

During active incidents or periods of elevated threat, organisations prioritise immediate response capability over style alignment. These crisis periods create what one participant called "leadership compression," where all CISOs must temporarily adopt action-oriented approaches regardless of their natural style or organisational phase.

The continuous evolution of the threat landscape creates ongoing adaptation pressure that distinguishes security leadership from other executive roles:

*"The threat has completely changed to a point now that we are against things like nation-states if you're in an important sort of organization… We're talking about organized crime, which is basically the traditional crime groups that basically bring in– they outsource all of their development and exploit development and everything else… You're actually dealing with an entity that's actually putting money in to do what they need to do to extract information from you or make it into a business on their end"* (Bank01)

This constant adversarial pressure means that: - Crisis overrides occur regularly, not exceptionally - Leaders must maintain readiness to shift styles rapidly - Organisational phases can regress suddenly due to incidents - The model represents idealised states between crises

As BuildingMat01 further elaborated:

*"Every organization is completely different, and every organization has different levels of cyber maturity, and threat profiles… Despite the differences in the organizations and the different threat profiles, you still face the same threat. I think the risk might be quantified in a different manner"* (BuildingMat01)

**Cultural Context.** Western business environments valuing structured governance and role specialisation form the model's cultural foundation. The model assumes cultural acceptance of functional specialisation typical in Anglo-sphere business environments.

**Organisational Scale.** The model applies most clearly to organisations with sufficient scale to support dedicated security leadership distinct from broader IT responsibilities. In smaller organisations, the CISO role often combines with other technology leadership functions, fundamentally altering the dynamics described by the model. As participants noted, lean organisations create different leadership challenges where visibility increases but resources constrain options. Aviation01 observed:

*"We're only less than 600 people. Given the spread of people on Aviation01 at any day could be up to 40,000 people working there. It's a one to many type set up. So that means we're quite lean, but also means that even as ahead of– and I'm about three layers below the CEO, you're quite visible."* (Aviation01)

This visibility in smaller organisations can compensate for limited resources but also intensifies pressure on leaders to perform multiple roles simultaneously.

**Regulatory Environment.** Whilst the model accommodates regulatory variation, extreme regulatory environments impose leadership requirements that override natural progression. Critical infrastructure, defence, and financial services may mandate stewardship approaches regardless of organisational phase.

### 6.3.2 Alternative Explanations

Methodological rigour demanded systematic consideration of competing explanations for observed effectiveness patterns. We subjected three alternative explanations to particular scrutiny through constant comparison and theoretical sampling.

#### *6.3.2.1 Individual Competence Explanation*

The most parsimonious alternative suggests that competent individuals succeed regardless of contextual alignment: a *great person* theory of security leadership.

**Evidence Against.** Multiple participants explicitly described how the same capabilities that drove success in one context led to challenges in another. As Tech01 observed:

*“The organisation understanding what it needs at that time and place, and a CISO with the correct, not just skills, but the correct mindset to work with the business to deliver on that outcome.”* (Tech01)

Our cross-case analysis revealed competent leaders struggling in misaligned contexts whilst thriving when alignment improved. The data consistently contradicted pure competence explanations.

**Partial Truth.** Individual competence operates as a hygiene factor rather than differentiator. All participants met professional standards, yet effectiveness varied dramatically with context. Competence enables but does not ensure effectiveness.

#### *6.3.2.2 Industry Determinism*

This explanation proposes that industry characteristics determine optimal leadership styles: banking requires one approach, manufacturing another.

**Evidence Against.** Within-industry variation exceeded between-industry differences. Financial services participants ranged from action-oriented change agents in digital challengers to stewardship-focused leaders in established institutions. As Beverage02 noted:

*“Industries that have high levels of regulation… banking sector is a classic.”* (Beverage02)

Yet even within highly regulated sectors, leadership requirements varied based on organisational maturity and specific context. Manufacturing showed similar variation, from compliance-light contexts to highly regulated pharmaceutical manufacturing.

**Nuanced Reality.** Industry provides contextual overlay: regulatory requirements, threat profiles, technology dependence, that influences but does not determine optimal leadership style. Industry factors moderate rather than replace fit considerations.

#### *6.3.2.3 Resource Availability Explanation*

Perhaps effectiveness simply reflects resource allocation; well-funded functions succeed whilst under-resourced ones struggle.

**Evidence Against.** Several participants succeeded with minimal resources through contextual alignment, whilst others struggled despite generous funding. Rather than citing specific budget figures, participants emphasised the importance of alignment over resources. As MgtConsult01 noted:

*"We're doing information security... using a whole bunch of spreadsheets... It's exceptionally manual... However, spreadsheets are cheaper than going with a cybersecurity tool, which doesn't necessarily give you a better outcome."* (MgtConsult01)

Yet this constraint did not prevent effectiveness when leadership style matched organisational needs. Another participant reflected on the resource challenge:

*"So I think a CISO role now is not just a technology focus, it's a business focus for us and risk focus. But being able to split between all of those things in a day, in a meeting, in a conversation, that can happen at any level."* (Insurance01)

Resource availability emerged as an outcome rather than cause of effectiveness. Aligned leaders proved more adept at securing resources by framing requests in organisationally resonant language.

### 6.3.3 Contextual Variations

The model's expression varies systematically across multiple contextual dimensions, revealing implementation nuances crucial for practical application.

#### *6.3.3.1 Organisational Size Effects*

Organisational scale profoundly influences how dual-fit dynamics manifest through structural and resource mechanisms.

**Large Enterprises (5000+ employees).** Large organisations support multiple simultaneous leadership styles through structural differentiation. As Super01 illustrated:

*"We've probably got one of the most mature control frameworks in the country, and that's simply because of our client base. Our client base is the top end of town."* (Super01)

These organisations create "leadership portfolios" where vision-oriented executives oversee action-oriented divisional leaders, achieving alignment through aggregation rather than individual fit.

**Mid-size Organisations (500-5000 employees).** Mid-size organisations require more precise individual alignment due to limited structural flexibility. These organisations showed the strongest fit-effectiveness relationships. As Bank01 explained:

*"To be honest, right, there is no single blueprint for a security team. The security team needs to be built based on the experience of the CISO."* (Bank01)

Without resources for elaborate workarounds, style-phase alignment becomes critical for effectiveness.

**Small Organisations (<500 employees).** Small organisations exhibit fundamentally different dynamics where security leadership combines with broader responsibilities. Pure security leadership models apply inconsistently at this scale, with leaders needing to span multiple styles simultaneously.

#### *6.3.3.2 Temporal Variations*

The model's validity fluctuates with temporal context, particularly during organisational transitions and external shocks.

**Organisational Lifecycles.** Mergers, acquisitions, and restructures rapidly shift phase requirements. Retail01 described the impact of organisational change:

*"In September, so what are we now 24? So I've been there two years this month. And so that would have been '22. So September '21 was when they appointed a CTO. Previous to that, the security team was reporting into infrastructure."* (Retail01)

These transitions create leadership challenges when previously aligned leaders suddenly face new contexts requiring different approaches.

**Crisis Response Dynamics.** Major incidents temporarily override normal fit dynamics. During crisis, organisations value decisive action regardless of phase alignment:

*"I've been working on. So as a CISO my role is to really restore and recover the services. So contain the incident, restore and recover the services as soon as possible."* (Auto01)

Post-crisis periods often catalyse phase transitions as organisations reassess security postures. The cyclical nature of crisis and stability in cybersecurity creates unique leadership challenges absent in other executive roles.

**Technological Disruption.** Fundamental technology shifts reset organisational maturity. Cloud adoption, digital transformation, and emerging technologies force mature organisations toward establishment dynamics. As Tech01 observed:

*"When you sort of consider that, then yeah, the role and the remit of the CISO increased, and the expectation of what the CISO should be able to understand and do has also increased with that level of complexity."* (Tech01)

These disruptions create windows where leadership transitions become necessary or where hybrid approaches prove temporarily effective.

#### *6.3.3.3 Geographic and Cultural Variations*

Whilst maintaining Australian empirical focus, participants with international experience highlighted systematic variations.

**Regulatory Harmonisation.** Global regulatory convergence creates universal pressure toward stewardship orientation:

*"Financial institutions, A, they already had a head start and a lot of getting that maturity in place. So for them it's been a case of, okay, where are we– [again?], we got some of these gaps that we need to fix that the regulator is calling out."* (Beverage02)

This convergence may compress the model's phase variation in highly regulated industries, creating baseline stewardship requirements regardless of organisational maturity.

**Cultural Dimensions.** Organisational culture profoundly influences how security leadership operates. Cultural factors emerged as critical moderators of the model's relationships, particularly in terms of hierarchy acceptance and uncertainty tolerance. High power distance cultures, where hierarchical differences are strongly respected, may limit security leaders' ability to challenge organisational direction regardless of their leadership style. Conversely, egalitarian cultures may provide greater flexibility for CISOs to influence strategic direction even from lower organisational positions.

### 6.3.4 The Complexity of Political Navigation

Organisational politics emerged as the pervasive medium through which all model relationships operate. Rather than a discrete variable, political dynamics constitute the organisational physics governing how influence propagates and decisions crystallise.

#### *6.3.4.1 Politics as Organisational Physics*

Every successful alignment story included sophisticated political navigation; every misalignment involved political barriers. As GovAgency01 observed:

*"The moment that we lose influence, you know then we can't execute. So it's about constant effort through the changes that organization goes through, through the shifting of powers the*

*organization goes through. We continue to maintain that influence is how politics can impact us."* (GovAgency01)

Another participant acknowledged the cost of neglecting political dynamics:

*"But I have definitely underplayed politics. And I think I have been less successful at Retail01 because I did not play the political game."* (Retail01)

Political dynamics operate at multiple levels simultaneously. Formal structures provide the skeleton, but informal networks constitute the nervous system through which influence flows. Successful security leaders map and navigate both systems.

#### *6.3.4.2 Political Capital Accumulation Strategies*

Different leadership styles accumulate political capital through distinct mechanisms:

**Action-Oriented Leaders** build capital through rapid wins and crisis response. Their political currency derives from tangible results and demonstrated competence under pressure:

*"I think uplift in the security posture. Measurable uplift. Tangible uplift that you can show. And I don't mean you've rolled out a new security control. You can actually show an uplift in the security and maturity of the business and maybe reduced incidents, reduced phishing, whatever else"* (Super02)

This crisis competence proves particularly valuable:

*"I think the fourth piece is ability to manage incidents confidently... when things happen... the business turns to the Chief Information Security Officer to assist in leading major incident response or incidents, whether they're ones of availability or confidentiality, you need to be able to step in and in some cases lead the incident response for the organization"* (Beverage02)

**Stewardship-Oriented Leaders** accumulate capital via consistent delivery and risk prevention. Their political strength emerges from reliability and trust:

*"...how I define success is that if I am able to support the business in meeting objectives, if I am able to ensure that we have implemented the right tools, and we are aware of the gaps, and we have a plan in place to manage that. And where we are accepting the risk, we are making informed decisions where I can actually bring the risk to the right level of visibility."* (Auto01)

**Vision-Oriented Leaders** develop capital through strategic alignment and business partnership:

*"In my second iteration of strategy, I was very focused on how cybersecurity is going to add value to what HigherEd02 is going to be trying to achieve. If the target is retention of students... if the target is opening up a campus in another geography, how cybersecurity is going to help."* (HigherEd02)

#### *6.3.4.3 The Hidden Work of Alignment*

Maintaining dual-fit requires continuous political labour invisible in formal role descriptions. CISOs invested substantial effort in:

- **Reading organisational dynamics**: Understanding formal and informal power structures, identifying key influencers, recognising political undercurrents
- **Building strategic coalitions**: Cultivating allies across functions, managing up and sideways simultaneously, creating support networks before needing them
- **Framing security resonantly**: Translating technical risks into business language, aligning security initiatives with organisational priorities, choosing battles strategically
- **Managing competing interests**: Balancing stakeholder demands, navigating resource competition, mediating between security and business needs

This political work intensifies during misalignment. As Super01 noted:

*"There's politics everywhere, right? That's corporate world. To be honest, the corporate world is pretty brutal, especially if you're an ASX listed company."* (Super01)

Misaligned leaders must expend enormous energy compensating for style-phase mismatch through political manoeuvring, partially explaining high CISO turnover rates.

### 6.3.5 Synthesis: Embracing Theoretical Complexity

The nuances and contradictions revealed through systematic analysis strengthen rather than weaken the Security Leadership Contingency Model. By acknowledging complexity, negative cases, boundary conditions, alternative explanations, and contextual variations, we develop robust theory capable of guiding practice whilst respecting empirical reality.

The model's dual-fit framework provides necessary but insufficient conditions for effectiveness. Political skill, organisational support structures, and environmental pressures create a dynamic system where alignment represents the optimal but not exclusive path to effectiveness. This complexity reflects the multifaceted nature of cybersecurity leadership: technical expertise operating within organisational politics, strategic vision implemented through operational excellence, and continuous adaptation to environmental change.

Rather than seeking simplistic prescriptions, the model embraces the inherent complexity of security leadership:

*"There's never a simplistic utopian sort of thing. And certainly across organisation and organisation, there's no uniformity. Everyone does it differently, et cetera, right? So what you're faced with as a CISO is lots of complexity and trying to manage the risk around, well, just understanding that complexity to begin with and then trying to manage the risk associated with that complexity at an IT and business level is difficult."* (Tech01)

This sophisticated understanding enables security leaders and organisations to make informed decisions about leadership selection, development, and transition. The model provides a framework for navigating complexity whilst acknowledging that effective security leadership remains as much art as science. The ongoing adversarial nature of cybersecurity, with its cycles of stability and crisis, ensures that this complexity will persist and likely intensify as digital transformation continues to reshape organisational landscapes.

Having examined how the model addresses complexity through negative cases, alternative explanations, and contextual variations, Section 6.4 synthesises Chapter 6's theoretical contribution and connects it to broader scholarly and practical implications.

## 6.4 CHAPTER SYNTHESIS: THE THEORETICAL CONTRIBUTION

This chapter transformed the empirical patterns revealed in Chapter 5 into the Security Leadership Contingency Model, a theoretical framework explaining how security leadership effectiveness emerges and evolves. Through systematic integration of multiple theoretical perspectives, we developed a dynamic dual-fit model that reconceptualises effectiveness as continuous process rather than static achievement. This synthesis examines what we accomplished theoretically, how the model operates, and what complexity we engaged with to strengthen its explanatory power.

### 6.4.1 The Theoretical Achievement

We built the Security Leadership Contingency Model through systematic integration of multiple theoretical perspectives, each addressing distinct limitations in existing frameworks. Traditional fit theories explain alignment but conceptualise it as static achievement rather than ongoing process. Contingency theory accounts for phase-dependent effectiveness yet lacks mechanisms for continuous evolution. Dynamic capabilities perspectives address adaptation but underspecify the political dimensions through which organisational change actually occurs. No single theoretical lens could adequately explain the patterns we observed in participant experiences.

Our synthesis demonstrates how person-organisation fit theory, person-environment fit theory, contingency theory, dynamic capabilities perspective, and role theory complement each other when integrated through the dual-fit framework. This integration represents genuine theoretical innovation rather than mere eclecticism. By reconceptualising fit as inherently processual, we transform understanding from static alignment to continuous maintenance. This reconceptualisation emerged directly from participants describing constant adaptation, evolution, and recalibration as central to their effectiveness. The theoretical innovation lies not in the individual perspectives but in their integration through the dynamic dual-fit concept.

### 6.4.2 How the Model Works

The Security Leadership Contingency Model operates through three contextual groupings that collectively generate two aggregate dimensions. Organisational Context, CISO Characteristics, and Environmental Context shape the dual-fit dynamics central to effectiveness. These groupings themselves contain multiple interacting elements, with organisational phases evolving over time, leadership orientations adapting to demands, and environmental pressures creating continuous turbulence.

The model's core insight centres on dual-fit maintenance. CISO-Organisation Fit captures alignment between leader characteristics and organisational needs, varying systematically with organisational phase. The phase-contingent patterns reveal how action-oriented leaders excel during establishment phases requiring rapid capability building, whilst stewardship-oriented leaders match maturation phases demanding process refinement, and vision-oriented leaders suit strategic phases enabling business integration. Yet this alignment proves dynamic rather than static, requiring continuous recalibration as organisations evolve through these phases.

CISO-Environment Fit captures alignment between leader capabilities and external demands including threat landscapes, regulatory requirements, technological change, and resource constraints. Environmental factors operate independently of organisational control, creating adaptation pressures that leaders must navigate whilst simultaneously maintaining organisational alignment. The dual-fit imperative means effectiveness requires maintaining both alignments concurrently, with weakness in either dimension undermining overall effectiveness regardless of strength in the other.

The model identifies adaptive mechanisms enabling ongoing fit maintenance. Leaders develop new capabilities individually through formal education and experiential learning. Organisations adapt structures and processes to support evolving security requirements. CISOs proactively navigate environmental changes through regulatory relationship building and threat intelligence participation. When dual-fit maintenance becomes untenable despite these efforts, strategic succession preserves organisational effectiveness through leadership transition. These mechanisms operate through political dynamics rather than purely technical or rational processes, with political capital moderating the ability to maintain fits during turbulent periods.

### 6.4.3 Engaging with Complexity

Rather than seeking simplistic prescriptions, we explicitly engaged with complexity to strengthen the model's explanatory power. Negative cases, where CISOs succeeded despite apparent misalignment, revealed compensatory mechanisms unavailable through conventional analysis. Political capital accumulated through prior relationships buffers temporary misalignment during transitions, whilst structural workarounds enable effectiveness through complementary team composition. These discoveries refined the model whilst establishing boundary conditions for its operation.

We subjected the dual-fit framework to systematic testing against alternative explanations. The individual competence explanation proved insufficient, with our data showing technically excellent leaders succeeding or struggling based on contextual alignment rather than absolute capability. Industry determinism similarly failed to account for within-industry variation that exceeded between-industry differences in leadership requirements. Resource availability emerged as outcome rather than cause of effectiveness, with aligned leaders proving more adept at securing resources by framing security investments in organisationally resonant language. These tests strengthened confidence in the dual-fit framework's explanatory power whilst revealing its boundary conditions.

Contextual variations across organisational size, temporal periods, and cultural settings revealed implementation nuances crucial for understanding how the model operates in practice. Large organisations can support multiple simultaneous leadership styles through structural differentiation, creating "leadership portfolios" that achieve alignment through aggregation. Mid-size organisations demonstrate the strongest fit-effectiveness relationships due to limited capacity for elaborate workarounds, making style-phase alignment critical. Crisis periods temporarily override normal fit dynamics as organisations prioritise immediate response capability, requiring all CISOs to adopt action-oriented approaches regardless of their natural style or organisational phase. These variations acknowledge real-world complexity whilst maintaining theoretical coherence.

Political dynamics emerged as the pervasive medium through which all model relationships operate. Rather than treating politics as discrete variable, we recognise political navigation as the organisational physics governing how influence propagates and decisions crystallise. Different leadership orientations accumulate political capital through distinct mechanisms, yet all effective security leaders demonstrate sophisticated political skill in maintaining dual fits whilst navigating the stakeholder complexity characterising modern organisations.

### 6.4.4 Reflecting on Theory Building

The progression from empirical patterns to theoretical model exemplifies rigorous grounded theory methodology. We maintained strong empirical grounding whilst achieving the conceptual abstraction necessary for theoretical contribution. The systematic movement from over sixty first-order concepts through sixteen second-order themes to three contextual groupings and finally two aggregate dimensions demonstrates disciplined analytical progression that honours both participant voices and theoretical requirements.

Critical analytical turning points shaped theoretical development throughout the data collection and analysis process. Early interviews crystallised recognition of leadership orientations extending beyond simple technical-strategic dichotomies, revealing action, stewardship, and vision as distinct approaches to security leadership. Subsequent analysis revealed temporal dynamics that necessitated processual rather than static conceptualisation of fit relationships. Later negative cases illustrated political capital’s moderating role, demonstrating how pre-existing relationships could buffer misalignment during transitional periods. These pivotal moments demonstrate how theoretical sensitivity develops through iterative engagement with data rather than predetermined frameworks.

The negative cases proved particularly valuable in refining theoretical boundaries. Rather than treating anomalies as exceptions requiring dismissal, we incorporated them to develop more nuanced understanding of how dual fits operate under real-world constraints. This engagement with complexity produces robust theory capable of explaining success despite imperfect conditions whilst acknowledging alignment as the optimal path to sustained effectiveness.

### 6.4.5 Positioning for Discussion

This chapter developed the Security Leadership Contingency Model as a theoretical framework explaining how effectiveness emerges from maintaining dual dynamic fits between security leaders and their contexts. The model synthesises multiple theoretical perspectives whilst remaining firmly grounded in the empirical patterns revealed through systematic grounded theory analysis. By reconceptualising effectiveness as continuous process requiring simultaneous maintenance of organisational and environmental alignments through sophisticated political navigation, we transform understanding of security leadership from static role descriptions to dynamic relationship management.

The journey from participant experiences through empirical patterns to theoretical integration demonstrates how qualitative research can generate novel frameworks whilst maintaining methodological rigour. The explicit engagement with complexity, negative cases, and alternative explanations strengthens rather than weakens the model by establishing clear boundary conditions and acknowledging the multifaceted reality of security leadership in modern organisations.

Chapter 7 examines what this theoretical contribution means for information security management scholarship and organisational practice. The discussion explores how the model advances academic understanding, what implications emerge for practitioners navigating security leadership challenges, and where future research might extend or challenge these insights. Having established what the model is and how it operates, we now turn to what it means and why it matters.

# CHAPTER 7: DISCUSSION

## 7.1 INTRODUCTION: FROM EMPIRICAL INSIGHTS TO THEORETICAL AND PRACTICAL IMPLICATIONS

### 7.1.1 Restating the Research Journey

Building on the empirical findings presented in Chapters 5-6, this chapter examines the theoretical and practical implications of the Security Leadership Contingency Model (SLCM). Through systematic investigation of 20 Australian security leaders, we have uncovered patterns that extend beyond descriptive accounts to reveal deeper theoretical mechanisms governing security leadership effectiveness. The SLCM (Figure 6.2, p. 177), as presented in Section 6.2, demonstrates that security leadership effectiveness challenges conventional understandings of executive effectiveness in boundary-spanning roles.

This discussion chapter serves three essential purposes in the thesis narrative. First, it positions the SLCM within existing theoretical conversations, demonstrating how our findings extend established theories of person-organisation fit and person-environment fit whilst revealing novel mechanisms unique to security leadership contexts. Second, it translates theoretical insights into actionable implications for security leaders, organisations, and the broader profession. Third, it establishes boundaries for our contributions whilst charting pathways for future investigation. The chapter thus bridges the empirical discoveries of Chapters 5-6 with the thesis's concluding synthesis in Chapter 8, transforming raw findings into lasting scholarly and practical contributions.

### 7.1.2 Bridging Findings to Discussion

The dual-fit requirement identified in Chapter 6 extends existing theory by revealing how boundary-spanning executives must navigate potentially conflicting alignment demands. This chapter explores what these emergent patterns mean for theory, practice, and future research. This interpretive leap demands rigorous engagement with existing literature, critical examination of alternative explanations, and careful consideration of boundary conditions that shape the applicability of our insights.

Our discussion unfolds across three interconnected levels of contribution. At the theoretical level, we demonstrate how the SLCM advances multiple theoretical conversations simultaneously. The model extends person-organisation fit theory by introducing temporal dynamics and phase-contingent requirements previously unrecognised in the literature. It enriches person-environment fit theory by revealing how external stakeholder relationships create independent fit demands that may conflict with organisational requirements. Most significantly, it proposes dual-fit maintenance as a novel theoretical construct, demonstrating that effectiveness in boundary-spanning executive roles requires simultaneous navigation of potentially conflicting fit demands through political capital and relational mechanisms.

At the practical level, our findings offer immediately actionable insights for multiple constituencies. Security leaders gain frameworks for assessing their current fit status and developing capabilities aligned with their organisational phase requirements. Organisations receive guidance for restructuring security leadership roles, recruitment practices, and support systems to enable dual-fit achievement. The security profession benefits from evidence-based foundations for evolving education programmes, certification frameworks, and career development pathways. These practical implications emerge directly from theoretical insights, ensuring recommendations rest on solid empirical and conceptual foundations.

At the methodological level, this research demonstrates the value of practitioner-researcher perspectives in uncovering nuanced organisational phenomena. The insider knowledge enabling access to elite participants and recognition of subtle political dynamics proved essential for theoretical development. Our adaptation of the Gioia method to incorporate temporal dimensions and negative case analysis offers methodological innovations for future organisational research. These contributions extend beyond the specific phenomenon studied to inform broader conversations about rigorous qualitative research in professional contexts.

## 7.2 THEORETICAL CONTRIBUTIONS AND POSITIONING

The Security Leadership Contingency Model (SLCM) represents a significant theoretical advancement in understanding executive effectiveness in boundary-spanning security roles. This section positions the SLCM within established theoretical frameworks, demonstrating how it extends, challenges, and integrates existing knowledge to create novel theoretical insights. The model's core contribution lies in revealing the dynamic, phase-contingent nature of security leadership effectiveness, moving beyond static competency frameworks to embrace temporal complexity and simultaneous fit requirements.

### 7.2.1 Extending Person-Organisation Fit Theory

The dual-fit requirement extends emerging understandings of P-O fit as a dynamic phenomenon (Kristof-Brown et al., 2023; Subramanian et al., 2022). Whilst traditional P-O fit theory conceptualised fit as a relatively stable alignment between individual characteristics and organisational attributes (Chatman, 1989; Kristof, 1996), recent research has begun to recognise its dynamic nature (Sales et al., 2023; Sun et al., 2023). The SLCM extends these insights specifically to security leadership contexts, revealing unique complexities in how fit evolves through organisational maturity phases.

**Theoretical extension**: Whilst recent scholarship recognises that P-O fit is dynamic, these studies examine how fit fluctuates within relatively stable role contexts. The SLCM reveals a fundamentally different pattern: fit requirements transform systematically through organisational maturity phases, where capabilities that create fit in one phase actively undermine fit in subsequent phases. This phase-based transformation, demonstrated in the SLCM (Figure 6.2, p. 177), creates more extreme fit transitions than traditional executive roles experience. The progression from technical expertise requirements in establishment phases to business integration capabilities in strategic phases illustrates how security leaders face fundamentally changing effectiveness criteria as organisations mature. Moreover, the security-specific contribution lies in demonstrating how CISO-Organisation fit must be maintained simultaneously with CISO-Environment fit, where improving one dimension may damage the other, creating compound dynamics absent from single-domain fit studies (Chalutz Ben-Gal, 2023; Chayka, 2025). This extends general observations about changing fit (Caldwell, 2011; Caldwell et al., 2004) to the particular demands of security leadership, where organisational maturity phases create distinct, phase-contingent fit requirements that shift fundamentally across phases, rather than gradual evolution within stable roles.

### 7.2.2 Extending Person-Environment Fit Theory

The environmental fit dimensions advance P-E fit theory by demonstrating how boundary-spanning executives face unique environmental pressures that operate independently from organisational demands. Traditional P-E fit research has often focused on general work environments or task-specific demands (Edwards, 2008; Kristof-Brown et al., 2005), whilst the SLCM reveals how security leaders must maintain fit with external stakeholder communities that impose their own effectiveness criteria.

**Theoretical extension**: The findings demonstrate that P-E fit in boundary-spanning executive roles requires political capital for maintenance (Ellen et al., 2013; Ferris et al., 2007). Unlike traditional conceptualisations where fit is achieved through alignment or matching (Chen et al., 2015; Chorn, 1991), security leaders must actively manage environmental relationships to maintain fit over time. Political capital serves as the mechanism enabling this active fit maintenance, allowing leaders to influence environmental demands rather than merely responding to them. Political capital enables individuals to shape decisions and outcomes through networks, skills, and authority (Campbell, 2021; Lake & Huckfeldt, 1998). This active, politically mediated view of P-E fit extends theory into new territory.

### 7.2.3 The Dual-Fit Imperative: An Integrated Theoretical Contribution

The SLCM's central theoretical contribution lies in identifying the dual-fit imperative: the requirement for security leaders to maintain simultaneous CISO-Organisation and CISO-Environment fit despite inherent tensions between these domains. This contribution transcends simple addition of existing fit theories to reveal a fundamentally new understanding of executive effectiveness in boundary-spanning roles. Boundary-spanning roles connect organisations to their external environments, facilitating knowledge exchange, collaboration, and adaptation to change (Aldrich & Herker, 1977; Bednarek et al., 2018).

**Theoretical extension**: The dual-fit maintenance requirement emerges as an effectiveness requirement rather than an ideal state. Unlike the information security employee context studied in existing literature where individuals may emphasise one fit dimension over another, security executives must achieve and maintain both organisational and environmental fit despite frequent conflicts between these domains (Chen & Li, 2018; Yin et al., 2024). This requirement creates a dynamic tension that defines the security leadership experience and explains many of the role's unique challenges, including managing evolving threats, resource constraints, and credibility issues (Chrzaszcz et al., 2024; Triplett, 2022).

Whilst existing multi-fit studies often focus on complementarity (Edwards & Billsberry, 2020), the SLCM demonstrates that in security leadership contexts, fit types frequently conflict. Political capital acts as a higher-order resource that enables and amplifies other dynamic capabilities within organisations (Kotabe et al., 2017; Nee & Opper, 2010), serving as the critical mechanism for navigating these tensions. Without sufficient political capital, sensing capabilities cannot translate into organisational awareness, seizing opportunities becomes politically infeasible, and reconfiguring faces organisational resistance. This meta-capability nature of political capital provides new insight into how individual capabilities translate into organisational outcomes (Ferris et al., 2007; Munyon et al., 2015).

The direct link between individual dynamic capabilities of security leaders and organisational security capabilities extends existing theory by demonstrating how personal dynamic capabilities enable organisational security capabilities, helping organisations become more agile and effective in their cybersecurity incident response (Ben Selma et al., 2024; Naseer et al., 2024; Palmié et al., 2023). The temporal evolution of these mechanisms requires continuous rebalancing, as phase transitions alter the relative importance of different fit dimensions. Leaders must adapt dynamically, maintaining internal harmony whilst periodically disrupting it to respond to external changes (Miller, 1992).

**Critical comparison with existing fit theories**: The dual-fit imperative differs fundamentally from existing theoretical frameworks. Kristof-Brown et al. (Kristof-Brown et al.) assume fit domains operate independently, allowing separate optimisation. The SLCM reveals deep interdependence where actions to improve one fit dimension may damage another. Whilst recent work recognises that organisational change frequently alters employee-environment compatibility (Caldwell, 2011), existing frameworks do not address how boundary-spanning security roles must maintain contradictory fits simultaneously. Recent temporal fit research (Sun et al., 2023) explores how fit evolves over time but considers single-domain evolution. The simultaneous evolution of multiple fits creates compound complexity that existing theories cannot adequately address.

### 7.2.4 Contributions to Contingency Theory in Security Leadership

The phase-contingent effectiveness patterns identified in the SLCM extend contingency theory by revealing unique contingency factors operating in security leadership contexts (Fiedler, 1964; Lawrence & Lorsch, 1967). These factors create a contingency landscape more complex and dynamic than traditional applications of contingency theory suggest. Building on foundational work that established how organisational structures should adapt to different environmental conditions (Burns & Stalker, 1961; Woodward, 1965), the SLCM demonstrates how security leadership faces unprecedented contingency complexity.

**Contingency factors unique to security leadership**: The incident-driven role evolution extends contingency theory by demonstrating how security incidents can instantly alter leadership requirements, demanding immediate crisis management and new decision-making approaches (Monehin & Diers-Lawson, 2022). Unlike the stable environments that suit mechanistic structures or the general turbulence requiring organic structures (Ellis et al., 2002; Liu, 2020), security leadership must navigate sudden, unpredictable shifts. Regulatory changes serve as contingency triggers, forcing leaders to quickly adapt strategies, compliance measures, and communication styles (Monehin & Diers-Lawson, 2022). Technology disruption creates new contingencies at an accelerating pace, with each emerging technology potentially redefining security leadership requirements.

The theoretical contributions outlined above position the SLCM as a significant advancement in understanding executive effectiveness in complex, boundary-spanning roles. By extending and integrating multiple theoretical perspectives, the model provides a more complete understanding of security leadership whilst offering insights applicable to other executive contexts facing similar complexity. The next section examines how these theoretical contributions relate to existing empirical literature.

## 7.3 RELATIONSHIP TO EXISTING LITERATURE: CONVERGENCES AND DIVERGENCES

The Security Leadership Contingency Model (SLCM) emerges from empirical investigation of CISO effectiveness in Australian organisations, offering both confirmatory and contradictory insights relative to existing security leadership literature. This section systematically examines how our findings relate to established knowledge, identifying where our research strengthens existing understanding, challenges conventional wisdom, and extends theoretical frontiers.

### 7.3.1 Confirmatory Findings: Strengthening Established Knowledge

Our research confirms several established patterns in security leadership literature whilst adding important nuance through the phase-based lens.

**Leadership complexity confirmed.** The findings align with Da Silva and Jensen (2022) characterisation of the CISO navigating a “dark art” where security is perceived as mystical and fearful, creating both elevated status and alienation. This complexity manifests through multiple stakeholder management challenges, role ambiguity, and legitimacy issues (Piazza et al., 2024; Sahin & Vance, 2025). However, whilst these studies present complexity as a universal characteristic of security leadership, the SLCM shows how this complexity varies systematically with organisational maturity. This phase-contingent view explains why some studies find CISOs overwhelmed by complexity whilst others document successful navigation; the nature and intensity of complexity depend fundamentally on organisational phase.

**Technical-business balance requirements.** Our research partially supports Karanja and Rosso (2017) on the necessity of hybrid technical-business competencies for CISO effectiveness. Recent competency frameworks emphasise this hybrid nature (Maynard et al., 2018; Ramezan, 2025; Whitten, 2008). However, the SLCM demonstrates a more nuanced relationship characterised by substitutive trade-offs rather than simple addition, reflecting role identity tensions and the challenge of transcending technical mindsets (Ashenden & Sasse, 2013; Short & Carandang, 2022).

**Environmental pressures validated.** The SLCM confirms environmental pressures documented extensively in existing literature (Adebola et al., 2024; AlGhamdi et al., 2020; Guan, 2024). Regulatory frameworks such as the GDPR and sector-specific mandates create non-negotiable requirements that security leaders must address (Adebola et al., 2024; AlGhamdi et al., 2020; Madnick & Marotta, 2021). Whilst these studies document how external compliance requirements, evolving threat landscapes, and industry standards shape security programmes, our research illuminates their interaction with organisational factors through the dual-fit mechanism (Edwards, 2008; Kristof-Brown et al., 2005). Environmental pressures do not simply impose requirements; they create fit demands that must be balanced against organisational fit requirements, sometimes creating irreconcilable tensions.

These confirmatory findings establish the SLCM's connection to existing scholarship whilst highlighting where our phase-based and dual-fit perspectives add explanatory power.

### 7.3.2 Contradictory Findings: Challenging Conventional Wisdom

Several findings directly contradict prevailing assumptions in security leadership literature, suggesting the need for theoretical refinement.

**Technical expertise as potential liability.** Whilst academic literature confirms that technical expertise is viewed as foundational for CISO credibility, particularly in early-stage organisations where technical knowledge establishes legitimacy and enables problem-solving (Adams & Makramalla, 2015; Haislip et al., 2021; Haqaf & Koyuncu, 2018; Khallaf & Majdalawieh, 2012), our findings indicate that technical expertise can become a significant hindrance to executive effectiveness, particularly in Maturation-phase and Strategic-phase organisations. This technical identity trap, where CISOs struggle to transcend their engineering mindset to embrace political and strategic thinking, aligns with role transition challenges (Ashforth, 2001) and cognitive entrenchment theory (Dane, 2010; Phan & Ngu, 2021; Zhang et al., 2022).

The explanation for this contradiction lies in organisational phase differences. Studies examining early security maturity stages focus on organisations where technical challenges dominate. Their participants faced fundamental questions about security architecture, tool selection, and implementation approaches that require deep technical knowledge (Dawson & Thomson, 2018; Shreeve et al., 2022). Our research, however, establishes how effectiveness requirements fundamentally shift across phases, explaining why capabilities essential in one phase become liabilities in another.

**Standardisation versus customisation paradox.** Conventional wisdom advocates for standardised security frameworks and controls to ensure consistency and regulatory compliance (Grigaliūnas et al., 2023; Karie et al., 2021). Yet standardisation often conflicts with organisational culture and business requirements, as our findings demonstrate. Unlike frameworks assuming universal applicability of standards, the dual-fit requirement shows how environmental demands for standardisation frequently clash with organisational preferences for customisation, forcing CISOs to navigate competing imperatives rather than simply implementing best practices.

**Transparency and communication constraints.** Contrary to widespread advocacy for security awareness, transparency, and metric-driven communication with executives (Brucciani, 2023; Corallo et al., 2023; Ramezan, 2025), effective CISOs deliberately limit security-specific metrics in executive communications, instead translating security outcomes into business impact narratives. Whilst previous research emphasises educating executives about security, our findings show successful CISOs minimise security-centric language, focusing instead on business outcomes and risk narratives that resonate with executive concerns.

These contradictory findings challenge fundamental assumptions about security leadership effectiveness, suggesting that conventional wisdom may reflect specific contexts rather than universal principles.

### 7.3.3 Novel Discoveries: Extending the Knowledge Frontier

Beyond confirming or contradicting existing literature, our research identifies several phenomena absent from current security leadership scholarship.

**Phase-based leadership requirements.** To our knowledge, no existing literature addresses security leadership through a phase-contingency lens. Whilst IT leadership evolution has been explored (Benmira & Agboola, 2021; Karakose et al., 2022; Patterson, 2015; Peppard, 2010), security's unique characteristics create distinct phase dynamics. The SLCM's identification of establishment, maturation, and strategic phases with distinct leadership requirements constitutes a novel theoretical contribution. This phase model explains previously puzzling variations in CISO effectiveness studies; researchers examining different phases naturally reach different conclusions about essential capabilities (Maynard et al., 2018; Ramezan, 2025; Whitten, 2008).

**Dual-fit simultaneity and tension.** Fit theory applications to leadership are well-established, yet existing studies assume multiple fits operate complementarily (Edwards & Billsberry, 2020). Person-organisation fit (Kristof-Brown et al., 2005) and person-environment fit (Edwards, 2008) theories traditionally examine fits independently. The affective components of P-E fit (Yu, 2009) and multi-domain fit considerations (Voydanoff, 2005) suggest complexity, yet the SLCM's discovery of inherent tension between CISO-Organisation and CISO-Environment fit requirements marks a significant theoretical advance.

This tension manifests when environmental demands for standardised security controls conflict with organisational preferences for flexibility. Unlike traditional fit theories assuming complementary relationships, CISOs face irreconcilable fit demands requiring continuous navigation rather than resolution.

**Political capital as mediating mechanism.** Political skill receives attention in leadership literature (Rosen & Levy, 2013; Silvester, 2008; Treadway et al., 2013), yet its role as a critical mediating mechanism for managing dual-fit tensions offers a novel finding. The SLCM extends beyond recognising political skill importance to demonstrating how political capital enables CISOs to navigate otherwise irreconcilable demands. This discovery advances understanding of how leaders manage conflicting stakeholder expectations in complex organisational environments.

### 7.3.4 Theoretical Positioning and Advancement

The SLCM advances security leadership theory through its integration of contingency thinking with fit theory dynamics. Contingency theory has informed IT leadership research (Fiedler, 1964; House, 1971), and fit theory has examined person-environment alignment (Edwards, 2008), yet their combination to explain security leadership effectiveness offers theoretical innovation.

The model extends Ashenden and Sasse (2013) security-business tension observations by providing a theoretical framework explaining how these tensions arise and can be managed. Where previous research identified problems, the SLCM offers explanatory mechanisms and practical pathways forward.

Furthermore, the SLCM addresses calls for more nuanced understanding of security leadership (Da Silva & Jensen, 2022; Sahin & Vance, 2025) by moving beyond static competency lists to dynamic, context-dependent effectiveness requirements. This shift from universal to contingent effectiveness criteria constitutes a fundamental reconceptualisation of security leadership theory.

### 7.3.5 Addressing Alternative Explanations

Critical examination of alternative explanations strengthens our theoretical contribution whilst acknowledging legitimate boundary conditions.

**Individual variance effects.** Critics might argue that observed patterns reflect individual differences rather than systematic phase-based requirements. However, the consistency of patterns across diverse individuals within similar phases, coupled with documented transitions as individuals moved between phases, supports the phase-based explanation. The SLCM does not deny individual differences but demonstrates how phase context shapes which individual capabilities become assets or liabilities.

**Industry selection effects.** Whilst the research sample is distributed approximately equally between heavily regulated and less regulated industries, the significant presence of the former might be perceived to overemphasise environmental pressures. However, participants from less regulated sectors reported similar dual-fit tensions arising from competitive rather than regulatory pressures. Research confirms that in less regulated sectors, market forces and internal organisational values can be as important as regulatory pressure in creating environmental demands (Graafland & Smid, 2017; He et al., 2018; Wang et al., 2020).

These rebuttals demonstrate the SLCM's robustness against alternative explanations whilst acknowledging legitimate boundary conditions.

### 7.3.6 Summary

This examination of convergences, divergences, and novel discoveries positions the SLCM within existing security leadership literature whilst demonstrating its theoretical advances. Confirmatory findings provide foundation and credibility, contradictory findings challenge assumptions requiring theoretical refinement, and novel discoveries extend knowledge frontiers. The phase-based perspective and dual-fit framework explain previously puzzling variations in security leadership effectiveness research whilst opening new avenues for investigation.

## 7.4 PRACTICAL IMPLICATIONS: FROM THEORY TO ACTION

The Security Leadership Contingency Model (SLCM) offers substantial practical guidance for multiple stakeholder groups. This section translates theoretical insights into actionable recommendations, demonstrating how the model improves security leadership effectiveness in organisational contexts. The theoretical contribution lies in the SLCM itself, which explains how dual-fit dynamics shape security leadership effectiveness. The practical contribution extends beyond theory through four practitioner-grade operational tools that translate these insights into immediately usable instruments for leadership selection, dual-fit assessment, and in-role health monitoring.

### 7.4.1 For Security Leaders: Navigating the Dual-Fit Challenge

#### *7.4.1.1 Understanding Your Leadership Orientation*

Security leaders should begin by understanding their natural leadership orientation, as effectiveness depends fundamentally on alignment between individual preferences and organisational context. The research identified three distinct orientations (action-oriented, stewardship-oriented, and vision-oriented) that reflect where leaders find energy, what work they gravitate toward, and which contexts enable their best performance.

Self-awareness about orientation proves critical for several practical reasons. First, it enables recognition of contexts where leaders will thrive versus struggle. Leaders who accept roles misaligned with their natural orientation, regardless of compensation or title, typically experience frustration and reduced effectiveness. Second, it informs development priorities by distinguishing capabilities leaders can readily develop from those requiring sustained effort against natural preferences. Third, it provides framework for career decisions, helping identify when organisational evolution necessitates transition rather than adaptation.

Leaders can identify their natural orientation through systematic reflection on career experiences. They should consider which work has consistently energised them: building capabilities from scratch signals action orientation, optimising existing processes suggests stewardship orientation, and positioning security strategically indicates vision orientation. They should equally examine which contexts frustrated them despite success, as maintenance operations may indicate action orientation, greenfield ambiguity may suggest stewardship orientation, and purely technical work may reveal vision orientation. Observing where they gravitate when unconstrained, whether toward hands-on implementation, systematic improvement, or strategic influence, provides additional confirmatory insight.

This self-knowledge enables informed career management. Action-oriented leaders should recognise when organisations mature beyond contexts suited to their strengths and plan transitions before experiencing diminished effectiveness from steady-state operations. Stewardship-oriented leaders should avoid perpetual transformation environments lacking the stability their orientation requires. Vision-oriented leaders should ensure organisations possess the maturity necessary for strategic security positioning before accepting executive roles.

The Leadership Orientation Self-Assessment provides systematic approach to this self-evaluation, though honest reflection on career patterns often reveals orientation without formal assessment.

#### *7.4.1.2 Assessing Organisational Fit*

Security leaders should systematically evaluate their current fit status across both organisational and environmental dimensions. This assessment approach builds on established person-organisation and person-environment fit measurement practices (Edwards, 2008; Kristof-Brown et al., 2005), adapted for the security leadership context.

For CISO-Organisation fit assessment, leaders should evaluate how well their leadership orientation aligns with organisational maturity phase requirements. A technically-oriented leader in a strategic phase organisation faces significant misalignment that requires attention. Similarly, leaders should assess their understanding of organisational culture, politics, and decision-making processes. The political capital dynamics demonstrate how organisational fit depends not only on technical competence but also on relationship networks, influence capacity, and cultural alignment.

For CISO-Environment fit assessment, leaders should evaluate their standing within industry security communities, relationships with regulators and auditors, and ability to translate external requirements into organisational language. This dual assessment reveals fit gaps that may not be apparent when considering only internal or external dimensions in isolation. Once assessment reveals misalignment areas, leaders should prioritise remediation efforts based on their specific context, identifying which fit dimension currently constrains their effectiveness most severely.

### *7.4.1.3 Navigating Career Transitions*

Security leaders should continuously monitor their fit status, as both organisational and environmental contexts evolve. The temporal dynamics demonstrate how previously effective leaders may experience performance decline without role evolution as organisations mature through security phases. This proactive relationship building contrasts with reactive approaches that attempt to build credibility during crisis periods when stakeholders are least receptive.

Leaders experiencing misalignment face several strategic options. They may develop compensatory strategies by building teams with complementary orientations, delegating activities outside their natural strengths, or developing capabilities in weaker areas. Alternatively, they may assess sustainability of continued misalignment, considering whether they possess sufficient political capital to buffer misfit or whether organisational evolution trends toward their natural orientation. Finally, they may plan transition timing, determining when to seek better-aligned roles or facilitate succession to more appropriately matched leaders.

Social capital theory demonstrates such resources as assets rooted in networks of social relations (Cook, 2017; Preston et al., 2017), requiring sustained cultivation rather than crisis deployment.

### *7.4.1.4 Operational Tool: Leadership Orientation Self-Assessment*

To operationalise these insights, we developed a Leadership Orientation Self-Assessment instrument grounded in the empirical patterns identified in Chapters 5 and 6. The assessment comprises 24 behavioural questions measuring action-oriented, stewardship-oriented, and vision-oriented tendencies through frequency-based scales.

The instrument operationalises the phase-leadership alignment patterns observed in Chapter 6, where distinct leadership orientations demonstrated differential effectiveness across organisational maturity phases. Each question reflects behavioural patterns observed across multiple interview participants and grounds itself in how successful CISOs described their approaches and preferences. The assessment avoids social desirability bias through behavioural frequency questions ("How often do you...") rather than self-evaluation statements ("I am good at...").

Security leaders can use this assessment for multiple purposes. For self-awareness, it clarifies natural preferences and blind spots, helping leaders recognise contexts where they will thrive versus struggle. For career decisions, it enables assessment of role-fit before accepting positions and recognition of when transition becomes appropriate. For development planning, it focuses attention on enhancing strengths in natural orientation whilst developing compensatory capabilities for likely contexts.

The complete Leadership Orientation Self-Assessment appears in Appendix D: SLCM Practical Implications Toolkit with detailed scoring interpretation guidance. Results prove most valuable when combined with organisational maturity assessment to evaluate alignment between leadership orientation and organisational phase requirements.

**Limitations:** This instrument represents an operational tool derived from academic research rather than a psychometrically validated scale. It has not undergone formal instrument validation procedures including test-retest reliability, construct validity assessment, or factor analysis. Its value lies in systematically operationalising grounded theory findings rather than in statistical validation. Results capture point-in-time snapshots of dynamic situations and should be reassessed periodically as leaders develop and contexts evolve.

#### *7.4.1.5 Operational Tool: Dual-Fit Health Diagnostic*

To enable ongoing monitoring of fit status, we developed a Dual-Fit Health Diagnostic instrument that operationalises the dual-fit dynamics identified as central to the SLCM. The diagnostic comprises 26 questions measuring CISO-Organisation fit (13 questions) and CISO-Environment fit (13 questions) through behavioural frequency scales.

The instrument directly addresses the dual-fit maintenance requirement detailed in Chapter 6, where security leaders must simultaneously achieve and maintain fit with both organisational context and external environment. Questions ground themselves in the first-order concepts and second-order themes (Chapter 5) that emerged from participant descriptions of effectiveness challenges. The diagnostic incorporates political capital dynamics, assessing both formal authority and informal influence mechanisms.

Security leaders should use this diagnostic quarterly or semi-annually to monitor fit evolution. Early warning indicators of deteriorating fit enable proactive intervention before effectiveness significantly declines. The diagnostic proves particularly valuable during organisational transitions, following leadership changes, or when experiencing persistent challenges in stakeholder engagement or programme advancement.

The complete Dual-Fit Health Diagnostic appears in Appendix [Practical Implication Toolkit] with scoring guidance and interpretation frameworks. Results should inform development priorities, relationship building strategies, and, where appropriate, career transition decisions.

**Limitations:** This diagnostic measures perceived fit rather than objective fit, as fit assessment inherently involves subjective judgment about alignment between individual capabilities and contextual requirements. Results may be influenced by recent experiences, current organisational climate, or leader mood states. The diagnostic informs but does not replace professional judgment about effectiveness or career decisions.

### 7.4.2 For Organisations: Enabling Security Leadership Success

Organisations seeking effective security leadership should move beyond traditional approaches focusing solely on technical competencies or reporting structures. Recent research challenges structural determinism in security governance (Granata et al., 2024; Madnick & Marotta, 2021), demonstrating that effectiveness depends on alignment between leadership capabilities and organisational context.

#### *7.4.2.1 Matching Leadership to Phase*

Organisations should assess candidate-phase fit by evaluating whether candidates' leadership orientations align with current organisational security maturity requirements. A maturation phase transformation organisation requires different leadership capabilities than an establishment phase organisation, yet traditional recruitment approaches rarely acknowledge these distinctions. The phase-contingent effectiveness patterns demonstrate how technical competence alone proves insufficient for executive success.

When evaluating candidates, organisations can assess political capital and adaptive capacity by examining candidates' success in previous role transitions, their ability to explain technical concepts to non-technical audiences, and their experience managing competing stakeholder demands. The political dynamics highlight how informal influence mechanisms often prove more critical than formal authority structures for security leadership effectiveness.

#### *7.4.2.2 Supporting Leadership Transitions*

For organisations undergoing phase transitions, security leaders require particular support. As organisations evolve through security maturity phases, yesterday's successful security leadership approach may constrain tomorrow's effectiveness. Organisations can support leadership transitions by acknowledging that phase evolution may necessitate leadership changes or significant capability development for existing leaders.

Organisations should ensure that formal structures are complemented by informal mechanisms enabling CISOs to build necessary political capital. Security leaders should have forums for regular interaction with key stakeholders, clear decision rights that acknowledge the CISO's expertise whilst respecting business ownership, and communication channels that enable both formal reporting and informal influence. These structural enablers address the political capital development challenges.

#### *7.4.2.3 Building Complementary Teams*

Organisations should recognise dual-fit maintenance as a performance indicator, valuing activities that build political capital or strengthen stakeholder relationships even when these activities do not produce immediate security outcomes. Research demonstrates the limitations of conventional security metrics (Brucciani, 2023; Corallo et al., 2023). For performance evaluation, organisations should account for phase-specific requirements. An establishment phase CISO delivering basic security capabilities deserves different evaluation than a strategic phase CISO enabling business transformation through security. Generic security metrics applied across different contexts mask important differences in leadership effectiveness.

When valuing political capital building activities, organisations can explicitly recognise and reward CISO participation in industry forums, regulator relationships, or internal stakeholder engagement as essential for dual-fit maintenance and long-term effectiveness. Building security teams with complementary leadership orientations proves particularly valuable, as action-oriented deputies can support vision-oriented CISOs whilst strategic advisers can complement stewardship-oriented leaders.

#### *7.4.2.4 Operational Tool: Organisational Security Maturity Phase Diagnostic*

To operationalise the phase model developed in Chapter 6, we created an Organisational Security Maturity Phase Diagnostic instrument that systematically assesses organisational positioning across establishment, maturation, and strategic phases. The diagnostic comprises 39 questions (13 per phase) measuring behavioural characteristics through frequency-based scales.

The instrument reflects the phase characteristics identified in Chapter 6, where organisations demonstrated distinct patterns of security programme maturity, stakeholder engagement, and strategic integration. Each question operationalises empirical observations from Chapter 5 about how security functions differently across organisational contexts. The assessment employs a 4-point frequency scale that proves gaming-resistant through behavioural focus ("How often does this happen?") rather than opinion elicitation ("Do you agree?").

Organisations should use this diagnostic during CISO recruitment to determine optimal leadership orientation requirements, during strategic security programme reviews to assess current positioning, and when experiencing persistent leadership challenges that may reflect phase-leadership misalignment. Multiple respondents across organisational levels can provide valuable perspective comparison, as divergent scores may indicate communication gaps or genuine variation across business units.

The complete Organisational Security Maturity Phase Diagnostic appears in Appendix [Practical Implication Toolkit] with detailed scoring interpretation guidance and phase threshold definitions.

**Limitations:** This diagnostic represents organisational self-assessment rather than external audit. Results depend on respondent organisational awareness and may be influenced by role perspective (security leaders may assess maturity differently than business stakeholders). The instrument captures point-in-time snapshots of evolving situations; organisations in transition between phases may demonstrate mixed patterns. Results inform but do not determine leadership selection decisions, which should also consider individual candidate factors and organisational strategic direction.

##### *7.4.2.5 Operational Tool: Phase-Leadership Alignment Matrix*

To translate diagnostic results into actionable guidance, we developed a Phase-Leadership Alignment Matrix that systematically maps organisational maturity phase assessment results against leadership orientation patterns. This decision-support tool operationalises the phase-leadership alignment findings from Chapter 6, providing practical guidance for recruitment, performance evaluation, and succession planning decisions.

The matrix operates as a comprehensive decision-support framework rather than a prescriptive rulebook. It identifies optimal alignment patterns (action-oriented leaders in establishment phase, stewardship-oriented in maturation phase, vision-oriented in strategic phase) whilst acknowledging that exceptional leaders may succeed despite apparent misalignment through extraordinary political capital, unique organisational circumstances, or individual adaptation capacity.

For each phase-orientation combination, the matrix provides guidance on expected effectiveness patterns, potential challenges, compensatory strategies, and succession planning considerations. Organisations experiencing leadership challenges can use the matrix to diagnose whether misalignment contributes to difficulties and to develop appropriate support strategies or transition plans.

The complete Phase-Leadership Alignment Matrix appears in Appendix [Practical Implication Toolkit] with detailed application guidance and decision frameworks for common organisational scenarios.

**Limitations:** The matrix provides informed guidance based on empirical patterns observed across 20 security leaders in Australian organisations, but cannot account for all contextual factors that influence effectiveness. Organisational culture, industry characteristics, resource availability, and individual leader capabilities all moderate the phase-leadership relationships. The matrix should inform rather than determine organisational decisions, with professional judgment considering factors beyond phase-leadership alignment.

#### *7.4.2.6 Integrated Tool Usage for Selection and Evaluation*

The three operational tools introduced above (Organisational Security Maturity Phase Diagnostic, Leadership Orientation Assessment, and Phase-Leadership Alignment Matrix) work in concert to support CISO selection and evaluation decisions. Rather than constituting a separate tool, this integrated approach combines the diagnostic instruments to address the complete leadership lifecycle from recruitment through ongoing performance assessment.

For recruitment decisions, organisations can use the tools sequentially: first, complete the Organisational Security Maturity Phase Diagnostic to determine current phase positioning; second, consult the Phase-Leadership Alignment Matrix to identify optimal leadership orientation requirements; third, assess candidates’ leadership orientations through interview questions derived from the Leadership Orientation Assessment; finally, evaluate candidate-phase alignment and structured comparison against phase-appropriate effectiveness criteria. This systematic approach operationalises the dual-fit requirements, ensuring evaluation considers both organisational and environmental fit dimensions.

For performance evaluation, the tools enable phase-appropriate assessment by providing success indicators that reflect contingent effectiveness criteria. Rather than applying universal metrics, organisations can adjust evaluation standards based on organisational maturity phase, acknowledging that establishment phase success looks fundamentally different from strategic phase success. The Dual-Fit Health Diagnostic provides additional monitoring capability for ongoing effectiveness assessment.

**Integration considerations:** This multi-tool approach provides structure for systematic decision-making but cannot replace organisational judgment about candidate fit or leader performance. Cultural fit, technical credibility, and stakeholder relationships all matter beyond what structured assessments capture. The integrated approach should complement rather than replace existing recruitment and evaluation processes, adding phase-contingency awareness to traditional assessment approaches.

### 7.4.3 For the Security Profession: Evolving Professional Development

The security profession should evolve its approach to leadership development, moving beyond technical certification paradigms to acknowledge the complex political and organisational dimensions of security leadership effectiveness. Current frameworks emphasise hybrid technical-business competencies (Karanja, 2017; Ramezan, 2025) but underestimate political requirements documented throughout this research.

Security education programmes should expand beyond technical knowledge to prepare future security leaders for the political realities of executive roles. Whilst boards readily approve technical security investments, they may question investments in leadership development, relationship building, or political capital development. The research demonstrates these investments are essential for security leadership effectiveness and ultimately organisational security outcomes.

For boards to successfully integrate security into strategic thinking, they should involve CISOs in strategic planning processes, recognise security as a business enabler rather than merely a cost centre, and understand how security capabilities influence competitive positioning. Boards that create contexts where security leaders can be most effective should ensure this integration reflects genuine recognition of security's role in organisational success. When boards understand and articulate security's strategic value, they create environments where CISOs can move beyond defensive positions to become true strategic partners.

Professional certification bodies can incorporate phase-contingency awareness into competency frameworks, recognising that different contexts require different leadership approaches. Rather than prescribing universal competencies, certification programmes could emphasise contextual awareness, adaptive capacity, and self-knowledge about natural leadership orientation. This evolution would better prepare security professionals for the contextual complexity of executive roles.

### 7.4.4 Synthesis: From Individual Action to Systemic Change

These practical implications operate synergistically rather than independently. Individual leaders who develop dual-fit capabilities find greater success in organisations that understand and support contextual requirements. Similarly, organisations investing in enabling security leadership effectiveness benefit from access to more capable security leaders who can navigate complex demands successfully. This systemic perspective aligns with calls for holistic approaches to security leadership development. The microfoundations of dynamic capabilities, spanning individual, structural, and processual elements, are central to enabling innovation across diverse organisational settings (Ben Selma et al., 2024; Palmié et al., 2023).

As education and certification programmes develop more sophisticated leaders, organisations gain access to candidates better prepared for executive challenges. This mutual reinforcement suggests that systemic change, whilst requiring coordinated effort across multiple stakeholder groups, offers substantial benefits for all involved. The operational tools facilitate this systemic change by providing shared language and structured frameworks that enable more informed dialogue between security leaders, organisations, and the profession.

When security leaders successfully maintain dual-fit, organisations achieve better security outcomes through improved stakeholder buy-in, more effective resource allocation, and enhanced ability to navigate external demands. Personal dynamic capabilities play a key role in enabling organisational security capabilities, helping organisations become more agile and effective in their cybersecurity incident response (Naseer et al., 2024). These improvements in security effectiveness benefit not only organisations but also their customers, partners, and broader ecosystems.

Implementation of these recommendations requires sustained commitment rather than one-time interventions. Security leaders should continuously monitor and adjust their fit configuration using the Dual-Fit Health Diagnostic. Organisations should evolve their support systems as security maturity advances, reassessing phase positioning and leadership requirements periodically. The profession should continually refine development approaches based on emerging challenges, incorporating phase-contingency awareness into certification frameworks. Governance bodies should maintain ongoing attention to security leadership effectiveness rather than episodic engagement during crises.

By understanding the dynamic interplay between individual capabilities, organisational contexts, and environmental demands, stakeholders can make more informed decisions that ultimately strengthen organisational security postures in an increasingly complex threat landscape. This translation from theoretical insight to practical action represents a critical contribution of academic research to professional practice, bridging the often-cited theory-practice gap (Patterson, 2015; Peppard, 2010). The four operational tools extend this practical contribution by providing immediately usable instruments that operationalise theoretical understanding, ensuring research insights inform real-world security leadership decisions.

## 7.5 METHODOLOGICAL CONTRIBUTIONS AND REFLECTIONS

The development of the Security Leadership Contingency Model (SLCM) through this research represents not merely a theoretical contribution but also advances methodological understanding in security leadership research. This section examines how practitioner-researcher positioning enhanced the research process, the specific methodological innovations employed, and the reflexive insights gained through navigating dual identities.

### 7.5.1 Advancing Grounded Theory in Security Research

The application of grounded theory methodology to security leadership research revealed both the power and complexity of practitioner-researcher positioning. This study demonstrates how insider knowledge, when systematically managed, enhances rather than compromises rigorous qualitative research.

#### *7.5.1.1 Practitioner-Researcher Advantages Demonstrated*

The practitioner-researcher position provided distinct advantages that would have been difficult for external researchers to replicate. Access to elite participants proved particularly valuable, as CISOs and security executives demonstrated remarkable openness when interviewed by a peer. Participants frequently prefaced sensitive disclosures with phrases such as "you understand this" or "as you know from experience", indicating a level of trust and candour that emerged from shared professional identity.

This insider status enabled nuanced understanding of technical contexts that pervaded participant narratives. When participants discussed complex security architectures, regulatory compliance challenges, or incident response scenarios, the ability to comprehend technical details without requiring extensive explanation allowed interviews to progress rapidly to deeper analytical levels. Participants could employ technical shorthand and industry-specific terminology whilst remaining confident in researcher comprehension.

Perhaps most critically, practitioner experience facilitated recognition of subtle political dynamics that external researchers might overlook. The unspoken tensions between security and business priorities, the delicate navigation of board relationships, and the informal influence mechanisms within organisations revealed themselves through seemingly casual comments and indirect references. Understanding these political undercurrents proved essential for developing the dual-fit framework that emerged as the study's core contribution.

#### *7.5.1.2 Methodological Innovations and Adaptations*

The integration of multiple theoretical lenses represented a departure from purely inductive approaches, acknowledging that practitioner knowledge inherently carries theoretical assumptions (see Section 4.2.2 for detailed methodology). This adaptation proved particularly valuable when navigating the complex interplay between technical competence and organisational politics that characterises security leadership.

Temporal analysis innovations emerged through tracking how participants' leadership approaches evolved across career stages and organisational contexts. Rather than treating leadership as static, the methodology captured dynamic adaptation processes, revealing how security leaders continuously recalibrate their fit configurations. This temporal dimension, often overlooked in cross-sectional leadership studies, proved crucial for understanding effectiveness patterns.

Negative case analysis contributed significantly to theoretical refinement. Instances where successful leaders departed from emerging patterns forced deeper examination of contextual factors. These anomalies revealed that effectiveness depends not on universal practices but on achieving appropriate fit between individual capabilities, organisational needs, and environmental demands. This insight fundamentally shaped the contingency nature of the SLCM.

Supervisory sessions proved invaluable for identifying instances where practitioner assumptions might overshadow participant perspectives, with supervisors consistently challenging interpretations that relied too heavily on researcher experience rather than empirical evidence.

Leveraging insider knowledge appropriately required careful calibration. Whilst technical comprehension and political awareness enhanced data collection and initial analysis, the abstraction to theoretical concepts demanded stepping back from practitioner frameworks. The distinction between "what practitioners know" and "what the data reveals" required constant vigilance, particularly during the transition from first-order concepts to second-order themes.

Maintaining analytical distance proved most challenging when participants described experiences that resonated strongly with personal practitioner history. The temptation to complete participant narratives based on similar personal experiences required active resistance, ensuring interpretations remained grounded in participant data rather than researcher projection.

#### *7.5.1.3 Benefits Realised*

The practitioner-researcher position yielded substantial benefits that validated the methodological approach. Participant openness and trust manifested in remarkably candid discussions of leadership failures, political missteps, and organisational frustrations. The depth of disclosure around sensitive topics such as board politics, peer conflicts, and career challenges suggested a level of comfort that emerged from shared professional understanding.

Rich contextual understanding emerged naturally from shared professional background. Participants could communicate complex organisational dynamics through brief allusions, confident that the researcher would grasp implications without extensive explanation. This efficiency allowed interviews to explore multiple dimensions of leadership experience within time constraints, contributing to the comprehensiveness of the final model.

Practical relevance remained assured throughout the research process. The practitioner lens ensured that emerging theoretical insights maintained applicability to real-world security leadership challenges. Participants frequently validated preliminary findings by recognising their own experiences within developing frameworks, confirming that theoretical abstractions retained practical grounding.

#### *7.5.1.4 Challenges Navigated*

The practitioner-researcher position also presented distinct challenges requiring active management. Assumption questioning demanded systematic approaches to surface and examine practitioner beliefs that might constrain theoretical development. Regular engagement with non-practitioner academics provided essential external perspective, challenging taken-for-granted assumptions about security leadership.

Bias mitigation through systematic method proved essential for research credibility (see Section 4.6.5 for detailed reflexivity discussion). Each analytical decision required explicit justification grounded in data rather than experience, creating an audit trail that demonstrated empirical rather than experiential foundations for theoretical claims.

Achieving balance between emic and etic perspectives represented perhaps the greatest challenge. The emic perspective (understanding security leadership from within the practitioner community) came naturally. Developing etic perspective (viewing the phenomenon from theoretical distance) required conscious effort and methodological discipline. The iterative movement between data immersion and theoretical abstraction facilitated this perspectival balance, though tension between insider and outsider viewpoints persisted throughout the analysis.

#### *7.5.1.5 Methodological Insights for Future Research*

This research demonstrates that practitioner-researchers can produce rigorous theoretical contributions when appropriate methodological safeguards are employed (Coghlan, 2001; Last et al., 2019; Warner, 2024). The combination of systematic analytical methods, external supervision, and explicit reflexivity enables the benefits of insider knowledge whilst mitigating potential limitations. Future security research might productively engage more practitioner-researchers, particularly for phenomena where technical complexity and political sensitivity limit external researcher access (Scott, 2024).

The adaptation of established qualitative methodologies for practitioner research contexts opens new possibilities for knowledge creation at the intersection of practice and theory (Sandberg & Tsoukas, 2011; Van de Ven, 2007). Rather than viewing practitioner knowledge as a source of bias to be eliminated, this study demonstrates how such knowledge can enhance research when properly managed. The key lies not in bracketing all practitioner understanding but in maintaining clear distinctions between experiential knowledge, empirical evidence, and theoretical abstraction.

The reflexive journey of conducting research within one's own professional domain ultimately enriched both practitioner and researcher identities (Cramp & Khan, 2019). The systematic analysis of familiar phenomena through theoretical lenses revealed patterns and relationships invisible from pure practitioner perspective. Simultaneously, maintaining connection to practitioner realities ensured that theoretical insights retained practical relevance. This dual enrichment suggests that practitioner research, when rigorously conducted, contributes not only to academic knowledge but also to enhanced professional practice (Dadds, 2008; Newman & Leggett, 2018).

## 7.6 BOUNDARY CONDITIONS AND LIMITATIONS: DEFINING THE CONTRIBUTION SCOPE

This section establishes the boundaries within which the Security Leadership Contingency Model (SLCM) operates and acknowledges the limitations that constrain its applicability. Understanding these boundaries is essential for appropriate theoretical application and empirical extension. The boundaries identified are theoretically justified rather than methodologically defensive, arising from the specific conditions required for the dual-fit dynamics to manifest and create the leadership challenges that the SLCM addresses. For methodological limitations of the research approach, see Section 4.8.

### 7.6.1 Empirical Boundaries (Theoretically Justified)

The SLCM's applicability is bounded by several empirical conditions that emerge from the theoretical requirements for dual-fit tensions to manifest. These boundaries are not arbitrary constraints but necessary conditions for the phenomena under investigation.

#### *7.6.1.1 Geographic Context: Australian Regulatory Environment as Critical Case*

The study's Australian context provides what Gerring (2007) terms a 'critical case' for understanding security leadership effectiveness in mature regulatory environments. This approach aligns with established critical case methodology (Easton, 2010; Hyett et al., 2014; Meyer, 2001), which emphasises selecting contexts with theoretical significance for broader inference. Australia's regulatory landscape combines the Westminster governance system with sophisticated privacy legislation (Privacy Act 1988) and emerging critical infrastructure protection requirements. This regulatory maturity creates specific accountability structures that generate the external environmental pressures necessary for CISO-Environment fit tensions to emerge, aligning with Lanivich et al. (2010) who demonstrate how high accountability environments amplify person-environment fit effects in the broader P-E fit literature.

The theoretical significance of this context lies in its combination of regulatory sophistication with cultural characteristics that influence organisational politics. Australian business culture, characterised by relatively collaborative decision-making processes and preference for consultative leadership, creates particular pathways for developing the relational capital that enables dual-fit navigation. These cultural factors shape how political capital is accumulated and deployed within organisations, consistent with Zibenberg (2017), Valle et al. (2019), Park et al.

(2021), and Den Hartog and De Hoogh (2024) who demonstrate how cultural characteristics influence organisational politics.

The SLCM's transferability extends to similar regulatory contexts, particularly other Westminster governance systems including the United Kingdom, Canada, and New Zealand. These jurisdictions share comparable accountability structures and regulatory sophistication that create analogous environmental pressures, as demonstrated by Moses et al. (2025) and Dimitrov (2025) in their analyses of regulatory frameworks across Westminster systems. However, adaptation would be required for jurisdictions with fundamentally different regulatory approaches, such as the prescriptive sector-specific mandates common in the European Union or the more decentralised regulatory environment in the United States.

The geographic boundary is theoretically justified because the dual-fit model requires sufficient environmental pressure to create meaningful CISO-Environment fit challenges. Jurisdictions lacking regulatory maturity or sophisticated governance expectations may not generate the environmental complexity necessary for dual-fit tensions to emerge as significant factors in security leadership effectiveness.

#### *7.6.1.2 Organisational Scope: Large, Complex Organisations with Political Dynamics*

The SLCM applies specifically to large, established organisations that possess sufficient internal complexity to generate meaningful organisational politics. The theoretical requirement aligns with Mintzberg (1983) professional bureaucracy characteristics, where political capital becomes a scarce resource and informal influence networks operate alongside formal authority structures. This complexity threshold is supported by organisational politics research (Andrews, 2010; Child, 1972; Mayes & Allen, 1977), which demonstrates how large organisations' internal complexity generates meaningful political dynamics essential for CISO-Organisation fit maintenance challenges.

Furthermore, the model applies across all three phases of organisational security maturity (Establishment, Maturation, and Strategic), though the nature of dual-fit tensions varies by phase. This phase-based boundary specification provides theoretical precision beyond simple organisational characteristics. In Establishment phase organisations, whilst action-oriented leadership and technical competence are critical, dual-fit tensions still emerge as leaders must balance internal capability building with external threat response. The model's explanatory power reveals different alignment patterns across phases: action-oriented leaders excel in Establishment, stewardship-oriented in Maturation, and vision-oriented in Strategic phases, where the competing demands of CISO-Organisation and CISO-Environment fit manifest differently but remain equally important.

The organisational boundary explicitly excludes flat organisational structures, technical startups where security remains primarily a technical function, and small organisations lacking the complexity necessary for significant internal political dynamics. In these contexts, traditional leadership effectiveness models may provide adequate explanatory power without requiring the additional complexity of dual-fit considerations, as suggested by Mumford et al. (2007) in their strataplex model of leadership skill requirements across organisational levels. This is supported by comparative leadership research (Deng et al., 2023; Marion & Uhl-Bien, 2001; Wang et al., 2014) demonstrating that simpler organisational contexts require less complex leadership frameworks.

Organisational complexity emerged through analysis as a critical factor influencing theoretical pattern manifestation. Very small organisations (typically those with limited departmental structures and minimal hierarchical layers) may lack the organisational complexity necessary for the political capital dynamics that characterise dual-fit navigation. However, size alone is insufficient; organisational complexity and political sophistication are the determining factors, consistent with Ferris et al. (2019) and Landells and Albrecht (2017) who demonstrate how organisational complexity generates political dynamics. The study included organisations ranging from several hundred to many thousands of employees, demonstrating that political capital dynamics can emerge in mid-sized organisations (500-1000 employees) when they possess sufficient structural complexity and political sophistication.

#### *7.6.1.3 Role Specification: Boundary-Spanning Executive Positions*

The SLCM specifically addresses boundary-spanning executive positions that require simultaneous accountability to internal organisational stakeholders and external environmental actors. This boundary-spanning criterion follows Aldrich and Herker (1977) on boundary-spanning roles, with subsequent research (Jemison, 2007; Korschun, 2015; Prysor & Henley, 2018) confirming the unique challenges of roles requiring simultaneous internal and external accountability. The theoretical requirement is for roles where incumbents must maintain legitimacy across multiple constituencies with potentially conflicting expectations and success criteria.

Chief Information Security Officers exemplify such boundary-spanning roles through their simultaneous accountability to internal executive teams, boards of directors, regulatory bodies, industry peers, and external stakeholders including customers and business partners. This multiplicity of accountability relationships creates the conditions necessary for dual-fit tensions to emerge as significant challenges, as documented in recent studies on CISO role complexity (Karanja, 2017; Sahin & Vance, 2025).

The model may extend to other C-suite boundary roles that exhibit similar characteristics, including Chief Risk Officers, Chief Compliance Officers, and potentially Chief Technology Officers in organisations where technology strategy requires significant external stakeholder management. However, the model's applicability to these roles would require empirical validation to confirm that similar dual-fit dynamics operate, following Malmendier et al. (2023) and Nath and Mahajan (2011) who demonstrate varying influence patterns across C-suite boundary roles.

The role boundary excludes purely internal leadership positions, technical specialist roles regardless of seniority, and positions where external stakeholder accountability remains minimal. The theoretical logic is that single-fit dynamics (either CISO-Organisation or CISO-Environment) may adequately explain effectiveness in these contexts without requiring the additional complexity of dual-fit considerations.

#### *7.6.1.4 Temporal Considerations: Cross-Sectional Design Capturing Evolution*

The study employed a cross-sectional interview design that captured participants' reflections on their career experiences rather than tracking evolution over time (see Section 4.8 for methodological limitations). This creates specific boundary conditions for the theoretical claims made by the SLCM. Participants provided retrospective accounts of their experiences across different roles and organisations, offering insights into how dual-fit requirements vary across contexts. The cross-sectional design included participants at different career stages and organisational phases, creating a composite picture of security leadership challenges, following Fortier et al. (2017) on retrospective data harmonisation approaches.

However, the temporal boundary means that real-time adaptation processes remain incompletely captured. The SLCM describes patterns of fit evolution and maintenance but cannot claim to fully explain the micro-processes through which leaders navigate dual-fit tensions in real-time. Future longitudinal research would be required to validate the temporal dynamics proposed by the model.

### 7.6.2 Theoretical Boundaries and Scope Conditions

The SLCM operates within specific theoretical boundaries that define the conditions under which its explanatory mechanisms function. These boundaries emerge from the theoretical requirements for the model's core propositions.

#### *7.6.2.1 Dual-Fit Model Applicability Requirements*

The dual-fit model requires sufficient role complexity to create meaningful tensions between CISO-Organisation and CISO-Environment fit demands. Simple roles with clear success criteria and minimal stakeholder complexity do not generate the competing pressures that necessitate dual-fit navigation. The theoretical threshold is reached when role incumbents face situations where actions that improve fit with one domain (organisation or environment) may compromise fit with the other domain, extending Etzel and Nagy (2021) on multidimensional fit interactions in the P-E fit literature. This is further supported by research on P-E fit complexity (Andela & van der Doef, 2018; Danese et al., 2020; Oh et al., 2014), which demonstrates the tensions inherent in managing multiple fit dimensions simultaneously.

Organisational maturity sufficient to create political dynamics represents another critical scope condition. The model assumes that informal influence networks operate alongside formal authority structures, creating multiple pathways for achieving organisational effectiveness. Organisations lacking such political sophistication may not provide the context necessary for relational capital to function as a dual-fit navigation mechanism, as suggested by Ferris et al. (2019) and Hochwarter et al. (2020) on organisational politics and effectiveness. Additional research (Elbanna, 2016; Mathieu et al., 2019) confirms that political dynamics significantly influence organisational effectiveness and leadership requirements.

External environmental pressure that exerts independent influence on role effectiveness constitutes the third major scope condition. The environment must create expectations and success criteria that are at least partially independent of internal organisational dynamics. Without such environmental pressure, CISO-Environment fit becomes merely an extension of CISO-Organisation fit rather than a distinct and potentially competing domain.

#### *7.6.2.2 Phase Model Prerequisites*

The organisational phase model requires that organisations experience evolution through security maturity phases driven by internal development and external pressures. Static industries or regulatory environments that do not create pressure for security evolution may not exhibit the phase transitions that generate changing dual-fit requirements.

Market or regulatory pressures must be sufficient to drive phase transitions rather than allowing organisations to remain in comfortable equilibrium states. The model assumes that security requirements evolve in response to threat landscapes, regulatory changes, and business model evolution. Stable environments may not create the dynamic conditions necessary for phase-based leadership requirement changes, consistent with Wilden and Gudergan (2015) and Pavlou and El Sawy (2011) on dynamic capabilities in stable versus turbulent environments. Recent research (Kalubanga & Gudergan, 2022; Tempelmayr et al., 2019; Zhang & Bang, 2023) further confirms that environmental turbulence significantly moderates the value of dynamic capabilities.

The phase model may not apply in jurisdictions or industries where security requirements remain static over extended periods. However, the increasing digitalisation of business processes and evolving threat landscapes suggest that such static environments are becoming increasingly rare.

#### *7.6.2.3 Relational Capital Mechanism Assumptions*

The relational capital mechanisms that enable dual-fit navigation are a form of bridging social capital Adler and Kwon (2002): ties that extend a CISO's influence beyond formal authority structures. The model requires that influence processes matter beyond formal authority structures and that relationships can be leveraged to achieve organisational outcomes. Highly bureaucratised environments that operate purely through formal authority may not provide the context necessary for relational capital to function effectively, as noted in Leana and Van Buren (1999). This is further supported by research demonstrating social capital’s context-dependent effects (Andrews, 2010; Kemper et al., 2013; Watson & Papamarcos, 2002).

The mechanisms assume Western business contexts characterised by relatively networked rather than purely hierarchical decision-making processes. The study tested these assumptions only within the Australian context, and their transferability to more hierarchical business cultures or different cultural contexts regarding relationship-building and influence processes remains to be empirically established, following Dickson et al. (2003) and Gerstner and Day (1994) on cross-cultural validity of leadership models.

Professional contexts where expertise and reputation influence organisational dynamics represent another scope condition. The model assumes that technical credibility and professional reputation can be converted into organisational influence through appropriate relationship management. Contexts where such professional capital is not valued or where purely positional authority dominates may not support the relational capital mechanisms described.

### 7.6.3 Theoretical Limitations and Alternative Explanations

Beyond the boundary conditions that define the SLCM's applicability, several theoretical limitations constrain the claims that can be made based on this research (see Section 4.8 for methodological limitations).

#### *7.6.3.1 Theoretical Generalisation Constraints*

The qualitative design enables theoretical rather than statistical generalisation, aligning with Yin (2018) on analytical generalisation approach. This approach is further supported by established case study methodology (Eisenhardt, 1991; Eisenhardt, 2021; Eisenhardt & Graebner, 2007), which emphasises theoretical development over statistical inference. The SLCM proposes causal mechanisms and theoretical relationships rather than probabilistic predictions about security leadership effectiveness. The theoretical propositions require empirical testing through quantitative methods to establish their statistical validity and predictive power.

Causal mechanisms are proposed rather than proven through the qualitative design. The study identifies plausible causal pathways linking dual-fit dynamics to leadership effectiveness but cannot establish causal certainty. This approach aligns with established qualitative methods for exploring causal mechanisms (Bonell et al., 2022; Falleti & Lynch, 2009; Jensen, 2021; Maxwell, 2004), which focus on understanding processes and contexts that generate outcomes rather than establishing statistical associations. The theoretical logic is supported by empirical evidence, but alternative explanations for the observed patterns cannot be definitively ruled out.

#### *7.6.3.2 Elite Perspective Limitations*

The focus on senior security leaders who have achieved executive positions creates an elite perspective that may not capture the full range of security leadership experiences. Unsuccessful leadership transitions, derailed careers, and alternative pathways to security influence remain underrepresented in the theoretical model. This elite bias is inherent in studying successful incumbents of senior positions, as noted in research on elite interviewing challenges (Harvey, 2011; Lancaster, 2017; Welch et al., 2002).

This limitation was addressed through diverse recruitment strategies that included participants who had experienced career setbacks and role transitions. The study explicitly explored negative cases and implementation challenges rather than focusing solely on success stories, consistent with established approaches for addressing elite bias in qualitative research (Furnas & Lapira, 2024; Natow, 2020; Ramač et al., 2022). Triangulation across multiple data sources helped mitigate the privileged perspectives inherent in elite interviews.

#### *7.6.3.3 Alternative Explanations Considered*

Selection bias toward articulate leaders was addressed through recruitment strategies that prioritised theoretical sampling over convenience sampling. Participants were selected based on their ability to provide insight into specific theoretical dimensions rather than their communication skills or career success, following Eisenhardt (2021) on theoretical sampling best practices. This approach aligns with established theoretical sampling methodology (Conlon et al., 2020; Draucker et al., 2007; Ligita et al., 2019), where participant selection is guided by emerging theoretical constructs rather than predetermined criteria.

Australian cultural factors that might influence the observed patterns were considered through analysis of participants' diverse cultural backgrounds and international experience. The pattern consistency across participants with varying cultural origins provides some confidence in the findings' broader applicability, though Dickson et al. (2003) on cross-cultural measurement invariance suggests further validation is needed.

Temporal period effects during data collection were considered through analysis of the stability of the regulatory and business environment during the research period. Data collection occurred during a relatively stable period, minimising the potential for external shocks to influence participant experiences systematically.

The theoretical boundaries establish the context within which the SLCM's contributions should be understood whilst providing a foundation for future research that can extend and validate the model's propositions across different contexts and through alternative methodological approaches. These boundaries are not weaknesses but rather honest acknowledgements of scope that strengthen rather than diminish the contribution, as boundary specification is essential for theoretical precision (Garzella et al., 2021; Mayrl & Quinn, 2016; Zhang, 2025). Clear boundary definition directly impacts the accuracy, reliability, and interpretability of research findings (Chen et al., 2015).

## 7.7 FUTURE RESEARCH AGENDA: BUILDING ON THE FOUNDATION

Future research could extend the SLCM through systematic investigation across multiple dimensions, from immediate empirical validation to long-term theoretical development and practical application.

### 7.7.1 Immediate Research Opportunities

#### *7.7.1.1 Quantitative Validation Studies*

Future research could develop scales to measure CISO-Organisation fit and CISO-Environment fit constructs. Existing person-organisation fit scales (Chatman, 1989; Kristof, 1996; O'Reilly et al., 1991 1991) and person-environment fit measures (Caplan, 1987; Edwards, 2008) provide starting points, but security leadership contexts require domain-specific adaptation. Recent scale development in leadership research demonstrates the complexity of capturing contextual nuances (Crawford & Kelder, 2019; Hassandoust & Johnston, 2023; Oc, 2018). Future studies might develop instruments that capture the interdependent nature of dual-fit whilst maintaining psychometric rigour.

Large-sample testing could establish the statistical significance and effect sizes of phase-contingent leadership effectiveness patterns. Methodological exemplars of qualitative-to-quantitative validation demonstrate the value of this progression (Eisenhardt, 1991; Gioia et al., 2013 2013). Quantitative validation might test whether the three-phase progression holds universally and whether specific leadership orientations demonstrate consistent effectiveness patterns within phases.

Longitudinal fit evolution studies could investigate temporal dynamics by following security leaders through phase transitions. Exemplary longitudinal studies in leadership research (Day, 2011; Ployhart & Vandenberg, 2010; Riggio & Mumford, 2011) provide methodological guidance for tracking leadership development over time. Future research might track how leaders adjust their approaches as organisational contexts evolve and identify predictors of successful navigation through transitions.

#### *7.7.1.2 Context Extensions*

Cross-cultural investigations could examine how the SLCM applies across different geographic contexts, particularly in North America, Europe, and Asia-Pacific regions. Cultural dimensions such as power distance, uncertainty avoidance, and individualism-collectivism (Hofstede, 1980) may influence how dual-fit dynamics manifest and how phase transitions occur. Cross-cultural leadership effectiveness studies demonstrate significant cultural contingencies (House et al., 2004). Future research might explore whether collaborative decision-making strategies prove less viable in more hierarchical cultures.

Different organisational types warrant investigation. Studies showing how organisational context affects leadership requirements (Peppard, 2010) suggest that startup versus established organisations and public versus private sector contexts may exhibit distinct patterns. Government contexts, particularly those with mandated cybersecurity frameworks, might demonstrate accelerated phase progression or unique effectiveness criteria. Future studies could explore whether startups, government agencies, and non-profit organisations exhibit distinct phase progressions or leadership requirements.

Adjacent roles offer extension opportunities. CTOs, CROs, and CIOs occupy similar boundary-spanning positions. Studies of boundary-spanning roles (Aldrich & Herker, 1977) and comparative leadership research across C-suite positions (Hiller & Peterson, 2019; Whysall & Bruce, 2023) suggest potential commonalities. Comparative studies could establish whether phase-contingent effectiveness patterns and dual-fit requirements represent broader phenomena or security-specific dynamics.

#### *7.7.1.3 Mechanism Investigation*

Political capital development processes require systematic examination. Studies investigating political capital development in organisational contexts (Ferris et al., 2007; Pfeffer, 1992) provide methodological exemplars. Future research could explore how security leaders build and leverage influence across different stakeholder groups, including specific strategies and timeframes for building relational capital.

Fit maintenance strategies warrant detailed investigation. Future studies might examine specific tactics successful leaders employ to monitor fit evolution and respond to misalignment, including how leaders recognise fit deterioration, assess the relative importance of CISO-Organisation versus CISO-Environment alignment in different circumstances, and implement corrective actions.

Phase transition navigation requires investigation of specific competencies, strategies, and organisational supports that facilitate successful navigation. Future research could examine succession planning, interim leadership arrangements, and organisational development interventions in supporting effective phase transitions.

These immediate research opportunities provide clear next steps for validating and extending the SLCM's core constructs through empirical investigation. The quantitative validation studies offer pathways for establishing statistical generalisability, whilst context extensions and mechanism investigations would deepen understanding of how the model operates across diverse settings. Additional research directions, including theoretical development pathways, methodological innovations, and practical research applications, are addressed in the concluding chapter's comprehensive research agenda.

## 7.8 CONCLUSION: THE SLCM'S CONTRIBUTION TO SECURITY LEADERSHIP UNDERSTANDING

This chapter has examined the theoretical, practical, and methodological implications of the Security Leadership Contingency Model (SLCM) that emerged from this research. The journey from empirical insights to theoretical contributions demonstrates how practitioner puzzles can generate meaningful academic knowledge whilst maintaining relevance for professional practice. This concluding section synthesises the core arguments, establishes the significance of the contributions, and reflects on the journey from practice-based questions to theoretically grounded insights.

### 7.8.1 Synthesis of Core Arguments

This research set out to answer a fundamental question: "How is the CISO role enacted across different organisational contexts in modern organisations?" The SLCM provides a systematic answer to this fundamental question by revealing that the CISO role is characterised by dual-fit imperatives, phase-contingent requirements, and political navigation mechanisms. Understanding the CISO role, as the research demonstrates, requires comprehending not merely technical responsibilities but the complex dynamics of maintaining simultaneous alignment with organisational contexts and external environments across varying maturity phases. The four interconnected theoretical arguments synthesised below demonstrate how the research moved from the primary research question to theoretically grounded insights about role requirements, effectiveness patterns, and contextual dependencies.

The SLCM advances our understanding of security leadership effectiveness through four interconnected theoretical arguments that challenge conventional wisdom whilst building on established theory. These arguments collectively position the model as both a theoretical contribution and a practical framework for understanding the complexities of security leadership in contemporary organisations.

**Security leadership effectiveness requires dual-fit achievement and maintenance**. Effective security leaders must simultaneously achieve and maintain fit with both their organisational context and the external environment. This dual-fit imperative represents a fundamental departure from traditional fit theories that examine person-organisation or person-environment alignment in isolation. The simultaneity requirement creates resource allocation dilemmas and necessitates sophisticated political skills for successful navigation.

This dual-fit imperative operates within dynamic organisational contexts that fundamentally shape its manifestation. **Phase-contingent requirements create dynamic effectiveness criteria**. The traditional view of leadership effectiveness as comprising stable competencies and behaviours is challenged by the finding that effectiveness criteria evolve systematically as organisations mature through security phases. This phase contingency means that previously effective leaders may experience performance decline without role evolution, whilst organisations may struggle with leaders whose capabilities matched earlier phases but not current requirements. The interaction between dual-fit demands and phase-specific requirements creates compound complexity: the nature of both organisational and environmental fit requirements changes as organisations progress through maturity phases, requiring continuous recalibration of leadership approaches.

Navigating these phase-contingent dual-fit demands requires specific organisational mechanisms. **Political dynamics mediate between organisational and environmental demands**. Political capital serves as the critical mechanism enabling dual-fit maintenance. Unlike traditional models that position politics as peripheral to leadership effectiveness, this research demonstrates that political processes serve essential functions in security leadership contexts. Political capital enables leaders to bridge competing demands between organisational preferences and external requirements and provides the influence necessary to shape organisational perspectives when external pressures demand security approaches that appear to conflict with business priorities. This mediating function of political capital becomes particularly critical during phase transitions, when both fit requirements and political dynamics shift simultaneously.

The complexity revealed through these three interconnected mechanisms demonstrates fundamental limitations in existing approaches. **Traditional approaches oversimplify security leadership complexity**. The dominant paradigm in security leadership research that focuses on identifying universal competencies or optimal organisational structures fails to capture the dynamic nature of these requirements. The SLCM provides a more nuanced understanding that accommodates this complexity whilst offering practical guidance for leaders and organisations.

### 7.8.2 Significance of Contributions

The SLCM makes substantive contributions across multiple domains, each offering distinct value to different research and practitioner communities whilst maintaining coherence within an integrated theoretical framework.

**Theoretical significance emerges from the extension and integration of multiple established theories**. The model extends person-organisation fit theory by demonstrating that fit requirements change systematically as organisations evolve, challenging static conceptualisations of fit relationships. Person-environment fit theory is advanced through the demonstration that environmental fit operates simultaneously with, rather than independently from, organisational fit, creating novel theoretical territories that existing frameworks do not adequately address. The integration of these perspectives with contingency theory and role theory generates new insights about the temporal dimensions of leadership effectiveness and the mechanisms through which contextual factors influence role performance. This theoretical synthesis represents a meaningful contribution to leadership theory beyond the security domain, offering insights relevant to other boundary-spanning executive roles that must balance internal and external demands.

**Practical significance lies in the actionable framework for leaders and organisations seeking to enhance security leadership effectiveness**. The SLCM provides security professionals with a diagnostic tool for assessing their current effectiveness and identifying development priorities based on organisational phase and environmental demands. Rather than prescribing universal best practices, the framework enables contextually appropriate decision-making about competency development, role positioning, and stakeholder engagement strategies. For organisations, the framework offers guidance for recruitment decisions, performance evaluation criteria, and support systems that enable security leadership success. This bridge between academic theory and professional practice represents a significant contribution to applied management research.

**Methodological significance is demonstrated through the successful application of practitioner-research approaches in elite professional contexts**. This research demonstrates the value of insider knowledge in accessing and interpreting the experiences of senior executives,

whilst showing how systematic qualitative methods can mitigate potential bias risks. The adaptation of Gioia methodology to accommodate temporal dimensions and multiple theoretical lenses provides a template for future research in dynamic organisational contexts. Furthermore, the research validates the utility of grounded theory approaches for developing contextually relevant frameworks in professional domains where existing theory provides insufficient guidance. These methodological contributions offer value for researchers studying other executive roles or professional contexts where practitioner access and contextual understanding are critical for meaningful insights.

**Professional significance establishes a foundation for evidence-based development in the security leadership domain**. The SLCM provides the profession with its first theoretically grounded framework for understanding leadership effectiveness, moving beyond anecdotal evidence and prescriptive frameworks to empirically derived insights. This foundation enables more sophisticated professional development programmes that account for contextual variation and dynamic requirements. It also provides a basis for evidence-based governance practices that recognise the complexity of security leadership roles and support appropriate development and evaluation approaches. The research contributes to the professionalisation of security leadership by establishing academic credibility for practitioner concerns and demonstrating the value of systematic research in addressing professional challenges.

### 7.8.3 Final Reflection

The journey from practitioner puzzle to theoretical insight illustrates the potential for practice-based research to generate meaningful academic contributions whilst addressing real-world professional challenges. This research began with observations about the apparent inconsistency in security leadership effectiveness across similar organisational contexts. Traditional explanations focusing on competency gaps or structural deficiencies seemed inadequate to explain the complex patterns observed in professional practice. The systematic investigation of these patterns through rigorous qualitative research has generated insights that advance both theoretical understanding and practical effectiveness.

**Bridging academic rigour with practical relevance has been a central challenge and achievement of this research**. The SLCM maintains theoretical sophistication whilst offering practical applicability, demonstrating that academic research can serve both scholarly and professional communities effectively. The framework's grounding in established theory ensures its contribution to academic knowledge, whilst its emergence from practitioner experiences

guarantees its relevance to professional practice. This balance required careful attention to methodological rigour, particularly in managing the researcher's dual identity as both academic investigator and industry practitioner. The successful navigation of this challenge demonstrates the potential for practitioner-researchers to make distinctive contributions that purely academic or purely practitioner perspectives might miss.

**Contributing to security leadership professionalisation represents a broader impact beyond immediate research contributions**. The establishment of evidence-based frameworks for understanding leadership effectiveness supports the maturation of security leadership as a professional discipline. This research provides a foundation for future empirical studies, professional development programmes, and governance practices that recognise the sophisticated nature of security leadership roles. The professionalisation impact extends beyond security to demonstrate how systematic research can enhance understanding of emerging executive roles that bridge technical and business domains.

**Opening pathways for continued investigation ensures that this research serves as a platform for ongoing development rather than a concluding statement**. The theoretical foundations established can support quantitative validation studies, cross-cultural investigations, and longitudinal examinations of leadership development processes. Furthermore, the research demonstrates methodological approaches that can be adapted for studying other professional domains where contextual complexity and practitioner insight are critical for understanding effectiveness.

### 7.8.4 Transition to Thesis Conclusion

This chapter has examined the implications of the SLCM for theory, practice, and methodology, establishing its significance across multiple domains whilst acknowledging its limitations and boundaries. Chapter 8 will synthesise the entire thesis journey and present final recommendations for both research and practice.

# CHAPTER 8: CONCLUSION

## 8.1 THE RESEARCH JOURNEY: FROM PARADOX TO UNDERSTANDING

The Chief Information Security Officer role presents a distinctive organisational paradox. Globally, organisations have elevated the CISO to executive prominence, investing substantial resources in a position commanding board attention and strategic authority. Yet, at the point this research commenced, systematic scholarly understanding of what enables security leaders to succeed in that position remained virtually absent. Organisations were making consequential decisions about recruitment, structure, and performance evaluation on the basis of assertion rather than evidence.

This thesis addressed that paradox through a single guiding question: how is the CISO role enacted across different organisational contexts in modern organisations? The question was framed deliberately broadly, reflecting both the nascent state of security leadership scholarship and the methodological commitment to discovery over confirmation. Twenty senior Australian security leaders participated in a grounded theory investigation designed to reveal patterns that neither practitioner discourse nor adjacent academic literatures had previously articulated.

The investigation established that security leadership effectiveness is not a fixed standard achievable through universal competencies. It is a contextually contingent achievement, shaped by organisational circumstances, sensitive to how those circumstances evolve over time, and dependent on forms of political sophistication that existing executive leadership frameworks had neither recognised nor theorised. This finding reframes the field's foundational assumptions: the question is not what a good CISO looks like, but what effectiveness requires in a given context at a given moment.

The sections that follow synthesise what this research contributes, what it means for practice, where its boundaries lie, and what questions it opens for continued scholarly investigation. The paradox with which this research began has not been resolved by assertion; it has been addressed through systematic empirical investigation producing theoretical understanding where previously only assumption existed.

## 8.2 ANSWERING THE RESEARCH QUESTIONS: WHAT THE INVESTIGATION REVEALED

The primary research question asked how the CISO role is enacted across different organisational contexts in modern organisations. The answer this investigation produces is both theoretically precise and practically consequential: the CISO role is not enacted as a stable configuration of responsibilities applied uniformly across settings, but as a continuously negotiated achievement whose character is fundamentally constituted by the contexts within which it operates. This finding challenges the foundational assumptions underlying most existing treatments of security leadership, which have proceeded as though the role possesses an essential definition awaiting empirical description. What systematic investigation reveals instead is that context does not merely shape the CISO role from the outside; it determines what the role is, and therefore what effectiveness within it requires, from the inside. The Security Leadership Contingency Model gives this insight theoretical form, but the deeper claim precedes the model: organisations cannot understand their security leadership without first understanding the organisational and environmental conditions that define what leadership in that context demands.

### 8.2.1 Organisational Context and Security Leadership

Organisational context does not produce variation in how a fixed role is performed; it produces variation in what the role itself demands, what legitimacy requires, and what constitutes effectiveness at any given moment. Prior scholarship had consistently acknowledged that context matters without systematically investigating how or why; the investigation resolved that gap in terms more fundamental than existing literature anticipated. Critically, this contextual variation interacts with what individual leaders naturally bring. The investigation identified distinct leadership orientations, characterised broadly as action-oriented, stewardship-oriented, and vision-oriented, each of which creates natural alignment with particular organisational phases whilst generating potential misalignment with others. Effectiveness therefore emerges not from the possession of any single orientation but from the degree of fit between what a leader is disposed to do and what the organisational phase demands. The theoretical implication is significant: frameworks presuming a universal CISO capability profile misspecify the nature of the phenomenon by treating as a fixed target what is in fact a moving and contextually constituted set of requirements.

### 8.2.2 Establishing and Maintaining Credibility

Credibility is neither established once through credential acquisition nor maintained through consistent performance against fixed standards; it is continuously renegotiated as the organisational conditions that originally conferred it evolve. Leaders whose legitimacy rested on technical responsiveness in early phases encountered those same capacities operating as liabilities when organisations matured toward expectations of strategic integration; the very source of prior credibility became an obstacle to sustaining it. This dynamic reveals credibility as phase-contingent rather than cumulative, and political sophistication, specifically the capacity to recognise when the prevailing terms of legitimacy have shifted and to reposition accordingly, as the mechanism through which security leaders sustain influence across transitions that static capability frameworks cannot navigate.

### 8.2.3 Technical Expertise and Business Acumen

The investigation reveals the longstanding debate between technical depth and business orientation as a misframing of a more complex and consequential dynamic. Technical credibility and business acumen are both necessary, but their relationship is neither fixed nor uniformly complementary; it is phase-contingent, and the consequences of miscalibration are not merely suboptimal but actively damaging to legitimacy. In Establishment phases, technical credibility constitutes the primary currency through which security leaders earn organisational standing; in Strategic phases, that same dominance of technical orientation can undermine the business integration that legitimacy then requires. Furthermore, different leadership orientations create natural dispositions toward one end of this spectrum, meaning that phase transitions demand not only capability development but deliberate recalibration of emphasis in ways that may run counter to a leader’s instinctive orientation. The tension between technical expertise and business acumen is therefore a permanent structural feature of the role, phase-sensitive in its expression and navigable through political sophistication rather than resolvable through any fixed developmental prescription.

### 8.2.4 The Integrated Answer

Considered in relation to one another, these findings converge on a theoretical insight that unifies what the guiding themes approached as separate concerns: the CISO role is enacted through the continuous management of two independent alignment relationships, between the leader and the internal organisational context, and between the leader and the external environment, across conditions that do not remain stable. Legitimacy, credibility, and capability orientation are not independent dimensions of security leadership; they are expressions of the same underlying dynamic, namely the degree to which a leader maintains dual-fit alignment as both dimensions evolve and as the fit demands of each phase place different pressures on different leadership orientations. The Security Leadership Contingency Model articulates this dynamic with theoretical precision, explaining why identical approaches produce divergent outcomes across contexts, why leaders with strong natural orientations sometimes fail precisely because those orientations create misalignment as organisations mature, and why the question of what makes a security leader effective cannot be answered without first specifying the organisational and environmental conditions within which effectiveness is being assessed.

## 8.3 CONTRIBUTIONS TO KNOWLEDGE: SIGNIFICANCE AND REACH

The Security Leadership Contingency Model constitutes this research's primary contribution to knowledge, but its significance extends beyond the model itself to the theoretical reorientation it produces, the methodological demonstration it offers, and the practical foundations it establishes. These contributions operate at different registers and serve different communities; taken together, they advance security leadership from a domain characterised by prescription without empirical foundation toward one capable of systematic, evidence-based inquiry.

### 8.3.1 Theoretical Contributions

The research advances theoretical understanding across three established frameworks. Person-organisation fit theory is extended by revealing that fit requirements do not evolve gradually but transform systematically and discontinuously across organisational maturity phases, with capabilities constituting fit in one phase actively undermining fit in another. This represents a more radical form of fit dynamics than the theory had previously conceptualised, and one with direct implications for how executive effectiveness in complex roles is theorised. Person-environment fit theory is advanced by demonstrating that boundary-spanning executives face independent external alignment demands that operate regardless of organisational preferences, requiring active political maintenance rather than passive adjustment. Most significantly, the integration of these perspectives with contingency theory and role theory generates the dual-fit construct, a theoretically novel mechanism explaining how security leaders navigate potentially conflicting alignment requirements simultaneously. The contribution is not the extension of any single theory but the demonstration that security leadership effectiveness cannot be explained within any single theoretical domain; it emerges from dynamic interactions across frameworks that existing scholarship had not previously synthesised into a coherent explanatory framework.

### 8.3.2 Methodological Contributions

The research demonstrates that practitioner-researcher positioning, when managed through systematic reflexivity, constitutes a methodological asset rather than a source of bias requiring mitigation. Insider knowledge enabled recognition of political dynamics and organisational subtleties that external researchers might not have identified; systematic supervisory challenge ensured that this knowledge informed rather than predetermined analytical interpretation. The research additionally demonstrates how temporal dynamics can be incorporated into the Gioia methodology systematically, extending its applicability to phenomena where phase-contingent variation is theoretically significant. These contributions offer templates for other professionally embedded researchers navigating the tension between contextual access and analytical distance, establishing that rigorous scholarship and practical embeddedness are complementary rather than competing research conditions.

### 8.3.3 Practical Contributions

This research establishes empirical foundations from which security leadership decisions can be made with theoretical precision rather than assumption. The Security Leadership Contingency Model reframes the criteria against which consequential organisational decisions are made. Recruitment can now be oriented toward phase-appropriate alignment rather than generic competency matching. Development can acknowledge political sophistication as a meta-capability rather than a peripheral attribute. Performance evaluation can account for the contextual conditions within which leaders operate, rather than applying universal standards that cannot distinguish effective leadership from favourable circumstances. The operational tools developed from the SLCM translate this understanding into instruments immediately applicable in organisational settings, representing the first systematic, empirically grounded practical guidance available to a profession that has until now relied on intuition and frameworks adapted from adjacent executive roles.

### 8.3.4 Broader Significance

Collectively, these contributions position security leadership as a legitimate and theoretically distinctive domain of scholarly inquiry. The research demonstrates that the adversarial context, prevention paradox, and asymmetric visibility characterising security leadership are not merely incidental features of the context but theoretically consequential dimensions demanding dedicated frameworks rather than those adapted from adjacent executive leadership frameworks. Beyond security leadership, the dual-fit construct and phase-contingent effectiveness model may offer theoretical perspectives informing investigation of other boundary-spanning executive roles facing analogous challenges of conflicting stakeholder demands and evolving contextual requirements. The research establishes that emerging professional domains offer opportunities for genuine theoretical innovation, and that the absence of empirical foundations is a condition amenable to systematic scholarly investigation rather than an insurmountable constraint.

## 8.4 RECOMMENDATIONS FOR PRACTICE: TRANSLATING UNDERSTANDING INTO ACTION

Security leadership fit is now diagnosable. Where organisations previously had no systematic basis for assessing whether a leader's orientation matched their organisational phase, or for evaluating the health of the dual-fit relationship as conditions evolved, the Security Leadership Contingency Model and its associated instruments provide that basis. This changes what is possible across every dimension of security leadership practice.

Appointment decisions can move beyond generic competency matching toward deliberate alignment between leadership orientation and phase-contingent organisational demands, accounting simultaneously for the independent pressures of the external environment. Phase transitions, rather than arriving as unexpected disruptions to established effectiveness, can be anticipated and planned for through developmental investment and succession frameworks that reflect the different orientations different phases require. When security leadership struggles, the diagnostic question is no longer solely whether the individual is capable but whether the alignment conditions for effectiveness have been established and maintained. Political sophistication, the capacity to navigate dual-fit tensions and sustain influence across evolving contexts, must be recognised as a selection and development criterion rather than treated as a peripheral attribute that leaders either possess or do not.

For the profession, these findings establish that preparing security leaders for executive roles requires more than technical and business capability development. Curricula and certification frameworks must cultivate phase awareness, leadership orientation recognition, and the adaptive capacity to sustain alignment as organisational conditions evolve, equipping practitioners with the conceptual foundations to manage their own effectiveness systematically rather than intuitively.

## 8.5 RESEARCH BOUNDARIES AND FUTURE DIRECTIONS

This research was conducted with senior security leaders in large, complex Australian organisations operating within mature regulatory environments. These conditions define the scope within which the model's propositions hold, and mark the boundaries from which future investigation can extend.

The dual-fit dynamics theorised through the SLCM establish a foundation for investigating executive effectiveness in other boundary-spanning roles facing analogous challenges of conflicting stakeholder demands and evolving contextual requirements. The model's core propositions, that effectiveness emerges from simultaneous alignment with independent internal and external demands mediated by political capital, are not inherently specific to security leadership. Systematic investigation of other C-suite roles operating under comparable conditions of irreconcilable accountability tensions and asymmetric visibility would establish the generalisability of these theoretical mechanisms and extend their explanatory reach.

The three-phase maturity progression was identified within Australian organisations operating under Westminster governance structures and common law regulatory frameworks. These institutional conditions shaped the phenomenon as it was observed, and cross-cultural investigation represents a natural and necessary extension. Examining whether the phase model and its associated leadership orientation requirements manifest consistently across different governance traditions and regulatory environments would establish the boundaries of the model's universality and deepen understanding of the institutional conditions that shape security leadership demands.

Political capital emerges from this research as the central mechanism through which dual-fit alignment is navigated and sustained. Its developmental trajectory, how it is built, what conditions enable or constrain its accumulation, and the degree to which it can be cultivated through deliberate practice, represents a significant direction for continued investigation. Establishing how political capital develops has direct implications for how the profession prepares security leaders and how organisations support them through phase transitions.

## 8.6 CONCLUDING STATEMENT

The theoretical frameworks that currently inform our understanding of executive effectiveness each began as an attempt to explain phenomena that existing frameworks could not adequately address. The Security Leadership Contingency Model stands in that tradition. It emerges from a domain that existing theories were not designed to explain, and in doing so, it offers theoretical language for a class of leadership challenges that those theories cannot fully capture: the simultaneous navigation of independent and potentially conflicting alignment demands, across organisational conditions that transform rather than merely evolve, mediated by political capital as the central enabling mechanism.

Security leadership provided the empirical setting for this theoretical development, but the model's reach is not confined to it. Boundary-spanning executive roles, characterised by irreconcilable accountability tensions, asymmetric visibility, and evolving contextual demands, are not unique to security. They are a defining feature of modern organisational life. The SLCM establishes the theoretical foundations from which that broader class of leadership challenges can be investigated systematically, offering a framework capable of generating cumulative knowledge rather than isolated insights.

The SLCM represents a theoretical foundation from which security leadership scholarship, and the broader investigation of boundary-spanning executive effectiveness, can develop systematically and cumulatively.

# REFERENCES

Abdallah, M., Naghizadeh, P., Hota, A. R., Cason, T., Bagchi, S., & Sundaram, S. (2020). Behavioral and Game-Theoretic Security Investments in Interdependent Systems Modeled by Attack Graphs. *IEEE Transactions on Control of Network Systems*, *7*(4), 1585-1596. https://doi.org/10.1109/tcns.2020.2988007

Abouelmehdi, K., Beni-Hessane, A., & Khaloufi, H. (2018). Big healthcare data: preserving security and privacy. *Journal of Big Data*, *5*(1), 1. https://doi.org/10.1186/s40537-017-0110-7

Abrahamsen, E. B., Selvik, J. T., Milazzo, M. F., Langdalen, H., Dahl, R. E., Bansal, S., & Abrahamsen, H. B. (2021). On the use of the 'Return Of Safety Investments' (ROSI) measure for decision-making in the chemical processing industry. *Reliability Engineering & System Safety*, *210*, 107537. https://doi.org/10.1016/j.ress.2021.107537

Abrardi, L., Comino, S., & Grassini, S. (2025). The economics of cyber risk: a survey of the literature. *Journal of Industrial and Business Economics*. https://doi.org/10.1007/s40812-025-00370-3

Adams, M., & Makramalla, M. (2015). Cybersecurity Skills Training: An Attacker-Centric Gamified Approach. *Technology Innovation Management Review*, *5*(1), 5-14. https://doi.org/10.22215/timreview/861

Adebola, F., Ifeoluwa, W., Bunmi, S., & Viqaruddin, M. (2024). Security compliance and its implication for cybersecurity. *World Journal of Advanced Research and Reviews*, *24*(1), 2105-2121. https://doi.org/10.30574/wjarr.2024.24.1.3170

Adekoya, O. A., Atlam, H. F., & Lallie, H. S. (2025). Quantifying the Multidimensional Impact of Cyber Attacks in Digital Financial Services: A Systematic Literature Review. *Sensors*, *25*(14), 4345.

Adler, P. S., & Kwon, S. W. (2002). Social capital: Prospects for a new concept. *Academy of Management Review*, *27*(1), 17-40. https://doi.org/10.2307/4134367

Aftabi, N., Moradi, N., Mahroo, F., & Kianfar, F. (2025). SD-ABM-ISM: An integrated system dynamics and agent-based modeling framework for information security management in complex information systems with multi-actor threat dynamics. *Expert Systems with Applications*, *263*, 125681. https://doi.org/10.1016/j.eswa.2024.125681

Agrafiotis, I., Nurse, J. R. C., Goldsmith, M., Creese, S., & Upton, D. (2018). A taxonomy of cyber-harms: Defining the impacts of cyber-attacks and understanding how they propagate. *Journal of Cybersecurity*, *4*(1). https://doi.org/10.1093/cybsec/tyy006

Ahmad, A., Desouza, K. C., Maynard, S. B., Naseer, H., & Baskerville, R. L. (2020). How integration of cyber security management and incident response enables organizational learning. *Journal of the Association for Information Science and Technology*, *71*(8), 939-953. https://doi.org/10.1002/asi.24311

Ahmad, A., Hadgkiss, J., & Ruighaver, A. B. (2012). Incident response teams – Challenges in supporting the organisational security function. *Computers & Security*, *31*(5), 643-652. https://doi.org/10.1016/j.cose.2012.04.001

Ahmad, A., Maynard, S. B., Desouza, K. C., Kotsias, J., Whitty, M. T., & Baskerville, R. L. (2021). How can organizations develop situation awareness for incident response: A case study of management practice. *Computers & Security*, *101*, 102122. https://doi.org/10.1016/j.cose.2020.102122

Ahmad, A., Maynard, S. B., & Shanks, G. (2015). A case analysis of information systems and security incident responses. *International Journal of Information Management*, *35*(6), 717-723. https://doi.org/10.1016/j.ijinfomgt.2015.08.001

Ahmad, A., Webb, J., Desouza, K., & Boorman, J. (2019). Strategically-motivated advanced persistent threat: Definition, process, tactics and a disinformation model of counterattack. *Computers & Security*, *86*, 402-418. https://doi.org/10.1016/j.cose.2019.07.001

Aidoo, A., Schiller, E., Fuhrer, J., Stahl, J., Ziorjen, M., & Stiller, B. (2022). Landscape of IoT security. *Computer Science Review*, *44*, 100467. https://doi.org/10.1016/j.cosrev.2022.100467

Akbar, M. A., Smolander, K., Mahmood, S., & Alsanad, A. (2022). Toward successful DevSecOps in software development organizations: A decision-making framework. *Information and Software Technology*, *147*, 106894. https://doi.org/10.1016/j.infsof.2022.106894

Akter, S., Uddin, M. R., Sajib, S., Lee, W. J. T., Michael, K., & Hossain, M. A. (2025). Reconceptualizing cybersecurity awareness capability in the data-driven digital economy. *Annals of Operations Research*, *350*(2), 673-698. https://doi.org/10.1007/s10479-022-04844-8

Al-Taie, M., Lane, M., & Cater-Steel, A. (2018). An Empirical Assessment of the CIO Role Expectations Instrument Using PLS Path Modelling. *Communications of the Association for Information Systems*, *42*(1), 1-20. https://doi.org/10.17705/1cais.04201

Alazab, M., Hong, S. H., & Ng, J. (2021). Louder bark with no bite: Privacy protection through the regulation of mandatory data breach notification in Australia. *Future Generation Computer Systems-the International Journal of Escience*, *116*, 22-29. https://doi.org/10.1016/j.future.2020.10.017

Aldaajeh, S., & Alrabaee, S. (2024). Strategic cybersecurity. *Computers & Security*, *141*, 103845. https://doi.org/10.1016/j.cose.2024.103845

Aldrich, H., & Herker, D. (1977). Boundary Spanning Roles and Organization Structure. *Academy of Management Review*, *2*(2), 217-230. https://doi.org/10.2307/257905

AlGhamdi, S., Win, K. T., & Vlahu-Gjorgievska, E. (2020). Information security governance challenges and critical success factors: Systematic review. *Computers & Security*, *99*, 102030. https://doi.org/10.1016/j.cose.2020.102030

Alhassan, I., Sammon, D., & Daly, M. (2016). Data governance activities: an analysis of the literature. *Journal of Decision Systems*, *25*, 64-75. https://doi.org/10.1080/12460125.2016.1187397

Alhassan, I., Sammon, D., & Daly, M. (2018). Data governance activities: a comparison between scientific and practice-oriented literature. *Journal of Enterprise Information Management*, *31*(2), 300-316. https://doi.org/10.1108/Jeim-01-2017-0007

Ali Milaat, F., & Lubell, J. (2024). Layered Security Guidance for Data Asset Management in Additive Manufacturing. *Journal of computing and information science in engineering*, *24*(7), 071001. https://doi.org/10.48550/arXiv.2309.16842

Ali, R. F., Dominic, P. D. D., Ali, S. E. A., Rehman, M., & Sohail, A. (2021). Information Security Behavior and Information Security Policy Compliance: A Systematic Literature Review for Identifying the Transformation Process from Noncompliance to Compliance. *Applied Sciences-Basel*, *11*(8), 3383. https://doi.org/10.3390/app11083383

Almalawi, A., Khan, A. I., Alsolami, F., Abushark, Y. B., & Alfakeeh, A. S. (2023). Managing Security of Healthcare Data for a Modern Healthcare System. *Sensors (Basel)*, *23*(7). https://doi.org/10.3390/s23073612

Almuqrin, A. (2024). How About Enhancing Organizational Security: Critical Success Factors in Information Security Management Performance. *Journal of Global Information Management*, *32*(1), 1-18. https://doi.org/10.4018/Jgim.358745

AlNuaimi, B. K., Singh, S. K., Ren, S., Budhwar, P., & Vorobyev, D. (2022). Mastering digital transformation: The nexus between leadership, agility, and digital strategy. *Journal of Business Research*, *145*, 636-648. https://doi.org/10.1016/j.jbusres.2022.03.038

Aloseel, A., He, H. M., Shaw, C., & Khan, M. A. (2021). Analytical Review of Cybersecurity for Embedded Systems. *IEEE Access*, *9*, 961-982. https://doi.org/10.1109/Access.2020.3045972

Alotaibi, Y., & Liu, F. (2017). Survey of business process management: challenges and solutions. *Enterprise Information Systems*, *11*(8), 1119-1153. https://doi.org/10.1080/17517575.2016.1161238

Alshaikh, M., Ahmad, A., Maynard, S. B., & Chang, S. (2014, 2014/01/01/2014). Towards a taxonomy of information security management practices in organisations. Proceedings of the 25th Australasian Conference on Information Systems, ACIS 2014,

Alshaikh, M., Maynard, S., Ahmad, A., & Chang, S. (2015). Information Security Policy: A Management Practice Perspective.

Alshamrani, A., Myneni, S., Chowdhary, A., & Huang, D. J. (2019). A Survey on Advanced Persistent Threats: Techniques, Solutions, Challenges, and Research Opportunities. *Ieee Communications Surveys and Tutorials*, *21*(2), 1851-1877. https://doi.org/10.1109/Comst.2019.2891891

Ameen, N., Tarhini, A., Shah, M. H., & Madichie, N. O. (2020). Employees' behavioural intention to smartphone security: A gender-based, cross-national study. *Computers in Human Behavior*, *104*, 106184. https://doi.org/10.1016/j.chb.2019.106184

Ampatzoglou, A., Bibi, S., Avgeriou, P., Verbeek, M., & Chatzigeorgiou, A. (2019). Identifying, categorizing and mitigating threats to validity in software engineering secondary studies. *Information and Software Technology*, *106*, 201-230. https://doi.org/10.1016/j.infsof.2018.10.006

Andela, M., & van der Doef, M. (2018). A Comprehensive Assessment of the Person–Environment Fit Dimensions and Their Relationships With Work-Related Outcomes. *Journal of Career Development*, *46*(5), 567-582. https://doi.org/10.1177/0894845318789512

Anderson, E. E., & Choobineh, J. (2008). Enterprise information security strategies. *Computers & Security*, *27*(1-2), 22-29. https://doi.org/10.1016/j.cose.2008.03.002

Andrews, R. (2010). Organizational social capital, structure and performance. *Human Relations*, *63*(5), 583-608. https://doi.org/10.1177/0018726709342931

Applegate, L. M., & Elam, J. J. (1992). New Information-Systems Leaders - a Changing-Role in a Changing World. *Mis Quarterly*, *16*(4), 469-490. https://doi.org/10.2307/249732

Argaw, S. T., Troncoso-Pastoriza, J. R., Lacey, D., Florin, M.-V., Calcavecchia, F., Anderson, D., Burleson, W., Vogel, J.-M., O'Leary, C., Eshaya-Chauvin, B., & Flahault, A. (2020). Cybersecurity of Hospitals: discussing the challenges and working towards mitigating the risks. *BMC Medical Informatics and Decision Making*, *20*(1), 146. https://doi.org/10.1186/s12911-020-01161-7

Armenia, S., Angelini, M., Nonino, F., Palombi, G., & Schlitzer, M. F. (2021). A dynamic simulation approach to support the evaluation of cyber risks and security investments in SMEs. *Decision Support Systems*, *147*, 113580. https://doi.org/10.1016/j.dss.2021.113580

Ashenden, D., & Sasse, A. (2013). CISOs and organisational culture: Their own worst enemy? *Computers & Security*, *39*, 396-405. https://doi.org/10.1016/j.cose.2013.09.004

Ashforth, B. (2001). Role Transitions in Organisational Life : An Identity-based Perspective. *Administrative Science Quarterly*, *46*, 778. https://doi.org/10.4324/9781410600035

Awan, U., Hannola, L., Tandon, A., Goyal, R. K., & Dhir, A. (2022). Quantum computing challenges in the software industry. A fuzzy AHP-based approach. *Information and Software Technology*, *147*, 106896. https://doi.org/10.1016/j.infsof.2022.106896

Axon, L., Fletcher, K., Scott, A. S., Stolz, M., Hannigan, R., El Kaafarani, A., Goldsmith, M., & Creese, S. (2022). Emerging Cybersecurity Capability Gaps in the Industrial Internet of Things: Overview and Research Agenda. *Digital Threats: Research and Practice*, *3*(4), 1-27. https://doi.org/10.1145/3503920

Badakhshan, P., Wurm, B., Grisold, T., Geyer-Klingeberg, J., Mendling, J., & vom Brocke, J. (2022). Creating business value with process mining. *The Journal of Strategic Information Systems*, *31*(4), 101745. https://doi.org/10.1016/j.jsis.2022.101745

Bailey, C. (2019). The Relationship Between Chief Risk Officer Expertise, ERM Quality, and Firm Performance. *Journal of Accounting, Auditing & Finance*, *37*(1), 205-228. https://doi.org/10.1177/0148558x19850424

Bajwa, D. S., Rai, A., & Brennan, I. (1998). Key antecedents of Executive Information System success: a path analytic approach. *Decision Support Systems*, *22*(1), 31-43. https://doi.org/10.1016/S0167-9236(97)00032-8

Balozian, P., Burns, A. J., & Leidner, D. E. (2023). An Adversarial Dance: Toward an Understanding of Insiders' Responses to Organizational Information Security Measures. *Journal of the Association for Information Systems*, *24*(1), 161-221. https://doi.org/10.17705/1jais.00798

Balzano, M., & Marzi, G. (2025). At the Cybersecurity Frontier: Key Strategies and Persistent Challenges for Business Leaders. *Strategic Change-Briefings in Entrepreneurial Finance*, *34*(2), 181-192. https://doi.org/10.1002/jsc.2622

Bansal, G., & Axelton, Z. (2025). She's worth IT: challenges for female CIOs in ensuring IT security compliance. *Information Technology & People*, *38*(3), 1305-1327. https://doi.org/10.1108/Itp-05-2023-0524

Bansal, V., & Agarwal, A. (2015). Enterprise resource planning: identifying relationships among critical success factors. *Business Process Management Journal*, *21*(6), 1337-1352. https://doi.org/10.1108/Bpmj-12-2014-0128

Barik, K., Misra, S., Fernandez-Sanz, L., & Koyuncu, M. (2023). RONSI: a framework for calculating return on network security investment. *Telecommunication Systems*, *84*(4), 533-548. https://doi.org/10.1007/s11235-023-01039-9

Barnes, S., Rutter, R. N., La Paz, A., & Scornavacca, E. (2021). Empirical identification of skills gaps between chief information officer supply and demand: a resource-based view using machine learning. *Industrial Management & Data Systems*, *121*(8), 1749-1766. https://doi.org/10.1108/Imds-01-2021-0015

Baskerville, R., Spagnoletti, P., & Kim, J. (2014). Incident-centered information security: Managing a strategic balance between prevention and response. *Information & Management*, *51*(1), 138-151. https://doi.org/10.1016/j.im.2013.11.004

Batyashe, T. N., & Iyamu, T. (2017). Examining IT Governance through Diffusion of Innovations: A Case of a South African Telecom. *Information Resources Management Journal (IRMJ)*, *30*(3), 26-40. https://doi.org/10.4018/IRMJ.2017070102

Beck, R., Avital, M., Rossi, M., & Thatcher, J. B. (2017). Blockchain Technology in Business and Information Systems Research. *Business & Information Systems Engineering*, *59*(6), 381-384. https://doi.org/10.1007/s12599-017-0505-1

Becker, J., Knackstedt, R., & Pöppelbuss, J. (2009). Developing Maturity Models for IT Management - A Procedure Model and its Application. *Business & Information Systems Engineering*, *1*(3), 213-+. https://doi.org/10.1007/s12599-009-0044-5

Bednarek, A. T., Wyborn, C., Cvitanovic, C., Meyer, R., Colvin, R. M., Addison, P. F. E., Close, S. L., Curran, K., Farooque, M., Goldman, E., Hart, D., Mannix, H., McGreavy, B., Parris, A., Posner, S., Robinson, C., Ryan, M., & Leith, P. (2018). Boundary spanning at the science-policy interface: the practitioners' perspectives. *Sustain Sci*, *13*(4), 1175-1183. https://doi.org/10.1007/s11625-018-0550-9

Ben Selma, M., Bouzinab, K., Papadopoulos, A., Chebbi, H., Labouze-Nasica, A., & Desmarteau, R. H. (2024). The effect of the dynamic capabilities' microfoundations on innovation: insights from crossing levels. *EuroMed Journal of Business*, *20*(3), 882-899. https://doi.org/10.1108/emjb-10-2023-0269

Benamati, J. H., Ozdemir, Z. D., & Smith, H. J. (2021). Information Privacy, Cultural Values, and Regulatory Preferences. *Journal of Global Information Management*, *29*(3), 131-164. https://doi.org/10.4018/Jgim.2021050106

Bendig, D., Wagner, R., Jung, C., & Nüesch, S. (2022). When and why technology leadership enters the C-suite: An antecedents perspective on CIO presence. *Journal of Strategic Information Systems*, *31*(1), 101705. https://doi.org/10.1016/j.jsis.2022.101705

Bendler, D., & Felderer, M. (2023). Competency Models for Information Security and Cybersecurity Professionals: Analysis of Existing Work and a New Model. *ACM Transactions on Computing Education*, *23*(2), 1-33. https://doi.org/10.1145/3573205

Benlian, A., & Haffke, I. (2016). Does mutuality matter? Examining the bilateral nature and effects of CEO–CIO mutual understanding. *The Journal of Strategic Information Systems*, *25*(2), 104-126. https://doi.org/10.1016/j.jsis.2016.01.001

Benmira, S., & Agboola, M. (2021). Evolution of leadership theory. *BMJ Leader*, *5*(1), 3. https://doi.org/10.1136/leader-2020-000296

Bentley, M., Stephenson, A., Toscas, P., & Zhu, Z. (2020). A Multivariate Model to Quantify and Mitigate Cybersecurity Risk. *Risks*, *8*(2), 61. https://www.mdpi.com/2227-9091/8/2/61

Benz, M., & Chatterjee, D. (2020). Calculated risk? A cybersecurity evaluation tool for SMEs. *Business Horizons*, *63*(4), 531-540. https://doi.org/10.1016/j.bushor.2020.03.010

Berger, R. (2015). Now I see it, now I don't: researcher's position and reflexivity in qualitative research. *Qualitative Research*, *15*(2), 219-234. https://doi.org/10.1177/1468794112468475

Bethlehem, J. (2010). Selection Bias in Web Surveys. *International Statistical Review*, *78*(2), 161-188. https://doi.org/10.1111/j.1751-5823.2010.00112.x

Bhamare, D., Zolanvari, M., Erbad, A., Jain, R., Khan, K., & Meskin, N. (2020). Cybersecurity for industrial control systems: A survey. *Computers & Security*, *89*, 101677. https://doi.org/10.1016/j.cose.2019.101677

Bhatt, S., Manadhata, P. K., & Zomlot, L. (2014). The Operational Role of Security Information and Event Management Systems. *Ieee Security & Privacy*, *12*(5), 35-41. https://doi.org/10.1109/Msp.2014.103

Biddle, B. J. (1986). Recent Developments in Role-Theory. *Annual Review of Sociology*, *12*, 67-92. https://doi.org/10.1146/annurev.so.12.080186.000435

Birks, M., Hoare, K., & Mills, J. (2019). Grounded Theory: The FAQs. *International Journal of Qualitative Methods*, *18*. https://doi.org/10.1177/1609406919882535

Blažič, B. J. (2022). Changing the landscape of cybersecurity education in the EU: Will the new approach produce the required cybersecurity skills? *Education and Information Technologies*, *27*(3), 3011-3036. https://doi.org/10.1007/s10639-021-10704-y

Bolden, J. E., Peart, M. J., & Johnstone, R. W. (2006). Anticancer activities of histone deacetylase inhibitors. *Nat Rev Drug Discov*, *5*(9), 769-784. https://doi.org/10.1038/nrd2133

Bonell, C., Warren, E., & Melendez-Torres, G. J. (2022). Methodological reflections on using qualitative research to explore the causal mechanisms of complex health interventions. *Evaluation*, *28*(2), 166-181. https://doi.org/10.1177/13563890221086309

Bouncken, R. B., Qiu, Y. X., & García, F. J. S. (2021). Flexible pattern pattern matching approach: Suggestions for augmenting theory evolvement. *Technological Forecasting and Social Change*, *167*, 120685. https://doi.org/10.1016/j.techfore.2021.120685

Bouraffa, T., & Hui, K. L. (2025). Regulating Information and Network Security: Review and Challenges. *Acm Computing Surveys*, *57*(5), Article 126. https://doi.org/10.1145/3711124

Brass, I., & Sowell, J. H. (2021). Adaptive governance for the Internet of Things: Coping with emerging security risks. *Regulation & Governance*, *15*(4), 1092-1110. https://doi.org/10.1111/rego.12343

Braun, V., & Clarke, V. (2019). Reflecting on reflexive thematic analysis. *Qualitative Research in Sport Exercise and Health*, *11*(4), 589-597. https://doi.org/10.1080/2159676x.2019.1628806

Brock, T. (2025). Ontology and interdisciplinary research in esports. *Sport Ethics and Philosophy*, *19*(1), 48-64. https://doi.org/10.1080/17511321.2023.2260567
Brucciani, P. (2023). Embracing an outcome-based approach to cyber security. *Network Security*, *2023*(10). https://doi.org/10.12968/s1353-4858(23)70046-9
Brummer, K., Young, M. D., Özdamar, Ö., Canbolat, S., Thiers, C., Rabini, C., Dimmroth, K., Hansel, M., & Mehvar, A. (2020). Forum: Coding in Tongues: Developing Non-English Coding Schemes for Leadership Profiling. *International Studies Review*, *22*(4), 1039-1067. https://doi.org/10.1093/isr/viaa001
Buis, E. E. G., Ashby, S. S. R., & Kouwenberg, K. K. P. A. (2023). Increasing the UX maturity level of clients: A study of best practices in an agile environment. *Information and Software Technology*, *154*, 107086. https://doi.org/10.1016/j.infsof.2022.107086
Burke, W., Stranieri, A., Oseni, T., & Gondal, I. (2024). The need for cybersecurity self-evaluation in healthcare. *BMC Med Inform Decis Mak*, *24*(1), 133. https://doi.org/10.1186/s12911-024-02551-x
Burns, T., & Stalker, G. M. (1961). *The management of innovation*. Tavistock Publications.
Bushee, B. J., Gow, I. D., & Taylor, D. J. (2017). Linguistic Complexity in Firm Disclosures: Obfuscation or Information? *Journal of Accounting Research*, *56*(1), 85-121. https://doi.org/10.1111/1475-679x.12179
Caglio, A., Dossi, A., & Van der Stede, W. A. (2018). CFO role and CFO compensation: An empirical analysis of their implications. *Journal of Accounting and Public Policy*, *37*(4), 265-281. https://doi.org/10.1016/j.jaccpubpol.2018.07.002
Caldwell, S. D. (2011). Bidirectional Relationships Between Employee Fit and Organizational Change. *Journal of Change Management*, *11*(4), 401-419. https://doi.org/10.1080/14697017.2011.590453
Caldwell, S. D., Herold, D. M., & Fedor, D. B. (2004). Toward an understanding of the relationships among organizational change, individual differences, and changes in person-environment fit: a cross-level study. *J Appl Psychol*, *89*(5), 868-882. https://doi.org/10.1037/0021-9010.89.5.868
Campbell, A. (2021). Spending Political Capital. *Economic Journal*, *131*(640), 3103-3121. https://doi.org/10.1093/ej/ueab040
Campbell, S., Greenwood, M., Prior, S., Shearer, T., Walkem, K., Young, S., Bywaters, D., & Walker, K. (2020). Purposive sampling: complex or simple? Research case examples. *J Res Nurs*, *25*(8), 652-661. https://doi.org/10.1177/1744987120927206
Caplan, R. D. (1987). Person-Environment Fit Theory and Organizations - Commensurate Dimensions, Time Perspectives, and Mechanisms. *Journal of Vocational Behavior*, *31*(3), 248-267. https://doi.org/10.1016/0001-8791(87)90042-X
Cassell, C., Radcliffe, L., & Malik, F. (2020). Participant Reflexivity in Organizational Research Design. *Organizational Research Methods*, *23*(4), 750-773. https://doi.org/10.1177/1094428119842640
Castleberry, A. (2014). NVivo 10 [software program]. Version 10. QSR International; 2012. *American Journal of Pharmaceutical Education*, *78*(1). https://doi.org/10.5688/ajpe78125
Cavusoglu, H., Cavusoglu, H., & Raghunathan, S. (2004). Economics of ITSecurity Management: Four Improvements to Current Security Practices. *Communications of the Association for Information Systems*, *14*, 65-75. https://doi.org/10.17705/1CAIS.01403
Cavusoglu, H., Cavusoglu, H., Son, J.-Y., & Benbasat, I. (2015). Institutional pressures in security management: Direct and indirect influences on organizational investment in information security control resources. *Information & Management*, *52*(4), 385-400. https://doi.org/10.1016/j.im.2014.12.004
Chalutz Ben-Gal, H. (2023). Person–Skill Fit: Why a New Form of Employee Fit Is Required. *Academy of Management Perspectives*. https://doi.org/10.5465/amp.2022-0024

Chan, C. M. L., Teoh, S. Y., Yeow, A., & Pan, G. (2018). Agility in responding to disruptive digital innovation: Case study of an SME. *Information Systems Journal*, *29*(2), 436-455. https://doi.org/10.1111/isj.12215

Chan, R. Y. K. (2021). Do chief information officers matter for sustainable development? Impact of their regulatory focus on green information technology strategies and corporate performance. *Business Strategy and the Environment*, *30*(5), 2523-2534. https://doi.org/10.1002/bse.2761

Chapman, A. L., Hadfield, M., & Chapman, C. J. (2015). Qualitative research in healthcare: an introduction to grounded theory using thematic analysis. *J R Coll Physicians Edinb*, *45*(3), 201-205. https://doi.org/10.4997/JRCPE.2015.305

Charmaz, K. (2014). *Constructing grounded theory* (2nd ed.). Sage.

Charmaz, K., & Thornberg, R. (2021). The pursuit of quality in grounded theory. *Qualitative Research in Psychology*, *18*(3), 305-327. https://doi.org/10.1080/14780887.2020.1780357

Chatman, J. A. (1989). Improving Interactional Organizational Research - a Model of Person-Organization Fit. *Academy of Management Review*, *14*(3), 333-349. https://doi.org/10.5465/Amr.1989.4279063

Chaudhary, S., Gkioulos, V., & Katsikas, S. (2023). A quest for research and knowledge gaps in cybersecurity awareness for small and medium-sized enterprises. *Computer Science Review*, *50*, 100592. https://doi.org/10.1016/j.cosrev.2023.100592

Chawla, R. N., Goyal, P., & Saxena, D. K. (2023). The role of CIO in digital transformation: an exploratory study. *Information Systems and E-Business Management*, *21*(4), 797-835. https://doi.org/10.1007/s10257-023-00651-1

Chayka, R. (2025). Types of fit in the context of organizational psychology: a systematic literature review. *Організаційна психологія Економічна психологія*, *34*(1), 151-160. https://doi.org/10.31108/2.2025.1.34.13

Chen, D. Q., Preston, D. S., & Xia, W. D. (2010). Antecedents and Effects of CIO Supply-Side and Demand-Side Leadership: A Staged Maturity Model. *Journal of Management Information Systems*, *27*(1), 231-272. https://doi.org/10.2753/Mis0742-1222270110

Chen, H., & Li, W. (2018). Understanding commitment and apathy in is security extra-role behavior from a person-organization fit perspective. *Behaviour & Information Technology*, *38*(5), 454-468. https://doi.org/10.1080/0144929x.2018.1539520

Chen, J., Dong, H., Wang, X., Feng, F., Wang, M., & He, X. (2023). Bias and Debias in Recommender System: A Survey and Future Directions. *ACM Transactions on Information Systems*, *41*(3), 1-39. https://doi.org/10.1145/3564284

Chen, Z. Z., Peng, S. P., Li, X. K., Qiu, H. B., Xiong, H. D., Gao, L., & Li, P. G. (2015). An important boundary sampling method for reliability-based design optimization using kriging model. *Structural and Multidisciplinary Optimization*, *52*(1), 55-70. https://doi.org/10.1007/s00158-014-1173-0

Cheng, L., Liu, F., & Yao, D. (2017). Enterprise data breach: causes, challenges, prevention, and future directions. *WIREs Data Mining and Knowledge Discovery*, *7*(5), e1211. https://doi.org/10.1002/widm.1211

Chidukwani, A., Zander, S., & Koutsakis, P. (2022). A Survey on the Cyber Security of Small-to-Medium Businesses: Challenges, Research Focus and Recommendations. *IEEE Access*, *10*, 85701-85719. https://doi.org/10.1109/Access.2022.3197899

Chidukwani, A., Zander, S., & Koutsakis, P. (2024). Cybersecurity preparedness of small-to-medium businesses: A Western Australia study with broader implications. *Computers & Security*, *145*, 104026. https://doi.org/10.1016/j.cose.2024.104026

Child, J. (1972). Organizational Structure, Environment and Performance: The Role of Strategic Choice. *Sociology*, *6*(1), 1-22. https://doi.org/10.1177/003803857200600101

Choi, M. (2016). Leadership of Information Security Manager on the Effectiveness of Information Systems Security for Secure Sustainable Computing. *Sustainability*, *8*(7), 638. https://doi.org/10.3390/su8070638

Choi, S., Martins, J. T., & Bernik, I. (2018). Information security: Listening to the perspective of organisational insiders. *Journal of Information Science*, *44*(6), 752-767. https://doi.org/10.1177/0165551517748288

Choi, Y. B., Capitan, K. E., Krause, J. S., & Streeper, M. M. (2006). Challenges associated with privacy in health care industry: implementation of HIPAA and the security rules. *J Med Syst*, *30*(1), 57-64. https://doi.org/10.1007/s10916-006-7405-0

Chorn, N. H. (1991). The “Alignment” Theory: Creating Strategic Fit. *Management Decision*, *29*(1). https://doi.org/10.1108/eum0000000000066

Chowdhury, N., & Gkioulos, V. (2021a). Cyber security training for critical infrastructure protection: A literature review. *Computer Science Review*, *40*, 100361. https://doi.org/10.1016/j.cosrev.2021.100361

Chowdhury, N., & Gkioulos, V. (2021b). Key competencies for critical infrastructure cyber-security: a systematic literature review. *Information and Computer Security*, *29*(5), 697-723. https://doi.org/10.1108/Ics-07-2020-0121

Christians, C. G. (2018). Ethics and Politics in Qualitative Research. In N. K. Denzin & Y. S. Lincoln (Eds.), *The SAGE handbook of qualitative research* (5th ed. ed., pp. pp.66-82). Sage.

Chrzaszcz, A., Ciekanowski, M., Zurawski, S., Zaloga, W., & Pietrzyk, S. (2024). Managing Organizational Security in the Context of Global Challenges. *European Research Studies Journal*, *XXVII*(Issue 3), 765-777. https://doi.org/10.35808/ersj/3464

Chun, M., & Mooney, J. (2009). CIO roles and responsibilities: Twenty-five years of evolution and change. *Information & Management*, *46*(6), 323-334. https://doi.org/10.1016/j.im.2009.05.005

Clougherty, J. A., Duso, T., & Muck, J. (2016). Correcting for Self-selection Based Endogeneity in Management Research:Review, Recommendations and Simulations. *Organizational Research Methods*, *19*(2), 286-347. https://doi.org/10.1177/1094428115619013

Coghlan, D. (2001). Insider Action Research Projects: Implications for Practising Managers. *Management Learning*, *32*(1), 49-60. https://doi.org/10.1177/1350507601321004

Colicchia, C., Creazza, A., & Menachof, D. A. (2018). Managing cyber and information risks in supply chains: insights from an exploratory analysis. *Supply Chain Management: An International Journal*, *24*(2), 215-240. https://doi.org/10.1108/scm-09-2017-0289

Collier, Z. A., Briglia, B., Finkelston, T., Manasco, M. C., Slutzky, D. L., & Lambert, J. H. (2023). On metrics and prioritization of investments in hardware security. *Systems Engineering*, *26*(4), 425-437. https://doi.org/10.1002/sys.21667

Collins, M. D., Dasborough, M. T., Gregg, H. R., Xu, C., Midel Deen, C., He, Y., & Restubog, S. L. D. (2023). Traversing the storm: An interdisciplinary review of crisis leadership. *The Leadership Quarterly*, *34*(1), 101661. https://doi.org/10.1016/j.leaqua.2022.101661

Compastié, M., López Martínez, A., Fernández, C., Gil Pérez, M., Tsarsitalidis, S., Xylouris, G., Mlakar, I., Kourtis, M. A., & Šafran, V. (2023). PALANTIR: An NFV-Based Security-as-a-Service Approach for Automating Threat Mitigation. *Sensors*, *23*(3), 1658. https://www.mdpi.com/1424-8220/23/3/1658

Conlon, C., Timonen, V., Elliott-O'Dare, C., O'Keeffe, S., & Foley, G. (2020). Confused About Theoretical Sampling? Engaging Theoretical Sampling in Diverse Grounded Theory Studies. *Qual Health Res*, *30*(6), 947-959. https://doi.org/10.1177/1049732319899139

Cook, K. (2017). *Social Capital: Theory and Research*. Taylor and Francis. https://doi.org/10.4324/9781315129457

Corallo, A., Lazoi, M., Lezzi, M., & Pontrandolfo, P. (2023). Cybersecurity Challenges for Manufacturing Systems 4.0: Assessment of the Business Impact Level. *IEEE*

*Transactions on Engineering Management*, *70*(11), 3745-3765. https://doi.org/10.1109/tem.2021.3084687
Corbin, J., & Strauss, A. (2015). *Basics of qualitative research: Techniques and procedures for developing grounded theory* (4th ed ed.). Sage.
Corradi, A., Di Modica, G., Foschini, L., Patera, L., & Solimando, M. (2022). SIRDAM4.0: A Support Infrastructure for Reliable Data Acquisition and Management in Industry 4.0. *Ieee Transactions on Emerging Topics in Computing*, *10*(3), 1605-1620. https://doi.org/10.1109/Tetc.2021.3111974
Council on Foreign Relations. (2024). *Cyber operations tracker*. Retrieved August 2025 from https://www.cfr.org/cyber-operations/
Coventry, L., & Branley, D. (2018). Cybersecurity in healthcare: A narrative review of trends, threats and ways forward. *Maturitas*, *113*, 48-52. https://doi.org/10.1016/j.maturitas.2018.04.008
Crabtree, A. (2025). H is for human and how (not) to evaluate qualitative research in HCI. *Human–Computer Interaction*, *2409*, 1-24. https://doi.org/10.1080/07370024.2025.2475743
Cramp, A., & Khan, S. (2019). The convivial space - exploring teacher learning through practitioner research. *Professional Development in Education*, *45*(3), 344-355. https://doi.org/10.1080/19415257.2018.1431957
Crawford, J. A., & Kelder, J.-A. (2019). Do we measure leadership effectively? Articulating and evaluating scale development psychometrics for best practice. *The Leadership Quarterly*, *30*(1), 133-144. https://doi.org/10.1016/j.leaqua.2018.07.001
Creazza, A., Colicchia, C., Spiezia, S., & Dallari, F. (2021). Who cares? Supply chain managers' perceptions regarding cyber supply chain risk management in the digital transformation era. *Supply Chain Management: An International Journal*, *27*(1), 30-53. https://doi.org/10.1108/scm-02-2020-0073
CrowdStrike. (2024). *2024 Global threat report*. https://www.crowdstrike.com/resources/reports/crowdstrike-2024-global-threat-report/
Culot, G., Nassimbeni, G., Podrecca, M., & Sartor, M. (2021). The ISO/IEC 27001 information security management standard: literature review and theory-based research agenda. *Tqm Journal*, *33*(7), 76-105. https://doi.org/10.1108/Tqm-09-2020-0202
Cumps, B., Viaene, S., & Dedene, G. (2006). Managing for Better Business-IT Alignment. *IT Professional*, *8*(5), 17-24. https://doi.org/10.1109/mitp.2006.115
Cutcliffe, J. R. (2000). Methodological issues in grounded theory. *J Adv Nurs*, *31*(6), 1476-1484. https://doi.org/10.1046/j.1365-2648.2000.01430.x
Da Silva, J., & Jensen, R. (2022). "Cyber security is a dark art": The CISO as Soothsayer. *Proceedings of the ACM on Human-Computer Interaction*, *6*(CSCW2), 1-31. https://doi.org/10.1145/3555090
da Veiga, A., Astakhova, L., Botha, A., & Herselman, M. (2020). Defining organisational information security culture—Perspectives from academia and industry. *Computers & Security*, *92*, 101713-undefined. https://doi.org/10.1016/j.cose.2020.101713
Dadds, M. (2008). Empathetic validity in practitioner research. *Educational Action Research*, *16*(2), 279-290. https://doi.org/10.1080/09650790802011973
Dahlin, E. (2021). Email Interviews: A Guide to Research Design and Implementation. *International Journal of Qualitative Methods*, *20*. https://doi.org/10.1177/16094069211025453
Dalal, R. S., Howard, D. J., Bennett, R. J., Posey, C., Zaccaro, S. J., & Brummel, B. J. (2022). Organizational science and cybersecurity: abundant opportunities for research at the interface. *J Bus Psychol*, *37*(1), 1-29. https://doi.org/10.1007/s10869-021-09732-9

Dalkin, S., Forster, N., Hodgson, P., Lhussier, M., & Carr, S. M. (2021). Using computer assisted qualitative data analysis software (CAQDAS; NVivo) to assist in the complex process of realist theory generation, refinement and testing. *International Journal of Social Research Methodology*, *24*(1), 123-134. https://doi.org/10.1080/13645579.2020.1803528

Dane, E. (2010). Reconsidering the Trade-Off between Expertise and Flexibility: A Cognitive Entrenchment Perspective. *Academy of Management Review*, *35*(4), 579-603. https://doi.org/10.5465/Amr.2010.53502832

Danese, P., Molinaro, M., & Romano, P. (2020). Investigating fit in supply chain integration: A systematic literature review on context, practices, performance links. *Journal of Purchasing and Supply Management*, *26*(5), 100634. https://doi.org/10.1016/j.pursup.2020.100634

Dang-Pham, D., Kautz, K., Hoang, A.-P., & Pittayachawan, S. (2022). Identifying information security opinion leaders in organizations: Insights from the theory of social power bases and social network analysis. *Computers & Security*, *112*, 102505. https://doi.org/10.1016/j.cose.2021.102505

Darem, A. A., Alhashmi, A. A., Alkhaldi, T. M., Alashjaee, A. M., Alanazi, S. M., & Ebad, S. A. (2023). Cyber Threats Classifications and Countermeasures in Banking and Financial Sector. *IEEE Access*, *11*, 125138-125158. https://doi.org/10.1109/Access.2023.3327016

Davies, H. T., & Crombie, I. K. (1995). Assessing the quality of care. *BMJ*, *311*(7008), 766. https://doi.org/10.1136/bmj.311.7008.766

Davis, J. M., Kettinger, W. J., & Kunev, D. G. (2009). When users are IT experts too: the effects of joint IT competence and partnership on satisfaction with enterprise-level systems implementation. *European Journal of Information Systems*, *18*(1), 26-37. https://doi.org/10.1057/ejis.2009.4

Dawson, J., & Thomson, R. (2018). The Future Cybersecurity Workforce: Going Beyond Technical Skills for Successful Cyber Performance [Review]. *Front Psychol*, *9*, 744. https://doi.org/10.3389/fpsyg.2018.00744

Day, D. V. (2011). Integrative perspectives on longitudinal investigations of leader development: From childhood through adulthood. *Leadership Quarterly*, *22*(3), 561-571. https://doi.org/10.1016/j.leaqua.2011.04.012

De Cássia De Faria, P., R, Da, S., M, M., Lapão, & L, V. (2021). Business/IT alignment through IT governance patterns in Portuguese healthcare. *International Journal of IT/Business Alignment and Governance*, *5*(1), 1-22.

De Haes, S., Van Grembergen, W., & Debreceny, R. S. (2013). COBIT 5 and Enterprise Governance of Information Technology: Building Blocks and Research Opportunities. *Journal of Information Systems*, *27*(1), 307-324. https://doi.org/10.2308/isys-50422

Dedeke, A. (2017). Cybersecurity Framework Adoption: Using Capability Levels for Implementation Tiers and Profiles. *Ieee Security & Privacy*, *15*(5), 47-54. https://doi.org/10.1109/Msp.2017.3681063

Demertzis, K., Tziritas, N., Kikiras, P., Sanchez, S. L., & Iliadis, L. (2019). The Next Generation Cognitive Security Operations Center: Adaptive Analytic Lambda Architecture for Efficient Defense against Adversarial Attacks. *Big Data and Cognitive Computing*, *3*(1), 6. https://doi.org/10.3390/bdcc3010006

Den Hartog, D. N., & De Hoogh, A. H. B. (2024). Cross-Cultural Leadership: What We Know, What We Need to Know, and Where We Need to Go. *Annual Review of Organizational Psychology and Organizational Behavior*, *11*(1), 535-566. https://doi.org/10.1146/annurev-orgpsych-110721-033711

Deng, C., Gulseren, D., Isola, C., Grocutt, K., & Turner, N. (2023). Transformational leadership effectiveness: an evidence-based primer. *Human Resource Development International*, *26*(5), 627-641. https://doi.org/10.1080/13678868.2022.2135938

Denner, J., Bean, S., Campe, S., Martinez, J., & Torres, D. (2019). Negotiating Trust, Power, and Culture in a Research-Practice Partnership. *Aera Open*, *5*(2). https://doi.org/10.1177/2332858419858635
Denzin, N. K., & Lincoln, Y. S. (Eds.). (1994). *Handbook of qualitative research*. Sage Publications.
Dhillon, G., Smith, K., & Dissanayaka, I. (2021). Information systems security research agenda: Exploring the gap between research and practice. *Journal of Strategic Information Systems*, *30*(4), 101693. https://doi.org/10.1016/j.jsis.2021.101693
Dhillon, G., & Torkzadeh, G. (2006). Value-focused assessment of information system security in organizations. *Information Systems Journal*, *16*(3), 293-314. https://doi.org/10.1111/j.1365-2575.2006.00219.x
Dhirani, L. L., Armstrong, E., & Newe, T. (2021). Industrial IoT, Cyber Threats, and Standards Landscape: Evaluation and Roadmap. *Sensors*, *21*(11), 3901. https://www.mdpi.com/1424-8220/21/11/3901
Diamantopoulou, V., Tsohou, A., & Karyda, M. (2020). From ISO/IEC 27002:2013 Information Security Controls to Personal Data Protection Controls: Guidelines for GDPR Compliance. *Computer Security, Esorics 2019*, *11980*, 238-257. https://doi.org/10.1007/978-3-030-42048-2_16
Dickson, M. W., Den Hartog, D. N., & Mitchelson, J. K. (2003). Research on leadership in a cross-cultural context: Making progress, and raising new questions. *Leadership Quarterly*, *14*(6), 729-768. https://doi.org/10.1016/j.leaqua.2003.09.002
Diesch, R., Pfaff, M., & Krcmar, H. (2020). A comprehensive model of information security factors for decision-makers. *Computers & Security*, *92*, 101747. https://doi.org/10.1016/j.cose.2020.101747
Dimitrov, D. (2025). The Institutional Role of Charity Regulators in Bringing Charities to Account: An International Comparative Study of Charity Regulators in New Zealand, Australia, Canada and England and Wales. *Nonprofit Policy Forum*, *0*. https://doi.org/10.1515/npf-2024-0071
Dirani, K. M., Abadi, M., Alizadeh, A., Barhate, B., Garza, R. C., Gunasekara, N., Ibrahim, G., & Majzun, Z. (2020). Leadership competencies and the essential role of human resource development in times of crisis: a response to Covid-19 pandemic. *Human Resource Development International*, *23*(4), 380-394. https://doi.org/10.1080/13678868.2020.1780078
Dobson, R. (2003). Industry sponsored studies twice as likely to have positive conclusions about costs. *BMJ*, *327*(7422), 1006. https://doi.org/10.1136/bmj.327.7422.1006-b
Dodgson, J. E. (2019). Reflexivity in Qualitative Research. *J Hum Lact*, *35*(2), 220-222. https://doi.org/10.1177/0890334419830990
Donalds, C., & Barclay, C. (2022). Beyond technical measures: a value-focused thinking appraisal of strategic drivers in improving information security policy compliance. *European Journal of Information Systems*, *31*(1), 58-73. https://doi.org/10.1080/0960085x.2021.1978344
Draucker, C. B., Martsolf, D. S., Ross, R., & Rusk, T. B. (2007). Theoretical sampling and category development in grounded theory. *Qual Health Res*, *17*(8), 1137-1148. https://doi.org/10.1177/1049732307308450
Dritsas, E., & Trigka, M. (2025). A Survey on Cybersecurity in IoT. *Future Internet*, *17*(1), 30. https://doi.org/10.3390/fi17010030
Ducas, E., & Wilner, A. (2017). The security and financial implications of blockchain technologies: Regulating emerging technologies in Canada. *International Journal*, *72*(4), 538-562. https://doi.org/10.1177/0020702017741909
Dupont, B. (2019). The cyber-resilience of financial institutions: significance and applicability. *Journal of Cybersecurity*, *5*(1). https://doi.org/10.1093/cybsec/tyz013

Dutta, A., & McCrohan, K. (2002). Management's role in information security in a cyber economy. *California Management Review*, *45*(1), 67-+. https://doi.org/10.2307/41166154

Dwyer, S. C., & Buckle, J. L. (2009). The Space Between: On Being an Insider-Outsider in Qualitative Research. *International Journal of Qualitative Methods*, *8*(1), 54-63. https://doi.org/10.1177/160940690900800105

Easton, G. (2010). Critical realism in case study research. *Industrial Marketing Management*, *39*(1), 118-128. https://doi.org/10.1016/j.indmarman.2008.06.004

Ebel, A., & Mitra, D. (2024). Economics and Optimal Investment Policies of Attackers and Defenders in Cybersecurity. In *Journal of Cybersecurity* (Vol. 10, pp. 1-14).

Edwards, J. A., & Billsberry, J. (2020). Testing a Multidimensional Theory of Person-Environment Fit. *Journal of Managerial Issues*, *32*(1), 8-25.

Edwards, J. R. (2008). Person–Environment Fit in Organizations: An Assessment of Theoretical Progress. *Academy of Management Annals*, *2*, 167-230. https://doi.org/10.1080/19416520802211503

Efthymiopoulos, M. P. (2019). A cyber-security framework for development, defense and innovation at NATO. *Journal of Innovation and Entrepreneurship*, *8*(1), 12. https://doi.org/10.1186/s13731-019-0105-z

Einola, K., & Alvesson, M. (2021). Behind the Numbers: Questioning Questionnaires. *Journal of Management Inquiry*, *30*(1), 102-114. https://doi.org/10.1177/1056492620938139

Eisenhardt, K. M. (1991). Better Stories and Better Constructs - the Case for Rigor and Comparative Logic. *Academy of Management Review*, *16*(3), 620-627. https://doi.org/10.2307/258921

Eisenhardt, K. M. (2021). What is the Eisenhardt Method, really? *Strategic Organization*, *19*(1), 147-160. https://doi.org/10.1177/1476127020982866

Eisenhardt, K. M., & Graebner, M. E. (2007). Theory building from cases: Opportunities and challenges. *Academy of Management Journal*, *50*(1), 25-32. https://doi.org/10.5465/Amj.2007.24160888

Eisenhardt, K. M., & Martin, J. A. (2000). Dynamic capabilities: what are they? *Strategic Management Journal*, *21*(10-11), 1105-1121. https://doi.org/10.1002/1097-0266(200010/11)21:10/11<1105::Aid-smj133>3.0.Co;2-e

Elbanna, S. (2016). Managers' autonomy, strategic control, organizational politics and strategic planning effectiveness: An empirical investigation into missing links in the hotel sector. *Tourism Management*, *52*, 210-220. https://doi.org/10.1016/j.tourman.2015.06.025

Eling, M., Elvedi, M., & Falco, G. (2023). The Economic Impact of Extreme Cyber Risk Scenarios. *North American Actuarial Journal*, *27*(3), 429-443. https://doi.org/10.1080/10920277.2022.2034507

Ellen, B. P., Ferris, G. R., & Buckley, M. R. (2013). Leader political support: Reconsidering leader political behavior. *Leadership Quarterly*, *24*(6), 842-857. https://doi.org/10.1016/j.leaqua.2013.10.007

Ellis, S., Almor, T., & Shenkar, O. (2002). Structural Contingency Revisited: Toward a Dynamic System Model. *Emergence*, *4*(4), 51-85. https://doi.org/10.1207/s15327000em0404_6

Esmaeili, R., Yazdi, M., Rismanchian, M., & Shakerian, M. (2025). Unveiling the dynamics of team cognition in emergency response teams. *Front Psychol*, *16*, 1534224. https://doi.org/10.3389/fpsyg.2025.1534224

Etzel, J. M., & Nagy, G. (2021). Challenging the Multidimensional Conception of Perceived Person-Environment Fit Are Specific Fit Dimensions Related to Educational Outcomes Beyond a Higher-Order Factor? *European Journal of Psychological Assessment*, *37*(5), 368-376. https://doi.org/10.1027/1015-5759/a000622

European Union Agency for Cybersecurity. (2021). *Threat landscape for supply chain attacks*. https://www.enisa.europa.eu/publications/threat-landscape-for-supply-chain-attacks

European Union Agency for Cybersecurity. (2023). *ENISA threat landscape 2023*. https://www.enisa.europa.eu/publications/enisa-threat-landscape-2023

Europol. (2023). *Internet organised crime threat assessment (IOCTA) 2023*. P. O. o. t. E. Union. https://www.europol.europa.eu/publication-events/main-reports/internet-organised-crime-threat-assessment-iocta-2023

Fabbri, A. (2020). Using research sponsorship to skew the evidence base towards policies and interventions that favour industry. *European Journal of Public Health*, *30*. https://doi.org/10.1093/eurpub/ckaa165.504

Falleti, T. G., & Lynch, J. F. (2009). Context and Causal Mechanisms in Political Analysis. *Comparative Political Studies*, *42*(9), 1143-1166. https://doi.org/10.1177/0010414009331724

Falowo, O. I., Ozer, M., Li, C. C., & Abdo, J. B. (2024). Evolving Malware and DDoS Attacks: Decadal Longitudinal Study. *IEEE Access*, *12*, 39221-39237. https://doi.org/10.1109/Access.2024.3376682

Farrow, S., & von Winterfeldt, D. (2020). Retrospective Benefit-Cost Analysis of Security-Enhancing and Cost-Saving Technologies. *Journal of Benefit-Cost Analysis*, *11*(3), 479-500. https://doi.org/10.1017/bca.2020.24

Fedele, A., & Roner, C. (2022). Dangerous games: A literature review on cybersecurity investments. *Journal of Economic Surveys*, *36*(1), 157-187. https://doi.org/10.1111/joes.12456

Felser, M., Rentschler, M., & Kleineberg, O. (2019). Coexistence Standardization of Operation Technology and Information Technology. *Proceedings of the Ieee*, *107*(6), 962-976. https://doi.org/10.1109/Jproc.2019.2901314

Ferdous, J., Islam, R., Mahboubi, A., & Islam, M. Z. (2023). A Review of State-of-the-Art Malware Attack Trends and Defense Mechanisms. *IEEE Access*, *11*, 121118-121141. https://doi.org/10.1109/Access.2023.3328351

Fernandez De Arroyabe, I., Arranz, C. F. A., Arroyabe, M. F., & Fernandez de Arroyabe, J. C. (2023). Cybersecurity capabilities and cyber-attacks as drivers of investment in cybersecurity systems: A UK survey for 2018 and 2019. *Computers & Security*, *124*, 102954. https://doi.org/10.1016/j.cose.2022.102954

Fernandez De Arroyabe, I., & Fernandez de Arroyabe, J. C. (2023). The severity and effects of Cyber-breaches in SMEs: a machine learning approach. *Enterprise Information Systems*, *17*(3), 1942997. https://doi.org/10.1080/17517575.2021.1942997

Ferreira, C., Park, A., Kietzmann, J., Demetis, D., Flostrand, A., Mccarthy, I., Pitt, L., & Dabirian, A. (2024). Cybercrime: Understanding the Current State of Literature and Issues Facing CISOs. *It Professional*, *26*(2), 83-89. https://doi.org/10.1109/Mitp.2024.3375571

Ferris, G. R., Ellen, B. P., McAllister, C. P., & Maher, L. P. (2019). Reorganizing Organizational Politics Research: A Review of the Literature and Identification of Future Research Directions. *Annual Review of Organizational Psychology and Organizational Behavior, Vol 6*, *6*, 299-323. https://doi.org/10.1146/annurev-orgpsych-012218-015221

Ferris, G. R., Perrewé, P. L., Ranft, A. L., Zinko, R., Stoner, J. S., Brouer, R. L., & Laird, M. D. (2007). Human resources reputation and effectiveness. *Human Resource Management Review*, *17*(2), 117-130. https://doi.org/10.1016/j.hrmr.2007.03.003

Fiedler, F. E. (1964). A Contingency-Model of Leadership Effectiveness. *Advances in Experimental Social Psychology*, *1*(1), 149-190. https://doi.org/10.1016/S0065-2601(08)60051-9

Fiedler, F. E. (1967). *A theory of leadership effectiveness*. McGraw-Hill.

Fielder, A., Panaousis, E., Malacaria, P., Hankin, C., & Smeraldi, F. (2016). Decision support approaches for cyber security investment. *Decision Support Systems*, *86*, 13-23. https://doi.org/10.1016/j.dss.2016.02.012

Firk, S., Hanelt, A., Oehmichen, J., & Wolff, M. (2021). Chief Digital Officers: An Analysis of the Presence of a Centralized Digital Transformation Role. *Journal of Management Studies*, *58*(7), 1800-1831. https://doi.org/10.1111/joms.12718

Fisher, C. D., & To, M. L. (2012). Using experience sampling methodology in organizational behavior. *Journal of Organizational Behavior*, *33*(7), 865-877. https://doi.org/10.1002/job.1803

Fleetwood, S. (2005). Ontology in organization and management studies: A critical realist perspective. *Organization*, *12*(2), 197-222. https://doi.org/10.1177/1350508405051188

Flemming, K., & Noyes, J. (2021). Qualitative Evidence Synthesis: Where Are We at? *International Journal of Qualitative Methods*, *20*. https://doi.org/10.1177/1609406921993276

Fletcher, G., & Griffiths, M. (2020). Digital transformation during a lockdown. *International Journal of Information Management*, *55*, 102185. https://doi.org/10.1016/j.ijinfomgt.2020.102185

Flowerday, S. V., & Tuyikeze, T. (2016). Information security policy development and implementation: The what, how and who. *Computers & Security*, *61*, 169-183. https://doi.org/10.1016/j.cose.2016.06.002

Foley, G., Timonen, V., Conlon, C., & O'Dare, C. E. (2021). Interviewing as a Vehicle for Theoretical Sampling in Grounded Theory. *International Journal of Qualitative Methods*, *20*, 1-12. https://doi.org/10.1177/1609406920980957

Folger, R., & Stein, C. (2017). Abduction 101: Reasoning processes to aid discovery. *Human Resource Management Review*, *27*(2), 306-315. https://doi.org/10.1016/j.hrmr.2016.08.007

Forster, C., Paparella, C., Duchek, S., & Guttel, W. H. (2022). Leading in the Paradoxical World of Crises: How Leaders Navigate Through Crises. *Schmalenbach Z Betriebswirtsch Forsch*, *74*(4), 631-657. https://doi.org/10.1007/s41471-022-00147-7

Fortier, I., Raina, P., Van den Heuvel, E. R., Griffith, L. E., Craig, C., Saliba, M., Doiron, D., Stolk, R. P., Knoppers, B. M., Ferretti, V., Granda, P., & Burton, P. (2017). Maelstrom Research guidelines for rigorous retrospective data harmonization. *Int J Epidemiol*, *46*(1), 103-105. https://doi.org/10.1093/ije/dyw075

Franco, M. F., Künzler, F., von der Assen, J., Feng, C., & Stiller, B. (2024). RCVaR: An economic approach to estimate cyberattacks costs using data from industry reports. *Computers & Security*, *139*, 103737. https://doi.org/10.1016/j.cose.2024.103737

Friesl, M., Stensaker, I., & Colman, H. L. (2021). Strategy implementation: Taking stock and moving forward. *Long Range Planning*, *54*(4), 102064. https://doi.org/10.1016/j.lrp.2020.102064

Furnas, A. C., & Lapira, T. M. (2024). The people think what I think: False consensus and unelected elite misperception of public opinion. *American Journal of Political Science*, *68*(3), 958-971. https://doi.org/10.1111/ajps.12833

Furnell, S. (2021). The cybersecurity workforce and skills. *Computers & Security*, *100*, 102080. https://doi.org/10.1016/j.cose.2020.102080

Gambrill, E. (2011). Evidence-based practice and the ethics of discretion. *Journal of Social Work*, *11*(1), 26-48. https://doi.org/10.1177/1468017310381306

Ganin, A. A., Quach, P., Panwar, M., Collier, Z. A., Keisler, J. M., Marchese, D., & Linkov, I. (2020). Multicriteria Decision Framework for Cybersecurity Risk Assessment and Management. *Risk Anal*, *40*(1), 183-199. https://doi.org/10.1111/risa.12891

Gao, X., Qiu, M., Wang, Y., & Wang, X. (2023). Information security investment with budget constraint and security information sharing in resource-sharing environments. *Journal of the Operational Research Society*, *74*(6), 1520-1535. https://doi.org/10.1080/01605682.2022.2096506

Garcia-Perez, A., Cegarra-Navarro, J. G., Sallos, M. P., Martinez-Caro, E., & Chinnaswamy, A. (2023). Resilience in healthcare systems: Cyber security and digital transformation. *Technovation*, *121*. https://doi.org/10.1016/j.technovation.2022.102583

Garnaev, A., Baykal-Gursoy, M., & Poor, H. V. (2016). Security Games With Unknown Adversarial Strategies. *IEEE Trans Cybern*, *46*(10), 2291-2299. https://doi.org/10.1109/TCYB.2015.2475243

Garretsen, H., Stoker, J. I., Soudis, D., & Wendt, H. (2022). The pandemic that shocked managers across the world: the impact of the COVID-19 crisis on leadership behavior. *Leadersh Q*, *33*(6), 101630. https://doi.org/10.1016/j.leaqua.2022.101630

Gartner. (2025). *Predicts 2025: Cybersecurity strategy and planning*. https://www.gartner.com/en/newsroom/press-releases/2025-03-03-gartner-identifiesthe-top-cybersecurity-trends-for-2025

Garzella, S., Fiorentino, R., Caputo, A., & Lardo, A. (2021). Business model innovation in SMEs: the role of boundaries in the digital era. *Technology Analysis & Strategic Management*, *33*(1), 31-43. https://doi.org/10.1080/09537325.2020.1787374

Gehman, J., Glaser, V. L., Eisenhardt, K. M., Gioia, D., Langley, A., & Corley, K. G. (2018). Finding Theory–Method Fit: A Comparison of Three Qualitative Approaches to Theory Building. *Journal of Management Inquiry*, *27*(3), 284-300. https://doi.org/10.1177/1056492617706029

Georgiadou, A., Mouzakitis, S., & Askounis, D. (2021). Assessing MITRE ATT&CK Risk Using a Cyber-Security Culture Framework. *Sensors (Basel)*, *21*(9), 3267. https://doi.org/10.3390/s21093267

Germann, F., Ebbes, P., & Grewal, R. (2015). The Chief Marketing Officer Matters! *Journal of Marketing*, *79*(3), 1-22. https://doi.org/10.1509/jm.14.0244

Gerring, J. (2007). Is There a (Viable) Crucial-Case Method? *Comparative Political Studies*, *40*(3), 231-253. https://doi.org/10.1177/0010414006290784

Gerstner, C. R., & Day, D. V. (1994). Cross-Cultural-Comparison of Leadership Prototypes. *Leadership Quarterly*, *5*(2), 121-134. https://doi.org/10.1016/1048-9843(94)90024-8

Gerth, A. B., & Peppard, J. (2016). The dynamics of CIO derailment: How CIOs come undone and how to avoid it. *Business Horizons*, *59*(1), 61-70. https://doi.org/10.1016/j.bushor.2015.09.001

Ghadge, A., Weiß, M., Caldwell, N. D., & Wilding, R. (2020). Managing cyber risk in supply chains: a review and research agenda. *Supply Chain Management: An International Journal*, *25*(2), 223-240. https://doi.org/10.1108/SCM-10-2018-0357

Ghafir, I., Saleem, J., Hammoudeh, M., Faour, H., Prenosil, V., Jaf, S., Jabbar, S., & Baker, T. (2018). Security threats to critical infrastructure: the human factor. *The Journal of Supercomputing*, *74*(10), 4986-5002. https://doi.org/10.1007/s11227-018-2337-2

Gioia, D. A., Corley, K. G., & Hamilton, A. L. (2013). Seeking Qualitative Rigor in Inductive Research: Notes on the Gioia Methodology. *Organizational Research Methods*, *16*(1), 15-31. https://doi.org/10.1177/1094428112452151

Glaser, B. G., & Strauss, A. L. (1967). *The discovery of grounded theory: strategies for qualitative research*. Aldine Pub. Co.

Goel, R., Kumar, A., & Haddow, J. (2020). PRISM: a strategic decision framework for cybersecurity risk assessment. *Information and Computer Security*, *28*(4), 591-625. https://doi.org/10.1108/Ics-11-2018-0131

Gordon, L. A., & Loeb, M. P. (2006). *Managing cybersecurity resources: a cost-benefit analysis*. McGraw-Hill.

Gordon, L. A., Loeb, M. P., & Zhou, L. (2020). Integrating cost-benefit analysis into the NIST Cybersecurity Framework via the Gordon-Loeb Model. *Journal of Cybersecurity*, *6*(1). https://doi.org/10.1093/cybsec/tyaa005

Gozman, D., & Currie, W. (2014). The Role of Investment Management Systems in Regulatory Compliance: A Post-Financial Crisis Study of Displacement Mechanisms. *Journal of Information Technology*, *29*(1), 44-58. https://doi.org/10.1057/jit.2013.16

Graafland, J., & Smid, H. (2017). Reconsidering the relevance of social license pressure and government regulation for environmental performance of European SMEs. *Journal of Cleaner Production*, *141*, 967-977. https://doi.org/10.1016/j.jclepro.2016.09.171

Graham, C. M., & Lu, Y. (2023). Skills Expectations in Cybersecurity: Semantic Network Analysis of Job Advertisements. *Journal of Computer Information Systems*, *63*(4), 937-949. https://doi.org/10.1080/08874417.2022.2115954

Granata, D., Mastroianni, M., Rak, M., Cantiello, P., & Salzillo, G. (2024). GDPR compliance through standard security controls: An automated approach. *Journal of High Speed Networks*, *30*(2), 147-174. https://doi.org/10.3233/Jhs-230080

Grigaliūnas, Š., Schmidt, M., Brūzgienė, R., Smyrli, P., & Bidikov, V. (2023). Leveraging Taxonomical Engineering for Security Baseline Compliance in International Regulatory Frameworks. *Future Internet*, *15*(10), 330. https://www.mdpi.com/1999-5903/15/10/330

Guan, Z. (2024). Difficulties and Solutions for Industrial Data Security and Compliance Governance. In M. Luo & L.-J. Zhang (Eds.), *Cloud Computing – CLOUD 2023* (Vol. 14204, pp. 66-75). Springer Nature Switzerland. https://doi.org/10.1007/978-3-031-51709-9_6

Guest, G., Bunce, A., & Johnson, L. (2006). How Many Interviews Are Enough? *Field Methods*, *18*(1), 59-82. https://doi.org/10.1177/1525822x05279903

Guest, G., Namey, E., & Chen, M. (2020). A simple method to assess and report thematic saturation in qualitative research. *PLoS One*, *15*(5), e0232076. https://doi.org/10.1371/journal.pone.0232076

Guhr, N., Lebek, B., & Breitner, M. H. (2019). The impact of leadership on employees' intended information security behaviour: An examination of the full-range leadership theory. *Information Systems Journal*, *29*(2), 340-362. https://doi.org/10.1111/isj.12202

Gupta, M., Shoja, A., & Mikalef, P. (2022). Toward the understanding of national culture in the success of non-pharmaceutical technological interventions in mitigating COVID-19 pandemic. *Ann Oper Res*, *319*(1), 1433-1450. https://doi.org/10.1007/s10479-021-03962-z

Haapamäki, E., & Sihvonen, J. (2019). Cybersecurity in accounting research. In *Managerial Auditing Journal* (Vol. 34, pp. 808-834): Emerald Publishing Limited.

Haislip, J., Lim, J. H., & Pinsker, R. (2021). The Impact of Executives' IT Expertise on Reported Data Security Breaches. *Information Systems Research*, *32*(2), 318-334. https://doi.org/10.1287/isre.2020.0986

Hambrick, D. C., & Mason, P. A. (1984). Upper Echelons - the Organization as a Reflection of Its Top Managers. *Academy of Management Review*, *9*(2), 193-206. https://doi.org/10.2307/258434

Hammi, B., Zeadally, S., & Nebhen, J. (2023). Security threats, countermeasures, and challenges of digital supply chains. *ACM Computing Surveys*, *55*(14s), 1-40. https://doi.org/10.1145/3588999

Handoyo, S. (2024). Purchasing in the digital age: A meta-analytical perspective on trust, risk, security, and e-WOM in e-commerce. *Heliyon*, *10*(8). https://doi.org/10.1016/j.heliyon.2024.e29714

Hannah, S. T., Uhl-Bien, M., Avolio, B. J., & Cavarretta, F. L. (2009). A framework for examining leadership in extreme contexts. *Leadership Quarterly*, *20*(6), 897-919. https://doi.org/10.1016/j.leaqua.2009.09.006

Hannousse, A., & Yahiouche, S. (2021). Securing microservices and microservice architectures: A systematic mapping study. *Computer Science Review*, *41*, 100415. https://doi.org/10.1016/j.cosrev.2021.100415

Haqaf, H., & Koyuncu, M. (2018). Understanding key skills for information security managers. *International Journal of Information Management*, *43*, 165-172. https://doi.org/10.1016/j.ijinfomgt.2018.07.013

Harvey, W. S. (2011). Strategies for conducting elite interviews. *Qualitative Research*, *11*(4), 431-441. https://doi.org/10.1177/1468794111404329

Harvey, W. S., Mitchell, V.-W., Almeida Jones, A., & Knight, E. (2021). The tensions of defining and developing thought leadership within knowledge-intensive firms. *Journal of Knowledge Management*, *25*(11), 1-33. https://doi.org/10.1108/jkm-06-2020-0431

Hasan, S., Ali, M., Kurnia, S., & Thurasamy, R. (2021). Evaluating the cyber security readiness of organizations and its influence on performance. *Journal of Information Security and Applications*, *58*, 102726. https://doi.org/10.1016/j.jisa.2020.102726

Hassandoust, F., & Johnston, A. C. (2023). Peering through the lens of high-reliability theory: A competencies driven security culture model of high-reliability organisations. *Information Systems Journal*, *33*(5), 1212-1238. https://doi.org/10.1111/isj.12441

Hausken, K. (2006). Returns to information security investment: The effect of alternative information security breach functions on optimal investment and sensitivity to vulnerability. *Information Systems Frontiers*, *8*(5), 338-349. https://doi.org/10.1007/s10796-006-9011-6

He, Y., Aliyu, A., Evans, M., & Luo, C. (2021). Health Care Cybersecurity Challenges and Solutions Under the Climate of COVID-19: Scoping Review. *J Med Internet Res*, *23*(4), e21747. https://doi.org/10.2196/21747

He, Z. X., Shen, W. X., Li, Q. B., Xu, S. C., Zhao, B., Long, R. Y., & Chen, H. (2018). Investigating external and internal pressures on corporate environmental behavior in papermaking enterprises of China. *Journal of Cleaner Production*, *172*, 1193-1211. https://doi.org/10.1016/j.jclepro.2017.10.115

Heidt, M., Gerlach, J. P., & Buxmann, P. (2019). Investigating the Security Divide between SME and Large Companies: How SME Characteristics Influence Organizational IT Security Investments. *Information Systems Frontiers*, *21*(6), 1285-1305. https://doi.org/10.1007/s10796-019-09959-1

Heinrich, C. J. (2002). Outcomes-based performance management in the public sector: Implications for government accountability and effectiveness. *Public Administration Review*, *62*(6), 712-725. https://doi.org/10.1111/1540-6210.00253

Hekkala, R., Stein, M. K., & Sarker, S. (2022). Power and conflict in inter-organisational information systems development. *Information Systems Journal*, *32*(2), 440-468. https://doi.org/10.1111/isj.12335

Hennink, M., & Kaiser, B. N. (2022). Sample sizes for saturation in qualitative research: A systematic review of empirical tests. *Soc Sci Med*, *292*, 114523. https://doi.org/10.1016/j.socscimed.2021.114523

Herath, T. C., Herath, H. S. B., & Cullum, D. (2023). An Information Security Performance Measurement Tool for Senior Managers: Balanced Scorecard Integration for Security Governance and Control Frameworks. *Information Systems Frontiers*, *25*(2), 681-721. https://doi.org/10.1007/s10796-022-10246-9

Herath, T. C., Herath, H. S. B., & D'Arcy, J. (2020). Organizational Adoption of Information Security Solutions: An Integrative Lens Based on Innovation Adoption and the Technology-Organization-Environment Framework. *Data Base for Advances in Information Systems*, *51*(2), 12-35. https://doi.org/10.1145/3400043.3400046

Hiller, N. J., & Peterson, S. J. (2019). Assessment and development first requires a deeper understanding of unique categories of senior leaders: A focus on CEOs and C-level executives. *Industrial and Organizational Psychology*, *12*(2), 211-214. https://doi.org/10.1017/iop.2019.24

Hochwarter, W. A., Rosen, C. C., Jordan, S. L., Ferris, G. R., Ejaz, A., & Maher, L. P. (2020). Perceptions of Organizational Politics Research: Past, Present, and Future. *Journal of Management*, *46*(6), 879-907. https://doi.org/10.1177/0149206319898506

Hofstede, G. (1980). Culture and Organizations. *International Studies of Management & Organization*, *10*(4), 15-41. https://doi.org/10.1080/00208825.1980.11656300

Hoitash, R., Hoitash, U., & Kurt, A. C. (2016). Do accountants make better chief financial officers? *Journal of Accounting and Economics*, *61*(2-3), 414-432. https://doi.org/10.1016/j.jacceco.2016.03.002

Holt, T. J., Leukfeldt, R., & van de Weijer, S. (2020). An Examination of Motivation and Routine Activity Theory to Account for Cyberattacks Against Dutch Web Sites. *Criminal Justice and Behavior*, *47*(4), 487-505. https://doi.org/10.1177/0093854819900322

Hooijberg, R. (1996). A multidirectional approach toward leadership: An extension of the concept of behavioral complexity. *Human Relations*, *49*(7), 917-946. https://doi.org/10.1177/001872679604900703

Hooper, V., & McKissack, J. (2016). The emerging role of the CISO. *Business Horizons*, *59*(6), 585-591. https://doi.org/10.1016/j.bushor.2016.07.004

Horkan, E., & Baker, M. R. (2025). Experimental trials provide insight to climate impacts on condition and over-winter survival in Pacific sand lance, Ammodytes personatus. *Behav Processes*, *226*, 105169. https://doi.org/10.1016/j.beproc.2025.105169

Horne, C. A., Ahmad, A., & Maynard, S. B. (2016). A theory on information security. Australasian Conference on Information Systems, Wollongong, Australia.

Horne, C. A., Maynard, S. B., & Ahmad, A. (2017). Organisational Information Security Strategy: Review, Discussion and Future Research. *Australasian Journal of Information Systems*, *21*(0). https://doi.org/10.3127/ajis.v21i0.1427

House, R. J. (1971). A Path Goal Theory of Leader Effectiveness. *Administrative Science Quarterly*, *16*(3), 321-339. https://doi.org/10.2307/2391905

House, R. J., Hanges, P. J., Javidan, M., Dorfman, P. W., & Gupta, V. (2004). *Culture, leadership, and organizations: The GLOBE study of 62 societies*. Sage Publications.

Hu, Q., Hart, P., & Cooke, D. (2007). The role of external and internal influences on information systems security - a neo-institutional perspective. *Journal of Strategic Information Systems*, *16*(2), 153-172. https://doi.org/10.1016/j.jsis.2007.05.004

Humayun, M., Niazi, M., Jhanjhi, N. Z., Alshayeb, M., & Mahmood, S. (2020). Cyber Security Threats and Vulnerabilities: A Systematic Mapping Study. *Arabian Journal for Science and Engineering*, *45*(4), 3171-3189. https://doi.org/10.1007/s13369-019-04319-2

Hyett, N., Kenny, A., & Dickson-Swift, V. (2014). Methodology or method? A critical review of qualitative case study reports. *Int J Qual Stud Health Well-being*, *9*, 23606. https://doi.org/10.3402/qhw.v9.23606

IBM Security. (2023). *Cost of a data breach report 2023*. P. Institute. https://www.ibm.com/reports/data-breach

Imran, F., Shahzad, K., Butt, A., & Kantola, J. (2021). Digital Transformation of Industrial Organizations: Toward an Integrated Framework. *Journal of Change Management*, *21*(4), 451-479. https://doi.org/10.1080/14697017.2021.1929406

Iqbal, W., Abbas, H., Daneshmand, M., Rauf, B., & Bangash, Y. A. (2020). An In-Depth Analysis of IoT Security Requirements, Challenges, and Their Countermeasures via Software-Defined Security. *Ieee Internet of Things Journal*, *7*(10), 10250-10276. https://doi.org/10.1109/Jiot.2020.2997651

Jack, E. P., & Raturi, A. S. (2006). Lessons learned from methodological triangulation in management research. *Management Research News*, *29*(6), 345-357. https://doi.org/10.1108/01409170610683833

Jaeger, L., & Eckhardt, A. (2021). Eyes wide open: The role of situational information security awareness for security-related behaviour. *Information Systems Journal*, *31*(3), 429-472. https://doi.org/10.1111/isj.12317

Jafri, J. A., Mohd Amin, S. I., Abdul Rahman, A., & Mohd Nor, S. (2024). A systematic literature review of the role of trust and security on Fintech adoption in banking. *Heliyon*, *10*(1), e22980. https://doi.org/10.1016/j.heliyon.2023.e22980

Jakóbik, A. (2020). Stackelberg game modeling of cloud security defending strategy in the case of information leaks and corruption. *Simulation Modelling Practice and Theory*, *103*, 102071. https://doi.org/10.1016/j.simpat.2020.102071

Jalali, M. S., & Kaiser, J. P. (2018). Cybersecurity in Hospitals: A Systematic, Organizational Perspective. *Journal of medical Internet research*, *20*(5), e10059. https://doi.org/10.2196/10059

Javid, T., Gupta, M. K., & Gupta, A. (2022). A hybrid-security model for privacy-enhanced distributed data mining. *Journal of King Saud University - Computer and Information Sciences*, *34*(6, Part B), 3602-3614. https://doi.org/10.1016/j.jksuci.2020.06.010

Jemison, D. B. (2007). The Importance of Boundary Spanning Roles in Strategic Decision-Making [I]. *Journal of Management Studies*, *21*(2), 131-152. https://doi.org/10.1111/j.1467-6486.1984.tb00228.x

Jensen, R. (2021). Exploring causal relationships qualitatively: An empirical illustration of how causal relationships become visible across episodes and contexts. *Journal of Educational Change*, *23*(2), 179-196. https://doi.org/10.1007/s10833-021-09415-5

Johnson, M. E., & Goetz, E. (2007). Embedding information security into the organization. *Ieee Security & Privacy*, *5*(3), 16-24. https://doi.org/10.1109/Msp.2007.59

Jones, M. C., Kappelman, L., Pavur, R., Nguyen, Q. N., & Johnson, V. L. (2020). Pathways to being CIO: The role of background revisited. *Information & Management*, *57*(5), 103234. https://doi.org/10.1016/j.im.2019.103234

Judge, T. A., Cable, D. M., Boudreau, J. W., & Bretz, R. D. (1995). An Empirical-Investigation of the Predictors of Executive Career Success. *Personnel Psychology*, *48*(3), 485-519. https://doi.org/10.1111/j.1744-6570.1995.tb01767.x

Kaloudi, N., & Li, J. (2020). The AI-Based Cyber Threat Landscape. *ACM Computing Surveys*, *53*(1), 1-34. https://doi.org/10.1145/3372823

Kalubanga, M., & Gudergan, S. (2022). The impact of dynamic capabilities in disrupted supply chains-The role of turbulence and dependence. *Industrial Marketing Management*, *103*, 154-169. https://doi.org/10.1016/j.indmarman.2022.03.005

Kamil, Y., Lund, S., & Islam, M. S. (2023). Information security objectives and the output legitimacy of ISO/IEC 27001: stakeholders' perspective on expectations in private organizations in Sweden. *Information Systems and e-Business Management*, *21*(3), 699-722. https://doi.org/10.1007/s10257-023-00646-y

Kammersgaard, T. (2021). Private security guards policing public space: using soft power in place of legal authority. *Policing & Society*, *31*(2), 117-130. https://doi.org/10.1080/10439463.2019.1688811

Karahanna, E., & Preston, D. S. (2013). The Effect of Social Capital of the Relationship Between the CIO and Top Management Team on Firm Performance. *Journal of Management Information Systems*, *30*(1), 15-55. https://doi.org/10.2753/Mis0742-1222300101

Karakose, T., Kocabas, I., Yirci, R., Papadakis, S., Ozdemir, T. Y., & Demirkol, M. (2022). The Development and Evolution of Digital Leadership: A Bibliometric Mapping Approach-Based Study. *Sustainability*, *14*(23). https://doi.org/10.3390/su142316171

Karanja, E. (2017). The role of the chief information security officer in the management of IT security. *Information and Computer Security*, *25*(3), 300-329. https://doi.org/10.1108/Ics-02-2016-0013

Karanja, E., Grant, D., & Zaveri, J. S. (2021). CIO reporting structure and firm strategic orientation – a content analysis approach. *Journal of Systems and Information Technology*, *23*(1), 20-52. https://doi.org/10.1108/jsit-02-2020-0022

Karanja, E., & Rosso, M. A. (2017). The Chief Information Security Officer: An Exploratory Study. *Journal of International Technology & Information Management*, *26*(2), 23-47. https://doi.org/10.58729/1941-6679.1299

Karie, N. M., Sahri, N. M., Yang, W. C., Valli, C., & Kebande, V. R. (2021). A Review of Security Standards and Frameworks for IoT-Based Smart Environments. *IEEE Access*, *9*, 121975-121995. https://doi.org/10.1109/Access.2021.3109886

Katz, D., & Kahn, R. L. (1978). *The social psychology of organizations*. Wiley.

Kazimierczak, M., Habib, N., Chan, J. H., & Thanapattheerakul, T. (2024). Impact of AI on the Cyber Kill Chain: A Systematic Review. *Heliyon*, *10*(24), e40699. https://doi.org/10.1016/j.heliyon.2024.e40699

Kelle, U. (2007). "Emergence" vs. "Forcing" of Empirical Data? A Crucial Problem of "Grounded Theory" Reconsidered. *Historical Social Research / Historische Sozialforschung. Supplement*(19), 133-156. http://www.jstor.org/stable/40981074

Kemper, J., Schilke, O., & Brettel, M. (2013). Social Capital as a Microlevel Origin of Organizational Capabilities. *Journal of Product Innovation Management*, *30*(3), 589-603. https://doi.org/10.1111/jpim.12004

Kersten, A. (2005). Crisis as usual: Organizational dysfunction and public relations. *Public Relations Review*, *31*(4), 544-549. https://doi.org/10.1016/j.pubrev.2005.08.014

Khallaf, A., & Majdalawieh, M. (2012). Investigating the Impact of CIO Competencies on IT Security Performance of the U.S. Federal Government Agencies. *Information Systems Management*, *29*(1), 55-78. https://doi.org/10.1080/10580530.2012.634298

Khan, R. A., Khan, S. U., Khan, H. U., & Ilyas, M. (2021). Systematic Mapping Study on Security Approaches in Secure Software Engineering. *IEEE Access*, *9*, 19139-19160. https://doi.org/10.1109/Access.2021.3052311

Khando, K., Gao, S., Islam, S. M., & Salman, A. (2021). Enhancing employees information security awareness in private and public organisations: A systematic literature review. *Computers & Security*, *106*, 102267. https://doi.org/10.1016/j.cose.2021.102267

Khayat, M., Barka, E., Adel Serhani, M., Sallabi, F., Shuaib, K., & Khater, H. M. (2025). Empowering Security Operation Center With Artificial Intelligence and Machine Learning—A Systematic Literature Review. *IEEE Access*, *13*, 19162-19197. https://doi.org/10.1109/access.2025.3532951

KhoKhar, F. A., Shah, J. H., Khan, M. A., Sharif, M., Tariq, U., & Kadry, S. (2022). A review on federated learning towards image processing. *Computers & Electrical Engineering*, *99*, 107818. https://doi.org/10.1016/j.compeleceng.2022.107818

Klaus, J. P., Kim, K., Masli, A., Guerra, K., & Kappelman, L. (2022). Prioritizing IT Management Issues and Business Performance. *Journal of Information Systems*, *36*(2), 83-99. https://doi.org/10.2308/Isys-2020-016

Koch, M., Forgues, B., & Monties, V. (2015). The Way to the Top: Career Patterns of Fortune 100 CEOS. *Human Resource Management*, *56*(2), 267-285. https://doi.org/10.1002/hrm.21759

Kok, A., Martinetti, A., & Braaksma, J. (2024). The Impact of Integrating Information Technology With Operational Technology in Physical Assets: A Literature Review. *IEEE Access*, *12*, 111832-111845. https://doi.org/10.1109/Access.2024.3442443

Kolukisa Tarhan, A., Garousi, V., Turetken, O., Soylemez, M., & Garossi, S. (2020). Maturity assessment and maturity models in health care: A multivocal literature review. *Digit Health*, *6*, 2055207620914772. https://doi.org/10.1177/2055207620914772

Koolen, C., Wuyts, K., Joosen, W., & Valcke, P. (2024). From insight to compliance: Appropriate technical and organisational security measures through the lens of cybersecurity maturity models. *Computer Law & Security Review*, *52*, 105914. https://doi.org/10.1016/j.clsr.2023.105914
Korschun, D. (2015). Boundary-Spanning Employees and Relationships with External Stakeholders: A Social Identity Approach. *Academy of Management Review*, *40*(4), 611-629. https://doi.org/10.5465/amr.2012.0398
Korzhyk, D., Yin, Z. Y., Kiekintveld, C., Conitzer, V., & Tambe, M. (2011). Stackelberg vs. Nash in Security Games: An Extended Investigation of Interchangeability, Equivalence, and Uniqueness. *Journal of Artificial Intelligence Research*, *41*, 297-327. https://doi.org/10.1613/jair.3269
Kotabe, M., Jiang, C. X., & Murray, J. Y. (2017). Examining the Complementary Effect of Political Networking Capability With Absorptive Capacity on the Innovative Performance of Emerging-Market Firms. *Journal of Management*, *43*(4), 1131-1156. https://doi.org/10.1177/0149206314548226
Kovács, G., & Spens, K. M. (2005). Abductive reasoning in logistics research. *International Journal of Physical Distribution & Logistics Management*, *35*(2), 132-144. https://doi.org/10.1108/09600030510590318
Kratzer, S., Westner, M., & Strahringer, S. (2023). Four Decades of Chief Information Officer Research: A Literature Review and Research Agenda Based on Main Path Analysis. *Data Base for Advances in Information Systems*, *54*(3), 37-74. https://doi.org/10.1145/3614178.3614182
Kraude, R., Narayanan, S., & Talluri, S. (2022). Evaluating the performance of supply chain risk mitigation strategies using network data envelopment analysis. *European Journal of Operational Research*, *303*(3), 1168-1182. https://doi.org/10.1016/j.ejor.2022.03.016
Kraus, S., Breier, M., & Dasí-Rodríguez, S. (2020). The art of crafting a systematic literature review in entrepreneurship research. *International Entrepreneurship and Management Journal*, *16*(3), 1023-1042. https://doi.org/10.1007/s11365-020-00635-4
Kraus, S., Durst, S., Ferreira, J. J., Veiga, P., Kailer, N., & Weinmann, A. (2022). Digital transformation in business and management research: An overview of the current status quo. *International Journal of Information Management*, *63*, 102466. https://doi.org/10.1016/j.ijinfomgt.2021.102466
Krishnan, P., Jain, K., Buyya, R., Vijayakumar, P., Nayyar, A., Bilal, M., & Song, H. (2022). MUD-Based Behavioral Profiling Security Framework for Software-Defined IoT Networks. *Ieee Internet of Things Journal*, *9*(9), 6611-6622. https://doi.org/10.1109/jiot.2021.3113577
Kristensen, K., & Andersen, K. N. (2023). C-suite Leadership of Digital Government. *Digital Government: Research and Practice*, *4*(1), 1-23. https://doi.org/10.1145/3580000
Kristof-Brown, A., Schneider, B., & Su, R. (2023). Person-organization fit theory and research: Conundrums, conclusions, and calls to action. *Personnel Psychology*, *76*(2), 375-412. https://doi.org/10.1111/peps.12581
Kristof-Brown, A. L., Zimmerman, R. D., & Johnson, E. C. (2005). Consequences of individuals' fit at work: A meta-analysis of person-job, person-organization, person-group, and person-supervisor fit. *Personnel Psychology*, *58*(2), 281-342. https://doi.org/10.1111/j.1744-6570.2005.00672.x
Kristof, A. L. (1996). Person-organization fit: An integrative review of its conceptualizations, measurement, and implications. *Personnel Psychology*, *49*(1), 1-49.
Kruse, C. S., Smith, B., Vanderlinden, H., & Nealand, A. (2017). Security Techniques for the Electronic Health Records. *J Med Syst*, *41*(8), 127. https://doi.org/10.1007/s10916-017-0778-4
Krutilla, K., Good, D., Toman, M., & Arin, T. (2021). Addressing Fundamental Uncertainty in Benefit-Cost Analysis: The Case of Deep Seabed Mining. *Journal of Benefit-Cost Analysis*, *12*(1), 122-151. https://doi.org/10.1017/bca.2020.28

Kunisch, S., Menz, M., & Langan, R. (2022). Chief digital officers: An exploratory analysis of their emergence, nature, and determinants. *Long Range Planning*, *55*(2), 101999. https://doi.org/10.1016/j.lrp.2020.101999

Kure, H. I., Islam, S., & Mouratidis, H. (2022). An integrated cyber security risk management framework and risk predication for the critical infrastructure protection. *Neural Computing and Applications*, *34*(18), 15241-15271. https://doi.org/10.1007/s00521-022-06959-2

Kwon, J., & Johnson, M. E. (2013). Health-Care Security Strategies for Data Protection and Regulatory Compliance. *Journal of Management Information Systems*, *30*(2), 41-65. https://doi.org/10.2753/Mis0742-1222300202

Lake, R. L., & Huckfeldt, R. (1998). Social capital, social networks, and political participation. *Political Psychology*, *19*(3), 567-584. https://doi.org/10.1111/0162-895x.00118

Lancaster, K. (2017). Confidentiality, anonymity and power relations in elite interviewing: conducting qualitative policy research in a politicised domain. *International Journal of Social Research Methodology*, *20*(1), 93-103. https://doi.org/10.1080/13645579.2015.1123555

Landells, E. M., & Albrecht, S. L. (2017). The Positives and Negatives of Organizational Politics: A Qualitative Study. *Journal of Business and Psychology*, *32*(1), 41-58. https://doi.org/10.1007/s10869-015-9434-5

Langlois, E. V., McKenzie, A., Schneider, H., & Mecaskey, J. W. (2020). Measures to strengthen primary health-care systems in low- and middle-income countries. *Bull World Health Organ*, *98*(11), 781-791. https://doi.org/10.2471/BLT.20.252742

Langner, R. (2011). Stuxnet: Dissecting a Cyberwarfare Weapon. *Ieee Security & Privacy*, *9*(3), 49-51. https://doi.org/10.1109/Msp.2011.67

Lanivich, S. E., Brees, J. R., Hochwarter, W. A., & Ferris, G. R. (2010). P-E Fit as moderator of the accountability - employee reactions relationships: Convergent results across two samples. *Journal of Vocational Behavior*, *77*(3), 425-436. https://doi.org/10.1016/j.jvb.2010.05.004

Last, D., Morris, T., & Dececchi, B. (2019). Preparing for future security challenges with practitioner research. *Security and Defence Quarterly*, *24*(2), 105-122. https://doi.org/10.35467/sdq/103345

Lavee, E., & Itzchakov, G. (2023). Good listening: A key element in establishing quality in qualitative research. *Qualitative Research*, *23*(3), 614-631. https://doi.org/10.1177/14687941211039402

Law, C. C. H., & Ngai, E. W. T. (2007). IT Infrastructure Capabilities and Business Process Improvements: Association with IT Governance Characteristics. *Information Resources Management Journal*, *20*(4), 25-47. https://doi.org/10.4018/irmj.2007100103

Lawrence, P. R., & Lorsch, J. W. (1967). *Organization and environment; managing differentiation and integration*. Division of Research, Graduate School of Business Administration, Harvard University.

Leana, C. R., & Van Buren, H. I. I. I. (1999). Organizational social capital and employment practices. *Academy of Management Review*, *24*(3), 538-555. https://doi.org/10.2307/259141

Lee, I. (2021). Cybersecurity: Risk management framework and investment cost analysis. *Business Horizons*, *64*(5), 659-671. https://doi.org/10.1016/j.bushor.2021.02.022

Lehto, M., & Limnéll, J. (2021). Strategic leadership in cyber security, case Finland. *Information Security Journal*, *30*(3), 139-148. https://doi.org/10.1080/19393555.2020.1813851

Levenson, A. R., Van der Stede, W. A., & Cohen, S. G. (2006). Measuring the relationship between managerial competencies and performance. *Journal of Management*, *32*(3), 360-380. https://doi.org/10.1177/0149206305280789

Lexchin, J. (2012). Those who have the gold make the evidence: how the pharmaceutical industry biases the outcomes of clinical trials of medications. *Sci Eng Ethics*, *18*(2), 247-261. https://doi.org/10.1007/s11948-011-9265-3

Li, Y., & Liu, Q. (2021). A comprehensive review study of cyber-attacks and cyber security; Emerging trends and recent developments. *Energy Reports*, *7*, 8176-8186. https://doi.org/10.1016/j.egyr.2021.08.126

Li, Y., & Tan, C. H. (2013). Matching business strategy and CIO characteristics: The impact on organizational performance. *Journal of Business Research*, *66*(2), 248-259. https://doi.org/10.1016/j.jbusres.2012.07.017

Li, Y., & Xu, L. (2021). Cybersecurity investments in a two-echelon supply chain with third-party risk propagation. *International Journal of Production Research*, *59*(4), 1216-1238. https://doi.org/10.1080/00207543.2020.1721591

Lichtenthaler, P. W., & Fischbach, A. (2018). A meta-analysis on promotion- and prevention-focused job crafting. *European Journal of Work and Organizational Psychology*, *28*(1), 30-50. https://doi.org/10.1080/1359432x.2018.1527767

Ligita, T., Harvey, N., Wicking, K., Nurjannah, I., & Francis, K. (2019). A practical example of using theoretical sampling throughout a grounded theory study A methodological paper. *Qualitative Research Journal*, *20*(1), 116-126. https://doi.org/10.1108/Qrj-07-2019-0059

Lincoln, Y. S., & Guba, E. G. (1985). *Naturalistic inquiry*. Sage Publications, Inc.

Ling, T. (2012). Evaluating complex and unfolding interventions in real time. *Evaluation*, *18*(1), 79-91. https://doi.org/10.1177/1356389011429629

Linnenluecke, M. K., Marrone, M., & Singh, A. K. (2019). Conducting systematic literature reviews and bibliometric analyses. *Australian Journal of Management*, *45*(2), 175-194. https://doi.org/10.1177/0312896219877678

Liu, C.-W., Huang, P., & Lucas Jr, H. C. (2020). Centralized IT Decision Making and Cybersecurity Breaches: Evidence from U.S. Higher Education Institutions. *Journal of Management Information Systems*, *37*(3), 758-787. https://doi.org/10.1080/07421222.2020.1790190

Liu, M., Shore, M., Yeoh, W., Jiang, F., & Zeadally, S. (2025). Toward effective cybersecurity management: a hierarchical process model with performance assessment. *Journal of Cybersecurity*, *11*(1). https://doi.org/10.1093/cybsec/tyaf020

Liu, M., Zhu, L., & Cionea, I. A. (2016). What Makes Some Intercultural Negotiations More Difficult Than Others? Power Distance and Culture-Role Combinations. *Communication Research*, *46*(4), 555-574. https://doi.org/10.1177/0093650216631096

Liu, X. (2020). Understanding the Classical Researches in Contingency Theory: A Review. https://doi.org/10.23977/ICEMGD2020.073

Liu, X., Ahmad, S. F., Anser, M. K., Ke, J., Irshad, M., Ul-Haq, J., & Abbas, S. (2022). Cyber security threats: A never-ending challenge for e-commerce. *Front Psychol*, *13*, 927398. https://doi.org/10.3389/fpsyg.2022.927398

Liu, Y., Gan, H. Q., & Karim, K. (2021). The effectiveness of chief financial officer board membership in improving corporate investment efficiency. *Review of Quantitative Finance and Accounting*, *57*(2), 487-521. https://doi.org/10.1007/s11156-020-00953-2

Locke, E. A. (2007). The case for inductive theory building. *Journal of Management*, *33*(6), 867-890. https://doi.org/10.1177/0149206307307636

Loonam, J., Zwiegelaar, J., Kumar, V., & Booth, C. (2022). Cyber-Resiliency for Digital Enterprises: A Strategic Leadership Perspective. *IEEE Transactions on Engineering Management*, *69*(6), 3757-3770. https://doi.org/10.1109/tem.2020.2996175

Luftman, J., & Brier, T. (1999). Achieving and sustaining business-IT alignment. *California Management Review*, *42*(1), 109-+. https://doi.org/10.2307/41166021

Lundh, A., Lexchin, J., Mintzes, B., Schroll, J. B., & Bero, L. (2017). Industry sponsorship and research outcome. *Cochrane Database Syst Rev*, *2*(2), MR000033. https://doi.org/10.1002/14651858.MR000033.pub3

Lundh, A., Lexchin, J., Mintzes, B., Schroll, J. B., & Bero, L. (2018). Industry sponsorship and research outcome: systematic review with meta-analysis. *Intensive Care Med*, *44*(10), 1603-1612. https://doi.org/10.1007/s00134-018-5293-7

Luo, Y. (2022). A general framework of digitization risks in international business. *Journal of International Business Studies*, *53*(2), 344-361. https://doi.org/10.1057/s41267-021-00448-9

Madnick, S., & Marotta, A. (2021). A Framework for Investigating GDPR Compliance Through the Lens of Security. In *Mobile Web and Intelligent Information Systems* (Vol. 12896, pp. 16–31). Springer-Verlag. https://doi.org/10.1007/978-3-030-83164-6_2

Magnani, G., & Gioia, D. (2023). Using the Gioia Methodology in international business and entrepreneurship research. *International Business Review*, *32*(2), 102097. https://doi.org/10.1016/j.ibusrev.2022.102097

Magnusson, L., Iqbal, S., Elm, P., & Dalipi, F. (2025). Information security governance in the public sector: investigations, approaches, measures, and trends. *International Journal of Information Security*, *24*(4), 177. https://doi.org/10.1007/s10207-025-01097-x

Malmendier, U., Pezone, V., & Zheng, H. (2023). Managerial Duties and Managerial Biases. *Management Science*, *69*(6), 3174-3201. https://doi.org/10.1287/mnsc.2022.4467

Mandiant. (2024). *M-Trends 2024 special report*. G. Cloud. https://services.google.com/fh/files/misc/m-trends-2024.pdf

Manfreda, A., & Stemberger, M. I. (2019). Establishing a partnership between top and IT managers A necessity in an era of digital transformation. *Information Technology & People*, *32*(4), 948-972. https://doi.org/10.1108/Itp-01-2017-0001

Mansfield-Devine, S. (2017). Coming of age: how organisations achieve security maturity. *Computer Fraud & Security*, *2017*(12), 16-20. https://doi.org/10.1016/S1361-3723(17)30110-0

Mantravadi, S., Schnyder, R., Moller, C., & Brunoe, T. D. (2020). Securing IT/OT Links for Low Power IIoT Devices: Design Considerations for Industry 4.0. *IEEE Access*, *8*, 200305-200321. https://doi.org/10.1109/Access.2020.3035963

Marican, M. N. Y., Abd Razak, S., Selamat, A., & Othman, S. H. (2023). Cyber Security Maturity Assessment Framework for Technology Startups: A Systematic Literature Review. *IEEE Access*, *11*, 5442-5452. https://doi.org/10.1109/Access.2022.3229766

Marion, R., & Uhl-Bien, M. (2001). Leadership in complex organizations. *Leadership Quarterly*, *12*(4), 389-418. https://doi.org/10.1016/S1048-9843(01)00092-3

Marsh, S., Atele-Williams, T., Basu, A., Dwyer, N., Lewis, P. R., Miller-Bakewell, H., & Pitt, J. (2020). Thinking about Trust: People, Process, and Place. *Patterns (N Y)*, *1*(3), 100039. https://doi.org/10.1016/j.patter.2020.100039

Martin, K. D., Borah, A., & Palmatier, R. W. (2017). Data Privacy: Effects on Customer and Firm Performance. *Journal of Marketing*, *81*(1), 36-58. https://doi.org/10.1509/jm.15.0497

Maschmeyer, L., Deibert, R. J., & Lindsay, J. R. (2021). A tale of two cybers - how threat reporting by cybersecurity firms systematically underrepresents threats to civil society. *Journal of Information Technology & Politics*, *18*(1), 1-20. https://doi.org/10.1080/19331681.2020.1776658

Mathieu, J. E., Gallagher, P. T., Domingo, M. A., & Klock, E. A. (2019). Embracing Complexity: Reviewing the Past Decade of Team Effectiveness Research. *Annual Review of Organizational Psychology and Organizational Behavior, Vol 6*, *6*, 17-46. https://doi.org/10.1146/annurev-orgpsych-012218-015106

Mattord, H., Kotwica, K., Whitman, M., & Battaglia, E. (2023). Organizational perspectives on converged security operations. *Information and Computer Security*, *32*(2), 218-235. https://doi.org/10.1108/ics-03-2023-0029

Maxwell, J. A. (2004). Using Qualitative Methods for Causal Explanation. *Field Methods*, *16*(3), 243-264. https://doi.org/10.1177/1525822x04266831

Mayes, B. T., & Allen, R. W. (1977). Toward A Definition of Organizational Politics. *Academy of Management Review*, *2*(4), 672-678. https://doi.org/10.2307/257520
Maynard, S. B., Onibere, M., & Ahmad, A. (2018). Defining the Strategic Role of the Chief Information Security Officer. *Pacific Asia Journal of the Association for Information Systems*, *10*(3), 61-85. https://doi.org/10.17705/1pais.10303
Mayrl, D., & Quinn, S. (2016). Defining the State from within: Boundaries, Schemas, and Associational Policymaking. *Sociological Theory*, *34*(1), 1-26. https://doi.org/10.1177/0735275116632557
Mbonihankuye, S., Nkunzimana, A., & Ndagijimana, A. (2019). Healthcare Data Security Technology: HIPAA Compliance. *Wireless Communications & Mobile Computing*, *2019*, 1927495. https://doi.org/10.1155/2019/1927495
McIntosh, T. R., Susnjak, T., Liu, T., Watters, P., Xu, D., Liu, D., Nowrozy, R., & Halgamuge, M. N. (2024). From COBIT to ISO 42001: Evaluating cybersecurity frameworks for opportunities, risks, and regulatory compliance in commercializing large language models. *Computers & Security*, *144*, 103964. https://doi.org/10.1016/j.cose.2024.103964
McLaughlin, S., Konstantinou, C., Wang, X. Y., Davi, L., Sadeghi, A. R., Maniatakos, M., & Karri, R. (2016). The Cybersecurity Landscape in Industrial Control Systems. *Proceedings of the Ieee*, *104*(5), 1039-1057. https://doi.org/10.1109/Jproc.2015.2512235
Mees-Buss, J., Welch, C., & Piekkari, R. (2020). From Templates to Heuristics: How and Why to Move Beyond the Gioia Methodology. *Organizational Research Methods*, *25*(2), 405-429. https://doi.org/10.1177/1094428120967716
Merhi, M. I., & Ahluwalia, P. (2019). Examining the impact of deterrence factors and norms on resistance to Information Systems Security. *Computers in Human Behavior*, *92*, 37-46. https://doi.org/10.1016/j.chb.2018.10.031
Meso, P., Negash, S., & Musa, P. F. (2021). Interactions Between Culture, Regulatory Structure, and Information Privacy Across Countries. *Journal of Global Information Management*, *29*(6), 1-14. https://doi.org/10.4018/JGIM.20211101.oa49
Meyer, C. B. (2001). A Case in Case Study Methodology. *Field Methods*, *13*(4), 329-352. https://doi.org/10.1177/1525822x0101300402
Mihelic, A., Vrhovec, S., Markelj, B., & Hovelja, T. (2024). Delegation-Based Agile Secure Software Development Approach for Small and Medium-Sized Businesses. *IEEE Access*, *12*, 189611-189635. https://doi.org/10.1109/Access.2024.3514889
Mijnhardt, F., Baars, T., & Spruit, M. (2016). Organizational Characteristics Influencing Sme Information Security Maturity. *Journal of Computer Information Systems*, *56*(2), 106-115. https://doi.org/10.1080/08874417.2016.1117369
Miller, D. (1992). Environmental Fit Versus Internal Fit. *Organization Science*, *3*(2), 159-178. https://doi.org/10.1287/orsc.3.2.159
Mintzberg, H. (1983). *Power in and around organizations*. Prentice-Hall.
Mir, R., & Watson, A. (2000). Strategic Management and the Philosophy of Science: The Case for a Constructivist Methodology. *Strategic Management Journal*, *21*(9), 941-953. http://www.jstor.org/stable/3094262
Mirza, N. A., Akhtar-Danesh, N., Noesgaard, C., Martin, L., & Staples, E. (2014). A concept analysis of abductive reasoning. *J Adv Nurs*, *70*(9), 1980-1994. https://doi.org/10.1111/jan.12379
Mishra, A., Alzoubi, Y. I., Gill, A. Q., & Anwar, M. J. (2022). Cybersecurity Enterprises Policies: A Comparative Study. *Sensors*, *22*(2), 538. https://www.mdpi.com/1424-8220/22/2/538
Mohammed, Y., Warkentin, M., & Beshah, T. (2025). Cultural drivers behind employees neutralizing deviant information systems security behaviors. *Journal of Knowledge Management*, *29*(4), 1129-1161. https://doi.org/10.1108/Jkm-09-2024-1108

Molok, N. N. A., Ahmad, A., & Chang, S. (2010, 2010/01/01/2010). Understanding the factors of information leakage through online social networking to safeguard organizational information. ACIS 2010 Proceedings - 21st Australasian Conference on Information Systems,

Monehin, D., & Diers-Lawson, A. (2022). Pragmatic optimism, crisis leadership, and contingency theory: A view from the C-suite. *Public Relations Review*, *48*(4). https://doi.org/10.1016/j.pubrev.2022.102224

Moon, Y. J., Choi, M., & Armstrong, D. J. (2018). The impact of relational leadership and social alignment on information security system effectiveness in Korean governmental organizations. *International Journal of Information Management*, *40*, 54-66. https://doi.org/10.1016/j.ijinfomgt.2018.01.001

Moses, O., Bui, B., Houqe, M. N., & Borghei, Z. (2025). Readiness for Mandatory Climate-Related Disclosures: A Tri-Jurisdictional Analysis of Governance Attributes in Australia, New Zealand and the United Kingdom. *Business Strategy and the Environment*, *34*(3), 3739-3763. https://doi.org/10.1002/bse.4154

Müller, S. D., Konzag, H., Nielsen, J. A., & Sandholt, H. B. (2024). Digital transformation leadership competencies: A contingency approach. *International Journal of Information Management*, *75*, 102734. https://doi.org/10.1016/j.ijinfomgt.2023.102734

Mullet, V., Sondi, P., & Ramat, E. (2021). A Review of Cybersecurity Guidelines for Manufacturing Factories in Industry 4.0. *IEEE Access*, *9*, 23235-23263. https://doi.org/10.1109/Access.2021.3056650

Mumford, T. V., Campion, M. A., & Morgeson, F. P. (2007). The leadership skills strataplex: Leadership skill requirements across organizational levels. *Leadership Quarterly*, *18*(2), 154-166. https://doi.org/10.1016/j.leaqua.2007.01.005

Munyon, T. P., Summers, J. K., Thompson, K. M., & Ferris, G. R. (2015). Political Skill and Work Outcomes: A Theoretical Extension, Meta-Analytic Investigation, and Agenda for the Future. *Personnel Psychology*, *68*(1), 143-184. https://doi.org/10.1111/peps.12066

Mura, P., & Wijesinghe, S. N. R. (2023). Critical theories in tourism - a systematic literature review. *Tourism Geographies*, *25*(2-3), 487-507. https://doi.org/10.1080/14616688.2021.1925733

Murphy, C., Klotz, A. C., & Kreiner, G. E. (2017). Blue skies and black boxes: The promise (and practice) of grounded theory in human resource management research. *Human Resource Management Review*, *27*(2), 291-305. https://doi.org/10.1016/j.hrmr.2016.08.006

Nadkarni, S., & Prügl, R. (2020). Digital transformation: a review, synthesis and opportunities for future research. *Management Review Quarterly*, *71*(2), 233-341. https://doi.org/10.1007/s11301-020-00185-7

Naseer, H., Desouza, K., Maynard, S. B., & Ahmad, A. (2024). Enabling cybersecurity incident response agility through dynamic capabilities: the role of real-time analytics. *European Journal of Information Systems*, *33*(2), 200-220. https://doi.org/10.1080/0960085X.2023.2257168

Naseer, H., Maynard, S. B., & Desouza, K. C. (2021). Demystifying analytical information processing capability: The case of cybersecurity incident response. *Decision Support Systems*, *143*, 113476. https://doi.org/10.1016/j.dss.2020.113476

Nath, D. (2017). The politics of information systems planning: The consultant versus the CIO. *ACADEMICIA: An International Multidisciplinary Research Journal*, *7*(2), 104-111. https://doi.org/10.5958/2249-7137.2017.00020.9

Nath, P., & Mahajan, V. (2008). Chief marketing officers: A study of their presence in firms' top management teams. *Journal of Marketing*, *72*(1), 65-81. https://doi.org/10.1509/jmkg.72.1.65

Nath, P., & Mahajan, V. (2011). Marketing in the C-Suite: A Study of Chief Marketing Officer Power in Firms' Top Management Teams. *Journal of Marketing*, *75*(1), 60-77. https://doi.org/10.1509/jmkg.75.1.60

Natow, R. S. (2020). The use of triangulation in qualitative studies employing elite interviews. *Qualitative Research*, *20*(2), 160-173. https://doi.org/10.1177/1468794119830077

Nee, V., & Opper, S. (2010). Political Capital in a Market Economy. *Social Forces*, *88*(5), 2105-2132. https://doi.org/10.1353/SOF.2010.0039

Nelson, J. (2017). Using conceptual depth criteria: addressing the challenge of reaching saturation in qualitative research. *Qualitative Research*, *17*(5), 554-570. https://doi.org/10.1177/1468794116679873

Nepal, S., Hernandez, J., Lewis, R., Chaudhry, A., Houck, B., Knudsen, E., Rojas, R., Tankus, B., Prafullchandra, H., & Czerwinski, M. (2024). Burnout in Cybersecurity Incident Responders: Exploring the Factors that Light the Fire. *Proceedings of the Acm on Human Computer Interaction*, *8*(Cscw1), Article 27. https://doi.org/10.1145/3637304

Newaz, A. K. M. I., Sikder, A. K., Rahman, M. A., & Uluagac, A. S. (2021). A Survey on Security and Privacy Issues in Modern Healthcare Systems: Attacks and Defenses. *Acm Transactions on Computing for Healthcare*, *2*(3), Article 27. https://doi.org/10.1145/3453176

Newman, L., & Leggett, N. (2018). Practitioner research: with intent. *European Early Childhood Education Research Journal*, *27*(1), 120-137. https://doi.org/10.1080/1350293x.2018.1556538

Ng, T. W. H., Eby, L. T., Sorensen, K. L., & Feldman, D. C. (2005). Predictors of objective and subjective career success: A meta-analysis. *Personnel Psychology*, *58*(2), 367-408. https://doi.org/10.1111/j.1744-6570.2005.00515.x

Nguyen, P. H., Kramer, M., Klein, J., & Traon, Y. L. (2015). An extensive systematic review on the Model-Driven Development of secure systems. *Information and Software Technology*, *68*, 62-81. https://doi.org/10.1016/j.infsof.2015.08.006

Nifakos, S., Chandramouli, K., Nikolaou, C. K., Papachristou, P., Koch, S., Panaousis, E., & Bonacina, S. (2021). Influence of Human Factors on Cyber Security within Healthcare Organisations: A Systematic Review. *Sensors*, *21*(15), 5119. https://www.mdpi.com/1424-8220/21/15/5119

NIST. (2024). The NIST Cybersecurity Framework (CSF) 2.0. https://doi.org/10.6028/NIST.CSWP.29

Noble, H., & Mitchell, G. (2016). What is grounded theory? *Evid Based Nurs*, *19*(2), 34-35. https://doi.org/10.1136/eb-2016-102306

O'Neil, S., & Koekemoer, E. (2016). Two decades of qualitative research in Psychology, Industrial and Organisational Psychology and Human Resource Management within South Africa: A critical review. *Sa Journal of Industrial Psychology*, *42*(1). https://doi.org/10.4102/sajip.v42i1.1350

O'Reilly, C. A., Chatman, J., & Caldwell, D. F. (1991). People and Organizational Culture: A Profile Comparison Approach to Assessing Person-Organization Fit. *The Academy of Management Journal*, *34*(3), 487-516. https://doi.org/10.2307/256404

Oc, B. (2018). Contextual leadership: A systematic review of how contextual factors shape leadership and its outcomes. *The Leadership Quarterly*, *29*(1), 218-235. https://doi.org/10.1016/j.leaqua.2017.12.004

Offner, K. L., Sitnikova, E., Joiner, K., & MacIntyre, C. R. (2020). Towards understanding cybersecurity capability in Australian healthcare organisations: a systematic review of recent trends, threats and mitigation. *Intelligence and National Security*, *35*(4), 556-585. https://doi.org/10.1080/02684527.2020.1752459

Oh, I. S., Guay, R. P., Kim, K., Harold, C. M., Lee, J. H., Heo, C. G., & Shin, K. H. (2014). Fit Happens Globally: A Meta-Analytic Comparison of the Relationships of Person-Environment Fit Dimensions with Work Attitudes and Performance across East Asia,

Europe, and North America. *Personnel Psychology*, *67*(1), 99-152. https://doi.org/10.1111/peps.12026

Okoli, J., & Watt, J. (2018). Crisis decision-making: the overlap between intuitive and analytical strategies. *Management Decision*, *56*(5), 1122-1134. https://doi.org/10.1108/Md-04-2017-0333

Onumo, A., Ullah-Awan, I., & Cullen, A. (2021). Assessing the Moderating Effect of Security Technologies on Employees Compliance with Cybersecurity Control Procedures. *ACM Transactions on Management Information Systems*, *12*(2), 1-29. https://doi.org/10.1145/3424282

Orts, E. W., & Strudler, A. (2002). The ethical and environmental limits of stakeholder theory. *Business Ethics Quarterly*, *12*(2), 215-233. https://doi.org/10.2307/3857811

Paarporn, K., & Xu, S. (2024). Preventive-Reactive Defense Tradeoffs in Resource Allocation Contests. *IEEE Control Systems Letters*, *8*, 2421-2426.

Palinkas, L. A., Horwitz, S. M., Green, C. A., Wisdom, J. P., Duan, N., & Hoagwood, K. (2015). Purposeful Sampling for Qualitative Data Collection and Analysis in Mixed Method Implementation Research. *Adm Policy Ment Health*, *42*(5), 533-544. https://doi.org/10.1007/s10488-013-0528-y

Palmié, M., Rüegger, S., & Parida, V. (2023). Microfoundations in the strategic management of technology and innovation: Definitions, systematic literature review, integrative framework, and research agenda. *Journal of Business Research*, *154*, 113351. https://doi.org/10.1016/j.jbusres.2022.113351

Palo Alto Networks Unit 42. (2024). *Incident Response Report 2024*. https://www.enablis.com.au/hubfs/2024%20ABM%20content/2024-unit42-incident-response-report.pdf

Papathanasiou, N., & Adey, B. T. (2020). Identifying the Input Uncertainties to Quantify When Prioritizing Railway Assets for Risk-Reducing Interventions. *Civileng*, *1*(2), 106-131. https://doi.org/10.3390/civileng1020008

Park, J. K., & Kim, I. (2014). A Study for Influencing Factors of Organizational Performance: The Perspective of the Mediating Effect of Information Security Maturity Level. *The Journal of Information Systems*, *23*(3), 99-125. https://doi.org/10.5859/kais.2014.23.3.99

Park, S., Lee, D. S., & Son, J. (2021). Regulatory reform in the era of new technological development: The role of organizational factors in the public sector. *Regulation & Governance*, *15*(3), 894-908. https://doi.org/10.1111/rego.12339

Parkin, S., Kuhn, K., & Shaikh, S. A. (2023). Executive decision-makers: a scenario-based approach to assessing organizational cyber-risk perception. *Journal of Cybersecurity*, *9*(1), undefined-undefined. https://doi.org/10.1093/cybsec/tyad018

Parry, D. A., Davidson, B. I., Sewall, C. J. R., Fisher, J. T., Mieczkowski, H., & Quintana, D. S. (2021). A systematic review and meta-analysis of discrepancies between logged and self-reported digital media use. *Nature Human Behaviour*, *5*(11), 1535-1547. https://doi.org/10.1038/s41562-021-01117-5

Pass, S. (2020). Benefits of Boredom: An 'Interlopers' Experience of Conducting Participant Observation on the Production Line. *European Management Review*, *17*(1), 285-295. https://doi.org/10.1111/emre.12393

Patacsil, F., & S. Tablatin, C. L. (2017). Exploring the importance of soft and hard skills as perceived by IT internship students and industry: A gap analysis. *Journal of Technology and Science Education*, *7*(3), 347-368. https://doi.org/10.3926/jotse.271

Patera, L., Garbugli, A., Bujari, A., Scotece, D., & Corradi, A. (2021). A Layered Middleware for OT/IT Convergence to Empower Industry 5.0 Applications. *Sensors (Basel)*, *22*(1). https://doi.org/10.3390/s22010190

Patterson, A. (2015). *Leader evolution : from technical expertise to strategic leadership*. Business Expert Press.

Patterson, C. M., Nurse, J. R. C., & Franqueira, V. N. L. (2023). Learning from cyber security incidents: A systematic review and future research agenda. *Computers & Security*, *132*, 103309. https://doi.org/10.1016/j.cose.2023.103309
Paul, J., & Criado, A. R. (2020). The art of writing literature review: What do we know and what do we need to know? *International Business Review*, *29*(4). https://doi.org/10.1016/j.ibusrev.2020.101717
Paul, J. A., & Zhang, M. (2021). Decision support model for cybersecurity risk planning: A two-stage stochastic programming framework featuring firms, government, and attacker. *European Journal of Operational Research*, *291*(1), 349-364. https://doi.org/10.1016/j.ejor.2020.09.013
Paul, M., Maglaras, L., Ferrag, M. A., & Almomani, I. (2023). Digitization of healthcare sector: A study on privacy and security concerns. *Ict Express*, *9*(4), 571-588. https://doi.org/10.1016/j.icte.2023.02.007
Pavlou, P. A., & El Sawy, O. A. (2011). Understanding the Elusive Black Box of Dynamic Capabilities. *Decision Sciences*, *42*(1), 239-273. https://doi.org/10.1111/j.1540-5915.2010.00287.x
Pearsall, M. J., Christian, J. S., Burgess, R. V., & Leigh, A. (2023). Preventing success: How a prevention focus causes leaders to overrule good ideas and reduce team performance gains. *J Appl Psychol*, *108*(7), 1121-1136. https://doi.org/10.1037/apl0000596
Pellegrini, M. M., Ciampi, F., Marzi, G., & Orlando, B. (2020). The relationship between knowledge management and leadership: mapping the field and providing future research avenues. *Journal of Knowledge Management*, *24*(6), 1445-1492. https://doi.org/10.1108/Jkm-01-2020-0034
Peppard, J. (2007). The conundrum of IT management. *European Journal of Information Systems*, *16*(4), 336-345. https://doi.org/10.1057/palgrave.ejis.3000697
Peppard, J. (2010). Unlocking the Performance of the Chief Information Officer (CIO). *California Management Review*, *52*(4), 73-99. https://doi.org/10.1525/cmr.2010.52.4.73
Peppard, J., Edwards, C., & Lambert, R. (2011). Clarifying the Ambiguous Role of the CIO. *MIS Quarterly Executive*, *10*(1), 31-44.
Perdana, A., Lee, H. H., Koh, S., & Arisandi, D. (2022). Data analytics in small and mid-size enterprises: Enablers and inhibitors for business value and firm performance. *International Journal of Accounting Information Systems*, *44*, 100547. https://doi.org/10.1016/j.accinf.2021.100547
Perera, S., Jin, X., Maurushat, A., & Opoku, D.-G. J. (2022). Factors Affecting Reputational Damage to Organisations Due to Cyberattacks. *Informatics*, *9*(1), 28. https://www.mdpi.com/2227-9709/9/1/28
Pfeffer, J. (1992). *Managing with power : politics and influence in organizations*. Harvard Business School Press.
Pfleeger, S. L., & Cunningham, R. K. (2010). Why Measuring Security Is Hard. *Ieee Security & Privacy*, *8*(4), 46-54. https://doi.org/10.1109/Msp.2010.60
Pham, H. C., Brennan, L., & Furnell, S. (2019). Information security burnout: Identification of sources and mitigating factors from security demands and resources. *Journal of Information Security and Applications*, *46*, 96-107. https://doi.org/10.1016/j.jisa.2019.03.012
Phan, H. P., & Ngu, B. H. (2021). A Case for Cognitive Entrenchment: To Achieve Optimal Best, Taking Into Account the Importance of Perceived Optimal Efficiency and Cognitive Load Imposition. *Front Psychol*, *12*, 662898. https://doi.org/10.3389/fpsyg.2021.662898
Philippou, E., Frey, S., & Rashid, A. (2020). Contextualising and aligning security metrics and business objectives: A GQM-based methodology. *Computers & Security*, *88*, 101634. https://doi.org/10.1016/j.cose.2019.101634

Piazza, A., Vasudevan, S., & Carr, M. (2024). *Am I Hired as a Firefighter? Exploring the Role Ambiguity and Board's Engagements on Job Stress and Perceived Organizational Support of CISOs*. https://doi.org/10.1109/ICCR61006.2024.10532977

Pita, J., Jain, M., Tambe, M., Ordóñez, F., & Kraus, S. (2010). Robust solutions to Stackelberg games: Addressing bounded rationality and limited observations in human cognition. *Artificial Intelligence*, *174*(15), 1142-1171. https://doi.org/10.1016/j.artint.2010.07.002

Ployhart, R. E., & Vandenberg, R. J. (2010). Longitudinal Research: The Theory, Design, and Analysis of Change. *Journal of Management*, *36*(1), 94-120. https://doi.org/10.1177/0149206309352110

Poeppelbuss, J., Niehaves, B., Simons, A., & Becker, J. (2011). Maturity Models in Information Systems Research: Literature Search and Analysis. *Communications of the Association for Information Systems*, *29*, 27. https://doi.org/10.17705/1cais.02927

Polakova, M., Suleimanova, J. H., Madzik, P., Copus, L., Molnarova, I., & Polednova, J. (2023). Soft skills and their importance in the labour market under the conditions of Industry 5.0. *Heliyon*, *9*(8), e18670. https://doi.org/10.1016/j.heliyon.2023.e18670

Pollini, A., Callari, T. C., Tedeschi, A., Ruscio, D., Save, L., Chiarugi, F., & Guerri, D. (2022). Leveraging human factors in cybersecurity: an integrated methodological approach. *Cogn Technol Work*, *24*(2), 371-390. https://doi.org/10.1007/s10111-021-00683-y

Poon, P., & Wagner, C. (2001). Critical success factors revisited: success and failure cases of information systems for senior executives. *Decision Support Systems*, *30*(4), 393-418. https://doi.org/10.1016/S0167-9236(00)00069-5

Pop, P., Zarrin, B., Barzegaran, M., Schulte, S., Punnekkat, S., Ruh, J., & Steiner, W. (2021). The FORA Fog Computing Platform for Industrial IoT. *Information Systems*, *98*, 101727. https://doi.org/10.1016/j.is.2021.101727

Posey, C., Roberts, T. L., Lowry, P. B., & Hightower, R. T. (2014). Bridging the divide: A qualitative comparison of information security thought patterns between information security professionals and ordinary organizational insiders. *Information & Management*, *51*(5), 551-567. https://doi.org/10.1016/j.im.2014.03.009

Pospisil, O., Blazek, P., Kuchar, K., Fujdiak, R., & Misurec, J. (2021). Application Perspective on Cybersecurity Testbed for Industrial Control Systems. *Sensors (Basel)*, *21*(23). https://doi.org/10.3390/s21238119

Preston, D. S., Chen, D., & Leidner, D. E. (2008). Examining the Antecedents and Consequences of CIO Strategic Decision-Making Authority: An Empirical Study. *Decision Sciences*, *39*(4), 605-642. https://doi.org/10.1111/j.1540-5915.2008.00206.x

Preston, D. S., Chen, D. Q., Swink, M., & Meade, L. (2017). Generating Supplier Benefits through Buyer-Enabled Knowledge Enrichment: A Social Capital Perspective. *Decision Sciences*, *48*(2), 248-287. https://doi.org/10.1111/deci.12220

Prümmer, J., van Steen, T., & van den Berg, B. (2024). A systematic review of current cybersecurity training methods. *Computers & Security*, *136*, 103585. https://doi.org/10.1016/j.cose.2023.103585

Prysor, D., & Henley, A. (2018). Boundary spanning in higher education leadership: identifying boundaries and practices in a British university. *Studies in Higher Education*, *43*(12), 2210-2225. https://doi.org/10.1080/03075079.2017.1318364

Purser, S. A. (2004). Improving the ROI of the security management process. *Computers & Security*, *23*(7), 542-546. https://doi.org/10.1016/j.cose.2004.09.004

Qammar, A., Karim, A., Ning, H., & Ding, J. (2023). Securing federated learning with blockchain: a systematic literature review. *Artificial Intelligence Review*, *56*(5), 3951-3985. https://doi.org/10.1007/s10462-022-10271-9

Qazi, A., Dickson, A., Quigley, J., & Gaudenzi, B. (2018). Supply chain risk network management: A Bayesian belief network and expected utility based approach for managing supply chain risks. *International Journal of Production Economics*, *196*, 24-42. https://doi.org/10.1016/j.ijpe.2017.11.008
Qollakaj, K., Larsson, L. E., & Memeti, S. (2025). Cybersecurity of remote work migration: A study on the VPN security landscape post Covid-19 outbreak. *Array*, *27*, 100437. https://doi.org/10.1016/j.array.2025.100437
Radanliev, P. (2024). Digital security by design. *Security Journal*, *37*(4), 1640-1679. https://doi.org/10.1057/s41284-024-00435-3
Ramač, R., Mandić, V., Taušan, N., Rios, N., Freire, S., Pérez, B., Castellanos, C., Correal, D., Pacheco, A., Lopez, G., Izurieta, C., Seaman, C., & Spinola, R. (2022). Prevalence, common causes and effects of technical debt: Results from a family of surveys with the IT industry. *Journal of Systems and Software*, *184*, 111114. https://doi.org/10.1016/j.jss.2021.111114
Ramezan, C. A. (2025). Understanding the chief information security officer: Qualifications and responsibilities for cybersecurity leadership. *Computers & Security*, *152*, 104363. https://doi.org/10.1016/j.cose.2025.104363
Ramos, A., Lazar, M., Holanda, R., & Rodrigues, J. J. P. C. (2017). Model-Based Quantitative Network Security Metrics: A Survey. *Ieee Communications Surveys and Tutorials*, *19*(4), 2704-2734. https://doi.org/10.1109/Comst.2017.2745505
Reis, J., & Melão, N. (2023). Digital transformation: A meta-review and guidelines for future research. *Heliyon*, *9*(1), e12834. https://doi.org/10.1016/j.heliyon.2023.e12834
Richardson, B., Kettles, D., Mazzola, D., & Li, H. (2024). Career Trajectory Analysis of Fortune 500 CIOs: A Linkedin Perspective. *Communications of the Association for Information Systems*, *55*, 625-650. https://doi.org/10.17705/1cais.05525
Rieger, K. L. (2019). Discriminating among grounded theory approaches. *Nurs Inq*, *26*(1), e12261. https://doi.org/10.1111/nin.12261
Rieger, M. O., Wang, M., & Hens, T. (2015). Risk Preferences Around the World. *Management Science*, *61*(3), 637-648. https://doi.org/10.1287/mnsc.2013.1869
Riggio, R. E., & Mumford, M. D. (2011). Introduction to the special issue: Longitudinal studies of leadership development. *Leadership Quarterly*, *22*(3), 453-456. https://doi.org/10.1016/j.leaqua.2011.04.002
Riggs, H., Tufail, S., Parvez, I., Tariq, M., Khan, M. A., Amir, A., Vuda, K. V., & Sarwat, A. I. (2023). Impact, Vulnerabilities, and Mitigation Strategies for Cyber-Secure Critical Infrastructure. *Sensors (Basel)*, *23*(8), 4060. https://doi.org/10.3390/s23084060
Rios, E., Rego, A., Iturbe, E., Higuero, M., & Larrucea, X. (2020). Continuous Quantitative Risk Management in Smart Grids Using Attack Defense Trees. *Sensors (Basel)*, *20*(16), 4404. https://doi.org/10.3390/s20164404
Romanosky, S. (2016). Examining the costs and causes of cyber incidents. *Journal of Cybersecurity*, *2*(2), 121-135. https://doi.org/10.1093/cybsec/tyw001
Roponen, J., Insua, D. R., & Salo, A. (2020). Adversarial risk analysis under partial information. *European Journal of Operational Research*, *287*(1), 306-316. https://doi.org/10.1016/j.ejor.2020.04.037
Rosen, C. C., & Levy, P. E. (2013). Stresses, Swaps, and Skill: An Investigation of the Psychological Dynamics That Relate Work Politics to Employee Performance. *Human Performance*, *26*(1), 44-65. https://doi.org/10.1080/08959285.2012.736901
Rowlands, T., Waddell, N., & McKenna, B. (2015). Are We There Yet? A Technique to Determine Theoretical Saturation. *Journal of Computer Information Systems*, *56*(1), 40-47. https://doi.org/10.1080/08874417.2015.11645799
Russell, C. J. (2001). A longitudinal study of top-level executive performance. *J Appl Psychol*, *86*(4), 560-573. https://doi.org/10.1037/0021-9010.86.4.560
Ryan, J. J. C. H., & Ryan, D. J. (2006). Expected benefits of information security investments. *Computers & Security*, *25*(8), 579-588. https://doi.org/10.1016/j.cose.2006.08.001

Saeed, S., Altamimi, S. A., Alkayyal, N. A., Alshehri, E., & Alabbad, D. A. (2023). Digital Transformation and Cybersecurity Challenges for Businesses Resilience: Issues and Recommendations. *Sensors (Basel)*, *23*(15). https://doi.org/10.3390/s23156666

Sætra, H. S. (2023). *Technology and Sustainable Development* (1 ed.). Routledge. https://doi.org/10.1201/9781003325086

Safa, N. S., Maple, C., Furnell, S., Azad, M. A., Perera, C., Dabbagh, M., & Sookhak, M. (2019). Deterrence and prevention-based model to mitigate information security insider threats in organisations. *Future Generation Computer Systems*, *97*, 587-597. https://doi.org/10.1016/j.future.2019.03.024

Sahi, M. A., Abbas, H., Saleem, K., Yang, X. D., Derhab, A., Orgun, M. A., Iqbal, W., Rashid, I., & Yaseen, A. (2018). Privacy Preservation in e-Healthcare Environments: State of the Art and Future Directions. *IEEE Access*, *6*, 464-478. https://doi.org/10.1109/Access.2017.2767561

Sahin, Z., & Vance, A. (2025). What do we need to know about the Chief Information Security Officer? A literature review and research agenda. *Computers & Security*, *148*, 104063. https://doi.org/10.1016/j.cose.2024.104063

Sailio, M., Latvala, O. M., & Szanto, A. (2020). Cyber Threat Actors for the Factory of the Future. *Applied Sciences-Basel*, *10*(12). https://doi.org/10.3390/app10124334

Sales, A., Mansur, J., & Roth, S. (2023). Fit for functional differentiation: new directions for personnel management and organizational change bridging the fit theory and social systems theory. *Journal of Organizational Change Management*, *36*(2), 273-289. https://doi.org/10.1108/Jocm-03-2022-0061

Sallos, M. P., Garcia-Perez, A., Bedford, D., & Orlando, B. (2019). Strategy and organisational cybersecurity: a knowledge-problem perspective. *Journal of Intellectual Capital*, *20*(4), 581-597. https://doi.org/10.1108/jic-03-2019-0041

Samtani, S., Kantarcioglu, M., & Chen, H. C. (2020). Trailblazing the Artificial Intelligence for Cybersecurity Discipline: A Multi-Disciplinary Research Roadmap. *Acm Transactions on Management Information Systems*, *11*(4), 1-19. https://doi.org/10.1145/3430360

Sandberg, J., & Tsoukas, H. (2011). Grasping the Logic of Practice: Theorizing through Practical Rationality. *Academy of Management Review*, *36*(2), 338-360. https://doi.org/10.5465/amr.2009.0183

Sankaran, S., Clegg, S. R., Killen, C. P., Smyth, H., & Scales, J. (2024). Enabling Collaborative Research in Project Management by Creating Gioia Data Structures as a Boundary Object. *Project Management Journal*, *55*(3), 281-296. https://doi.org/10.1177/87569728231212411

Santos-Neto, J. B. S. d., & Costa, A. P. C. S. (2019). Enterprise maturity models: a systematic literature review. *Enterprise Information Systems*, *13*(5), 719-769. https://doi.org/10.1080/17517575.2019.1575986

Schatz, D., & Bashroush, R. (2017). Economic valuation for information security investment: a systematic literature review. *Information Systems Frontiers*, *19*(5), 1205-1228. https://doi.org/10.1007/s10796-016-9648-8

Schinagl, S., Shahim, A., & Khapova, S. (2022). Paradoxical tensions in the implementation of digital security governance: Toward an ambidextrous approach to governing digital security. *Computers & Security*, *122*, 102903. https://doi.org/10.1016/j.cose.2022.102903

Schlette, D., Vielberth, M., & Pernul, G. (2021). CTI-SOC2M2 – The quest for mature, intelligence-driven security operations and incident response capabilities. *Computers & Security*, *111*, 102482. https://doi.org/10.1016/j.cose.2021.102482

Schmitt, M. (2023). Securing the digital world: Protecting smart infrastructures and digital industries with artificial intelligence (AI)-enabled malware and intrusion detection.

*Journal of Industrial Information Integration*, *36*, 100520. https://doi.org/10.1016/j.jii.2023.100520
Schmitz, C., Schmid, M., Harborth, D., & Pape, S. (2021). Maturity level assessments of information security controls: An empirical analysis of practitioners assessment capabilities. *Computers & Security*, *108*, 102306. https://doi.org/10.1016/j.cose.2021.102306
Schobel, K., & Denford, J. S. (2013). The Chief Information Officer and Chief Financial Officer Dyad in the Public Sector: How an Effective Relationship Impacts Individual Effectiveness and Strategic Alignment. *Journal of Information Systems*, *27*(1), 261-281. https://doi.org/10.2308/isys-50321
Schoemaker, P. J. H., Heaton, S., & Teece, D. (2018). Innovation, Dynamic Capabilities, and Leadership. *California Management Review*, *61*(1), 15-42. https://doi.org/10.1177/0008125618790246
Scott, S. (2024). Insights, Considerations, and Suggestions for the Practitioner-Researcher in Voice Studies. *Voice and Speech Review*, *18*(2), 146-165. https://doi.org/10.1080/23268263.2022.2137972
Shahim, A. (2021). Security of the digital transformation. *Computers & Security*, *108*, 102345. https://doi.org/10.1016/j.cose.2021.102345
Shaikh, F. A., & Siponen, M. (2024). Organizational Learning from Cybersecurity Performance: Effects on Cybersecurity Investment Decisions. *Information Systems Frontiers*, *26*(3), 1109-1120. https://doi.org/10.1007/s10796-023-10404-7
Sharma, A., Gupta, B. B., Singh, A. K., & Saraswat, V. K. (2023). Advanced Persistent Threats (APT): evolution, anatomy, attribution and countermeasures. *Journal of Ambient Intelligence and Humanized Computing*, *14*(7), 9355-9381. https://doi.org/10.1007/s12652-023-04603-y
Shaw, I. (2005). Practitioner research: Evidence or critique? *British Journal of Social Work*, *35*(8), 1231-1248. https://doi.org/10.1093/bjsw/bch223
Shen, S., Zhu, T., Wu, D., Wang, W., & Zhou, W. (2022). From distributed machine learning to federated learning: In the view of data privacy and security. *Concurrency and Computation: Practice and Experience*, *34*(16), e6002. https://doi.org/10.1002/cpe.6002
Shet, S. V., Patil, S. V., & Chandawarkar, M. R. (2019). Competency based superior performance and organizational effectiveness. *International Journal of Productivity and Performance Management*, *68*(4), 753-773. https://doi.org/10.1108/Ijppm-03-2018-0128
Short, A., & Carandang, R. (2022). The modern CISO: where marketing meets security. *Computer Fraud & Security*, *2022*(2). https://doi.org/10.12968/s1361-3723(22)70021-8
Shreeve, B., Hallett, J., Edwards, M., Ramokapane, K. M., Atkins, R., & Rashid, A. (2022). The Best Laid Plans or Lack Thereof: Security Decision-Making of Different Stakeholder Groups. *IEEE Transactions on Software Engineering*, *48*(5), 1515-1528. https://doi.org/10.1109/tse.2020.3023735
Silvester, J. (2008). The Good, the Bad and the Ugly: Politics and Politicians at Work. In *International Review of Industrial and Organizational Psychology 2008* (pp. 107-148). https://doi.org/10.1002/9780470773277.ch4
Simsek, Z., Fox, B., & Heavey, C. (2023). Systematicity in Organizational Research Literature Reviews: A Framework and Assessment. *Organizational Research Methods*, *26*(2), 292-321. https://doi.org/10.1177/10944281211008652
Skafi, M., Yunis, M. M., & Zekri, A. (2020). Factors Influencing SMEs' Adoption of Cloud Computing Services in Lebanon: An Empirical Analysis Using TOE and Contextual Theory. *IEEE Access*, *8*, 79169-79181. https://doi.org/10.1109/access.2020.2987331
Skeoch, H. R. K. (2022). Expanding the Gordon-Loeb model to cyber-insurance. *Computers & Security*, *112*, 102533. https://doi.org/10.1016/j.cose.2021.102533

Skopik, F., Settanni, G., & Fiedler, R. (2016). A problem shared is a problem halved: A survey on the dimensions of collective cyber defense through security information sharing. *Computers & Security*, *60*, 154-176. https://doi.org/10.1016/j.cose.2016.04.003

Smaga, P. (2025). Profiling the victim - cyber risk in commercial banks. *Computers & Security*, *150*, 104274. https://doi.org/10.1016/j.cose.2024.104274

Smaltz, D. H., Sambamurthy, V., & Agarwal, R. (2006). The antecedents of CIO role effectiveness in Organizations: An empirical study in the healthcare sector. *IEEE Transactions on Engineering Management*, *53*(2), 207-222. https://doi.org/10.1109/TEM.2006.872248

Smith, J. A., & Foti, R. J. (1998). A pattern approach to the study of leader emergence. *Leadership Quarterly*, *9*(2), 147-160. https://doi.org/10.1016/S1048-9843(98)90002-9

Smith, T., Tadesse, A. F., & Vincent, N. E. (2021). The impact of CIO characteristics on data breaches. *International Journal of Accounting Information Systems*, *43*, 100532. https://doi.org/10.1016/j.accinf.2021.100532

Sonnenreich, W., Albanese, J., & Stout, B. (2006). Return On Security Investment (ROSI) -- A Practical Quantitative Model. *Journal of Research & Practice in Information Technology*, *38*(1), 45-56. https://research.ebsco.com/linkprocessor/plink?id=a71656c4-1121-3a6c-96a9-efc6bd515aaf

Soomro, Z. A., Shah, M. H., & Ahmed, J. (2016). Information security management needs more holistic approach: A literature review. *International Journal of Information Management*, *36*(2), 215-225. https://doi.org/10.1016/j.ijinfomgt.2015.11.009

Spiekermann, S., Korunovska, J., & Langheinrich, M. (2019). Inside the Organization: Why Privacy and Security Engineering Is a Challenge for Engineers. *Proceedings of the IEEE*, *107*(3), 600-615. https://doi.org/10.1109/jproc.2018.2866769

Sriharan, A., Sekercioglu, N., Mitchell, C., Senkaiahliyan, S., Hertelendy, A., Porter, T., & Banaszak-Holl, J. (2024). Leadership for AI Transformation in Health Care Organization: Scoping Review. *J Med Internet Res*, *26*, e54556. https://doi.org/10.2196/54556

Srivastava, S., & Prakash, S. (2020). Security Enhancement of IoT Based Smart Home Using Hybrid Technique. In A. Bhattacharjee, S. K. Borgohain, B. Soni, G. Verma, & X.-Z. Gao (Eds.), *Machine Learning, Image Processing, Network Security and Data Sciences* (pp. 543-558). Springer.

Stavru, S. (2014). A critical examination of recent industrial surveys on agile method usage. *Journal of Systems and Software*, *94*, 87-97. https://doi.org/10.1016/j.jss.2014.03.041

Steinbart, P. J., Raschke, R. L., Gal, G., & Dilla, W. N. (2018). The influence of a good relationship between the internal audit and information security functions on information security outcomes. *Accounting Organizations and Society*, *71*, 15-29. https://doi.org/10.1016/j.aos.2018.04.005

Stevens, G. W. (2013). A Critical Review of the Science and Practice of Competency Modeling. *Human Resource Development Review*, *12*(1), 86-107. https://doi.org/10.1177/1534484312456690

Stone, T. H., & Jawahar, I. M. (2021). A leadership model for high-intensity organizational contexts. *Management Research Review*, *44*(8), 1199-1216. https://doi.org/10.1108/Mrr-06-2020-0324

Subramanian, S., Billsberry, J., & Barrett, M. (2022). A bibliometric analysis of person-organization fit research: significant features and contemporary trends. *Management Review Quarterly*, *73*(4), 1971-1999. https://doi.org/10.1007/s11301-022-00290-9

Sun, Y., Zhuang, F. Z., Zhu, H. S., Song, X., He, Q., & Xiong, H. (2023). Modeling the Impact of Person-Organization Fit on Talent Management With Structure-Aware Attentive Neural Networks. *Ieee Transactions on Knowledge and Data Engineering*, *35*(3), 2809-2822. https://doi.org/10.1109/Tkde.2021.3115620

Sutton, A., & Tompson, L. (2025). Towards a cybersecurity culture-behaviour framework: A rapid evidence review. *Computers & Security*, *148*, 104110. https://doi.org/10.1016/j.cose.2024.104110
Szczepaniuk, E. K., Szczepaniuk, H., Rokicki, T., & Klepacki, B. (2020). Information security assessment in public administration. *Computers & Security*, *90*, 101709. https://doi.org/10.1016/j.cose.2019.101709
Tabesh, P., & Vera, D. M. (2020). Top managers' improvisational decision-making in crisis: a paradox perspective. *Management Decision*, *58*(10), 2235-2256. https://doi.org/10.1108/Md-08-2020-1060
Tam, T., Rao, A., & Hall, J. (2021). The good, the bad and the missing: A Narrative review of cyber-security implications for australian small businesses. *Computers & Security*, *109*. https://doi.org/10.1016/j.cose.2021.102385
Tan, Z. R., Parambath, S. P., Anagnostopoulos, C., Singer, J., & Marnerides, A. K. (2025). Advanced Persistent Threats Based on Supply Chain Vulnerabilities: Challenges, Solutions, and Future Directions. *Ieee Internet of Things Journal*, *12*(6), 6371-6395. https://doi.org/10.1109/Jiot.2025.3528744
Tassabehji, R., Hackney, R., & Popovic, A. (2016). Emergent digital era governance: Enacting the role of the 'institutional entrepreneur' in transformational change. *Government Information Quarterly*, *33*(2), 223-236. https://doi.org/10.1016/j.giq.2016.04.003
Tawse, A., & Tabesh, P. (2021). Strategy implementation: A review and an introductory framework. *European Management Journal*, *39*(1), 22-33. https://doi.org/10.1016/j.emj.2020.09.005
Tazi, F., Nandakumar, A., Dykstra, J., Rajivan, P., & Das, S. (2024). SoK: Analyzing Privacy and Security of Healthcare Data from the User Perspective. *Acm Transactions on Computing for Healthcare*, *5*(2), 1-31. https://doi.org/10.1145/3650116
Teece, D. J., Pisano, G., & Shuen, A. (1997). Dynamic capabilities and strategic management. *Strategic Management Journal*, *18*(7), 509-533. https://doi.org/10.1002/(Sici)1097-0266(199708)18:7<509::Aid-Smj882>3.0.Co;2-Z
Tempelmayr, D., Ehrlinger, D., Stadlmann, C., Überwimmer, M., Mang, S., & Biedersberger, A. (2019). The Performance Effect of Dynamic Capabilities in Servitizing Companies. *Journal of International Business Research and Marketing*, *4*(6), 42-48. https://doi.org/10.18775/jibrm.1849-8558.2015.46.3005
Thornberg, R. (2012). Informed Grounded Theory. *Scandinavian Journal of Educational Research*, *56*(3), 243-259. https://doi.org/10.1080/00313831.2011.581686
Thurmond, V. A. (2001). The point of triangulation. *J Nurs Scholarsh*, *33*(3), 253-258. https://doi.org/10.1111/j.1547-5069.2001.00253.x
Timmermans, S., & Tavory, I. (2012). Theory Construction in Qualitative Research: From Grounded Theory to Abductive Analysis. *Sociological Theory*, *30*(3), 167-186. https://doi.org/10.1177/0735275112457914
Timonen, V., Foley, G., & Conlon, C. (2018). Challenges When Using Grounded Theory:A Pragmatic Introduction to Doing GT Research. *International Journal of Qualitative Methods*, *17*(1), 1609406918758086. https://doi.org/10.1177/1609406918758086
Tom, E., Aurum, A., & Vidgen, R. (2013). An exploration of technical debt. *Journal of Systems and Software*, *86*(6), 1498-1516. https://doi.org/10.1016/j.jss.2012.12.052
Tongco, M. D. C. (2007). Purposive Sampling as a Tool for Informant Selection. *Ethnobotany Research and Applications*, *5*, 147-158. https://doi.org/10.17348/era.5.0.147-158
Topping, C., Dwyer, A., Michalec, O., Craggs, B., & Rashid, A. (2021). Beware suppliers bearing gifts!: Analysing coverage of supply chain cyber security in critical national infrastructure sectorial and cross-sectorial frameworks. *Computers & Security*, *108*, 102324. https://doi.org/10.1016/j.cose.2021.102324

Toussaint, M., Krima, S., & Panetto, H. (2024). Industry 4.0 data security: A cybersecurity frameworks review. *Journal of Industrial Information Integration*, *39*, 100604. https://doi.org/10.1016/j.jii.2024.100604
Treadway, D. C., Breland, J. W., Williams, L. M., Cho, J., Yang, J., & Ferris, G. R. (2013). Social Influence and Interpersonal Power in Organizations: Roles of Performance and Political Skill in Two Studies. *Journal of Management*, *39*(6), 1529-1553. https://doi.org/10.1177/0149206311410887
Triplett, W. J. (2022). Addressing Human Factors in Cybersecurity Leadership. *Journal of Cybersecurity and Privacy*, *2*(3), 573-586. https://doi.org/10.3390/jcp2030029
Tu, C. Z., Yuan, Y., Archer, N., & Connelly, C. E. (2018). Strategic value alignment for information security management: a critical success factor analysis. *Information and Computer Security*, *26*(2), 150-170. https://doi.org/10.1108/ICS-06-2017-0042
Tufford, L., & Newman, P. (2012). Bracketing in Qualitative Research. *Qualitative Social Work*, *11*(1), 80-96. https://doi.org/10.1177/1473325010368316
Turner, C., & Astin, F. (2021). Grounded theory: what makes a grounded theory study? *Eur J Cardiovasc Nurs*, *20*(3), 285-289. https://doi.org/10.1093/eurjcn/zvaa034
Uchendu, B., Nurse, J. R. C., Bada, M., & Furnell, S. (2021). Developing a cyber security culture: Current practices and future needs. *Computers & Security*, *109*, 102387. https://doi.org/10.1016/j.cose.2021.102387
Uhl-Bien, M., & Arena, M. (2018). Leadership for organizational adaptability: A theoretical synthesis and integrative framework. *Leadership Quarterly*, *29*(1), 89-104. https://doi.org/10.1016/j.leaqua.2017.12.009
Urquhart, C., Lehmann, H., & Myers, M. D. (2010). Putting the ‘theory’ back into grounded theory: guidelines for grounded theory studies in information systems. *Information Systems Journal*, *20*(4), 357-381. https://doi.org/10.1111/j.1365-2575.2009.00328.x
Vaessen, J., & Todd, D. (2008). Methodological challenges of evaluating the impact of the Global Environment Facility's biodiversity program. *Eval Program Plann*, *31*(3), 231-240. https://doi.org/10.1016/j.evalprogplan.2008.03.002
Valdés-Rodríguez, Y., Hochstetter-Diez, J., Diéguez-Rebolledo, M., Bustamante-Mora, A., & Cadena-Martínez, R. (2024). Analysis of Strategies for the Integration of Security Practices in Agile Software Development: A Sustainable SME Approach. *IEEE Access*, *12*, 35204-35230. https://doi.org/10.1109/access.2024.3372385
Valdez-Juárez, L. E., Gallardo-Vázquez, D., & Ramos-Escobar, E. A. (2021). Online Buyers and Open Innovation: Security, Experience, and Satisfaction. *Journal of Open Innovation: Technology, Market, and Complexity*, *7*(1), 37. https://www.mdpi.com/2199-8531/7/1/37
Valle, M., Kacmar, K. M., & Zivnuska, S. (2019). Understanding the Effects of Political Environments on Unethical Behavior in Organizations. *Journal of Business Ethics*, *156*(1), 173-188. https://doi.org/10.1007/s10551-017-3576-5
Van de Ven, A. (2007). *Engaged Scholarship: A Guide For Organizational and Social Research*. Oxford University Press.
van der Raadt, B., Bonnet, M., Schouten, S., & van Vliet, H. (2010). The relation between EA effectiveness and stakeholder satisfaction. *Journal of Systems and Software*, *83*(10), 1954-1969. https://doi.org/10.1016/j.jss.2010.05.076
van Kemenade, E. (2019). Emergence in TQM, a concept analysis. *The TQM Journal*, *32*(1), 143-161. https://doi.org/10.1108/tqm-04-2019-0100
Varadharajan, V., Karmakar, K., Tupakula, U., & Hitchens, M. (2019). A Policy-Based Security Architecture for Software-Defined Networks. *Ieee Transactions on Information Forensics and Security*, *14*(4), 897-912. https://doi.org/10.1109/Tifs.2018.2868220
Varga, S., Brynielsson, J., & Franke, U. (2021). Cyber-threat perception and risk management in the Swedish financial sector. *Computers & Security*, *105*, 102239. https://doi.org/10.1016/j.cose.2021.102239

Venkatesan, K., & Rahayu, S. B. (2024). Blockchain security enhancement: an approach towards hybrid consensus algorithms and machine learning techniques. *Scientific Reports*, *14*(1), 1149. https://doi.org/10.1038/s41598-024-51578-7

Verhagen, M., de Reuver, M., & Bouwman, H. (2023). Implementing Business Models Into Operations: Impact of Business Model Implementation on Performance. *Ieee Transactions on Engineering Management*, *70*(1), 173-183. https://doi.org/10.1109/Tem.2020.3046365

Verizon. (2024). *2024 Data Breach Investigations Report*. https://www.verizon.com/business/resources/reports/dbir/

Vielberth, M., Böhm, F., Fichtinger, I., & Pernul, G. (2020). Security Operations Center: A Systematic Study and Open Challenges. *IEEE Access*, *8*, 227756-227779. https://doi.org/10.1109/Access.2020.3045514

Volberda, H. W., Khanagha, S., Baden-Fuller, C., Mihalache, O. R., & Birkinshaw, J. (2021). Strategizing in a digital world: Overcoming cognitive barriers, reconfiguring routines and introducing new organizational forms. *Long Range Planning*, *54*(5), 102110. https://doi.org/10.1016/j.lrp.2021.102110

Volckmann, R. (2005). Assessing executive leadership: an integral approach. *Journal of Organizational Change Management*, *18*(3), 289-302. https://doi.org/10.1108/09534810510599434

von Solms, S. H. (2005). Information Security Governance – Compliance management vs operational management. *Computers & Security*, *24*(6), 443-447. https://doi.org/10.1016/j.cose.2005.07.003

Voydanoff, P. (2005). Toward a conceptualization of perceived work-family fit and balance: A demands and resources approach. *Journal of Marriage and Family*, *67*(4), 822-836. https://doi.org/10.1111/j.1741-3737.2005.00178.x

Waldman, D. A., Javidan, M., & Varella, P. (2004). Charismatic leadership at the strategic level: A new application of upper echelons theory. *Leadership Quarterly*, *15*(3), 355-380. https://doi.org/10.1016/j.leaqua.2004.02.013

Walker-Roberts, S., Hammoudeh, M., Aldabbas, O., Aydin, M., & Dehghantanha, A. (2020). Threats on the horizon: understanding security threats in the era of cyber-physical systems. *Journal of Supercomputing*, *76*(4), 2643-2664. https://doi.org/10.1007/s11227-019-03028-9

Walker, D., & Myrick, F. (2006). Grounded theory: an exploration of process and procedure. *Qual Health Res*, *16*(4), 547-559. https://doi.org/10.1177/1049732305285972

Wang, D., Waldman, D. A., & Zhang, Z. (2014). A meta-analysis of shared leadership and team effectiveness. *J Appl Psychol*, *99*(2), 181-198. https://doi.org/10.1037/a0034531

Wang, J., Chaudhury, A., & Rao, H. R. (2008). Research Note—A Value-at-Risk Approach to Information Security Investment. *Information Systems Research*, *19*(1), 106-120. https://doi.org/10.1287/isre.1070.0143

Wang, L., Li, W., & Qi, L. (2020). Stakeholder Pressures and Corporate Environmental Strategies: A Meta-Analysis. *Sustainability*, *12*(3). https://doi.org/10.3390/su12031172

Wang, S., Asif, M., Shahzad, M. F., & Ashfaq, M. (2024). Data privacy and cybersecurity challenges in the digital transformation of the banking sector. *Computers & Security*, *147*, 104051. https://doi.org/10.1016/j.cose.2024.104051

Wang, W. Q., Xu, J. J., & Wang, M. (2018). Effects of Recommendation Neutrality and Sponsorship Disclosure on Trust vs. Distrust in Online Recommendation Agents: Moderating Role of Explanations for Organic Recommendations. *Management Science*, *64*(11), 5198-5219. https://doi.org/10.1287/mnsc.2017.2906

Wang, Y., Han, J. H., & Beynon-Davies, P. (2019). Understanding blockchain technology for future supply chains: a systematic literature review and research agenda. *Supply Chain*

*Management: An International Journal*, *24*(1), 62-84. https://doi.org/10.1108/scm-03-2018-0148
Warner, B. E. (2024). Clinical Colleague and Clinical Academic: A Physician's Autoethnographical Reflexive Account of Opportunities and Challenges as an 'Insider' Researcher. *International Journal of Qualitative Methods*, *23*. https://doi.org/10.1177/16094069241256278
Warner, K. S. R., & Wäger, M. (2019). Building dynamic capabilities for digital transformation: An ongoing process of strategic renewal. *Long Range Planning*, *52*(3), 326-349. https://doi.org/10.1016/j.lrp.2018.12.001
Wasserman, L., & Wasserman, Y. (2022). Hospital cybersecurity risks and gaps: Review (for the non-cyber professional) [Review]. *Front Digit Health*, *4*, 862221. https://doi.org/10.3389/fdgth.2022.862221
Watson, G. W., & Papamarcos, S. D. (2002). Social capital and organizational commitment. *Journal of Business and Psychology*, *16*(4), 537-552. https://doi.org/10.1023/A:1015498101372
Weishäupl, E., Yasasin, E., & Schryen, G. (2018). Information security investments: An exploratory multiple case study on decision-making, evaluation and learning. *Computers & Security*, *77*, 807-823. https://doi.org/10.1016/j.cose.2018.02.001
Welch, C., Marschan-Piekkari, R., Penttinen, H., & Tahvanainen, M. (2002). Corporate elites as informants in qualitative international business research. *International Business Review*, *11*(5), 611-628. https://doi.org/10.1016/S0969-5931(02)00039-2
Wen, J., Zhang, Z., Lan, Y., Cui, Z., Cai, J., & Zhang, W. (2023). A survey on federated learning: challenges and applications. *International Journal of Machine Learning and Cybernetics*, *14*(2), 513-535. https://doi.org/10.1007/s13042-022-01647-y
Wendler, R. (2012). The maturity of maturity model research: A systematic mapping study. *Information and Software Technology*, *54*(12), 1317-1339. https://doi.org/10.1016/j.infsof.2012.07.007
White, G. (2009). Strategic, Tactical, & Operational Management Security Model. *Journal of Computer Information Systems*, *49*(3), 71-75.
White, G. L. (2024). Security literacy model for strategic, tactical, & operational management levels. *Information Security Journal*, *33*(6), 626-634. https://doi.org/10.1080/19393555.2024.2307632
Whitten, D. (2008). The Chief Information Security Officer: An Analysis of the Skills Required for Success. *Journal of Computer Information Systems*, *48*(3), 15-19. https://doi.org/10.1080/08874417.2008.11646017
Whysall, Z., & Bruce, A. (2023). Changing the C-suite: opportunities and threats for leadership diversity and equality. *Management Decision*, *61*(4), 975-995. https://doi.org/10.1108/Md-07-2021-0875
Wicks, A. C., Gilbert, D. R., & Freeman, R. E. (2015). A Feminist Reinterpretation of The Stakeholder Concept. *Business Ethics Quarterly*, *4*(4), 475-497. https://doi.org/10.2307/3857345
Wilden, R., & Gudergan, S. (2015). The impact of dynamic capabilities on operational marketing and technological capabilities: investigating the role of environmental turbulence. *Journal of the Academy of Marketing Science*, *43*, 181-199. https://doi.org/10.1007/S11747-014-0380-Y
Willis, S., Clarke, S., & O'Connor, E. (2017). Contextualizing leadership: Transformational leadership and Management-By-Exception-Active in safety-critical contexts. *Journal of Occupational and Organizational Psychology*, *90*(3), 281-305. https://doi.org/10.1111/joop.12172
Wolgemuth, J. R., Erdil-Moody, Z., Opsal, T., Cross, J. E., Kaanta, T., Dickmann, E. M., & Colomer, S. (2015). Participants' experiences of the qualitative interview: considering the importance of research paradigms. *Qualitative Research*, *15*(3), 351-372. https://doi.org/10.1177/1468794114524222

Wood, L. E., & Ford, J. M. (1993). Structuring Interviews with Experts during Knowledge Elicitation. *International Journal of Intelligent Systems*, *8*(1), 71-90. https://doi.org/10.1002/int.4550080106

Woodward, J. (1965). *Industrial organization : theory and practice*. Oxford University Press.

Wylde, V., Rawindaran, N., Lawrence, J., Balasubramanian, R., Prakash, E., Jayal, A., Khan, I., Hewage, C., & Platts, J. (2022). Cybersecurity, Data Privacy and Blockchain: A Review. *SN Computer Science*, *3*(2), 127. https://doi.org/10.1007/s42979-022-01020-4

Xenofontos, C., Zografopoulos, I., Konstantinou, C., Jolfaei, A., Khan, M. K., & Choo, K. K. R. (2022). Consumer, Commercial, and Industrial IoT (In)Security: Attack Taxonomy and Case Studies. *Ieee Internet of Things Journal*, *9*(1), 199-221. https://doi.org/10.1109/Jiot.2021.3079916

Xu, F., Hsu, C., Wang, T., & Lowry, P. B. (2024). The antecedents of employees' proactive information security behaviour: The perspective of proactive motivation. *Information Systems Journal*, *34*(4), 1144-1174. https://doi.org/10.1111/isj.12488

Xu, F., Luo, X., Zhang, H., Liu, S., & Huang, W. (2017). Do Strategy and Timing in IT Security Investments Matter? An Empirical Investigation of the Alignment Effect. *Information Systems Frontiers*, *21*(5), 1069-1083. https://doi.org/10.1007/s10796-017-9807-6

Yang, D., & Li, M. (2018). Evolutionary Approaches and the Construction of Technology-Driven Regulations. *Emerging Markets Finance and Trade*, *54*(14), 3256-3271. https://doi.org/10.1080/1540496x.2018.1496422

Yaqoob, I., Hashem, I. A. T., Ahmed, A., Kazmi, S. M. A., & Hong, C. S. (2019). Internet of things forensics: Recent advances, taxonomy, requirements, and open challenges. *Future Generation Computer Systems-the International Journal of Escience*, *92*, 265-275. https://doi.org/10.1016/j.future.2018.09.058

Yeoh, W., Liu, M., Shore, M., & Jiang, F. (2023). Zero trust cybersecurity: Critical success factors and A maturity assessment framework. *Computers & Security*, *133*, 103412. https://doi.org/10.1016/j.cose.2023.103412

Yin, R. K. (2018). *Case study research and applications* (Vol. 6). Sage Thousand Oaks, CA.

Yin, Y., Hsu, C. R., & Zhou, Z. Y. (2024). Understanding employees' responses to information security management practices: a person-environment fit perspective. *Behaviour & Information Technology*, *43*(12), 2987-3009. https://doi.org/10.1080/0144929x.2023.2266024

Yu, K. Y. (2009). Affective influences in person-environment fit theory: exploring the role of affect as both cause and outcome of P-E fit. *J Appl Psychol*, *94*(5), 1210-1226. https://doi.org/10.1037/a0016403

Yucel, R. (2018). Scientists' Ontological and Epistemological Views about Science from the Perspective of Critical Realism. *Science & Education*, *27*(5-6), 407-433. https://doi.org/10.1007/s11191-018-9983-x

Yukl, G., Mahsud, R., Prussia, G., & Hassan, S. (2019). Effectiveness of broad and specific leadership behaviors. *Personnel Review*, *48*(3), 774-783. https://doi.org/10.1108/Pr-03-2018-0100

Żebrowski, P., Couce-Vieira, A., & Mancuso, A. (2022). A Bayesian Framework for the Analysis and Optimal Mitigation of Cyber Threats to Cyber-Physical Systems. *Risk Analysis*, *42*(10), 2275-2290. https://doi.org/10.1111/risa.13900

Zhang, C., & Bang, H. (2023). How Does Dynamic Capability Adjust Chinese Firms' Capabilities to Adapt to Environment Changes? *Korea International Trade Research Institute*, *10*. https://doi.org/10.16980/jitc.19.1.202302.103

Zhang, R. (2025). Optimization of multiple sampling for solving network boundary specification problem. *Sci Rep*, *15*(1), 4221. https://doi.org/10.1038/s41598-025-87760-8

Zhang, T., Harrington, K. B., & Sherf, E. N. (2022). The errors of experts: When expertise hinders effective provision and seeking of advice and feedback. *Curr Opin Psychol*, *43*, 91-95. https://doi.org/10.1016/j.copsyc.2021.06.011

Zhang, X. J., Khan, F., Wang, X. G., & Tang, C. L. (2023). Exploring the impact of national culture on the development of open government data: A cross-cultural analysis. *Big Data & Society*, *10*(2). https://doi.org/10.1177/20539517231206809

Zhang, Y., & Malacaria, P. (2025). Dealing with uncertainty in cybersecurity decision support. *Computers & Security*, *148*, 104153. https://doi.org/10.1016/j.cose.2024.104153

Zhang, Y., Weng, Q. X., & Zhu, N. (2018). The relationships between electronic banking adoption and its antecedents: A meta-analytic study of the role of national culture. *International Journal of Information Management*, *40*, 76-87. https://doi.org/10.1016/j.ijinfomgt.2018.01.015

Zhao, D. W., Wang, L. H., Wang, Z., & Xiao, G. X. (2019). Virus Propagation and Patch Distribution in Multiplex Networks: Modeling, Analysis, and Optimal Allocation. *Ieee Transactions on Information Forensics and Security*, *14*(7), 1755-1767. https://doi.org/10.1109/Tifs.2018.2885254

Zibenberg, A. (2017). Perceptions of Organizational Politics: A Cross-cultural Perspective. *Global Business Review*, *18*(4), 849-860. https://doi.org/10.1177/0972150917692211

Zorn, D. M. (2004). Here a chief, there a chief: The rise of the CFO in the American firm. *American Sociological Review*, *69*(3), 345-364. https://doi.org/10.1177/000312240406900302

Zwilling, M. (2022). Trends and Challenges Regarding Cyber Risk Mitigation by CISOs-A Systematic Literature and Experts' Opinion Review Based on Text Analytics. *Sustainability*, *14*(3). https://doi.org/10.3390/su14031311

# APPENDIX A: HUMAN ETHICS APPROVAL

In accordance with University of Melbourne policy on thesis deposit in Minerva Access, ethics approval documentation has been removed from this deposited copy.

# APPENDIX B: METHODOLOGICAL ARTEFACTS

## CISO INTERVIEW QUESTIONS

### Opening Question:

- Can you describe your role as the CISO from the strategic context of your organisation?

#### Requirements and drivers

- Looking at your role as a CISO, what would you say are some of the requirements of the role within your organisation? [What does your organisation require of the role?]
- What would you say were the drivers for establishing your role, within the context of your organisation? [Why did your organisation establish the role?]
- How would you describe the implication of these drivers on how you perform your duties as CISO?
- From your experience in several organisations, would you say your priorities and focus varied from one organisation to the other, or were they the same? Can you elaborate?

#### Firm Performance

- Talking about performance, what or how does your organisation benefit from having your role as a CISO?
- How does your role impact on or contribute to your organisation's performance?
- In your experience, how would you describe your organisation's perception of the value of your role?
- How have you been able to demonstrate the value of the CISO role to your organisation?
- How would you describe 'being successful' as a CISO?

#### Constraints and challenges

- What would you say are some of the constraints you face in performing your role as a CISO in your organisation?
- If you were to pick your 3 biggest challenges, what would they be?
- Would you say these challenges are peculiar to your organisation or would other organisations in the same industry face similar?

#### Business vs IT

- In your experience, do you think your organisation perceives cyber security as an IT issue?
- How does this perception affect or impact your role as the CISO?

#### Organisational Structure and Positional Power

- Can you describe how Cyber security is structured within your organisation?
- In your experience, do you think the positioning of the CISO role within your organisation's organisational structure has any impact on how your performance?
- In your experience do you as the CISO need power (whether formal or informal) to succeed?
- How would you describe the effect of internal politics on how you perform as the CISO?
- Does the support of other your organisation executives affect your success as CISO?
- Have you been in an organisation where you did not have the support of the other executives? How did this impact on your role?

**Strategic vs operational**

- In terms of the actual work that you do, how would you describe the time you spend on day to day operational activities versus activities that have strategic implications?

**Prevention vs response, Threat Landscape**

- Looking back to the last few years - 10, 15, would you say the role of the CISO has changed or transformed?
- Is there a relationship between the evolving cyber threat landscape and the role of the CISO?

**Bonus**

- Can you describe your role in your organisation's response to a major cyber incident?

*[Short explanation of the four faces of the CISO – Strategist, Guardian, Technologist, Advisor]*

- If you were to apportion percentages, how would you rate the time you spend as each face?

## CONSENT FORM

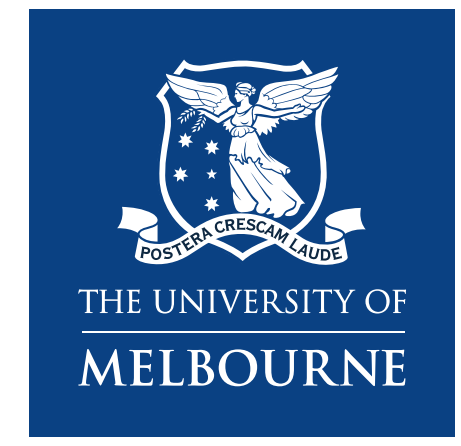

## SCHOOL OF COMPUTING AND INFORMATION SYSTEMS

## Faculty of Engineering and Information Technology

## Project: The Role of the CISO in Modern Organisations

**Responsible Researcher:** Associate Professor Atif Ahmad

**Additional Researchers:** A/Professor Sean Maynard, Mazino Onibere

**Name of Participant:** ______________________________

1. I consent to participate in this project, the details of which have been explained to me, and I have been provided with a written plain language statement to keep.
2. I understand that the purpose of this research is to investigate the evolution, the drivers, constraints, and impact on firm performance of the role of the CISO.
3. I understand that my participation in this project is for research purposes only.
4. I acknowledge that the possible effects of participating in this research project have been explained to my satisfaction.
5. In this project I will be required to provide responses to interview questions.
6. I understand that my interviews may be recorded.
7. I understand that my participation is voluntary and that I am free to withdraw from this project anytime without explanation or prejudice and to withdraw any unprocessed data that I have provided.
8. I understand that the data from this research will be stored at the University of Melbourne and will be destroyed 5 years after publication.
9. I have been informed that the confidentiality of the information I provide will be safeguarded subject to any legal requirements; my data will be password protected and accessible only by the named researchers.
10. I understand that after I sign and return this consent form, it will be retained by the researcher.

**Participant Signature:** ______________________ **Date:** ______________

## PLAIN LANGUAGE STATEMENT

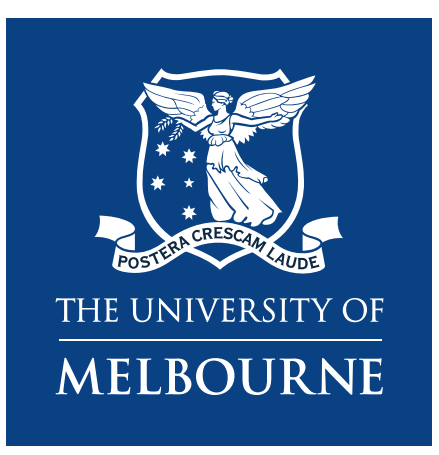

## SCHOOL OF COMPUTING AND INFORMATION SYSTEMS

## Faculty of Engineering and Information Technology

## Project: The Role of the CISO in Modern Organisations

Associate Professor Atif Ahmad (Responsible Researcher)

Tel: [redacted] Email: [redacted]

Associate Professor Sean Maynard, email: [redacted]

Mazino Onibere (PhD student); email: [redacted]

**Introduction**

Thank you for your interest in participating in this research project. The following few pages will provide you with further information about the project, so that you can decide if you would like to take part in this research.

Please take the time to read this information carefully. You may ask questions about anything you don't understand or want to know more about.

Your participation is voluntary. If you don't wish to take part, you don't have to. If you begin participating, you can also stop at any time.

**What is this research about?**

In today's fast-paced business world, information security is a critical business issue that affects the bottom line. With the increasing frequency, impact, and sophistication of cyber-attacks, organisations are facing an unprecedented level of risk to their sensitive data and critical infrastructure. As a result, the role of the Chief Information Security Officer (CISO) is more important than ever. Our research aims to understand the evolution of the CISO role, the drivers and requirements for the role, and how it impacts the overall performance of the organisation. By using an inductive grounded theory-based approach, we will provide valuable insights into how organisations can effectively utilize the CISO role to enhance organisational performance and achieve strategic goals, as well as protecting themselves from the ever-evolving cyber-threat landscape.

**What will I be asked to do?**

Should you agree to participate, the student researcher will interview you about your role as a CISO. The duration of the interview is capped at one hour. The interview will be audio recorded and later transcribed for the purpose of data analysis.

**What are the possible benefits?**

This research aims to provide organisations with insights on how to effectively utilize the CISO role to enhance performance, achieve strategic goals and protect against cyber threats by understanding the role's evolution, drivers, constraints, and impact on firm performance, as well as the effect of the organisation's type and industry and the CISO's position in the organisational structure.

**What are the possible risks?**

We recognise that breach of confidentiality is a possible risk in this research study, but we want to assure you that we have taken comprehensive measures to minimise this risk. All the information gathered from the discussions would be kept secure with access restricted to authorised personnel only. This gathered information will be anonymised and summarised before being included in any research findings or reports. We will not share your personal information in any published materials. Further, we will take stringent steps to protect your privacy and the confidentiality of the information we have collected.

**Do I have to take part?**

No. Participation is completely voluntary. You can withdraw at any time. You can also ask us to delete any shared information if it has not been processed.

**Will I hear about the results of this project?**

You will receive a copy of any publications based on the results of this project.

**What will happen to information about me?**

We will protect the confidentiality of the information that you have shared. To ensure complete anonymity, any reference that may reveal your identity or that of your organisation will be removed. All information you have shared will only be used for the purposes of this study. No personal information or reference to the name of your organisation would be included in any publications arising from this study. At the conclusion of this research, the information will be stored for five years from the last date of publication. At the expiration of the retention period, the information will be destroyed under university regulations.

**Who is funding this project?**

This study does not require funding.

**Where can I get further information?**

If you would like more information about the project, please contact the researchers:

- Associate Professor Atif Ahmad [redacted] or
- Associate Professor Sean Maynard [redacted] or
- Mazino Onibere [redacted]

**Who can I contact if I have any concerns about the project?**

This project has human research ethics approval from The University of Melbourne. If you have any concerns or complaints about the conduct of this research project, which you do not wish to discuss with the research team, you should contact the Research Integrity Administrator, Office of Research Ethics and Integrity, University of Melbourne, VIC 3010. Tel: +61 8344 1814 or Email: research-integrity@unimelb.edu.au. All complaints will be treated confidentially. In any correspondence please provide the name of the research team and/or the name or ethics ID number of the research project.

# APPENDIX C: REPRESENTATIVE QUOTES

This appendix presents the representative quotes underpinning the first-order concepts identified through the Gioia methodology data analysis process. Each quote is drawn verbatim from the CISO interview transcripts and is attributed to the corresponding participant organisation. The full data structure, showing the progression from first-order concepts through second-order themes to aggregate dimensions, is presented in Chapter 5.

## ORGANISATIONAL CONTEXT

| **Representative Quotes** | **First Order Concepts** | **Second order themes** |
|---|---|---|
| "... it was all purely based around, I guess, the evolving threat landscape and the risk that cyber posed the organization. There was very little focus there or drive from a compliance perspective." (BuildingMat01)<br><br>"One of the primary reasons is that, when customers ask and if a regulator asks, they can be completely transparent and proud as an organization of how they're going about protecting the information." (MgtConsult01)<br><br>"For me in the IT space, I always think it comes down to integrity. Right. How can you operate systems without high levels of integrity? And then availability and then confidentiality." (Beverage02) | Drivers for security | Organisational Security needs |
| "I think the role is ultimately responsible and in some ways accountable for managing cyber risk. I think cyber risk in our organization sits on the enterprise risk register. And it's within our top ten risks, but it's not in our-- in other organizations I've worked in, cyber has been the number-one risk." (BuildingMat01)<br><br>"I think the evolution of this role has been probably due to the size of the business of Food01 growing. And like I said, I've been around long enough to see growth from a really small team to a pretty decent team now." (Food01)<br><br>"Oh, I think growing awareness across industry, particularly critical infrastructure, of the risks and threats associated with cyber and technology resilience and technology risk and the need for identified senior leaders in this space." (Aviation01) | Drivers for establishing the CISO role | |
| "I am... critical about... people who race to that level without having a really good technical understanding of security threats and vulnerabilities who are just pure managers. I don't think that is a good trend for the industry." (GovAgency01).<br><br>"I guess, like anything else, it's probably more on the lines of ensuring that they have a trusted partner in security who can advise and consult with them and work with them in their business and be at the table..." (AgedCare1). | CISO role requirements | |
| "So I think in Beverage02, first and foremost, I would say it's always about the protection of information because that's the thing that's going to have-- if you look at it really simply, that's the thing that's going to have the biggest impact on the reputation." (Beverage02)<br><br>"I know coming into [an organization] that was hit with ransomware, that senior management are very aware of what weak cyber security controls look like." (Beverage02) | History of security incidents | |
| "I think there is a realization of a health insurance business acknowledging what I mentioned before about the personally identifiable information, health records. There came an overlay of that needs to be protected. We need to protect that." (Insurance01) | Sensitivity of data | |
| "We've got a lot of personal information. In fact, we've got 100-point data. So we've got medical records, IDs, bank accounts. We've got everything." (Super01). | | |
| "I've been quite fortunate in that I got senior leadership buy in from day one. So the traditional constraints around budget and time from the top and cut through weren't there." (Aviation01)<br><br>"I do have the support of the board and the executive... I report into a CIO and whilst a CISO reports into a CIO, it's always going to be seen as technology problem or issue or a function of technology only." (BuildingMat01)<br><br>"I align myself strategically with the CRO quite a bit. Whilst I do report to the CIO, I do have a dotted line to the CRO and I was pretty adamant of the beginning of my role is to have a relationship in terms of that." (AgedCare1) | Executive support | |
| "Our organization perceives cyber security has an IT problem or business? Not at all. I've worked in a lot of places, and I think Insurance01 is probably one of the most mature in terms of understanding it as an operational risk, and it's not necessarily linked to anything specific to IT." (Insurance01)<br><br>"...depending on which organization that you go to, you're going to come to organizations that have been through the war room and have gone through the hard elements there, and then it's always like, "Okay, we need to invest in all these procedures and we got to test." And they've gone through the battle scars in terms of the lessons learned. And generally, those organizations are more susceptible from a business perspective, to ensure that all the continuity plans are well understood. People have clear accountability for their roles within that, which is super key, as opposed to going to an organization where unfortunately people don't get it, right?" (AgedCare1)<br><br>"I think they think of it as a business issue. They treat it as an IT issue... But they ended up putting the cybersecurity function within an IT department... organization thinks it's a business issue. They treat it like a technical issue." (HigherEd02). | Perception of security | Organisational Culture |
| "I would say 18 months ago, yes, definitely. I would say 18 months on, i.e., now, to a certain extent, yes, but not as much. But there's still a long way to go again that's dependent on the organization." (BuildingMat01)<br><br>"Particularly that I report to the national IT operations manager, who reports to the CIO that said, it would be much more beneficial and, I mean, easier to deal directly with the CIO, for example." (Food01)<br><br>"I think it is perceived as an IT problem. I think it's perceived as something that's just too hard, that people don't understand, and that they're glad they've got a team in place who they see it's our job to do things." (Beverage01) | Business vs IT view | |
| "The structure of where the CISO role fits within Insurance01 is basically there's a board, then underneath the board, there are basically functional groups. And one of those functional groups is led by a senior executive who's part of the executive leadership group with technology and operations. And the CISO role reports directly to that person." (Insurance01)<br><br>"So my role probably changed a lot in the 18 months that I've been in my position... My role reports into the CIO, which forms part of the finance function. So reporting line is CEO, CFO, CIO, CISO." (BuildingMat01)<br><br>"Yeah, information security director at Beverage02 is the title. My role currently is to provide the cybersecurity- or the security-related controls to protect information systems and data at Beverage02. Organizationally, I sit under the chief technology officer." (Beverage02) | Reporting structure | Organisational structure |
| "Absolutely. I think you need a level of decision-making authority, right? So whether that's through setting a policy, having that endorsed, and then being able to enforce that policy without having to have everything approved every single time... then yes, right?" (MgtConsult02)<br><br>"So power is good in the sense that you need it for visibility at the board and Exec level. That's as much power as I need. The rest is actually influence." (Bank01)<br><br>"So in terms of power, I think one area where it's critical that they have formal power is that explicit agreement to do whatever needs to be done at speed during a major incident." (Beverage01) | Formal power | |
| "I think for many organizations if the CISO needs formal power, it means they've chosen the wrong CISO." (Aviation01) | | |
| "Instead of power, I would call it influence... With power you get influence. But you can influence without power... If you can't influence, you can't get people to change." (MgtConsult01)<br><br>"...security is always about influence rather than power mainly because it's one thing to compel people to do things, but that is only a temporary change in their behavior if you compel people to do changes. Certainly in a corporate environment, influence is more important. Being able to understand and have the empathy of understanding why certain things are done a certain way, that is more important." (Bank01)<br><br>"You create your own influence and your own stakeholders, and you manage and influence them, regardless of the size of stick that you've got." (Super02) | Influence | |
| "Thankfully, the board are very switched on and supportive of cybersecurity. They recognize how important it is for the organization and how important my role is to deliver on what we need to protect our most precious information. The approach that I took, one being thankful that I already had the support of the board for my role and I didn't have to convince them that I had to exist. So that was a huge-- I've been in other places where I haven't been as supported." (MgtConsult01)<br><br>"I've had situations where, in some instances, the CIO has been really happy for me to present to Audit risk, for example. And then in other situations where they're not, they want to do it themselves." (HigherEd01)<br><br>"So really my role is really the interface to the board and to the exec to help quantify and manage cyber risk." (BuildingMat01) | Board access | |
| "I'm probably in a quite unique position as a CISO because I came into Super01 just up to nine years ago. And it was no secret, but there was no security. It was pretty loose, to say it politely. So I had a green field there." (Super01)<br><br>"So, yeah. Look, my first element within AgedCare01 is-- I was security person number one, which is refreshing. First time I haven't had a team in 20 years, which was great at the time. And really is just to develop, okay, two things." (AgedCare1)<br><br>"...the cybersecurity function didn't even exist before that. My boss was hired two weeks before me. Up until then we had a compliance manager, but that was about it. So it is a very new function." (Beverage01) | Greenfield | Organisational Security maturity |
| "In my second iteration of strategy, I was very focused on how cybersecurity is going to add value to what university is going to be trying to achieve. If the target is retention of students... if the target is increased research revenue... that's when I shifted the focus of cybersecurity from building basic blocks to adding value." (HigherEd02)<br><br>"Our client base is the top end of town...and we're holding all their data. And we've got 50 million active personal records, but probably over 200 million personal records. That just changes the mindset of everything. You always have to be doing more than most. You have to be proactive and ahead of everyone else. And that's because you need the trust." (Super01)<br><br>"We benchmark our security capabilities... and then we benchmark against some of the best of industry frameworks. For example, there is a framework called C2M2. There is a framework called NIST. So what we do is we benchmark ourselves and then we identify whether we have already developed those practices..." (Food02)<br><br>"The security strategy for Bank01 is well documented, is actually well received. It is also board-approved as well. And it's actually seen as a way to deliver a service that people can trust." (Bank01) | Well-defined security posture | |
| "I'm shortly to be granted a group mandate for cybersecurity. And what does that mean? So Beverage02 as a business consists predominantly of brewing businesses in Australia and New Zealand, which I have accountability for. That's very clear that I have accountability there." (Beverage02) | Undergoing significant changes | |
| “We acquire businesses. Each two to three years, we acquire some large businesses... And so that’s how we kind of morphed, that the business grew, the industry grew, and the landscape changed drastically to create this role really.” (Super01)<br><br>“So my role probably changed a lot in the 18 months that I’ve been in my position... But we have since divested all of our US and Asian asset businesses and a number of local businesses, so it’s now very much an Australian based role.” (BuildingMat01). | | |
| “I think a good strategy is really around clear articulation of how we’re going to manage and control risk going forward as it relates to where the business is going” (Beverage02).<br><br>"We started showing results. You know We report to them on a quarterly basis. We show them how we are progressing, and they are able to see the progress within the first three months, six months." (GovAgency01)<br><br>"I've spent a lot of time in my role here doing those types of transformative projects where we can take processes or even security capabilities that we had previously, which were quite manual and cumbersome, and turn them into automated, cost-effective, and far more effective approaches for managing security." (Tech01) | Optimizing existing infrastructure | |

## CISO CHARACTERISTICS

| Representative Quotes | First order concepts | Second order themes |
|---|---|---|
| "I'm more of an agent of change, head of cyber. I'm someone you bring in to uplift or change directions." (Aviation01)<br><br>"I usually go to organizations when they have little maturity in terms of information security and I look to turn that around, and turn that around all within using a risk-based approach when it comes to security." (AgedCare1)<br><br>"I came into the role about 18 months ago... the team was three and three quite junior resources without any real structure, function in absence of a defined operating model. So one of the first things I did initially was build that operating model, deploy that operating model, and obtain approved funding..." (BuildingMat01) | Drive change | Action-Oriented CISO Type |
| “...one area where it’s critical that they have formal power is that explicit agreement to do whatever needs to be done at speed during a major incident.” (Beverage01)<br><br>"If we do this then, it's not a matter of solving this problem. There's going to be these other changes and things that are going to be potential impact. So let's think about those and have responses for them." (Insurance01)<br><br>"But I think that it's an important skill of leaders as well to actually make a decision when it calls for it. That's sometimes a challenge." (MgtConsult02) | Quick decisions | |
| “And of course, there was a bunch of certain technical technology and technical controls that we should be implementing that I don't need a strategy to tell me we should be doing. So worked on the program, developed a cyber program for really, 9 to 12 months, and we went really hard in terms of implementing that program.” (AgedCare1)<br><br>“And seeing as I came into essentially quarter two of the financial year, I had to figure out how to get the board engaged, how to come up with a strategy very quickly so that I had the authority to then go and make changes in the organization and draw on the resources I've been talking about previously from other teams to to support these initiatives.” (MgtConsult01) | Rapid implementation | |
| “So I kind of got that deep technical knowledge. And so I tend to, in incidents, dive in and out from, you know, that high-level oversight right down to solutioning technical stuff” (Super02)<br><br>“I think at the moment, this point in time, I will actually lead the investigation. I will lead the response mainly because I'm the only one that actually has the most experience in that space.” (Bank01) | Hands-on approach | |

| "My day starts off with literally checking each of our security products. I see if there's any risky sign-ons overnight, what are the alerts that have come through Sentinel." (Food01) | | |
|---|---|---|
| "So working in the industry that we work in, so an industrials organization, in manufacturing, there are very little regulatory compliance requirements for cybersecurity outside of ASX obligations, the Australian privacy act, PCI, DSS, where there isn't a regulator like APRA in finance or AEMO in energy." (BuildingMat01)<br><br>"PCI compliance was a big one. Our businesses revolves around taking payment. One of the compliance items was security, definitely. And a lot of the items from the PCI compliance was being failed as well." (Food01)<br><br>"Being in the financial services industry, that's a requirement. (Super01) | Ensuring compliance | |
| "I'm not really a steady state one, but I mean, some of it's simplistic, don't get breached, mitigate any impact of any attack." (Aviation01)<br><br>"The businesses want to do lots of things. We need sufficient prioritization of the particular supply chain engineers and also the it infrastructure staff to improve our security posture." (Beverage01)<br><br>"Identifying what our current level of capability maturity is helps us to identify what our gaps are. We can then formulate our strategy to fill these gaps and improve our capability for cybersecurity services over time." (AgedCare1) | Maintaining steady state | Stewardship-Oriented CISO Type |
| "I worked very much on it using a maturity model... it's easier for the executive and the board to understand– and palatable, of talking risk, but also talking about maturity and what maturity levels that you want to go, and between what domains." (AgedCare1).<br><br>"And that's where I've actually gone and put OKRs in for myself, so objectives and key results... And then I use CMMI to measure about 35 metrics in terms of security... I'll be able to show where the organisation was and then where the organisation is now, and then show a tangible uplift." (Super02) | Optimizing efficiency | |
| "And people would look at the security team and say, "Look, just protect us. Just do whatever you can, but live within your means." You've only got a percentage of budgets" (Super02)<br><br>"Funding and resources is always a problem. Look but I wouldn't say that it is preventing me from delivering what I want to deliver...Could I have had more money and do more things? Yeah." (GovAgency01) | Resource management | |
| "But in terms of the strategic context, Insurance01 has a vision of what they're trying to achieve, which is-- it's got multiple things. They're very focused on customers, very focused on health. So the strategic vision of where security fits into that is to basically enable but also protect the initiatives and operations of all elements of the business." (Insurance01)<br><br>"So a successful security leader articulates a clear strategy. A lot of this comes down to the CI. So you've got to have a strategy and a vision and you've got to be able to articulate that vision. All right? So put a strategy together with or without help, but then be able to clearly articulate that strategy piece to management, to the board, to peers, etc." (Beverage02) | Setting vision | |
| "My role in that from my previous experience was to really sell it to the executive board... So I really am responsible for the entire information security strategy and roadmap, as we call it." (MgtConsult02)<br><br>"Obviously being able to go through advising different levels of people through the organization, and that even means you're empowering and engaging the security team and doing whatever they need to do, helping them through any challenges that they need to make themselves look successful in representing the brand of security" (Insurance01)<br><br>"...the CISO play a role of an evangelist who can talk to the business, talk to external stakeholders, and give them the understanding of what security means and what kind of impact it could have on the organization, how it could add value..." (HigherEd02) | Inspiring others | Vision-Oriented CISO Type |
| "I've spent a lot of time in my role here doing those types of transformative projects where we can take processes or even security capabilities that we had previously, which were quite manual and cumbersome, and turn them into automated, cost-effective, and far more effective approaches..." (Tech01) | Leading transformation | |
| “So my role is to ensure that I have a direct relationship with these CIOs, understand what is their coming year or current year looks like. And that means whether ... we moving towards digitizing our environment from that geography perspective and how we are going to engage with businesses ... to transform the business into next level?” (Food02)<br><br>“...you can see that overnight with all that activity, you needed some helicopter view, someone to give that strategic advice. There are mergers acquisition integration, transformation without being bogged down by the day-to-day activities of BAU and project work. And so that's how we kind of morphed, that the business grew, the industry grew, and the landscape changed drastically to create this role really.” (Super01) | | |
| “...just like the bad guys are really good at innovating and automating, we can only survive if we go through a similar process ourselves, right? So anywhere we can look to not rely on people to do repetitive tasks, and they can be automated at scale...” (Tech01)<br><br>“We just built a little side project... to give exemptions to certain security controls... We built a little chatbot into Slack... Whereas... most organisations... want a Word document where the Word document is emailed to the manager.” (FinServ01)<br>“I prefer to make security sort of disappear. Disappear in the sense that people touch on it but don't realize the security. And the performance is that if a user gets a good user experience and stays secure at the same time, that's when I actually get the best sort of performance out of the organization.” (Bank01) | Prioritize creativity | |
| "I think we're coming into end of financial year, we're moving into the new financial year, there's budget, big restructure going on, etc. So right now, I'd say I'm probably 80% strategic, 20% operational." (BuildingMat01)<br><br>"In my second iteration of strategy, I was very focused on how cybersecurity is going to add value to what university is going to be trying to achieve. If the target is retention of students... if the target is increased research revenue... that’s when I shifted the focus of cybersecurity from building basic blocks to adding value." (HigherEd02)<br><br>“I think specific to CISO, they should always be a strategist. That's the only way you can keep up with the organization because if you're not a strategist, you're just going to be reacting to things all the time.” (Bank01) | Strategic focus | |
| “As a security person, you need to be fairly well abreast in all other aspects of IT to really be able to have enough insight into how to secure things.” (Tech01).<br><br>"I do lead the incident response. And the way the organization works today, if there is an incident, we do set up what we could essentially the CSIRT. So that's the Computer Security Incident Response Team." (MgtConsult01)<br><br>"So my role in the organization is pretty much end-to-end security. My coming along, I took on the risk assessment results and I was mainly focused on-- along with my BAU work with everyday checking all our security platforms." (Food01)<br><br>“Yes, having a technology background and being a top technologist is important, but I think you're you'll still start seeing a lot of roles now and where people are very successful in CISO or CSO roles, as we start to move from CISO to CSO and taking on more responsibility outside of cyber and information security, typically for enterprise resilience, privacy, fraud, etc., those roles, and people who are very successful in those roles aren't what you'd consider technologists. I think it's important. I've got a technical technology background. It definitely helps me in my role.” (BuildingMat01)<br><br>“But also, it's interesting because I've seen security that security people who are not really strong in technology also don't last very long. So you may still be in a big company, but if you don't understand technology and can do that, then also you'll struggle” (FinServ01) | Technical knowledge | CISO Technical Skills |
| "The tool that we are using to enable real-time analysis in our cybersecurity incident response process is SPLUNK. What it enables us to do is real-time analysis, search and monitoring of cybersecurity events." (MgtConsult01) | Hands-on tools experience | |
| "I've got products like Trend Micro that I actually have to touch on, visit, see their loading, what's happening, etc. And also dealing with ad hoc requests that come through, whether it be someone wants to get out of geo-blocking because they're going overseas, things like that." (Food01)<br><br>"We're building a bank. Very certainly, there's a lot of how do I put things together How do I set up a framework that works for the organization? And the reality is that, yes, it's easy enough to go, 'Well, we're going to use this cybersecurity framework or ISO 27,000,' or whatever else that's out there in the market." (Bank01) | | |
| "If a CISO is not reporting into the into the C suite or to the executive without having that seat at the board. I don't think you will be, you won't be truly. You can't be effective." (BuildingMat01)<br><br>"I think that's probably the main way to describe it in a more policy is, again, using this framework. The bottom line is that if we have avoided a major cyber incident, then we've been successful." (Beverage01)<br><br>" Well, I think the heart of everything in Insurance01 is focused on customers and customer information. So when we say like service availability, and what was the other thing you mentioned there?" (Insurance01) | Demonstrating expertise and value | CISO Experience |
| "I think the fourth piece is ability to manage incidents confidently. And that's because often when things happen, and I've got one that's currently going on at the moment, the business turns to the Chief Information Security Officer to assist in leading major Internet response or incidents." (Beverage02)<br><br>"My Incident response plan says I am actually the incident response manager. So I run it. I run the incident response. I also have a dovetail into the business crisis management plan." (AgedCare1)<br><br>"My role in responding to a major cyber incident would be the Incident Manager or the Incident Commander, using the example of a severity 1, P1, critical ransomware event." (BuildingMat01)<br><br>“I've sat across my career... across about 150 incidents, right? So I'm, I'm used to dealing with a big incident, dealing with incidents at scale, dealing with incidents that go on for weeks, sometimes months. And so I would be actively in that.” (Super02) | Incident management | |
| “Fundamentally, the business requires someone who can translate the threats that are obviously active in the environment, the capabilities of the organization at an operational level and align them based on risk to the organization's objectives. And usually to be a good CISO, in my opinion, means that you need to be able to translate between those two layers.” (Tech01).<br><br>"Because the CISO in the C Suite might have the board appointed responsibility, unless they've actually got everyone else in the C Suite on board, they can't actually implement what they're trying to achieve." (MgtConsult01) | Communicating effectively | CISO Personality |
| "So I do little things, I wear a tie, as you can see, I'm one of the few people who does wear a tie. I'm quite vocal, I'm quite obvious and present because, to be in it-- not so much an attention seeker, but a nexus point around that cultural change." (Aviation01)<br><br>"I'm a massive believer of-- if we're making a big decision or making a change in some space that's important to the business, then the executive leadership group needs to understand it." (Insurance01)<br><br>"I think where I'm at at the moment, it makes a lot of sense. However, I've been to organizations where it absolutely does not work because the CIO does push you to one side, and don't really understand the security function." (Bank01)<br><br>“You know I'm very much an introvert person. And I've kind of really stick to my foundations. But I can feel like the difference it makes when the personality of the person can make a lot of difference in communicating and engagement kind of thing.” (Auto01)<br><br>“And so that's why I think from a personality perspective, a highly political environment doesn't really suit me because, yeah, it's not that kind of game I want to play.” (Retail01) | Self-awareness of personality type and its impact on their leadership style | |
| "I'd fail without [other executives' support]. I'd fail. There's no point in my role without it. I strongly believe that. We might as well just be in IT and stick to our little worlds." (AgedCare1) | Adaptability, adjust their | |
| "CISOs need to be humble to understand that everyone's got their own pressures, everyone's got their own KPIs, and their own challenges. And the CISOs can help those executives with their challenges." (Super01)<br><br>"Being in a role, it's all about managing expectations of a very difficult audience. And it's across to your entire business." (BuildingMat01) | communication and approach | |
| "Any business needs to make trade-offs. That's a normal thing that you do. So I don't think the word conflict the word conflict, to me implies that somehow how you're almost going to do one thing, because-- Kelly's role, she is responsible for lots of things." (Beverage01)<br><br>"I think the best places that I've seen at work and then very much in particular with Insurance01, the senior executive, were the CISO reports to, I mean, basically [inaudible] a CISO [inaudible]. He's got just as much interest, just as much appetite for managing risk." (Insurance01) | Building relationships | |

## ENVIRONMENTAL CONTEXT

| Representative Quotes | First Order Concepts | Second order themes |
|---|---|---|
| “The threat has completely changed to a point now that we are against things like nation-states if you’re in an important sort of organization, but that hacker name has changed, right, because now it’s actually not driven by one person... We’re talking about organized crime...” (Bank01)<br><br>“...threat landscape has definitely played a part. We know that’s always ever evolving and whatnot, and with that, the CISO has to pivot and sort of move with that, as does an organization with that ever-changing threat landscape.” (Beverage01) | Prevailing threats | Threat Landscape |
| "The adoption of cloud, mobile, social media technologies, and focus on outsourcing is blurring the boundaries of the technology ecosystem. This has resulted in [a] much broader 'attack surface' for the threat actors to exploit." (BuildingMat01)<br><br>“You know, when I started out in security like 20 years ago, we were worried about script kiddies... then it's kind of evolved into, you know, basic cyber criminals that were, you know, using your site to sell malware or, you know, cracked games or something to, you know, where you've now got basically organised crime and nation-state actors.” (Super02)<br><br>"The threat has completely changed to a point now that we are against things like nation-states if you’re in an important sort of organization... we’re talking about organized crime, which is basically the traditional crime groups that basically bring in they outsource all of their development and exploit development and everything else..." (Bank01) | Evolving threat landscape | |
| "I mean, we've probably been in over 100 incidents in technology, probably more. So I've used elements around-- I think there's good understanding around how technologists actually use incident management as a whole." (AgedCare1)<br><br>"The call center dropping out for a day is probably, what would I say, a less important thing than the whole customer database being compromised and Insurance01 being stripped open and everything goes out on the dark web or whatever." (Insurance01)<br><br>“It could be, for instance, ransomware that effectively stops your business operating, which is what happened to Beverage02 in 2020, it could happen with regards to a data breach, it could be someone walks-- well, data breaches, because there's a couple of different types of that one there. All of those events, when they occur, come with a cost that the business has not budgeted for.” (Beverage02) | Cyber-attacks and data breaches | |
| "We use the new cybersecurity framework and we are assessed independently by a third party every 12 months. And that maturity score is also KPI for myself and the team and the wider technology team in part." (BuildingMat01) | Industry standards | Industry & Regulations |
| "Identifying what our current level of capability maturity is helps us to identify what our gaps are. We can then formulate our strategy to fill these gaps and improve our capability for cybersecurity services over time." (AgedCare1)<br><br>"Being a health insurer, there's requirements around PCI compliance because credit cards are taken. So there's some obligations out that we have to be really mindful of our own storage and use and management of credit card information." (Insurance01)<br><br>"Industries that have high levels of regulation and oversight, such as the banking sector is a classic, right? ...what that has driven over the last decade has been an information security function that is very focused and orientated towards satisfying the regulator's requirements for cybersecurity." (Beverage02) | | |
| "the focus is slightly different from organization to organization, particularly because in the past, I worked mainly for MSSPs, managed service providers. That said, we touch on many different organizations. Many different clients have differing needs. Financial institutions would have very different attack vectors to, for example, what we experience in the food and beverage organization, typically financial organizations would be heavily securing their payment environments, their emails, etc. Whereas we tend to focus on email security because that's how main attack vector." (Food01)<br><br>"So disruption is our key cyber risk." (Aviation01)<br><br>"The environment is very volatile today. We have a lot of political. We have a lot of other motives for the people to use the non-physical parts I mean, to launch attacks" (Food02) | Industry specific threats | |
| "...there are very little regulatory compliance requirements for cybersecurity outside of ASX obligations, the Australian privacy act, PCI, DSS, where there isn't a regulator like APRA in finance or AEMO in energy." (BuildingMat01).<br><br>"One of the primary reasons is that, when customers ask and if a regulator asks, they can be completely transparent" (MgtConsult01)<br><br>"For manufacturing organizations, actually, it's safety. ... Safety is [become?] - -they have an SIA. And then after safety, it's the integrity of the operations." (Beverage02)<br><br>"Our main concern is availability. Our main concern is the uptime of the stores. At the end of the day, the stores can sell their products, right?" (Food01) | Industry specific security requirements | |
| "So working in the industry that we work in, so an industrials organization, in manufacturing, there are very little regulatory compliance requirements for cybersecurity outside of ASX obligations, the Australian privacy act, PCI, DSS, where there isn't a regulator like APRA in finance or AEMO in energy." (BuildingMat01)<br><br>"Industries that have high levels of regulation and oversight, such as the banking sector is a classic, right? So you've got your requirements with ASIC that they're looking at [directives?] to do. You've then got APRA the Prudential regulator, which is very concerned about this." (Beverage02)<br><br>"Being an organization where you have data protection legislations across Europe, across Asia, like China, and in APAC, and also in Latin America and the US" (Food02) | Degree of regulation | |
| "So the CISO role is really well placed. I haven't seen any point where it's not given the right representation across everything that's happening in the business, from a risk level down to tactical discussions with all of the different functions within the business." (Insurance01)<br><br>"PCI compliance was a big one. Our businesses revolves around taking payment. One of the compliance items was security, definitely. And a lot of the items from the PCI compliance was being failed as well." (Food01)<br><br>"In the bank, it was regulated and it's been heavily regulated for a long time... people understand if you don't properly meet the regulations and obligations and you don't protect, then you will lose your banking license. There's a very clear consequence." (Retail01) | Compliance requirements | |
| "My job was to prepare the organization to be able to interact with other law enforcement agencies... they don't want to share information unless we are at a reasonable level and we show that we are there." (GovAgency01) | | |
| "I've been quite fortunate in that I got senior leadership buy in from day one. So the traditional constraints around budget and time from the top and cut through weren't there." (Aviation01)<br><br>"Let me make the point that I'm not sure that I would say you need to have power to be successful. I think you do need the ability to influence, and I think you need to be able to influence different levels of the organization." (MgtConsult01)<br><br>"The impacts to the sector in relation to budgets were very severe... students numbers, which is a loss in revenue... this desire to shape security and get it resourced hit at the same time that COVID hit. And it was quite negative." (HigherEd01)<br><br>"We are not profit-oriented. We are a public sector. You know, we don't have deep pockets, but we do have enough to make sure that we are able to run our regulatory business..." (GovAgency01) | securing buy-in for security initiatives | Available Funding |
| "I want to make sure that any decision around buying things or getting more headcount is very, very calculated...We've gone through, we've maximized, we've got these five tools, and we're really comfortable with the way those tools are working. We don't need to go and spend a million dollars on another one because, I don't know, they told us that we need to. Or some latest thing is telling us it's required." (Insurance01)<br><br>"We are definitely entering a cycle of tightening our belts and you know low margins and things... we've got to make sure that every dollar we spend equates to security improvement, risk reduction." (Retail01)<br><br>"I guess it's a constraint [manual spreadsheets], but I don't know what a plausible alternative is at this point in time... spreadsheets are cheaper than going with a cybersecurity tool, which doesn't necessarily give you a better outcome." (MgtConsult01) | Doing more with less | |
| "The biggest challenge I have is technology hygiene and tech debt and legacy infrastructure, poor lifecycle management for applications, hardware, software and the risk that presents to cyber. It's huge." (BuildingMat01)<br><br>"So with security, risk, there should be that tension. If it's all lovely-dovely, you've got the wrong model. People should be squeezing the CTOs and the technology directors and business units. "Why are you carrying this unnecessary technology debt risk?" (Super01)<br><br>"When I picked up this role, for example, we had different ways of managing, for example, endpoints on our laptops, different types of agents" (FinServ01) | Technology debt | Existing Technology |
| "The reality is that as 100% digital bank as well, we have a very different sort of landscape and threats and risks that we're exposed to." (Bank01)<br><br>"There's never a simplistic utopian sort of thing. And certainly across organisation...there's no uniformity. Everyone does it differently, et cetera, right? So what you're faced with as a CISO is lots of complexity and trying to manage the risk around, well, just understanding that complexity to begin with and then trying to manage the risk associated with that complexity at an IT and business level is difficult. And just when you think that maybe you've gotten a handle on it and you got under control, suddenly something new comes along like AI and you're doing something, you're having to take that into account now, right?" (Tech01) | Complexity of managing existing IT infrastructure | |
| "And just when you think that maybe you've gotten a handle on it and you got under control, suddenly something new comes along like AI and you're doing something, you're having to take that into account now, right?" (Tech01)<br><br>"You should always be looking both forward and in your rearview mirror. If you're not, you're just going to get caught. Because it's constantly changing." (Super01) | Shifting IT environment | |
| "As IT is evolving and the threat landscape of people trying to compromise that is evolving, you need to sort of stay across enough of it to be able to be confident, to talk through and understand what's going on…" (Insurance01) | | |
| "The biggest challenge for me is people. Not the technology, not the process, not the government, it's not the board, not the executive, it's the people." (AgedCare1)<br>"…we can take processes or even security capabilities that we had previously, which were quite manual and cumbersome, and turn them into automated, cost-effective, and far more effective approaches for managing security." (Tech01). | Ease of adopting Innovation | |
| "Fairly small compared to what I've sort of managed or built in the past, but that's in line with the organization size, the organization, so. Obviously, since divesting the US and Asian assets, I've lost members of the team in our Asian and North American businesses." (BuildingMat01)<br>"When I started, there was nobody, including IT security. Now, I think we're about 12 people, and with plans to grow next year" (HigherEd01).<br>"Being in large organizations, for instance, at the airport, I had about 160 people in my team. And my role, I would say 80% of it was strategic" (AgedCare1). | People management | Size of security team |
| "There is no single blueprint for a security team. The security team needs to be built based on the experience of the CISO. So if the CISO is more GRC focused, then you get more technical people. If the CISO is more technical focus, you get more sort of GRC compliance type of people, and then there's a mix in between" (Bank01)<br>"I could not tell you what was happening on the ground. I had 200 security people under me. I didn't know any of their names… there's no requirement at that level. But I think when you start looking at some of these startups and I've worked for fintechs and things like that, then I think it really pays to have a technical CISO." (Super02)<br>"The security team would be the SAS, and the technology team would be the army. And so you've got a specialist unit focused on key areas, highly skilled, highly trained, and have a vested interest to work at pace." (Super01)<br>"And now we're leading other projects that move towards our meeting the critical structure legislation requirements. For example, business continuity, insider risk, and stuff like that. Which may not be a traditional technology-led issue. But because of the risk that we manage and our approach, we've been asked to do more or to lead more of these…" (Aviation01) | Specialisation of security team | |
| "Whilst in AgedCare01, it's much smaller team, it's a much organization. There's more hands-on in certain aspects of people leadership and in terms of getting into some of the weeds." (AgedCare1)<br>" And whilst I'm technologically able to describe stuff, if I have to be on the tools, there is something wrong." (Aviation01)<br>"I think the smaller the company, probably the more technical the CISO needs to be." (Super02) | Hands-on work | |
| "And one of my roles in the past, I was actually joint CISO, joint CTO. I had two hats. That changed my entire landscape because I understood the complexities of running a large IT shop, all of the financial constraints, all of the fact that you're squeezed about keeping the lights on, P1s, P2s, but also that there's innovation and products that they want to go to market, they want agile and speed. And security can sometimes be like, "Oh, no, you must have this." You've just got to push up a decision to the CEOs or the business owners to say that you're not quite ready to go live with a product." (Super01)<br>"So my role two hats I actually have in there in an incident. I'm in the IT process side of things, providing recommendations, but I'm in the conductor in the actual incident response." (AgedCare1)<br>"That's been a real struggle, to be honest with you, especially when we've had some of that attrition and had some gaps in quite a small team. I feel like I've been dragged into sort of more tactical tasks more often." (HigherEd01) | Wearing multiple hats | |
| "So when I started 18 months ago, the team was three and three quite junior resources without any real structure, function in absence of a defined operating model. So one of the first things I did initially was build that operating model, deploy that operating model" (BuildingMat01)<br><br>"Because you need someone that can stand up and understand the business. Like I mentioned, the business strategy, what's happening from a technology perspective and a security perspective to support that strategy and making sure that stuff happens in the business with all the best efforts of protected as much as it can be." (Insurance01)<br><br>"I'll delegate based on the actual-- whatever the technology may be that is affected, I'll delegate that to the SMEs." (Food01) | Guidance and direction | |
| "In terms of vendors, we structure ourselves-- at least with our cyber team, we don't really have functions that are very much outsourced. We'll have a lot of vendors and also contractors who really work reporting to us, as opposed to being sort of a proper outsourced function." (Beverage01)<br><br>"You're effectively, you're leading that instant response, whether it's yourself or an outsourced third party through an IR retainer, between yourself and a third-party IR specialist, you're sharing the responsibility of an incident manager, an incident commander, depending on the nature or severity of the incident." (BuildingMat01)<br><br>"I do have an outsourced SOC as well, which we're just onboarding at the moment" (Bank01)<br><br>"I also have a bunch of vendors who provide lots of services." (Beverage01) | Outsourcing | |
| "I regularly meet with two board members every month outside of cycle. And it just makes my key messaging really clear and crystal in terms of not only education elements to the board." (AgedCare1)<br><br>"Being able to work through that not only is CISO level, but empowering to change and things like that. And also like I mentioned, even though it's not as strong as-- you need to sort of understand technology and understand what's happening in the space." (Insurance01) | navigating stakeholder relationships | Size of organization |
| "We're only less than 600 people. Given the spread of people on Aviation01 at any day could be up to 40,000 people working there. It's a one to many type set up. So that means we're quite lean, but also means that even as ahead of-- and I'm about three layers below the CEO, you're quite visible." (Aviation01)<br><br>"I'd say it's around about 30% strategy. Probably 10 to 15% of it would be governance and processes and a whole bunch of stuff like that, and policy pieces. I would say 20, 30% would be in people's stakeholder element around that." (AgedCare1)<br><br>"I find that that can be a little bit of a struggle where people don't want to make a decision. And I think making a decision is more important than not making a decision, right?" (MgtConsult02) | Bureaucracy | |
| "Instead of influencing and working with your other technology managers as a peer, right, and within the-- effectively the technology org unit, as opposed to being an outside party coming in and telling you what to do." (Beverage02)<br><br>"Well, actually, that's been one of our success stories because we've broken down the silos. Because of how shiny cyber is at the moment, we've actively gone out and done, 'We're here to help. We've got some threat warning. How can we help?'" (Aviation01)<br><br>"There's things where outside of information relevant to the insurance, it's a health insurance, there's other health-related services. So records from a GP, records from an Optometrist or whatever, they're all related to customer's health." (Insurance01) | Silos | |
| "The adoption of cloud, mobile, social media technologies, and focus on outsourcing is blurring the boundaries of the technology ecosystem. This has resulted in [a] much broader 'attack surface' for the threat actors to exploit." (BuildingMat01)<br><br>"The proliferation of number of technologies and the lack of integration between vendors and those technologies ended up creating a complexity rather than a simplicity of managing security." (HigherEd02) | Complexity in organisation | |
| "I've worked in two organizations which are large family-owned businesses. And in large family-owned businesses, politics-- I'm going to be very careful what I say here. The political agenda is very strong." (BuildingMat01)<br><br>"There's politics everywhere, right? That's corporate world. To be honest, the corporate world is pretty brutal, especially if you're an ASX listed company." (Super01)<br><br>"It's huge. So right now, we do not have a lot of internal politics, and that actually makes it much easier to perform the role. Going back to it, if you came back 12 months ago and you asked this, in fact, if you spoke to my predecessor and his predecessor, the answer would have been the opposite. They would actually tell you that they cannot perform the role very well at all. So absolutely. It's huge." (FinServ01)<br><br>"Yes. That's no different to any other... Any other senior manager. Everyone's victim to politics, right?" (Beverage02) | Internal politics | |

# APPENDIX D: SLCM PRACTICAL IMPLICATIONS TOOLKIT

The four instruments presented in this appendix operationalise the Security Leadership Contingency Model (SLCM) developed through this research. Each instrument was developed as a practical contribution grounded in the empirical findings and is intended for use by security leaders, executive teams, and organisations seeking to apply the SLCM in practice. The instruments are presented here in static form, with question items, response scales, and scoring guidance.

Three instruments take the form of self-administered diagnostics employing a four-point behavioural frequency scale (1 = Never, 2 = Sometimes, 3 = Often, 4 = Always). Scores for each section or dimension are calculated as a percentage of the maximum possible score for that section. The fourth instrument is a decision-support matrix and does not involve scored items. Reverse-scored items, where indicated, should be recoded before scoring (1 becomes 4, 2 becomes 3, 3 becomes 2, 4 becomes 1).

## TOOL 1: ORGANISATIONAL SECURITY MATURITY PHASE DIAGNOSTIC

This instrument enables organisations to determine their current security maturity phase as defined by the SLCM: Establishment, Maturation, or Strategic. The diagnostic comprises 39 items distributed equally across the three phases (13 items per phase). It is designed for completion by the CISO, the executive to whom security reports, or senior security stakeholders with sufficient organisational awareness to assess the frequency of the described behaviours. Results provide an empirical basis for leadership-related decisions, including the recruitment of a security leader whose orientation aligns with the dominant phase, the evaluation of incumbent CISO performance against phase-appropriate expectations, and the planning of succession as the organisation evolves.

**Scoring**: calculate the sum of responses for each phase section and divide by the maximum possible score for that section (13 items × 4 = 52 points). Express the result as a percentage. The phase with the highest percentage score is the dominant phase.

Response scale:

| 1 = Never | 2 = Sometimes | 3 = Often | 4 = Always |
|---|---|---|---|

| No. | Statement |
| --- | --- |
| **Section A: Establishment Phase Items (E1–E13)** | |
| **1** | When a new security technology or control is needed, implementation typically begins within days rather than weeks of the decision. |
| **2** | The security leader spends more time in technical discussions and hands-on problem-solving than in governance meetings. |
| **3** | Security initiatives are often adjusted or reprioritised based on recent incidents or emerging threats rather than following a pre-planned roadmap. |
| **4** | When asked about the security posture, different teams would likely give quite different descriptions of capabilities and priorities. |
| **5** | Security investments focus primarily on establishing baseline protections rather than optimising existing controls. |
| **6** | Executive leadership asks for frequent updates on security progress but rarely challenges the strategic direction or priorities. |
| **7** | When security issues arise, the initial response is usually to contact IT rather than to engage business process owners. |
| **8** | Security budget requests require extensive justification and multiple rounds of approval compared to other operational investments. |
| **9** | Staff awareness of security responsibilities varies significantly across different parts of the organisation. |
| **10** | Security policies exist but are not consistently referenced in business decisions or project planning. |
| **11** | The security function has undergone significant restructuring or leadership changes in the past three years. |
| **12** | The security leader personally reviews or approves most significant security decisions rather than delegating to established processes. |
| **13** | Security decisions are often made pragmatically based on immediate needs rather than assessed against formal risk criteria. |
| **Section B: Maturation Phase Items (M1–M13)** | |
| **14** | When someone new joins the security team, there are documented procedures and playbooks they can follow for most situations. |
| **15** | The organisation regularly compares security practices against industry frameworks and adjusts its approach based on gaps identified. |
| **16** | Security metrics are reported consistently each period, and trends over time inform improvement priorities. |
| **17** | Business units understand how to engage with security for different types of requests, and the process works consistently. |
| **18** | Most security improvement efforts focus on making existing processes more efficient or comprehensive rather than building new capabilities. |
| **19** | External audit findings or regulatory requirements are significant drivers of security priorities and investment decisions. |
| **20** | When discussing security with executives, conversations focus heavily on compliance status and control effectiveness metrics. |
| **21** | The organisation has achieved or is actively pursuing formal security certifications that demonstrate controls to external parties. |
| **22** | Security reports to leadership emphasise risk reduction, incident metrics, and compliance adherence rather than business enablement. |
| **23** | Business stakeholders view security primarily as ensuring the organisation meets its obligations rather than as a source of competitive advantage. |
| **24** | The security team has clearly defined roles with specialists responsible for distinct areas such as governance, risk and compliance, operations, and architecture. |
| **25** | The security leader spends more time on programme optimisation and team development than on technical implementation. |
| **26** | The main challenges faced are about scaling security capabilities consistently across the organisation rather than building foundational controls. |
| **Section C: Strategic Phase Items (S1–S13)** | |
| **27** | Security participates in strategic planning discussions for major business initiatives from the earliest stages, not just for technical review. |
| **28** | When launching new products or services, security capabilities are positioned as customer value propositions alongside other features. |
| **29** | Business cases for major initiatives include security as a potential differentiator or enabler, not just a cost or constraint. |
| **30** | Senior leadership regularly references security capabilities when discussing competitive positioning or market opportunities. |
| **31** | Security metrics reported to leadership emphasise business outcomes, innovation enablement, and strategic alignment as much as risk metrics. |
| **32** | The security leader has direct access to the CEO and board, and their input is sought on major organisational decisions beyond security. |
| **33** | When significant business decisions are being made, security considerations are raised naturally by business leaders without security needing to insert itself. |
| **34** | Security recommendations carry substantial weight in organisational decisions, comparable to input from other executive functions. |
| **35** | The security leader allocates more time to business partnership and strategic planning than to operational security oversight. |
| **36** | Security actively shapes organisational strategy and direction rather than primarily responding to business requirements with security constraints. |
| **37** | The security programme regularly explores emerging technologies and innovative approaches rather than primarily focusing on proven, established controls. |
| **38** | Security is viewed as an ongoing capability requiring continuous evolution and investment rather than a problem to be solved once. |
| **39** | Security improvement initiatives focus on strategic advancement and business enablement as much as closing gaps or improving efficiency. |

## Scoring Guide: Organisational Security Maturity Phase Diagnostic

Apply the following thresholds to each phase section score:

| Score Range | Alignment Level | Interpretation |
|---|---|---|
| **70% and above** | Strong Alignment | Phase characteristics are clearly and consistently present. This is the dominant phase. |
| **55–69%** | Moderate Alignment | Phase characteristics are recognisable. Some elements may still be developing. |
| **40–54%** | Emerging Alignment | Phase characteristics are beginning to appear. Mixed patterns are evident. |

| **Below 40%** | Weak Alignment | Phase characteristics are not yet present. The organisation is more likely in a different phase. |
|---|---|---|

Profile interpretation: where one phase scores more than 25 percentage points above the others, the phase classification is unambiguous. A spread of 15 to 25 points indicates a transitional state in which hybrid leadership approaches may be appropriate. Where no phase is clearly dominant, deeper contextual analysis is required.

## TOOL 2: CISO LEADERSHIP ORIENTATION ASSESSMENT

This self-assessment instrument enables Chief Information Security Officers and senior security leaders to identify their natural leadership orientation as defined by the SLCM: Action-Oriented, Stewardship-Oriented, or Vision-Oriented. The three orientations emerged inductively from the empirical data and represent distinct patterns in how security leaders approach their role, allocate their time, and derive professional satisfaction. The assessment comprises 24 items distributed equally across the three orientations (eight items per orientation).

The instrument supports self-awareness regarding ideal contexts, potential blind spots, and development opportunities. It may also be used by organisations during recruitment and succession planning to assess the orientation of candidates relative to the current or anticipated maturity phase of the organisation. When combined with Tool 1, the assessment enables evaluation of orientation-phase alignment as conceptualised in the SLCM.

Scoring: calculate the sum of responses for each orientation section and divide by the maximum possible score for that section (8 items × 4 = 32 points). Express the result as a percentage.

Response scale:

| 1 = Never | 2 = Sometimes | 3 = Often | 4 = Always |
|---|---|---|---|

| No. | Statement |
|---|---|
| **Section A: Action-Oriented Items (A1–A8)** | |
| **1** | When facing a security challenge, I prefer to implement a working solution quickly rather than wait for the perfect approach. |
| **2** | I find the most satisfaction in building new security capabilities from the ground up. |
| **3** | When security incidents occur, I naturally gravitate toward hands-on involvement in the response. |
| **4** | I make security decisions with incomplete information when speed is important. |
| **5** | I prefer working in environments where I can drive visible change rather than maintain existing operations. |
| **6** | When priorities shift due to emerging threats or incidents, I adjust plans readily rather than following established roadmaps. |
| **7** | I spend more time on implementation and execution than on documentation and process design. |
| **8** | I feel energised by transformation challenges and frustrated by steady-state operations. |
| **Section B: Stewardship-Oriented Items (S1–S8)** | |
| **9** | I prioritise improving existing security processes over building new capabilities. |
| **10** | I find satisfaction in achieving measurable improvements in security metrics and control effectiveness. |
| **11** | When evaluating security solutions, compliance and auditability are primary considerations for me. |
| **12** | I prefer establishing repeatable, documented processes for security activities. |
| **13** | I naturally focus on identifying and closing gaps in the security control framework. |
| **14** | I feel accomplished when security operations run smoothly and predictably. |
| **15** | I dedicate significant time to measurement, metrics, and demonstrating security programme maturity. |
| **16** | I approach security decisions systematically using risk assessment frameworks rather than intuition. |
| **Section C: Vision-Oriented Items (V1–V8)** | |
| **17** | I spend considerable time thinking about how security can enable business objectives rather than just reduce risk. |
| **18** | I find myself naturally positioning security in strategic business terms rather than technical terms. |
| **19** | I prioritise building relationships with business leaders over technical security community engagement. |
| **20** | When planning security initiatives, I focus primarily on long-term strategic impact rather than immediate tactical gains. |
| **21** | I feel most effective when influencing organisational direction rather than implementing specific controls. |
| **22** | I dedicate substantial effort to articulating the business value of security investments to non-technical stakeholders. |
| **23** | I actively seek opportunities to position security as a competitive differentiator. |
| **24** | I prefer exploring innovative security approaches over implementing proven, established solutions. |

## Scoring Guide: CISO Leadership Orientation Assessment

Apply the following thresholds to each orientation section score:

| Score Range | Alignment Level | Interpretation |
|---|---|---|
| **70% and above** | Strong Orientation | This orientation is a defining characteristic of your leadership style. You naturally gravitate toward contexts that align with this orientation. |
| **55–69%** | Moderate Orientation | This orientation is present but not dominant. You can operate in this mode when needed. |
| **40–54%** | Emerging Orientation | Limited natural inclination toward this orientation. It may require conscious effort to operate in this mode. |
| **Below 40%** | Weak Orientation | This orientation does not align with your natural style. Contexts requiring this orientation are likely to create significant strain. |

Profile interpretation: where one orientation scores more than 20 percentage points above the others, the leader has a clear primary orientation. A spread of 10 to 20 points indicates a hybrid profile with two strong orientations and greater versatility across contexts. A spread below 10 points indicates a balanced profile with maximum adaptability but less clear primary identity.

## TOOL 3: DUAL-FIT HEALTH DIAGNOSTIC

This instrument enables CISOs to assess the current health of the two fit relationships central to the SLCM: fit between the CISO and the internal organisational context (CISO-Organisation Fit), and fit between the CISO and the external environment (CISO-Environment Fit). In addition to these two fit dimensions, the diagnostic includes measures of leadership sustainability and political capital, which the SLCM identifies as the primary navigating mechanism through which leaders manage fit and misfit conditions.

The diagnostic comprises 26 items across four sections: CISO-Organisation Fit (7 items), CISO-Environment Fit (7 items), Sustainability Indicators (7 items), and Political Capital (5 items). Three items in the Sustainability section are reverse-scored, as indicated. It is designed for individual completion by the CISO and should be treated as a reflective tool rather than a performance measure. Results are most useful when revisited periodically, particularly following significant changes in organisational context, external environment, or leadership structure.

Scoring: calculate the sum of responses for each section (after recoding reverse-scored items) and divide by the maximum possible score for that section. Express the result as a percentage. Reverse-scored items: SUS2, SUS4, and SUS6 (recode 1→4, 2→3, 3→2, 4→1 before summing).

Response scale:

| 1 = Never | 2 = Sometimes | 3 = Often | 4 = Always |
|---|---|---|---|

| **No.** | **Statement** |
|---|---|

| **Section A: CISO-Organisation Fit (COF1–COF7)** | |
|---|---|
| **1** | My natural leadership approach aligns well with what my organisation currently needs from its security leader. |
| **2** | I find myself energised rather than drained by the type of work my role currently requires. |
| **3** | My strengths are regularly utilised and valued in my current organisational context. |
| **4** | The organisation's security maturity level is appropriate for my preferred leadership approach. |
| **5** | I feel like I am operating in my optimal range regarding the type of security leadership needed. |
| **6** | My leadership style is appreciated and effective within this organisational culture. |
| **7** | The pace of change my organisation requires matches my natural comfort level. |
| **Section B: CISO-Environment Fit (CEF1–CEF7)** | |
| **8** | I have the capabilities needed to address the current threat landscape facing my organisation. |
| **9** | I effectively navigate the regulatory and compliance requirements in my industry. |
| **10** | The available budget and resources are adequate for what the environment demands. |
| **11** | I successfully balance external pressures (threats, regulations, audits) with internal needs. |
| **12** | My technical skills remain relevant for the evolving technology landscape we face. |
| **13** | I maintain effective relationships with external stakeholders including auditors, regulators, vendors, and peers. |
| **14** | I feel capable of responding to the sophistication level of threats targeting our sector. |
| **Section C: Sustainability Indicators (SUS1–SUS7)** | |
| **15** | I feel a sense of accomplishment and satisfaction from my work. |
| **16** | I experience persistent fatigue or exhaustion despite adequate rest. (Reverse-scored) |
| **17** | I look forward to challenges in my role rather than dreading them. |
| **18** | I find myself becoming cynical about my role or organisation. (Reverse-scored) |
| **19** | My work-life balance feels sustainable long-term. |
| **20** | I am considering leaving this role or organisation. (Reverse-scored) |
| **21** | I feel I can maintain my current level of effort and engagement indefinitely. |
| **Section D: Political Capital (POL1–POL5)** | |
| **22** | I have strong, trusting relationships with key executives and board members. |
| **23** | My recommendations are taken seriously and usually adopted. |
| **24** | I can secure resources and support when needed for security initiatives. |
| **25** | I am consulted on organisational decisions beyond just security matters. |
| **26** | I have the formal authority and informal influence to be effective in my role. |
| **22** | I have strong, trusting relationships with key executives and board members. |
| **23** | My recommendations are taken seriously and usually adopted. |

## Scoring Guide: Dual-Fit Health Diagnostic

Apply the following thresholds to each section score:

| Score Range | Alignment Level | Interpretation |
|---|---|---|

| 70% and above | Strong Fit / High Capital | Alignment is robust. Conditions support sustained effectiveness. |
|---|---|---|
| **55–69%** | Moderate Fit / Adequate Capital | Alignment is functional but may show strain. Monitor for developing misalignment. |
| **40–54%** | Emerging Misalignment / Diminishing Capital | Meaningful gaps are evident. Compensatory strategies or intervention are warranted. |
| **Below 40%** | Significant Misalignment / Limited Capital | Alignment is substantially impaired. Sustainability risk is high. Structured intervention or transition planning is recommended. |

Interpretation note: low scores on the Sustainability Indicators section in conjunction with low scores on either or both fit dimensions are indicative of misfit-induced strain as described in the SLCM. Low political capital scores suggest that compensatory buffering strategies are unlikely to be sustainable. In such circumstances, structured intervention or transition planning is recommended.

## TOOL 4: PHASE-LEADERSHIP ALIGNMENT MATRIX

This decision-support matrix translates the core contingency relationships of the SLCM into actionable guidance for organisational and individual decision-making. It does not involve scored items; rather, it is intended to be consulted in conjunction with the results from Tools 1 and 2. The matrix specifies the alignment between each of the three organisational maturity phases and each of the three CISO leadership orientations, and provides guidance on the conditions under which each combination is likely to support or undermine effectiveness.

The matrix may be applied across several practical contexts: before recruiting or appointing a security leader, to match candidate orientation to the current phase; during performance evaluation, to assess effectiveness against phase-appropriate rather than universal expectations; when planning leadership development, to identify compensatory strategies in conditions of misalignment; and during succession planning, to anticipate the orientation requirements that will accompany future phase transitions.

### Core Alignment Matrix

| Organisational Phase | Leadership Type | Alignment | Key Considerations |
|---|---|---|---|
| **Establishment Phase** | Action-Oriented | Strong Alignment | Rapid capability building; hands-on involvement; pragmatic risk decisions |
| | Stewardship-Oriented | Challenging | Struggles with ambiguity and rapid change; requires significant adaptation |
| | Vision-Oriented | Poor Fit | Organisation lacks maturity for strategic positioning; technical credibility at risk |
| **Maturation Phase** | Action-Oriented | Challenging | Frustration with optimisation over transformation; must adapt to process discipline |
| | Stewardship-Oriented | Strong Alignment | Process optimisation; compliance excellence; systematic improvement; scalability |
| | Vision-Oriented | Moderate Fit | Viable in late maturation transitioning toward strategic phase; may feel constrained |
| **Strategic Phase** | Action-Oriented | Limited Fit | Organisation needs strategic positioning; hands-on approach less valued |
| | Stewardship-Oriented | Moderate Fit | Can maintain operations; struggles with innovation and business integration |
| | Vision-Oriented | Strong Alignment | Business integration; strategic influence; innovation enablement; competitive positioning |

## Managing Misalignment

Where the matrix indicates challenging or poor fit between a CISO's orientation and the organisational phase, the following compensatory strategies reflect the mechanisms observed across the empirical sample.

Structural deputisation involves building a leadership team with orientations complementary to the CISO's own. An action-oriented deputy serves a vision-oriented or stewardship-oriented CISO operating in an establishment-phase organisation; a vision-oriented deputy supports a stewardship-oriented CISO in a strategic-phase organisation. This approach enables the leadership collective to fulfil the full range of phase requirements without requiring the CISO to operate consistently outside their natural orientation.

Selective domain focus involves concentrating the CISO's personal effort on activities aligned with their orientation while delegating misaligned activities to team members. An action-oriented CISO in a strategic-phase organisation may lead transformation and innovation initiatives within the broader strategic agenda while delegating business integration and executive relationship management to a vision-oriented deputy.

Political capital buffering reflects the finding that accumulated political capital, in the form of credibility, trust, and institutional influence, provides temporary flexibility for style-phase misalignment. Organisations and CISOs should recognise, however, that political capital depreciates without reinforcement through contextually aligned performance. The SLCM findings suggest that strong political capital combined with structural workarounds may sustain misalignment for 18 to 36 months; moderate political capital without structural support reduces this to 12 to 18 months.

Phase transition planning involves anticipating the orientation requirements that accompany organisational maturity transitions and planning succession accordingly. The SLCM findings indicate that CISOs who have successfully led an organisation from one phase to the next often face declining effectiveness as the organisation's needs evolve beyond their natural orientation. Recognising this dynamic proactively, and framing transition as organisational evolution rather than individual failure, is consistent with the sustainability considerations embedded in the SLCM.

***

These instruments operationalise findings from grounded theory research examining 20 Chief Information Security Officers across diverse Australian organisations. Each item reflects behavioural patterns observed across multiple cases and grounded in how participants described their approaches, preferences, and organisational contexts. The instruments are provided as a practical contribution of this thesis and should be interpreted in conjunction with the theoretical framing of the SLCM presented in